\documentclass[binding=0.6cm, noexaminfo, oneside]{sapthesis}
\usepackage{microtype} 
\usepackage[english]{babel}
\usepackage[utf8]{inputenc} 
\usepackage{hyperref}
\usepackage{braket}
\usepackage{amssymb}
\usepackage{float}
\usepackage[export]{adjustbox}
\usepackage{graphicx}
\usepackage{amsthm}
\usepackage{amsmath}
\usepackage{bm}
\usepackage{booktabs}
\usepackage{bbding}
\usepackage{xcolor,tikz}
\usepackage{siunitx}
\usepackage{graphicx}
\usepackage{xcolor}
\usepackage{empheq}
\usepackage{microtype}
\usepackage{physics}
\usepackage{framed}
\usepackage{cancel}
\usepackage{slashed}

\usepackage[numbers,sort&compress]{natbib}

\usepackage{mathrsfs}

\usepackage{tikz}

\usepackage[normalem]{ulem} 

\hypersetup{pdftitle={Manuscript}, pdfauthor={Claudio Gambino}}

\title{The Reissner-Nordstr\"om-Tangherlini solution from graviton and photon emission processes} 

\author{Claudio Gambino}
\IDnumber{1793203}
\course{Laurea Magistrale in Fisica}
\courseorganizer{Facoltà di Scienze Matematiche, Fisiche e Naturali} \AcademicYear{2021/2022}
\advisor{Prof. Fabio Riccioni}
\authoremail{gambino.1793203@studenti.uniroma1.it}
\copyyear{2022}
\thesistype{Master Thesis}

\newcommand{\de}{\partial}

\versiondate{\today}

\begin{document}

\frontmatter
\maketitle

\begin{abstract}
It is known that the canonical quantization of general relativity leads to a non-renormalizable theory, which however with the modern tools of effective field theories it is possible to make well-defined at low energies. What is emerging nowadays is that recovering general relativity from the classical limit of scattering amplitudes seems more efficient than actually solve the Einstein equations.
The thesis then aims to explore the state-of-the-art strategies to recover classical gravitational observables from quantum computations and apply them in order to reconstruct the Reissner-Nordstr\"om-Tangherlini solution, which is the generalization to arbitrary dimensions of the Reissner-Nordstr\"om one, from the classical limit of scattering amplitudes of charged scalars. 
In the first part of the thesis, we recover the most generic energy-ordered effective action which is compatible with the symmetries of the theory, and then we discuss the classical limit from which it is possible to obtain back classical gravitational observables out of quantum computations.
In order to show how such limit works, an explicit example is made, and following the historical papers it is shown how to recover stationary space-time solutions from the classical limit of graviton emission processes in four dimensions.
Then the attention is focused on the Reissner-Nordstr\"om-Tangherlini solution and on computing it by using quantum scattering amplitudes in the de Donder gauge. Since the main objective will be to compare the quantum computation with the classical one, a procedure to rewrite such solution in the de Donder gauge is carried out.
Finally, the state-of-the-art techniques to extract the classical contributions out of graviton emission amplitudes are reviewed, and they are applied to the context of a theory which is also coupled to the electromagnetic field. Since divergences arise in four and five dimensions it is crucial to define non-minimal couplings with the purpose of canceling such singularities and obtaining finite results.
In the end, the considerations employed in the metric case are used to recover the electromagnetic potential associated to the black hole solution through the classical limit of photon emission processes. Also in this case there appear divergences that have to be renormalized. One of the most important outcomes is that the counter-terms that have to be fixed in order to eliminate the poles are the same throughout the theory, showing an extraordinary coherence of the procedure. 

\medskip

The thesis project has produced the publication of a paper in which the results obtained in this work have been presented [S. D’Onofrio, F. Fragomeno, C. Gambino and F. Riccioni, \textit{The Reissner-Nordström-Tangherlini solution from scattering amplitudes of charged scalars}, JHEP \textbf{09} (2022) 013, \texttt{2207.05841}].
\end{abstract}

\begin{acknowledgments}
First I would like to thank my advisor Fabio Riccioni for the precious help and inspiration during the thesis project. 

\medskip

I would like also to thank Simone D'Onofrio and Federica Fragomeno which with their thesis inspired the present work, and whose collaboration was very pleasant and productive. 

\medskip

A special thank to the Physics Department of the University of Rome "La Sapienza" for the years of formation which culminate with this thesis. 

\medskip

Finally, I am very grateful for the support I received during these years from my colleagues, which without them none of this would have been possible, and from friends and family, who helped me during this journey in many ways.
\end{acknowledgments}

\tableofcontents
\mainmatter

\chapter{Introduction}

Since the formulation of quantum mechanics and general relativity, attempts have been made to unify these two theories to provide a quantum description of gravity. However, to date, the formulation of a quantum theory of gravity remains one of the greatest challenges of modern physics. One of the reasons lies in the fact that a canonical quantization of gravity, in which the interaction is mediated by a massless spin-2 boson called graviton, gives rise to a non-renormalizable theory, which has pushed the scientific community to search for alternative approaches to field theories. 
However, by exploiting the principles of effective field theories, a non-renormalizable theory can be reformulated in such a way to have predictive power as long as the energy of the processes considered is below a certain threshold, above which a more fundamental theory is necessary. For the gravitational interaction, we believe that the energy scale of the so called heavy degrees of freedom is identified by the Planck scale. In the language of effective field theories therefore, quantum gravity is described in terms of an action composed of an infinite series of terms weighted by inverse powers of the Planck mass, in which the Einstein-Hilbert action corresponds to the first order of this expansion \cite{Feynman:1963ax, DeWitt:1967yk, DeWitt:1967ub, DeWitt:1967uc, tHooft:1974toh}. This scheme was employed by Donoghue in 1993 in order to recover the quantum corrections of the Newton gravitational potential, showing how the effective field theory approach to gravity could actually lead to relevant conclusions  \cite{Donoghue:1993eb, Donoghue:1994dn}.

\medskip

Later, it was starting to be clear how the most important outcome of this result was that besides the quantum corrections of gravitational observables\footnote{In this thesis, for gravitational observables we mean the set of objects related to gravitational phenomena which are associated to measurable quantities. For example, the metric in general relativity is not a directly measurable quantity, however, from it, experimental predictions can be extracted, such as the attraction strength between two bodies, the light-bending angle around some mass, the gravitational-wave emission amplitude generated by a merger of a binary system, etc... Therefore, in this thesis, the metric is considered a gravitational observable.}, also classical terms at zeroth-order in $\hbar$ could be extracted through a loop quantum expansion \cite{Donoghue:1994dn, Donoghue:2001qc, Bjerrum-Bohr:2002fji, Donoghue:1996mt, Iwasaki:1971vb, Holstein:2004dn, Bjerrum-Bohr:2018xdl, Kosower:2018adc, Cheung:2018wkq, Bjerrum-Bohr:2004qcf}. In particular the awareness that classical physics could be efficiently obtained from the classical limit of quantum loop scattering amplitudes was emerging. 
In fact since quantum gravity as an effective field theory is well-defined at low energies, in that regime the classical limit in which $\hbar\rightarrow 0$ must reproduce the original classical theory, which in this case corresponds to general relativity. 
In \cite{Bjerrum-Bohr:2002fji} it was shown how besides $\hbar$-dependent contributions, the first post-Minkowskian term of the Schwarzschild metric could be derived from the tree-level amplitude in which a massive scalar emits a graviton, while the second post-Minkowskian order is obtained by considering a 1-loop amplitude which contains a 3 graviton interaction vertex. In the same context it was shown that it is possible to reconstruct the Kerr metric by considering graviton emission from spinors, as well as the terms proportional to the electric charge of the Reissner-Nordstr\"om and Kerr-Newman metrics by considering the 1-loop scattering of charged particles with photons running in the loop \cite{Donoghue:2001qc}.
The knowledge on how the classical limit of quantum gravity occurs was taking shape, and in particular that general relativity is reconstructed by multi-loop scattering amplitudes. This phenomenology differs from the classical limit of the other known long-range force, the electromagnetic interaction, in which classical electromagnetism is recovered by the $\hbar\rightarrow 0$ limit of tree-level processes in QED. The reason why gravity needs loop amplitudes to be reconstructed is traceable to its strong non-linear behavior, which is translated in the quantum context by graviton self-interactions terms, that make loop amplitudes necessary for recovering classical physics. 

\medskip

More recently, a systematic procedure to extract the classical contribution of graviton emission loop amplitudes in any dimension\footnote{In this thesis we will only consider space-time dimensions $D>2$.} was given in \cite{Bjerrum-Bohr:2018xdl}, in which it is shown how such terms correspond to the interaction of external sources through a quantum-tree graph, originally observed by Duff in \cite{Duff:1973zz}. Applying these techniques, in \cite{Mougiakakos:2020laz, Mougiakakos:2021ngd} the Schwarzschild-Tangherlini metric \cite{Tangherlini:1963bw}, that generalizes the Schwarzschild solution to arbitrary dimensions, is reconstructed at fourth order in the post-Minkowskian expansion from amplitude calculations up to 3-loops, showing the existence of a precise relation between the loop order in the amplitude computation and the post-Minkowskian one in the classical formulation. Such derivation is performed in de Donder gauge, and in order to compare the results with the classical metric one has to write down the latter in the same gauge, in which logarithmic terms appear in the post-Minkowskian expansion starting from the second order in five dimensions and third order in four dimensions. Correspondingly the amplitudes develop divergences whose renormalization needs the addition of non-minimal couplings in the original action, originally noticed by Goldberger and Rothstein in the context of world-line formalism in \cite{Goldberger:2004jt}, as it is suggested by general principles of the effective field theories. Such renormalization process gives rise to logarithmic terms that are exactly the ones that appear in the classical metric in de Donder gauge. Moreover in \cite{Jakobsen:2020ksu, Jakobsen:2020diz} it is shown how to recover the Schwarzschild-Tangherlini metric in a generic gravitational gauge up to 1 loop, in which the same divergences are found, but it is proved that there exists a particular gauge choice in which the metric is well-defined in all space-time dimensions. 

\medskip

In this thesis it is shown how the analysis of \cite{Mougiakakos:2020laz} applied to the scattering of charged scalars, whose interact with both gravitons and photons, systematically leads to the post-Minkowskian expansion of the Reissner-Nordstr\"om-Tangherlini solution (the generalization to arbitrary dimensions of the Reissner-Nordstr\"om one), recovering the metric from graviton emission and the electromagnetic potential from photon emission. The analysis of the metric is a review of the theses \cite{Fragomeno, DOnofrio:2021tap}, in which it is presented how following the prescription of \cite{Bjerrum-Bohr:2018xdl} it is possible to isolate the classical terms of scattering amplitudes of graviton emission with photons running in the loops, in order to reconstruct the classical Reissner-Nordstr\"om-Tangherlini metric. The calculations are performed up to 2 loops and a perfect agreement with the classical solution rewritten in the de Donder gauge is obtained. The renormalization procedure follows directly from the one outlined in \cite{Mougiakakos:2020laz}, in which the same non-minimal coupling is considered but generalized in order to satisfy the electromagnetic gauge symmetry. The addition of a new coupling constant in the theory, namely the electric charge, makes possible to define other counter-terms that are needed in order to cancel the divergences that appear in the electromagnetic sector of the theory. In fact while the pure gravitational diagrams in which only gravitons circulate in the loops are renormalized in \cite{Mougiakakos:2020laz}, processes in which also internal photons are present need a mixed counter-term which contains both gravitational and electromagnetic coupling constant. So then, it is proved that the graviton emission amplitudes in which photons circulate in the loops are renormalized by considering the same counter-term of \cite{Mougiakakos:2020laz} combined with a new one proportional to the electric charge.

\medskip

However in order to fully reconstruct the Reissner-Nordstr\"om-Tangherlini solution, besides the metric, one has to specify also the electromagnetic potential associated to the black hole. So then the original work which aims to reconstruct the electromagnetic potential of the solution from scattering amplitudes is presented. The idea is that if the metric is recovered from graviton emission, which reconstructs the gravitational source, then the electromagnetic potential is reconstructed from photon emission processes from which the electromagnetic current is obtained. 
The analysis is again performed up to 2 loops and leads to a perfect agreement with the classical potential associated to the Reissner-Nordstr\"om-Tangherlini solution in de Donder gauge at third post-Minkowskian order. 
As it was for the metric case, the amplitudes develop divergences, which in particular arise in five dimensions at 2 loops, and the most important outcome is that they are renormalized by the insertion of the same counter-terms that are needed for the renormalization of graviton emission processes. Then the logarithmic terms that arise are compared with the ones in the classical solution, and again a perfect match is found.
The thesis project has produced the publication of a paper in which the results obtained in this work have been presented \cite{DOnofrio:2022cvn}.

\medskip

The reason why in the recent years the scientific community is becoming more and more interested in recovering general relativity from quantum scattering amplitudes, is because of the idea that this scheme allows to recover gravitational observables in a more efficient and precise way than directly solve the Einstein equations, in addition to the fact that these methods are completely analytic. This idea has been tested in the context of gravitational waves emitted from compact binary systems, in which emission templates are recovered from the Newton potential between the binary constituents, that is calculated employing scattering amplitudes techniques \cite{Buonanno:2022pgc, Bjerrum-Bohr:2022blt}. In fact the precision demanded by the future generation of gravitational-wave detectors \cite{https://doi.org/10.48550/arxiv.1702.00786, Saleem:2021iwi, Reitze:2019iox, Punturo_2010}, is beyond our theoretical capabilities prediction, which make use of standard techniques like effective-one-body methods and numerical relativity \cite{Nagar:2018zoe, Varma:2018mmi, Varma:2019csw, Gamba:2021ydi}. Depending on the source, gravitational-waves from binary systems will be observed with a signal-to-noise ratio that is one or two orders of magnitude higher (more precise) than with current detectors \cite{LIGOScientific:2014pky, VIRGO:2014yos, KAGRA:2020agh}. So then the accuracy of current waveform models, which are built from combining analytic and numerical relativity, needs to be improved by two order of magnitudes. Therefore the necessity of new techniques which will improve theoretical predictions is urgent, and the scattering amplitudes program is candidate to be one of them, as well as a very promising analytic alternative to the set of all gravitational processes which can be predicted only by numerical methods. 

\medskip

\begin{figure}[h]
\centering
\includegraphics[width=0.7\textwidth, valign=c]{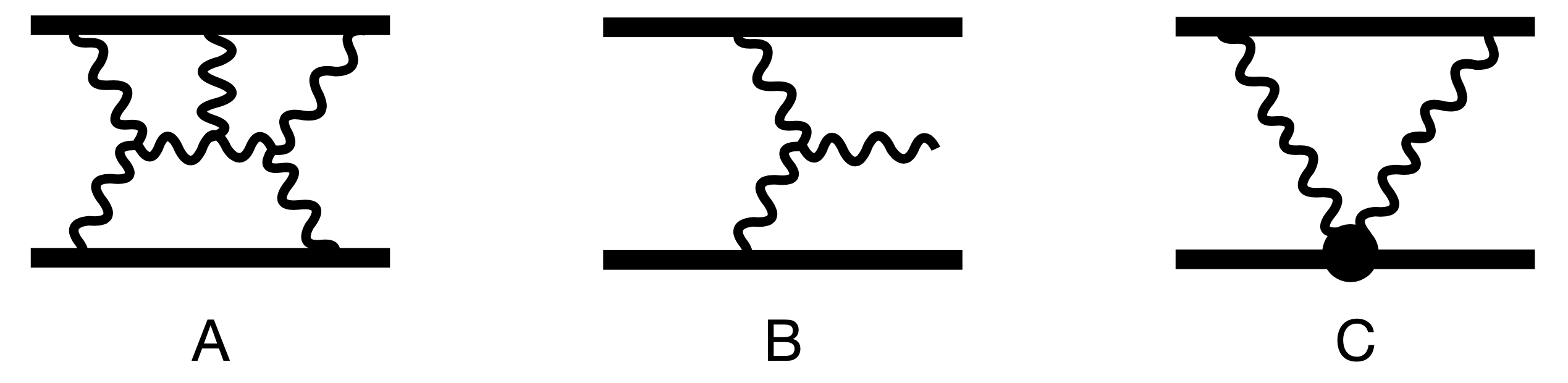}
\caption{2-to-2 scattering amplitudes taken from \cite{Buonanno:2022pgc}, which reconstruct the binary dynamics by means of interaction between massive particles (thick lines) and gravitons (wiggly lines). The amplitudes (A) reconstruct the conservative sector of the dynamics, (B) encodes all the radiative effects due to gravitational-wave emission represented by a five point scattering amplitude in which an external graviton is produced, while (C), through higher-dimensional vertices (solid circle), take into account all the tidal effects deriving from the fact that the binary constituents are not point particles. }
\label{fig:2-in-2_Scattering}
\end{figure}

\begin{figure}[h]
\centering
\includegraphics[width=0.8\textwidth, valign=c]{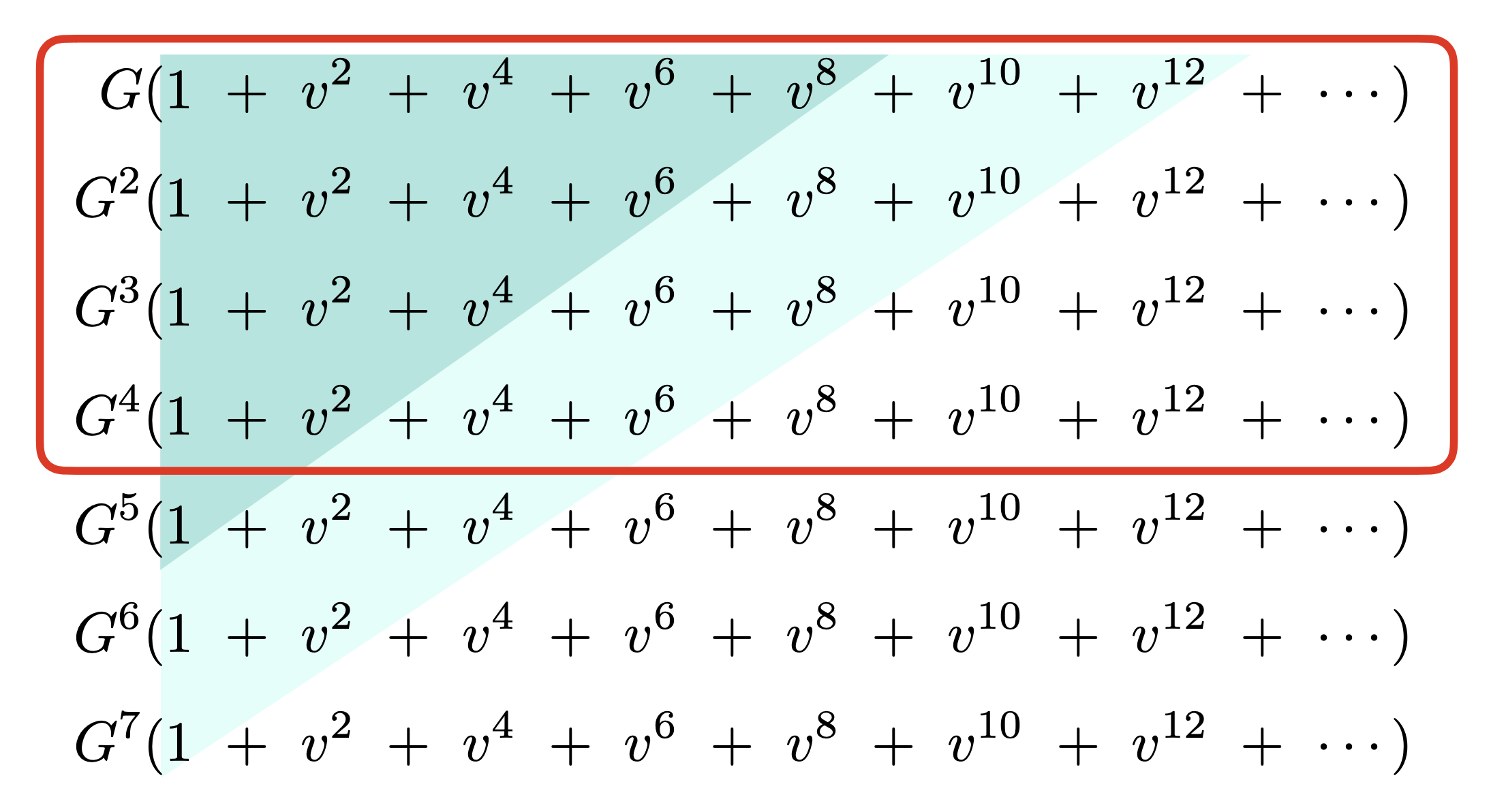}
\caption{Map of perturbative corrections to the Newton potential taken from \cite{Buonanno:2022pgc}, where $G_N$ is the universal gravitational Newton constant and $v$ is the relative velocity of the binary constituents. The red box contains the state-of-the-art calculations in the post-Minkowskian order, while the dark triangle the ones in the post-Newtonian expansion. The light triangle shows the contributions required by future detectors \cite{Favata:2013rwa, Samajdar:2018dcx, Purrer:2019jcp, Huang:2020pba, Gamba:2020wgg}.}
\label{fig:PMvsPN_Expansion}
\end{figure}

Up to now, the state-of-the-art predictions by means of scattering amplitudes methods are able to compute the gravitational-wave emission template in the inspiral phase through the classical limit of 2-to-2 scattering amplitudes between scalar particles mediated by gravitons, up to fourth order in the post-Minkowskian expansion, as shown in figure \ref{fig:2-in-2_Scattering}. 
Combining these techniques with the effective-one-body formalism, we expect that the required precision would be reached with calculations at $O(G_N^7)$ order in the post-Minkowskian expansion, or at $O(v^{12})$ in the post-Newtonian one\footnote{While the post-Minkowskian expansion is relativistic order by order, namely it is true for all velocities, the post-Newtonian series is an expansion in velocities, which is non-relativistic.}, as sketched in figure \ref{fig:PMvsPN_Expansion}.
So then, the work presented in this thesis aims to be a step forward in the understanding of how the next generation of theoretical gravitational predictions could be made employing scattering amplitudes techniques. 

\medskip

The thesis is organized as follows. In chapter \ref{QG_as_EFT} it is first reviewed how to interpret general relativity as a gauge theory. Then an attempt in order to quantize gravity is made, proving how such strategy leads to a non-renormalizable theory. The modern tools of effective field theories are employed and the most generic quantum effective action compatible with the symmetries of the theory is written. Finally it is discussed how to impose the classical limit in this theory, and replacement rules that state how to reinsert back $\hbar$ factors in the theory in order to perform the limit $\hbar\rightarrow 0 $ are obtained. In chapter \ref{Classical_Chapter} we present the review of the original papers \cite{Donoghue:1994dn, Donoghue:2001qc, Bjerrum-Bohr:2002fji} as an explicit example of how to employ the classical limit to obtain classical physics out of quantum computations. Moreover the chapter gives the opportunity to introduce some fundamental concepts and ideas very important in the following of the thesis. Then the attention is focused on the derivation of the Reissner-Nordstr\"om-Tangherlini solution from scattering amplitudes. In chapter \ref{chap:RNT_Classical_Solution} we report the classical derivation of the solution by solving the Einstein and Maxwell equations. Since the scattering amplitudes calculation we have in mind will be performed in the de Donder gauge, in order to make possible the comparison of the two derivations, the procedure to rewrite the Reissner-Nordstr\"om-Tangherlini solution in such gauge is reviewed. Finally in chapter \ref{chap:Metric_from_Amplitudes} the state-of-the-art strategy to extract the classical limit out of graviton emission loop amplitudes in generic dimensions, with photons and gravitons running in the loops, is discussed and then applied to recover the Reissner-Nordstr\"om-Tangherlini metric. In particular in section \ref{sec:Renorm_Metric} the non-minimal action that has to be added to the original theory in order to renormalize the divergences that appear is defined. Then in chapter \ref{chap:Photon_Emission_Processes} the same analysis is performed in order to consider the classical limit of photon emission process with both gravitons and photons running in the loops. It is shown how such processes reconstruct the electromagnetic potential associated to the Reissner-Nordstr\"om-Tangherlini solution, and we discuss the relevant outcome by which the divergences that appear in this case are renormalized by the very same counter-terms that enter in the metric case. In the end chapter  \ref{chap:Conclusions} contains a discussion of the obtained results and an overview of future possible perspectives.

\medskip

The thesis contains also three appendices. In appendix \ref{Appendix_FeynRules} all the expressions for the propagators and vertices that are used in this work are given. In appendix \ref{App:masterintegral} all the Fourier transforms employed to determine the metric and the potential from the amplitudes are listed. Finally, in appendix \ref{App:LoopRed} we list all the loop integrals reduction formulas needed to manage the expressions that arise from amplitude computations. 

\medskip

Notice that, unless it is specified the contrary, all the conclusions of this thesis are carried out in generic space-time dimensions.

\chapter{Quantum gravity as an effective field theory}
\label{QG_as_EFT}

In the first section of this chapter it is briefly reviewed how to construct scalar quantum electrodynamics by gauging the $\text{U}(1)$ global symmetry of the scalar action, then by analogy it is shown how to recover general relativity as a non-abelian gauge theory by gauging the space-time global translational symmetry. Once one has a classical field theory in which gravity is coupled to other fields, an attempt is made in order to quantize gravitational interaction in the old fashion way employing the path integral formalism, which however leads to a non-renormalizable theory. Then the modern interpretation of effective field theories can make the theory consistent at low energies, and the most generic quantum gravity effective Lagrangian is written. Finally it is shown how to recover the classical limit of such theory by explicitly restoring the $\hbar$'s and then performing the limit $\hbar\rightarrow 0$. In this context the low-energy limit implies specific rules of how vertices and propagators depend on the Planck constant, and they are applied in order to recover classical physics from quantum computations in an efficient way. 

\section{Scalar quantum electrodynamics}\label{Scalar_quantum_electrodynamics}

The action that describes a scalar complex massive field has the form
\begin{equation}\label{Free_Scalar_Action}
S=\int d^D x\Bigl( \de_\mu \phi^* \de^\mu \phi - m^2 \phi^* \phi\Bigl)\ ,
\end{equation}
where we neglect the self-interaction term and where we consider from here on a $D$-dimensional flat background space-time with signature $(+, -, ..., -)$ and natural units in which $\hbar=c=1$. 
This theory has an abelian $\text{U}(1)$ global symmetry, which means that the action is invariant under the transformation
\begin{equation}
\phi(x) \rightarrow e^{-i Q} \phi(x)\ ,
\end{equation}
with $Q=const$. Since this is a continuous symmetry, due to the Noether theorem there exists a conserved current, whose expression is
\begin{equation}\label{Global_EM_Noether_Current}
j^\mu(x) = i Q \Big(\phi^*(x) \de^\mu \phi(x) - \phi(x) \de^\mu \phi^*(x)\Big)\ ,
\end{equation}
and the relation 
\begin{equation}
\de_\mu j^\mu(x)=0
\end{equation}
is satisfied.
Now we want gauge the theory, which means make the original global symmetry local. However the local field transformation
\begin{equation}
\phi(x) \rightarrow e^{-i Q\alpha(x)} \phi(x)\ ,
\end{equation}
is not a symmetry of the theory since the space-time dependency in $\alpha(x)$ spoils it. In order to make the theory symmetric under this transformation, we need to define a new vector field, which in turn it makes possible to define the so called covariant derivative as
\begin{equation}
D_\mu = \de_\mu + iQ A_\mu(x)\ ,
\end{equation}
where $A_{\mu}(x)$ is the gauge field of the theory, called electromagnetic field, and $Q$ assumes the meaning of electric charge. Imposing that under the symmetry transformation we are looking for, the gauge field transforms as
\begin{equation}\label{EM_Gauge_Transform}
A_\mu(x)\rightarrow A_\mu(x)+\de_\mu\alpha(x)\ ,
\end{equation}
the theory described by the action
\begin{equation}
S = \int d^Dx \Bigl((D_\mu \phi)^* (D^\mu \phi) - m^2 \phi^* \phi\Bigl)
\end{equation}
is now symmetric under the local gauge transformation.

\medskip

Since we have introduced a new field, we have to consider its dynamics. The kinetic term can be constructed from the so called strength tensor, defined starting from the commutator between the covariant derivatives. Considering $F_{\mu\nu}$ as the strength tensor of the theory, here called electromagnetic tensor, schematically we can state
\begin{equation}
\left[D_\mu, D_\nu\right]\thicksim F_{\mu\nu}\ ,  
\end{equation}
where the actual definition is
\begin{equation}
F_{\mu\nu}=\de_\mu A_\nu - \de_\nu A_\mu \ .
\end{equation}
Finally, in order to construct the kinetic term of the electromagnetic field, we have to consider all the gauge invariant Lorentz scalars that can be made by contractions of $F_{\mu\nu}$ and that give rise to terms quadratic in the first derivative of the gauge field. From the symmetries of the electromagnetic tensor the only contraction that one must take into account is $F_{\mu\nu}F^{\mu\nu}$, from which the complete scalar electrodynamics action reads \cite{Srednicki}
\begin{equation}\label{Complete_SQED_no_Gauge}
S=\int d^Dx\Biggl(-\frac{1}{4}F_{\mu\nu}F^{\mu\nu}+(D_\mu \phi)^* (D^\mu \phi) - m^2 \phi^* \phi\Biggl)\ ,
\end{equation}
leading to the Maxwell equations for the electromagnetic potential.

\medskip

From the classical action recovered in \eqref{Complete_SQED_no_Gauge} we want to move to a quantum field theory. The standard approach to do this, is to define a generating functional through the path integral formalism. In order to quantize the scalar electrodynamics then, we need to define 
\begin{equation}
Z=\int [d\phi] [d\phi^*] [dA_\mu]\, e^{iS}\ .
\end{equation} 
However it is known, that apply this approach to a gauge invariant theory is problematic, in fact since we are integrating even over those configurations of the electromagnetic field which are related through a gauge transformation, we are considering infinite copies of the same physical system, which will give a divergent generating functional. The way to properly consider gauge theories in this context was originally outlined by Faddeev and Popov \cite{Faddeev:1967fc}, in which they formalized the gauge fixing procedure in order to not integrate over gauge equivalent orbits. The outcome is that the generating functional can be rewritten as
\begin{equation}\label{Generatinf_Functional_EM}
Z=\int [d\phi] [d\phi^*] [dA_\mu]\, \Delta[A_\mu]\, e^{iS-\frac{i}{2\xi}\int d^Dx \, f[A_\mu]^2}\ ,
\end{equation} 
where $f[A_\mu]$ is the gauge fixing function that eliminates the gauge freedom in the theory and $\Delta[A_\mu]$ is the so called Faddeev-Popov determinant, that in terms of the gauge transformation in eq. \eqref{EM_Gauge_Transform} has the explicit form
\begin{equation}\label{Faddeev_Popov_Determinant}
\Delta[A_\mu] = \det(\frac{\de f'}{\de \alpha})=\det(\frac{\de f[A_\mu+\de_\mu \alpha]}{\de \alpha})\ .
\end{equation}
The above expression can be seen as an integral over Grassman variables and takes the form  
\begin{equation}\label{Faddeev_Popov_Determinant_Expanded}
\Delta[A_\mu]=\int [d c][d \bar{c}]\, e^{i\int d^D x\, \bar{c} \frac{\de f'}{\de \alpha}c}\ , 
\end{equation}
where $c$ is called ghost field. It does not correspond to any physical field since it violates the spin-statistic theorem, however in order to preserve the gauge invariance in the quantized version of the theory, it is necessary to consider the ghost Lagrangian implicitly defined in \eqref{Faddeev_Popov_Determinant_Expanded} in addition to the complete theory. Although we considered explicitly the electromagnetic field as an example, this procedure holds even for more complicated non-abelian gauge theories.  

\medskip

In the case of electromagnetism, the standard choice for the gauge fixing function is $f[A_\mu] = \de_\mu A^\mu$, which corresponds to the Lorentz gauge and it is motivated by the fact that the field equations take a particularly simple form. This choice leads to the Faddeev-Popov determinant 
\begin{equation}
\Delta[A_\mu]=\det(\frac{\de}{\de \alpha(y)}(\de_\mu A^\mu(x)+\Box\, \alpha(x)))=\det(\Box\,  \delta^{(D)}(x-y))=const \ ,
\end{equation}
which shows the well-known result that an abelian theory does not require ghost fields since they do not interact with the gauge field. Notice that in our conventions $\Box=\de_t^2-\nabla^2$, where $\nabla^2 = \de_i\de_i$. Then fixing the so called Feynman gauge  by making the choice $\xi=1$, which leads to the gauge fixing Lagrangian term 
\begin{equation}
\mathcal{\mathcal{L}}^{GF}_A=-\frac{1}{2}(\de_\mu A^\mu)^2\ ,
\end{equation} 
considering its explicit addition in the action in \eqref{Complete_SQED_no_Gauge}, one finally obtains
\begin{equation}\label{Expanded_GF_SQED_Action}
S=\int d^Dx\Biggl(\frac{1}{2}\de_\mu A_\nu\de^\mu A^\nu+\de_\mu \phi^* \de^\mu \phi - m^2 \phi^* \phi +Q^2A_\mu A^\mu\phi^*\phi-j^\mu A_\mu\Biggl)\ ,
\end{equation}
where $j^\mu(x)$ is exactly the Noether current associated to the global symmetry in eq. \eqref{Global_EM_Noether_Current}. The action in \eqref{Expanded_GF_SQED_Action} is the very well-known scalar quantum electrodynamics action, which is renormalizable up to the addition of the self-interaction term of the scalar field. From eq. \eqref{Expanded_GF_SQED_Action} it is then possible to recover the electromagnetic vertices in our conventions, as the photon propagator and the 2 scalars - 1 photon vertex, reported in appendix \ref{Appendix_FeynRules}. It is also possible to write down the electromagnetic field equations in the Feynman gauge as
\begin{equation}
\Box A^\mu(x) = j^\mu_{cov}(x)\ ,
\end{equation}
where $j^\mu_{cov}(x)$ is exactly the covariantized version of the global Noether current as
\begin{equation}\label{EM_Current_Covariantized}
j^\mu_{cov}(x) = i Q \Big(\phi^*(x) D^\mu \phi(x) - \phi(x) D^\mu \phi^*(x)\Big)\ . 
\end{equation}
From now on the subscript $"cov"$ will be omitted and it will be clear from the context whether or not an expression is meant to be covariant. 

\medskip

Even if the procedure reviewed is performed considering the electromagnetic field, it can be easily generalized to more complicated theories. For example by gauging other symmetries of the theory it is possible to carry out different gauge fields and interactions vertices. 

\section{General relativity as a gauge theory}

Following a procedure similar to the one discussed above, it is possible to recover general relativity by gauging the space-time translations, which are a global symmetry of any field theory which is Poincaré invariant. This approach is presented in \cite{Donoghue:2017pgk}, and differs from other approaches, like the one in \cite{Ortin}.
Starting from the free action in \eqref{Free_Scalar_Action}, the theory is invariant under global space-time translations
\begin{equation}
x^\mu\rightarrow x^\mu+a^\mu\ ,
\end{equation}
where $a^\mu=const$. Due to the Noether theorem there exists a conserved current associated to the continuous symmetry, defined up to a divergenceless object, which has the expression
\begin{equation}\label{SET_Global_Symmetry}
T_\phi^{\mu\nu}=\de^\mu \phi^* \de^\nu \phi + \de^\nu \phi^* \de^\mu \phi-\eta^{\mu\nu} \left( \de_\alpha \phi^* \de^\alpha \phi - m^2 \phi^* \phi\right)\ ,
\end{equation}
and satisfies
\begin{equation}
\de_\mu T_\phi^{\mu\nu}(x)=0\ .
\end{equation}
The current in \eqref{SET_Global_Symmetry} is exactly the stress-energy tensor of the scalar field, in fact one can check that $T_\phi^{00}=\mathcal{H}_\phi$, where $\mathcal{H}_\phi$ is the Hamiltonian of the massive field. 

\medskip

Now following the same steps as in the electromagnetic case, we can gauge the translations making the transformation local. This corresponds to considering a coordinate transformation like
\begin{equation}\label{Coordinate_Transformation_Under_GCT}
x^\mu\rightarrow x^\mu+a^\mu(x)\ ,
\end{equation}
which since $a^\mu(x)$ is a generic function of space-time coordinates, it can be rewritten as
\begin{equation}
x^\mu\rightarrow \Lambda^\mu{}_{\nu}(x)x^\nu\ ,
\end{equation}
where one recognizes the structure of general coordinate transformations. 
So then, imposing the symmetry under the gauge transformation, means considering a theory invariant under general coordinate transformations, which is exactly what the machinery of general relativity does. In fact introducing a new gauge field in terms of a rank-2 symmetric tensor $h_{\mu\nu}(x)$ called graviton, and defining the object
\begin{equation}
g_{\mu\nu}(x)= \eta_{\mu\nu}+\kappa h_{\mu\nu}(x)\ ,
\end{equation}
which corresponds to the metric, we can build a covariantizing procedure as usual in gauge theories. Imposing that under the local symmetry transformation the gauge field implicitly transforms as
\begin{equation}\label{Metric_Transformation_Under_GCT}
g_{\mu\nu}(x)\rightarrow\Lambda^\alpha{}_{\mu}(x)\Lambda^\beta{}_{\nu}(x)g_{\alpha\beta}(x)\ ,
\end{equation}
by defining the usual Christoffel symbols as
\begin{equation}
\Gamma^{\rho}_{\mu\nu}=\frac{1}{2}g^{\rho \sigma}\left( \de_\mu g_{\sigma \nu}+\de_\nu g_{\sigma \mu}-\de_\sigma g_{\mu \nu}\right)\ ,
\end{equation}
and the covariant derivative
\begin{equation}
D_\mu V^\nu = \de_\mu V^\nu + \Gamma^\nu_{\mu\sigma}V^\sigma\ ,
\end{equation}
the covariantization of the theory is now straightforward, and it is done by the following replacements \cite{GR_Gualtieri}
\begin{align}\label{Gravitational_Covariantization_Procedure}
\eta_{\mu\nu} &\rightarrow g_{\mu\nu} & \de_\mu &\rightarrow D_\mu & d^Dx & \rightarrow d^D x \sqrt{-g}\ ,
\end{align}
which leads to the matter action 
\begin{equation}\label{Matter_Action_Gravity_Implicit}
S=\int d^D x \sqrt{-g} \, \Biggl( (D_\mu \phi)^* (D_\nu \phi) \ g^{\mu\nu} - m^2 \phi^* \phi\Biggl)\ .
\end{equation}

\medskip

Now since we introduced a new gauge field, we have to make it dynamical. As before we can define the gravitational strength tensor starting from the commutator between covariant derivatives, which leads to the definition
\begin{equation}
[D_\alpha, D_\beta]V^\mu = R^{\mu}_{\ \nu \alpha \beta}V^{\nu}\ ,
\end{equation}
where $R^{\mu}_{\ \nu \alpha \beta}$ is exactly the Riemann tensor. From it, it is possible to define the Ricci tensor and the Ricci scalar as
\begin{equation}
R_{\mu\nu}=R^\alpha{}_{\mu\alpha\nu} \quad \text{and} \quad R=g_{\mu\nu}R^{\mu\nu}\ .
\end{equation}
Now in order to construct the kinetic term of the gravitational field we have to consider all the possible scalars of the theory made by contractions of the Riemann tensor which are quadratic in the first derivative of the field. The term we are looking for is just the Ricci scalar, and the kinematic term is given by the well-known Einstein-Hilbert action
\begin{equation}\label{Einstein_Hilbert_Action}
S=-\frac{2}{\kappa^2}\int d^Dx \sqrt{-g}\, R\ ,
\end{equation}
where 
\begin{equation}
\kappa^2 = 32\, \pi \, G_N\ ,
\end{equation}
with $G_N$ the Newton universal gravitational constant. We then reconstruct general relativity as a gauge theory, and in particular since the general coordinate transformations that play the role of gauge transformations are non-abelian, gravity is placed in the category of non-abelian gauge theories. 

\medskip

Finally we can consider a classical action in which a complex scalar field transforms under both electromagnetic and gravitational gauge transformations. In such a theory a covariant derivative is meant to take into account all the symmetries\footnote{For example if we consider a charged vector field $W^\nu$, then 
\begin{equation*}
D_\mu W^\nu=\de_\mu W^\nu+iQ A_\mu W^\nu+\Gamma^\nu_{\mu\sigma}W^\sigma\ .
\end{equation*}}
and the gravitational covariantization procedure in eq. \eqref{Gravitational_Covariantization_Procedure} has to be carried out also in the electromagnetic sector of the theory.
This gives rise to interaction terms between the electromagnetic and gravitational field, as it is clear from the complete action 
\begin{equation}\label{Classical_Complete_Action}
S=\int d^Dx\, \sqrt{-g}\, \Biggl( -\frac{2}{\kappa^2}R-\frac{1}{4}F_{\mu\nu}F_{\alpha\beta}\, g^{\alpha\mu}g^{\beta\nu}+(D_\mu \phi)^* (D_\nu \phi) g^{\mu\nu}- m^2 \phi^* \phi\Biggl)\ .
\end{equation}
In this theory the graviton field is considered electrically neutral, while for the electromagnetic tensor holds its original expression since
\begin{equation}
F_{\mu\nu}=D_\mu A_\nu - D_\nu A_\mu = \de_\mu A_\nu - \de_\nu A_\mu -\Gamma_{\mu\nu}^{\sigma}A_\sigma+\Gamma_{\mu\nu}^\sigma A_\sigma = \de_\mu A_\nu - \de_\nu A_\mu\ .
\end{equation} 
From the action \eqref{Classical_Complete_Action} one derives the Einstein equations as the field equations of the gravitational field
\begin{equation}\label{Einstein_Eq}
G_{\mu\nu}=R_{\mu\nu}-\frac{1}{2}g_{\mu\nu}R=\frac{\kappa^2}{4}T_{\mu\nu}\ ,
\end{equation}
where $T_{\mu\nu}$ is the stress-energy tensor of the complete system, which takes into account both scalar and electromagnetic fields. In fact we can state that
\begin{equation}
T^{\mu\nu}=T_{\phi}^{\mu\nu}+T_{A}^{\mu\nu}+T_{\phi A}^{\mu\nu}\ ,
\end{equation}
where $T_{A}^{\mu\nu}$ is the electromagnetic stress-energy tensor, which corresponds exactly to the Noether current associated to the global translational symmetry of the electromagnetic field, and takes the expression\footnote{The tensor in \eqref{EM_StressEnergyTensor} corresponds to the Noether conserved current up to a symmetrization and a divergenceless object, since these operations do not spoil the conservation condition. Moreover, since the Einstein equations have to be symmetric under gauge transformations, the Noether current has to be covariantized, and $T_{A}^{\mu\nu}$ must be written in terms of the metric $g_{\mu\nu}$, exactly what we already commented in eq. \eqref{EM_Current_Covariantized}.}
\begin{equation}\label{EM_StressEnergyTensor}
T_{A}^{\mu\nu}=F^{\mu\alpha}F_{\alpha}{}^\nu+\frac{1}{4}g^{\mu\nu}F^{\alpha\beta}F_{\alpha\beta}\ ,
\end{equation}
often called in literature Belinfante–Rosenfeld stress-energy tensor,
while $T_{\phi A}^{\mu\nu}$ is the mixed term between scalar and electromagnetic field.
Equations \eqref{Einstein_Eq} respect the general behavior of gauge theories, in which the covariantized version of the current associated to the original global symmetry enters in the field equations of the gauge field as a source. Moreover eq. \eqref{Einstein_Eq} shows the characteristic non-linearity of gravity, which becomes relevant in the quantization process. In the end we notice that in a similar way, everything we have done considering a scalar field can be generalized to massive spinors \cite{Donoghue:2017pgk}.  

\section{Quantization of gravity}\label{sec:Quantization_of_Gravity}

The complete classical action in \eqref{Classical_Complete_Action} is constructed in terms of the metric, which in turn contains the graviton field in its definition. In order to make the physical content of the theory explicit in terms of graviton field interactions, one has to consider the expansions
\begin{equation}\label{h_Expansion_1}
g^{\mu\nu}=\eta^{\mu\nu}-\kappa h^{\mu\nu}+\kappa^2h^\mu{}_\alpha h^{\alpha\nu}+O(\kappa^3) 
\end{equation}
and
\begin{equation}\label{h_Expansion_2}
\sqrt{-g}=1+\frac{\kappa}{2}h+\frac{\kappa^2}{4}\left(\frac{1}{2}h^2-h^{\alpha}{}_\beta h^{\beta}{}_\alpha\right)+O(\kappa^3)\ ,
\end{equation}
by which it is possible to expand the theory perturbatively around the coupling $\kappa$, which makes the strongly non-linearity of gravity manifest. Considering the expansion of the Christoffel symbols in powers of coupling constant, explicitly derived by the insertion of \eqref{h_Expansion_1} and \eqref{h_Expansion_2} in its definition, schematically one obtains
\begin{equation}
\Gamma^{\mu}_{\alpha\beta} = \Gamma^{\mu\ (1)}_{\alpha\beta}+\Gamma^{\mu\ (2)}_{\alpha\beta}+O(\kappa^3)\ ,
\end{equation}
where explicitly
\begin{equation}
\begin{gathered}
\Gamma^{\mu\ (1)}_{\alpha\beta}=\frac{\kappa}{2}\eta^{\mu\nu}\Bigl( \de_\beta h_{\nu\alpha}+\de_\alpha h_{\nu\beta}-\de_\nu h_{\alpha\beta} \Bigl) \\
\Gamma^{\mu\ (2)}_{\alpha\beta}=-\frac{\kappa^2}{2}h^{\mu\nu}\Bigl( \de_\beta h_{\nu\alpha}+\de_\alpha h_{\nu\beta}-\de_\nu h_{\alpha\beta} \Bigl)\ .
\end{gathered}
\end{equation}
From this expansion it is possible to consider also 
\begin{equation}
R_{\mu\nu}=R_{\mu\nu}^{(1)}+R_{\mu\nu}^{(2)}+O(\kappa^3) \quad \text{and} \quad R=R^{(1)}+R^{(2)}+O(\kappa^3)\ ,
\end{equation}
where for example at first order they explicitly read
\begin{equation}
\begin{gathered}
R_{\mu\nu}^{(1)}=\frac{\kappa}{2}\Bigl(\de_\mu \de_\nu h+\Box h_{\mu\nu}- \de_\mu\de_\lambda h^{\lambda}{}_\nu-\de_\nu\de_\lambda h^{\lambda}{}_\mu\Bigl)\\
R^{(1)}=\kappa\Bigl(\Box h- \de_\mu\de_\nu h^{\mu\nu}\Bigl)\ ,
\end{gathered}
\end{equation}
where $h=\eta^{\mu\nu} h_{\mu\nu}$. 

\medskip

This expansion can be performed for each term in the action  \eqref{Classical_Complete_Action}. Rewriting it as
\begin{equation}
S=\int d^Dx\, \sqrt{-g}\, \Biggl( \mathcal{L}_h+\mathcal{L}_{A}+\mathcal{L}_\phi+\mathcal{L}_{\phi^2 A}+\mathcal{L}_{\phi^2 A^2}\Biggl)\ ,
\end{equation}
where respectively 
\begin{equation}\label{All_Terms_NotExpanded}
\begin{gathered}
\mathcal{L}_h = -\frac{2}{\kappa^2}  R \\
\mathcal{L}_A = -\frac{1}{4} F_{\mu\nu}F_{\alpha\beta}\, g^{\alpha\mu}g^{\beta\nu} \\
\mathcal{L}_\phi = (\de_\mu \phi)^* (\de_\nu \phi) g^{\mu\nu}- m^2 \phi^* \phi\\
\mathcal{L}_{\phi^2 A} = -j_\mu A_\nu g^{\mu\nu}\\
\mathcal{L}_{\phi^2 A^2}=Q^2A_\mu A_\nu g^{\mu\nu}\phi^*\phi\ ,
\end{gathered}
\end{equation}
one gets the expansions
\begin{equation}\label{All_Terms_Expanded}
\begin{gathered}
\sqrt{-g}\mathcal{L}_h = -\frac{2}{\kappa^2}  \Bigl(R^{(1)}+R^{(2)}+\frac{\kappa}{2}hR^{(1)}\Bigl)+O(\kappa) \\
\sqrt{-g}\mathcal{L}_A = -\frac{1}{4} F_{\mu\nu}F^{\mu\nu}-\frac{\kappa}{2}h_{\mu\nu}T^{\mu\nu}_{A}+O(\kappa^2)\\
\sqrt{-g}\mathcal{L}_\phi = (\de_\mu \phi)^* (\de^\mu \phi) - m^2 \phi^* \phi-\frac{\kappa}{2}h_{\mu\nu}T^{\mu\nu}_\phi + O(\kappa^2)\\
\sqrt{-g}\mathcal{L}_{\phi^2 A} = -j_\mu A^\mu +\kappa j_\mu A_\nu \Bigl(  h^{\mu\nu}-\frac{1}{2}h  \, \eta^{\mu\nu}\Bigl)+O(\kappa^2)\\
\sqrt{-g}\mathcal{L}_{\phi^2 A^2}=Q^2A_\mu A^\mu \phi^*\phi-\kappa Q^2 A_\mu A_\nu \phi^*\phi\Bigl(h^{\mu\nu}-\frac{1}{2}h\, \eta^{\mu\nu}\Bigl) +O(\kappa^2) \ ,
\end{gathered}
\end{equation}
where the indices are raised and lowered by the flat Minkowskian metric $\eta_{\mu\nu}$. The above expansion reproduces the terms at zeroth order in the graviton as well as terms that describe the interaction between electromagnetic, scalar and gravitational fields. For what concerns the pure gravity sector, the term linear in $h_{\mu\nu}$ in the expansion of the pure gravitational Lagrangian does not contribute since it is a total derivative, then it will be no longer considered in this thesis. The term quadratic in the graviton represents the kinetic term of the field, while higher terms in the expansion of $\mathcal{L}_h$ represent the self-interaction of the graviton. Then the pure scalar and electromagnetic sector, as usual in any gauge theory, give rise to an interaction term linear in the graviton which couples the gauge field directly with the original global Noether current. Again, due to the non-linearity of gravity, there are infinitely many other interaction terms that are higher order in the graviton field. Finally there exist terms which involve the simultaneous interaction of all the fields, with an arbitrary number of graviton fields involved, that correspond to the classical mixed contributions to the stress-energy tensor. 

\medskip

Now that the field content of the theory is manifest, it is possible to quantize the gravitational field \cite{Feynman:1963ax, DeWitt:1967yk, DeWitt:1967ub, DeWitt:1967uc}. Following the procedure outlined in section \ref{Scalar_quantum_electrodynamics}, since general relativity is a gauge theory, in order to quantize it we have to work out the Faddeev-Popov procedure for the gauge fixing, where now the quantum gauge field that mediates the interaction is the graviton field $h_{\mu\nu}$ \cite{tHooft:1974toh}. For simplicity we first consider gravity only coupled to the scalar field, and then after the quantization procedure we put back the electromagnetic interaction. 
The generating functional associated to the action in eq. \eqref{Classical_Complete_Action}, neglecting the electromagnetic interaction, is expected to be 
\begin{equation}\label{Gravity_Generating_Func_NoGauge}
Z=\int [d\phi] [d\phi^*] [dh_{\mu\nu}]\, e^{iS}\ ,
\end{equation}
where explicitly the kinetic term of the graviton field reads
\begin{equation}\label{Graviton_Kinetic_Term}
\mathcal{L}_h=\frac{1}{2}\de_\sigma h_{\mu\nu}\de^\sigma h^{\mu\nu}-\de^\nu h_{\mu\nu}\de_\sigma h^{\mu\sigma}+\de^\nu h_{\mu\nu}\de_\mu h-\frac{1}{2}\de_\mu h\de^\mu h+O(\kappa)\ .
\end{equation}
As already noticed in the electromagnetic case, eq. \eqref{Gravity_Generating_Func_NoGauge} diverges since we are integrating over infinite equivalent copies of the same physical systems due to gauge invariance. In order to remove such redundancy, we must choose a gauge, rewriting the generating functional in the form 
\begin{equation}
Z=\int [d\phi] [d\phi^*] [dh_{\mu\nu}]\, \Delta[h_{\mu\nu}]e^{iS+i\xi\int d^Dx\, F_\lambda F^\lambda} \ ,
\end{equation}
where $F_\lambda[h_{\mu\nu}]$ is the gauge fixing functional and $\Delta[h_{\mu\nu}]$ is the Faddeev-Popov determinant. Differently from electromagnetism, gravity is non-abelian, this means that the ghost Lagrangian that arises by the Faddeev-Popov determinant becomes relevant for the gauge invariance of the theory. Considering that under an infinitesimal gauge transformation\footnote{In literature one usually refers to this kind of transformation as the Lie derivative of the metric.} the transformed graviton field reads \cite{Hamber}
\begin{equation}\label{Infinitesimal_Graviton_Gauge_Transformation}
\kappa\, h'_{\mu\nu}=\kappa\, h_{\mu\nu}-\de_\mu a_\nu-\de_\nu a_\mu-\kappa\, h_{\lambda \nu}\de_\mu a^\lambda -\kappa\, h_{\lambda \mu}\de_\nu a^\lambda-\kappa\, a^\lambda \de_\lambda h_{\mu\nu}+O(a^2)\ ,
\end{equation}
one can consequently obtain 
\begin{equation}\label{Gravitational_Ghost_Lagrangian}
\Delta[h_{\sigma\rho}]=\det(\frac{\de F^\mu[h'_{\sigma\rho}]}{\de a^{\nu}})=\int [d c_{\mu}][d \bar{c}^\nu]e^{i\int d^Dx\, \bar{c}^\nu \frac{\de F'^\mu}{\de a^{\nu}} c_\mu}\ ,
\end{equation}
where $c_\mu$ and $\bar{c}^\nu$ are the anticommuting $"\text{fermionic vector}"$ ghost fields.

\medskip

In the context of quantization a particular convenient gauge choice is the de Donder gauge, whose condition is defined by
\begin{equation}
\eta^{\mu\nu}\Gamma_{\mu\nu}^\alpha=0\ ,
\end{equation}
from which in terms of the graviton field can be rewritten as
\begin{equation}
\de^\mu h_{\mu\alpha}-\frac{1}{2}\de_\alpha h=0\ .
\end{equation}
Now we can consider the gauge fixing function associated to the de Donder gauge, which is exactly 
\begin{equation}
F_\lambda = \de^\mu h_{\mu\lambda}-\frac{1}{2}\de_\lambda h\ ,
\end{equation}
from which it is possible to compute explicitly the ghost Lagrangian. Considering the infinitesimal gauge transformation in eq. \eqref{Infinitesimal_Graviton_Gauge_Transformation}, then one gets
\begin{equation}
F'_\lambda=F_\lambda-\Box a_\lambda+O(\kappa)\ ,
\end{equation}
that leads to  
\begin{equation}
\mathcal{L}_{ghost}=\de_\mu\bar{c}_\nu\de^\mu c^\nu+O(\kappa)\ .
\end{equation}
From the above Lagrangian one can derive the Feynman rules that involve ghosts in the old fashion way, in particular from the kinetic term one computes the propagator
\begin{equation}
\includegraphics[width=0.18\textwidth, valign=c]{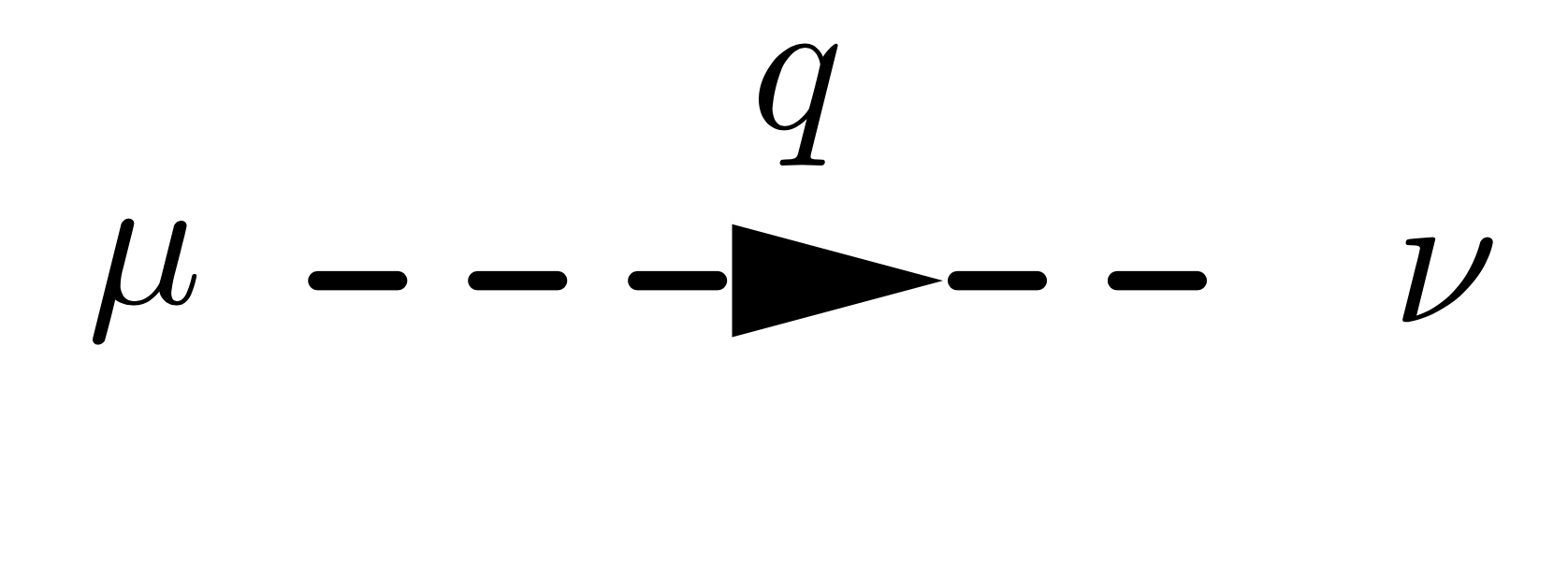}=-\frac{i\eta^{\mu\nu}}{q^2 +i\epsilon}\ ,
\end{equation}
while an explicit calculation of the interaction term will give rise to the vertex (for the explicit derivation see \cite{Hamber}). 

\medskip

Now it is possible to exploit the gauge fixing procedure to recover the graviton propagator. In fact considering the addition of 
\begin{equation}
\mathcal{L}_h^{GF}=F_\lambda F^\lambda
\end{equation} 
to the Lagrangian in eq. \eqref{Graviton_Kinetic_Term}, choosing $\xi= 1$, then the kinematical term of the graviton reads
\begin{equation}
S=\int d^Dx\, \Bigl( \mathcal{L}_h+\mathcal{L}_h^{GF}\Bigl)=\int d^Dx\, \Biggl( \frac{1}{2}\de_\sigma h_{\mu\nu}\de^\sigma h^{\mu\nu}-\frac{1}{4}\de_\mu h\de^\mu h + O(\kappa)\Biggl)\ ,
\end{equation}
by which after an integration by parts one gets
\begin{equation}
S=\int d^Dx\, \Biggl( h_{\mu\nu} V^{\mu\nu\alpha\beta}h_{\alpha \beta} + O(\kappa)\Biggl)\ ,
\end{equation}
where 
\begin{equation}
V^{\mu\nu\alpha\beta} = \left(\frac{1}{4}\eta^{\mu\nu}\eta^{\alpha\beta}-\frac{1}{2}\eta^{\mu\alpha}\eta^{\nu\beta}\right)\Box\ .
\end{equation}
The operator above can be inverted in order to determine the graviton propagator in the de Donder gauge, which in generic space-time dimension reads \cite{Donoghue:2017pgk}
\begin{equation}
\includegraphics[width=0.20\textwidth, valign=c]{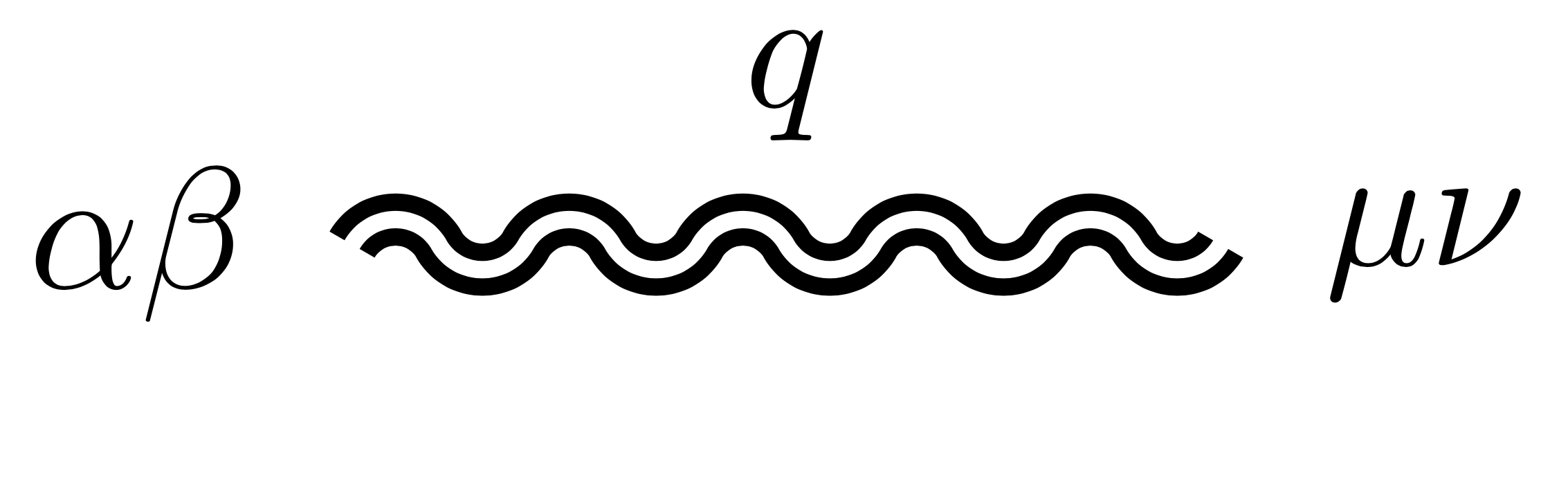}=i\frac{P_{\alpha\beta,\mu\nu}}{q^2 +i\epsilon} \ ,
\end{equation}
with $P_{\alpha\beta,\mu\nu}$ defined by
\begin{equation}
    P_{\alpha\beta,\mu\nu} =  \frac{1}{2} \Big( \eta_{\mu\alpha}\eta_{\nu\beta}+\eta_{\mu\beta}\eta_{\nu\alpha}-\frac{2}{D-2}\eta_{\mu\nu}\eta_{\alpha\beta} \Big) \ .
\end{equation}
Notice that here and in the following the double wiggly lines stand for gravitons. 
In the same gauge one can find the field equations of the graviton field, which read
\begin{equation}
\Box\Bigl(h_{\mu\nu}-\frac{1}{2}\eta_{\mu\nu}h\Bigl)=-\frac{\kappa}{2}\Bigl(T_{\mu\nu}+t_{\mu\nu}\Bigl)\ ,
\end{equation}
where $t_{\mu\nu}$ is the stress-energy tensor associated to the gravitational field itself. In fact the non-linearity of general relativity gives rise to gravitational self-interaction terms, that can be computed from the higher order terms in $\sqrt{-g}\mathcal{L}_h$. From here on we will write only $T_{\mu\nu}$ as a unique stress-energy tensor, and it will be clear from the context if a gravitational contribution is implied. Finally, by tracing the field equations, one obtains 
\begin{equation}\label{Graviton_Equation_Motion_deDonder}
\Box h_{\mu\nu} = -\frac{\kappa}{2} \Biggl(T_{\mu\nu}-\eta_{\mu\nu}\frac{1}{D-2}T\Biggl)\ .
\end{equation}

\medskip

Now putting back the electromagnetic interaction, since the gauge we choose is second order in the first derivative of the field, it does not affect the expressions of the other terms in eq. \eqref{All_Terms_Expanded} rather than the kinetic one. So it is possible to write down all the Feynman rules of the theory, keeping in mind that the interaction terms in \eqref{All_Terms_Expanded} are just the first pieces of an infinite expansion in the gravitational coupling constant, since the non-linearity of gravity gives rise to vertices with an arbitrary number of gravitons attached to them \cite{Bjerrum-Bohr:2014lea}. Some of these, in particular the ones useful for further analysis in this thesis, are reported in appendix \ref{Appendix_FeynRules}. 
It is worth to comment the 3 graviton vertex that arises from the $O(\kappa)$ contribution of the Einstein-Hilbert action expansion. The explicit expression of it was originally provided by de Witt \cite{DeWitt:1967uc}, in which the vertex is recovered in the old fashion way where the expression is symmetric under the exchange of any leg, and is listed in \eqref{3Graviton_Vertex_deWitt_AppEq}.
Since such object can be painful to handle due to the proliferation of Lorentz indices, one can derive a simplified expression that describes the interaction of two quantum fields and an external field. The expression was firstly given by Donoghue \cite{Donoghue:1994dn}, which used the background field method and obtained the vertex reported in eq. \eqref{3Graviton_Vertex_Donoghue}. The idea is to expand the metric rather than on a flat background space-time, around a generic $\bar{g}_{\mu\nu}$ metric, such as
\begin{equation}
g_{\mu\nu}=\bar{g}_{\mu\nu}+\kappa\, h_{\mu\nu}\ .
\end{equation}
Imposing than $\bar{g}_{\mu\nu}=\eta_{\mu\nu}+\kappa\, H_{\mu\nu}$, obtaining the relation
\begin{equation}
g_{\mu\nu}=\eta_{\mu\nu}+\kappa\, H_{\mu\nu}+\kappa\, h_{\mu\nu}\ ,
\end{equation}
where $H_{\mu\nu}$ is an external gravitational field, by replacing this expansion inside the Einstein-Hilbert action, the $O(\kappa)$ piece with two quantum fields and an external one gives rise to the aforementioned vertex. Notice that by construction such vertex is not symmetric under the exchange of all the legs, but only between the internal ones, therefore it is not possible to use such expression as an internal vertex.

\medskip

Once the quantization procedure is done and the Feynman rules are carried out, the natural question is whether this theory is renormalizable or not. The gravitational Einstein-Hilbert action, which gives rise to the propagator and the self-interaction terms of the graviton, is quadratic in the first derivatives of the metric. This means that taking $q$ as a characteristic momentum, the propagator goes like $q^{-2}$ and each vertex goes like $q^2$. Moreover in a $D$-dimensional space-time the measure of a loop integral is $d^Dq$, from which considering a diagram with $V$ vertices, $I$ internal lines and $L$ loops, the superficial degree of divergence of the theory reads
\begin{equation}
\mathfrak{D}=DL+2V-2I\ .
\end{equation}
Replacing the topological relation 
\begin{equation}
L=1+I-V\ ,
\end{equation}
one obtains the relation
\begin{equation}
\mathfrak{D}=2 + (D-2)L\ .
\end{equation}
It is easy to see that for dimensions $D>2$ the superficial degree of divergence of the theory is positive and increases with the number of loops, therefore the theory is non-renormalizable since an infinite number of processes produce divergent amplitudes. This can be seen also from the dimension of gravitational coupling. In fact in terms of the canonical mass dimension in which $[M]=1$, the dimension of the gravitational Newton constant is 
\begin{equation}
[G_N]=2-D\ ,
\end{equation} 
which is negative for $D>2$, and then leads to a non-renormalizable theory. 

\medskip

What happens when a theory is non-renormalizable, is that since each physical process is divergent, one should in principle add to the original Lagrangian an infinite number of terms in order to cancel the infinite amount of divergences that appear, for this reason such kind of theories are classified as non-fundamental ones. However in the modern way of thinking, non-renormalizable theories can be interpreted as effective field theories, which can be used to extrapolate predictions below some energy cut-off, in which above of it the effective theory is not useful anymore. The idea is that at high energies the unknown fundamental theory possesses some heavy degrees of freedom, which however at low energy are expected to not contribute, and their effect can be parametrized by the addition in the Lagrangian of all the possible higher-derivative non-minimal couplings compatible with the symmetry of the theory, whose ordered in an energy expansion mimic the fundamental theory and separates high-energy effects from the low-energy ones. 

\medskip

Embracing this idea and considering for the moment only the purely gravitational action, one can then rewrite it in terms of an infinite series of all the possible Riemann tensor contractions as
\begin{equation}\label{Effective_PureGravity_Action}
S=\int d^Dx \,\sqrt{-g}\Biggl(-\Lambda -\frac{2}{\kappa^2}R+c_1 R^2+c_2R_{\mu\nu}R^{\mu\nu}+c_3 R_{\alpha\beta\sigma\rho}R^{\alpha\beta\sigma\rho}+\cdots\Biggl)\ ,
\end{equation}
where $\Lambda, c_1, c_2, c_3...$ are effective couplings that have to be fixed experimentally. Since we are going to talk about their experimental values, it is useful to switch to $D=4$ until the end of this section, where in particular in four space-time dimensions the term proportional to $c_3$ disappears due to the Gauss-Bonnet identity. From the fact that the Riemann tensor is quadratic in the derivative of the metric, schematically written as $R_{\mu\nu\alpha\beta}\thicksim \de^2$, in momentum space it is considered to be quadratic in energy. Therefore the energy expansion is then organized as follows
\begin{equation}
\Lambda\thicksim O(1) \quad R\thicksim O(E^2) \quad R^2,  R_{\mu\nu}R^{\mu\nu}\thicksim O(E^4)\ ,
\end{equation}
meaning that the contribution of higher-order terms in energy becomes negligible when low-energy processes are considered. In fact schematically the coefficients $\Lambda, c_1, c_2, ...$ are not dimensionless in $D$ dimension, and they can be rewritten making explicit the cut-off energy scale of the gravitational interaction. We believe that such a scale is identified by the Planck scale, above which a fundamental theory of quantum gravity is needed. Including the Planck mass $M_p \simeq 10^{19}\ \text{GeV}$ and the dimensionless coefficient $\lambda, d_1, d_2, d_3, ...$, then it is possible to rewrite the effective action \eqref{Effective_PureGravity_Action} in a more manifest way as
\begin{equation}
S=\int d^Dx \,\sqrt{-g}\Biggl(-M_p^{D}\lambda -\frac{2}{\kappa^2}R+\frac{d_1}{M_p^{4-D}} R^2+\frac{d_2}{M_p^{4-D}}R_{\mu\nu}R^{\mu\nu}+\frac{d_3}{M_p^{4-D}}R_{\alpha\beta\sigma\rho}R^{\alpha\beta\sigma\rho}+\cdots\Biggl)\ ,
\end{equation}
which in $D=4$ reads
\begin{equation}\label{Effective_PureGravity_Action}
S=\int d^4x \,\sqrt{-g}\Biggl(-M_p^{4}\lambda -\frac{2}{\kappa^2}R+d_1 R^2+d_2R_{\mu\nu}R^{\mu\nu}+\cdots\Biggl)\ .
\end{equation}

\medskip

Once the most generic effective Lagrangian is defined, one can think about fixing the effective coupling constants. 
\begin{itemize}
\item The constant $\Lambda$ is called cosmological constant, and it is believed to be responsible for the current accelerated expansion of the Universe. By observations we inferred that its value should approximately be $\Lambda\simeq 10^{-47}\ \text{GeV}^4$. The energy of such object is very tiny compared to ordinary scales in which gravity occurs, so then we will neglect this term any further in this thesis. 
\item The parameter $\kappa^2$ defines the strength of the gravitational field at long distances and its value is fixed by matching the classical non-relativistic potential. In fact the contribution to the Newton potential only due to the Einstein-Hilbert action is 
\begin{equation}
V(r)=-\frac{\kappa^2}{32 \pi}\frac{m_1m_2}{r}=-G_N\frac{m_1m_2}{r}\ .
\end{equation}
\item The constants $d_1, d_2, ...$ produce Yukawa–type corrections to the gravitational potential which become relevant at short distances.  
Indeed by measuring the classical contribution to the gravitational potential of such pieces \cite{Stelle:1977ry}, by experiments one gets $|d_i|< 10^{-56}$. The length scale at which these terms are relevant is well below the astrophysical distances we are interested in, so any further in this thesis we will neglect any high-energy Riemann coupling. 
\end{itemize}

\medskip

Finally we can consider the non-minimal coupling between the gravitational field, the massive scalar and the photon. Considering the terms in eq. \eqref{All_Terms_NotExpanded}, one should consider also the addition of terms like
\begin{equation}
\begin{gathered}
\mathcal{L}_\phi \thicksim m^2 R \phi^*\phi, \,  R^{\mu\nu}\de_\mu \phi^*\de_\nu \phi, \,  R g^{\mu\nu}\de_\mu \phi^*\de_\nu \phi, \, ...\\
\mathcal{L}_A \thicksim F_{\mu\nu}F^{\mu\nu}R, \, F_{\mu\nu}R^{\mu\nu}, \, F_{\mu\alpha}F^{\mu}{}_{\beta}R^{\alpha\beta},\,  ...
\end{gathered}
\end{equation}
In order to organize an energy expansion of these terms, one has to notice that in this effective field theory approach, the energy of the processes, namely the typical transferred momentum between interacting particles, must be much smaller than every energy scale of the theory. This means that since in our theory also a massive field is present, this gives rise to another energy scale rather than the Planck one, and in the effective approach the typical transferred momentum must be $q \ll M_p, \, m$, which means that the derivative of massive fields is grater in energy than the derivative of massless fields, since in momentum space $m\thicksim \de \phi$. In virtue of this consideration it is crucial to differentiate the quantum effects of massless particles from the ones which have mass. While massless particles can propagate to long distances at low energies, massive particles cannot. In fact considering schematically a massive propagator in momentum space, since $q^2\ll m^2$ one can expand it as
\begin{equation}\label{Massive_Propagator_Expansion_EFT_Chapter}
\frac{1}{q^2-m^2+i\epsilon}=-\frac{1}{m^2}-\frac{q^2}{m^4}-\frac{q^4}{m^6}+O(q^6)\ .
\end{equation} 
Then going back to coordinate space, one gets
\begin{equation}
-\frac{1}{m^2}\thicksim \delta^{(D)}(x) \quad -\frac{q^2}{m^4}\thicksim \de^2 \quad -\frac{q^4}{m^6}\thicksim \de^4\ ,
\end{equation}
meaning that the leading order in the expansion of the massive propagator gives a local contribution, which in turn generates an effective interaction vertex, while higher order terms can be just reabsorbed in high-derivative couplings in the effective action. Obviously such expansion cannot be performed on massless propagators which go like $1/q^2$, and whose contribution cannot be integrated out.
\begin{figure}[h]
\centering
\begin{equation*}
\includegraphics[width=0.25\textwidth, valign=c]{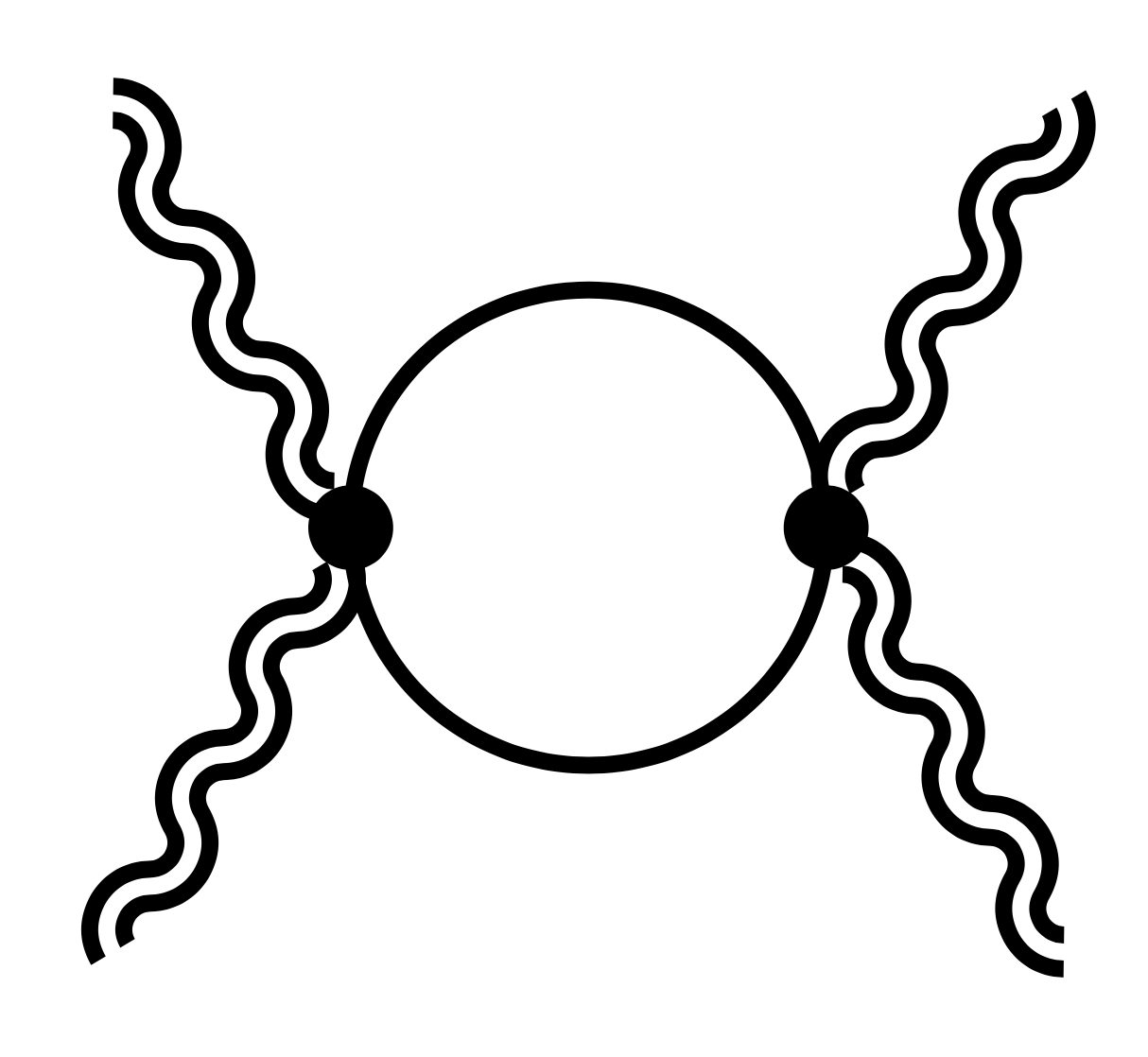} \underrightarrow{q^2\ll m^2}\includegraphics[width=0.25\textwidth, valign=c]{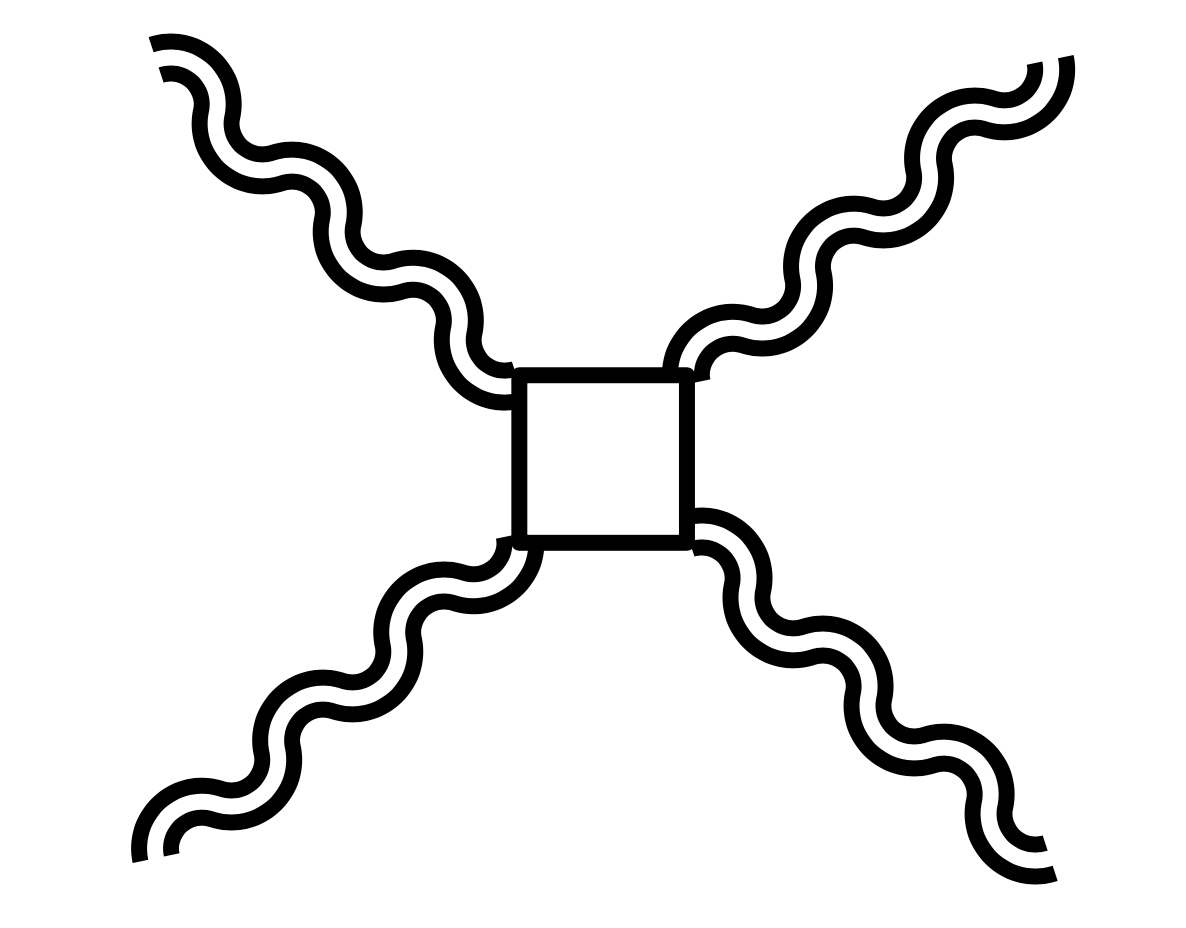}
\end{equation*}
\caption{In the left a 2-to-2 graviton scattering process which is mediated by a massive loop, which in the low-energy limit degenerate in an effective 4 graviton vertex on the right.}
\label{fig_Effective_4Graviton_Vertex}
\end{figure}
Then the final outcome is that a process in which the interaction is mediated by massive lines will collapse in an effective vertex as in figure \ref{fig_Effective_4Graviton_Vertex}, as usually happens when the heavy degrees of freedom are integrated out in an effective field theory.

\medskip

So in the very end, quantum gravity as an effective field theory is a well-defined quantum theory in the low-energy regime, in which at any fixed loop order it is possible to obtain finite predictions, as it is shown in \cite{Goroff:1985th}, where the 2-loop effective pure gravitational Lagrangian is obtained. 
Since this approach only enables us to make predictions when the exchanged momentum between virtual particles is well below all the energy scales of the theory, in coordinate space this is translated to the fact that we can only infer the long-range physics of the systems.

\section{Classical limit}\label{Sec_Classical_Limit}

In the last section we reviewed how the effective field theory approach to quantum gravity gives a perfectly well-defined quantum theory, able to make predictions in the low-energy regime. Then as any other quantum theory, the classical limit, in which $\hbar\rightarrow 0$, must reconstruct the original classical phenomenology, that in this case corresponds to general relativity. 
In order to do that, we have to restore the explicit presence of $\hbar$ in the theory. Generalizing to arbitrary space-time dimensions the procedure outlined in \cite{Brodsky:2010zk}, we still consider relativistic natural units in which $c=1$, while the Planck constant is considered to have dimensions
\begin{equation}
[\hbar]=L\cdot M\ ,
\end{equation}
where $L$ is a length dimension and $M$ is a mass dimension (which is equivalent to an energy). In order to recover the dimension of both fields and couplings, now the dimensional analysis must be performed keeping in mind that the dimensionless object is $S/\hbar$ and therefore the generating functional of the theory becomes
\begin{equation}
Z=\int [d\phi][d\phi^*][dA_\mu][dh_{\alpha\beta}]e^{iS/\hbar}\ .
\end{equation}

\medskip

From this consideration in a $D$-dimensional space-time the Lagrangian satisfies
\begin{equation}\label{Lagrangian_Dimension_hbar_Restored}
[\mathcal{L}]=[\hbar]\, L^{-D}\ .
\end{equation}
Studying the kinetic term of the fields, each boson of the theory has the same dimension, which is given by
\begin{equation}
[\phi]=[A_\mu]=[h_{\mu\nu}]=[\hbar]^{\frac{1}{2}}\, L^{-\frac{D}{2}+1}\ .
\end{equation}
Then considering the interaction terms of the theory and imposing eq. \eqref{Lagrangian_Dimension_hbar_Restored}, it is possible to recover whether it is needed or not to add an extra $\hbar$ in the Lagrangian by studying the dimension of the couplings.
\begin{itemize}
\item For what concerns the mass term, adding an unknown dimensional constant, the equation to solve is
\begin{equation}
[x\, m^2\phi^2]=[\hbar]\, L^{-D}\ ,
\end{equation}
which leads to 
\begin{equation}
[x]=[\hbar]^{-2}\ .
\end{equation}
This means that whenever $m$ appears in the theory, we have to consider the replacement
\begin{equation}\label{Mass_Replacement_hbar}
m\rightarrow \frac{m}{\hbar}\ .
\end{equation}
\item Considering the electromagnetic interaction term which involves the charge $Q$, the equation to solve is
\begin{equation}
[x\, QA_\mu \phi\de^\mu\phi]=[\hbar]\, L^{-D}\ .
\end{equation}
However one has also to keep in mind that the fine structure constant defined by
\begin{equation}
\alpha=\frac{Q^2}{4\pi\, \hbar}\ ,
\end{equation}
has to be dimensionless in $D=4$. So then the only solution that satisfies both relations is 
\begin{equation}
[Q]=[\hbar]^{\frac{1}{2}}\, L^{\frac{D}{2}-2} \quad \text{and} \quad [x]=[\hbar]^{-1}\ ,
\end{equation}
which in turn implies the replacement
\begin{equation}\label{Charge_Replacement_hbar}
Q\rightarrow \frac{Q}{\hbar}\ ,
\end{equation}
whenever a charge is present. 
\item The last coupling to take into account is the gravitational one. As it was for the charge we need to satisfy two dimensional relations. The first is due by considering the dimension of a coupling that involves $\kappa$. For example taking the 2 scalars - 1 graviton vertex the dimensional equation is
\begin{equation}
[x\, \kappa h_{\mu\nu}T_{\phi}^{\mu\nu}]=[\hbar]\, L^{-D}\ ,
\end{equation}
with 
\begin{equation}
[T_{\phi}^{\mu\nu}]=[\hbar]\, L^{-D}\ .
\end{equation}
The other relation to be satisfied is the one that defines the Planck mass, which is 
\begin{equation}
M_p=\sqrt{\frac{\hbar}{G_N}}\ ,
\end{equation}
and in $D=4$ imposes 
\begin{equation}
[G_N]=[\hbar]^{-1}\, L^2\ .
\end{equation}
Keeping in mind that $[\kappa]=[G_N]^{\frac{1}{2}}$, than the solution is 
\begin{equation}
[G_N]=[\hbar]^{-1}L^{D-2} \quad \text{and} \quad [x]=1 \ ,
\end{equation}
which means that terms involving $\kappa$ do not need an explicit insertion of $\hbar$ factors.
\end{itemize}

\medskip

Now that the explicit presence of $\hbar$'s is restored in the theory, the idea is to study how Feynman rules change, in order to consider the explicit dependence of the Planck constant in the diagrams. 
As in any other approach like this one, the fact that the ration $S/\hbar$ enters in the generating functional, means that for each vertex we have to add an extra $\hbar^{-1}$, and for each propagator we have to consider a $\hbar$ factor in front. Moreover, whenever one encounters a momentum, this is implied to be a wave-number $\tilde{q}$ instead of a physical one, where the relation that links each other is $q=\tilde{q}\, \hbar$, with $[\tilde{q}]=L^{-1}$. Making this explicit, for example the scalar propagator in coordinate space reads
\begin{equation}
i\Delta(x-y)=\int d^D \tilde q\, e^{i\tilde{q}\cdot (x-y)} \frac{i\, \hbar}{\tilde{q}^2-\frac{m^2}{\hbar^2}+i\epsilon}\ .
\end{equation}

\medskip

In this context it is important to remind the considerations of the section before, in which we underlined the differences between massive and massless particles since we constantly consider the low-energy limit. 
Even though we said that the only processes which give long-distance contribution are the ones with no massive particles circulating in the loops, an internal massive line may be attached to an external massive state, since the constraint of low-energy transferred momentum would be still valid.  
Consider then a massive propagator with momentum $q=p-k$, where $p$ is the momentum of the external particle, which satisfies the mass-shell relation $p^2=m^2$, and $k$ is an exchanged momentum that respects $k^2\ll m^2$. In this case the massive propagator can be expanded as
\begin{equation}
\frac{\hbar}{(\tilde{p}-\tilde{k})^2-\frac{m^2}{\hbar^2}+i\epsilon}=\frac{\hbar}{\tilde{k}^2-2\, \tilde{p}\cdot\tilde{k}+i\epsilon}\ ,
\end{equation}
where we used $\tilde{p}^2=m^2/\hbar^2$. Now considering $\tilde{p}\thicksim \frac{m}{\hbar}\gg \tilde{k}$, one gets schematically
\begin{equation}
\frac{\hbar}{\tilde{k}^2-2\, \tilde{p}\cdot\tilde{k}+i\epsilon}=-\frac{\hbar^2}{2m\, \tilde{k}}\propto \hbar^2\ .
\end{equation}
It is important to underline that this is just the leading order behavior, and a complete expansion of the propagator would lead to terms $O(\hbar^3)$.
Finally one can consider a massless propagator which simply goes like
\begin{equation}
\frac{\hbar}{\tilde{q}^2+i\epsilon}\propto \hbar \ . 
\end{equation}

\medskip

Similar arguments hold for the study of $\hbar$ dependency of vertices, which has a different behavior whether the interaction involves a massive particle or not. 
Vertices that involve only massless lines do not depend on the charge or the mass, and they simply go like
\begin{equation}
\tau_{massless} \propto \hbar^{-1}\ .
\end{equation}
To this category belong the 2 photons - 1 graviton vertex, 2 photons - 2 gravitons vertex, 3 gravitons vertex, and so on. On the other hand, vertices that are attached to external massive lines behave differently. For the 2 scalars - 1 photon vertex, the expression in eq. \eqref{2scalars_1photon} becomes
\begin{equation}
\left(\tau_{\phi^2A}\right)^\mu(p,p')=-i\frac{1}{\hbar}\frac{Q}{\hbar}\left(\tilde{p}+\tilde{p}\, '\right)^\mu \ , 
\end{equation}
where considering again $\tilde{p}\thicksim m/\hbar$ one gets
\begin{equation}
\left(\tau_{\phi^2A}\right)^\mu(p,p')\propto \hbar^{-3} \ . 
\end{equation}
Analogously for the 2 scalars - 1 graviton vertex one firstly has
\begin{equation}
\left(\tau_{\phi^2 h}\right) ^{\mu\nu}(p,p')=-\frac{1}{\hbar}\frac{i\kappa}{2}\Big(\big( \tilde{p} ^\mu \tilde{p}\, '{}^\nu + \tilde{p} ^\nu \tilde{p}\, '{}^\mu\big) - \eta ^{\mu\nu}\big(\tilde{p}\cdot \tilde{p}\, ' - \frac{m^2}{\hbar^2} \big)\Big) \ ,
\end{equation}
which leads to 
\begin{equation}
\left(\tau_{\phi^2 h}\right) ^{\mu\nu}(p,p')\propto \hbar^{-3}\ .
\end{equation}
From these two computations one can infer then that
\begin{equation}
\tau_{massive} \propto \hbar^{-3}\ .
\end{equation} 
Again this is just a leading order contribution, while a higher-order expansion contains pieces $O(\hbar^{-2})$.

\medskip

So then we obtain some replacement rules, which given a specific amplitude, allow us to see immediately to what power of $\hbar$ the process is proportional to. 
Then since a generic gravitational quantum observable can be described perturbatively in terms of Feynman diagrams, exploiting these considerations we are able to infer before calculations how each contribution of the observables depends on the Planck constant. 
In fact, since the main scope of this machinery is to recover classical physics from quantum processes, if we know a priori that some contributions of a specific observable depend on $\hbar$, then we can conclude that the Feynman diagrams associated to that specific piece do not contribute to the classical limit, and only reconstruct the so called quantum corrections that are vanishing in the limit $\hbar\rightarrow 0$. 
This procedure is crucial in order to perform actual calculations of classical gravitational observables from scattering amplitudes, because it allows us to focus only on those diagrams we know will contribute to the classical regime, avoiding all other processes which would lead just to $\hbar$-dependent contributions.
The idea then, is to define some selection rules which make the choice of classical processes systematical.  
To see explicitly how this machinery can be exploited, in the next chapter we pick a specific classical gravitational observable in order to see how it can be recovered from quantum calculations.   

\chapter{Classical metrics from scattering amplitudes in four dimensions}\label{Classical_Chapter}

In this section we want to exploit the procedure outlined so far in order to recover classical metrics from the $\hbar\rightarrow 0$ limit of quantum gravity as an effective field theory \cite{Donoghue:2001qc, Bjerrum-Bohr:2002fji}.
In first place it will be reconstructed the Schwarzschild metric. The strategy will be firstly identify the quantum counterpart of the classical gravitational observable, and then organize a perturbative expansion in terms of Feynman diagrams, picking up only those pieces we know will contribute to the classical limit, following the original work of \cite{Donoghue:1994dn}. 
Such observable, obtained from scattering amplitudes, will be compared with the one computed by directly solving the Einstein equations, whose however is expressed in the harmonic gauge. 
Then we will outline a procedure to systematically move from one gauge to the other, originally discussed in \cite{Bjerrum-Bohr:2002fji}, and we will prove the coherence of the procedure by matching the two derivations. 
Finally it will be computed the Reissner–Nordstr\"om metric from graviton emission processes with photons circulating in the loops, and it will be commented the procedure to recover the Kerr-Newman space-time from spinors-scattering in a completely analogous way.
The chapter follows the historical path of how we learned to employ the quantum field theory formalism in order to recover classical physics in gravity context, and gives the chance to introduce some important concepts and ideas later useful in this thesis. 

\section{Schwarzschild metric from graviton emission}
\label{sec:Schwarzschild_Metric_from_Graviton_Emission}

The Schwarzschild metric describes the four-dimensional space-time geometry associated to a spherically symmetric and static source, which is also electrically neutral. The quantum counterpart of the observable is the graviton field $h_{\mu\nu}(x)$, since by its evaluation it is possible to reconstruct the metric as $g_{\mu\nu}=\eta_{\mu\nu}+\kappa h_{\mu\nu}$. Such a solution of the Einstein equations is associated to an action that couples an electrically neutral scalar field to gravity, and where such scalar field is static, meaning that $\de_t\phi(x)=0$.
Consider then the graviton field, whose field equations in de Donder gauge and $D=4$ read
\begin{equation}\label{Box_Graviton_Example}
\Box h_{\mu\nu}(x)=-\frac{\kappa}{2}\left(T_{\mu\nu}(x)-\frac{1}{2}\eta_{\mu\nu}T(x)\right)\ ,
\end{equation} 
with $T^{\mu\nu}(x)=T^{\mu\nu}_{\phi}(x)+t^{\mu\nu}(x)$, where $T^{\mu\nu}_{\phi}(x)$ is simply the stress-energy tensor of the scalar field, while $t^{\mu\nu}(x)$ contains inside all the self-interaction terms of the gravitational field.
Imposing the staticity of the field and an explicit insertion of $\hbar$ factors, the solution of the above equation takes the form
\begin{equation}\label{Graviton_Example}
h_{\mu\nu}(\vec{x})=-\frac{\kappa}{2}\int\frac{d^3\tilde{q}}{(2 \pi)^3}\frac{e^{i\vec{\tilde{q}}\cdot \vec{x}}}{\vec{\tilde{q}}^2}\left(T_{\mu\nu}(\vec{\tilde q}^2)-\frac{1}{2}\eta_{\mu\nu}T(\vec{\tilde q}^2)\right)\ ,
\end{equation}
where $T_{\mu\nu}(\vec{\tilde q}^2)$ is the Fourier transform of the stress-energy tensor and again $\tilde{q}=q/\hbar$ has the meaning of wave-number. 
From the expression above we see that there are no explicit $\hbar$ inside the integral, meaning that eventual terms proportional to the Planck constant come from the stress-energy tensor itself. By definition $T_{\mu\nu}(q^2)$ is associated to the matrix element
\begin{equation}\label{StressEnergyTensor_as_MatrixElement_Example}
T_{\mu\nu}(q^2)=\bra{p'}T_{\mu\nu}(x=0)\ket{p}\ ,
\end{equation}
with $q=p-p'$ the transferred momentum, $p$ and $p'$ the external physical momenta associated to on-shell particles in which $p^2=p'^2=m^2$, and where the normalization of the particle states is
\begin{equation}
\bra{p'}\ket{p}=(2\pi)^3\delta^{(3)}(\vec{p}\, '-\vec{p})=(2\pi)^3\frac{1}{\hbar^3}\delta^{(3)}(\vec{\tilde p}\, '-\vec{\tilde p})\ .
\end{equation}
From \eqref{StressEnergyTensor_as_MatrixElement_Example} it is clear how $T_{\mu\nu}(q^2)$ can be expressed perturbatively in terms of Feynman diagrams which reconstruct the generic interaction vertex between two scalars with an emission of a graviton, as
\begin{equation}\label{Generic_Example_Process_chap3}
\includegraphics[width=0.3\textwidth, valign=c]{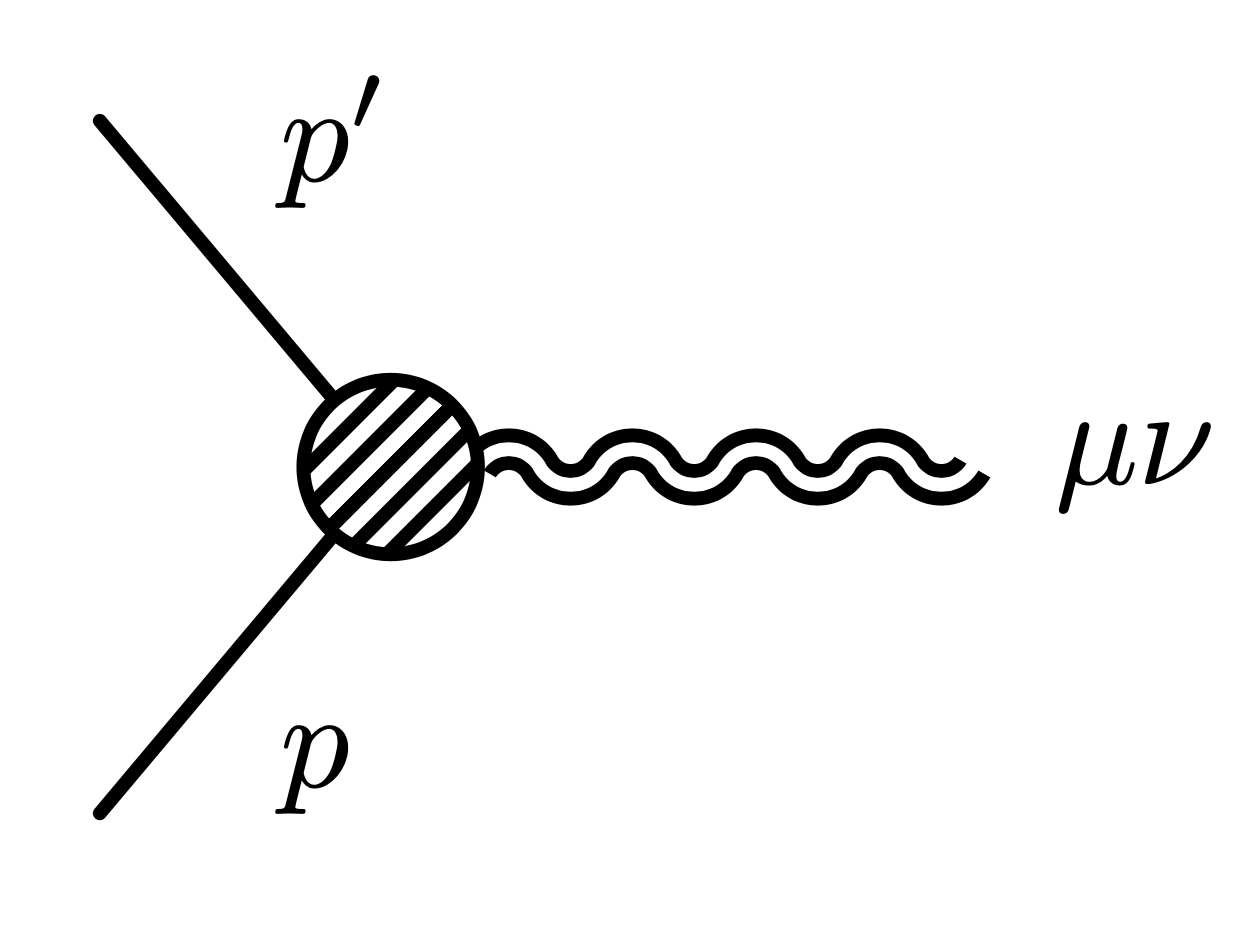}=-i\frac{1}{\hbar^3}\frac{\kappa}{2}\sqrt{4E \, E'}\, T_{\mu\nu}(q^2)\ ,
\end{equation}
where the energy-dependent factor is just a normalization choice with $E$ and $E'$ the energy of the external particles, and where the power of $\hbar$ is due to dimensional reasons. It is possible then to expand the stress-energy tensor in terms of form factors, and from the convention chosen the decomposition reads 
\begin{equation}\label{FormFactors_Example}
T_{\mu\nu}(q^2)=\frac{1}{\sqrt{4E \, E'}}\Biggl(2P_\mu P_\nu F_1(q^2)+(q_\mu q_\nu-\eta_{\mu\nu}q^2)F_2(q^2)\Biggl)\ ,
\end{equation}
with $P_\mu = \frac{1}{2}(p+p')_\mu$ and where the normalization condition imposes $F_1(0)=1$. It is easy to check the consistency of the above form factor expansion with the normalization given. Considering $\mathscr{P}^\mu$ as the momentum operator that acts on particle states and defined by the Noether current as
\begin{equation}
\mathscr{P}^\mu = \int d^3 x\,  T^{\mu 0}(x)\ ,
\end{equation} 
then it follows
\begin{equation}\label{Consistency_of_FormFactors}
\bra{p'}\mathscr{P}^\mu \ket{p}=p^\mu\bra{p'}\ket{p}=p^{\mu}(2 \pi )^3\delta^{(3)}(\vec{p}\, '-\vec{p})\ .
\end{equation}
On the other hand, it is possible to compute the same object considering
\begin{equation}
\bra{p'}\mathscr{P}^\mu \ket{p}=\int d^3 x 	\bra{p'}T^{\mu 0}(x)\ket{p}= (2\pi)^3 \delta^{(3)}(\vec{p}\, '-\vec{p})T^{\mu 0}(q^2=0)\ ,
\end{equation}
that using the normalization condition $F_1(0)=1$ leads to the same conclusion of \eqref{Consistency_of_FormFactors}.

\medskip

So then to compute the graviton field in eq. \eqref{Graviton_Example}, it is sufficient to evaluate the form factors defined in \eqref{FormFactors_Example}. The standard way to do that is to expand perturbatively the resummed vertex in \eqref{Generic_Example_Process_chap3}, and assign a value to $F_1$ and $F_2$ order by order in the coupling constant. However we are not interested in all the processes that reconstruct a given order in the quantum computation, but we only care about the ones which are not vanishing in the limit $\hbar\rightarrow 0$. Moreover, it is important to take into account that from the low-energy limit imposed by the effective approach, each diagram organizes itself in an energy expansion, and since the power of the momentum is related to the power of $\hbar$, we conclude that even if the leading contribution of a diagram contributes classically to an observable, its next-to-leading-order does not \cite{Donoghue:1994dn}. Making this clear, in the limit in which the momentum of the external particles satisfies ${m}^2\gg{\vec{p}}^2$ (imposed by staticity), the energy of the external particles reads
\begin{equation}
E=\sqrt{m^2-\vec{\tilde p}^2\hbar^2}=m\left(1-\frac{\vec{\tilde p}^2\hbar^2}{2m^2}+O(\vec{\tilde p}^4\hbar^4)\right)\ .
\end{equation}
Exploiting this analysis, in the following we will only consider the leading order, since terms $O(\vec{\tilde p}^2\hbar^2)$ contribute to quantum corrections. Finally considering the static limit in momentum space, which reads $q^0=0$, by replacing \eqref{FormFactors_Example} inside eq. \eqref{Graviton_Example} one gets
\begin{equation}\label{Metric_with_FormFactors_Example}
\kappa\, h_{\mu\nu}(\vec{x})=-16G_N \pi \int \frac{d^3\tilde q}{(2 \pi )^3}\frac{e^{i\vec{\tilde{q}}\cdot \vec{x}}}{\vec{\tilde{q}}^2} \frac{1}{2m}\Biggl(\left(\delta_\mu^0\delta_\nu^0-\frac{1}{2}\eta_{\mu\nu}\right)F_1(\vec{\tilde q}^2)+\frac{\tilde{q}_\mu \tilde{q}_\nu-\frac{1}{2}\eta_{\mu\nu}\vec{\tilde{q}}^2}{2m^2}\hbar^2F_2(\vec{\tilde q}^2)\Biggl)\ .
\end{equation}

\subsubsection*{Tree-level}

\begin{figure}[h]
\centering
\includegraphics[width=0.25\textwidth, valign=c]{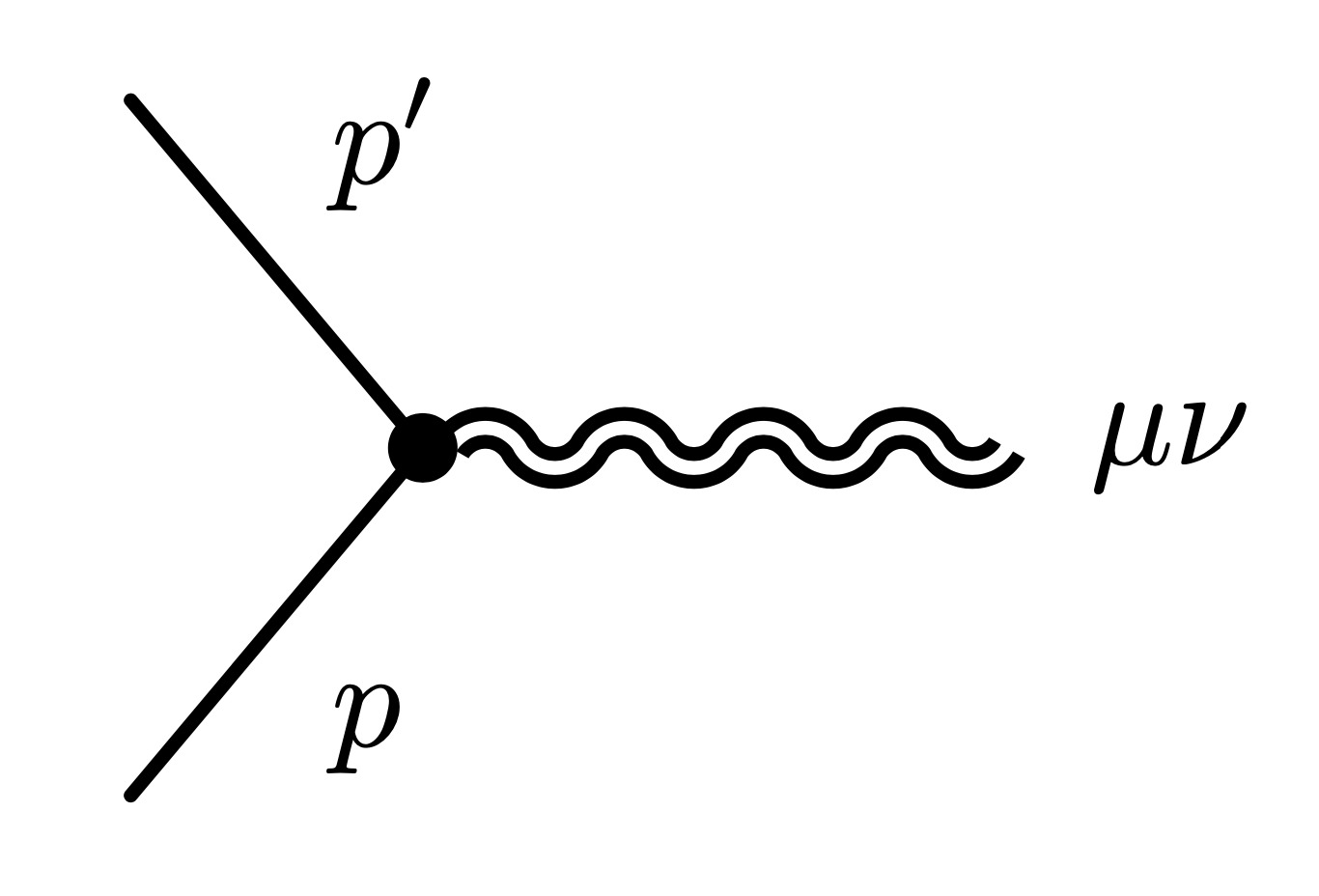} 
\caption{Tree-level of the 2 scalars - 1 graviton process.}
\label{Tree_Level_Example}
\end{figure}

The first order of the perturbative expansion is trivially the tree-level process, reported in figure \ref{Tree_Level_Example}.
From eq. \eqref{Generic_Example_Process_chap3} one obtains
\begin{equation}
-i\frac{1}{\hbar^3}\frac{\kappa}{2}\sqrt{4E\, E'}\, T^{tree}_{\mu\nu}(\tilde {q}^2)=(\tau_{\phi^2h})_{\mu\nu}(\tilde{p}, \tilde{p}')\ ,
\end{equation}
where $(\tau_{\phi^2h})_{\mu\nu}$ is the 2 scalars - 1 graviton vertex in eq. \eqref{2scalars_1graviton}. The leading order of the resulting stress-energy tensor, remembering that $(\tau_{massive})\propto \hbar^{-3}$, is then 
\begin{equation}\label{StressEnergy_TreeLevel_Example}
T^{tree}_{\mu\nu}(\vec{\tilde{q}}^2) = m\, \delta_{\mu}^0\delta_{\nu}^0+O(\hbar)\ ,
\end{equation}
which leads to the form factors
\begin{equation}
F_1^{tree}(\vec{\tilde{q}}^2)=1+O(\hbar) \quad \text{and} \quad F_2^{tree}(\vec{\tilde{q}}^2)=O(\hbar)\ .
\end{equation}
From prior considerations, since \eqref{StressEnergy_TreeLevel_Example} is classical, we can conclude that the metric contribution that comes from the leading-order of the tree-level amplitude is classical, and will be non-vanishing in the limit $\hbar\rightarrow 0$ since it is independent of the Planck constant.  
Replacing the form factors inside eq. \eqref{Metric_with_FormFactors_Example} and using the Fourier transforms in appendix \ref{App:masterintegral}, in particular eq. \eqref{D4_Identity1}, one finally obtains
\begin{equation}\label{Tree_Level_Chap3_Calc}
\begin{aligned}
\kappa \, h_{00}^{tree} &= -2\frac{G_N\, m}{r}+O(\hbar)\\
\kappa \, h_{i0}^{tree} &= O(\hbar)\\
\kappa \, h_{ij}^{tree} &= -2\frac{G_N\, m}{r}\, \delta_{ij}+O(\hbar)\ ,
\end{aligned}
\end{equation}
where $r=\sqrt{x^ix^i}$. 

\subsubsection*{1-loop}

\begin{figure}[h]
\centering
\includegraphics[width=0.48\textwidth, valign=c]{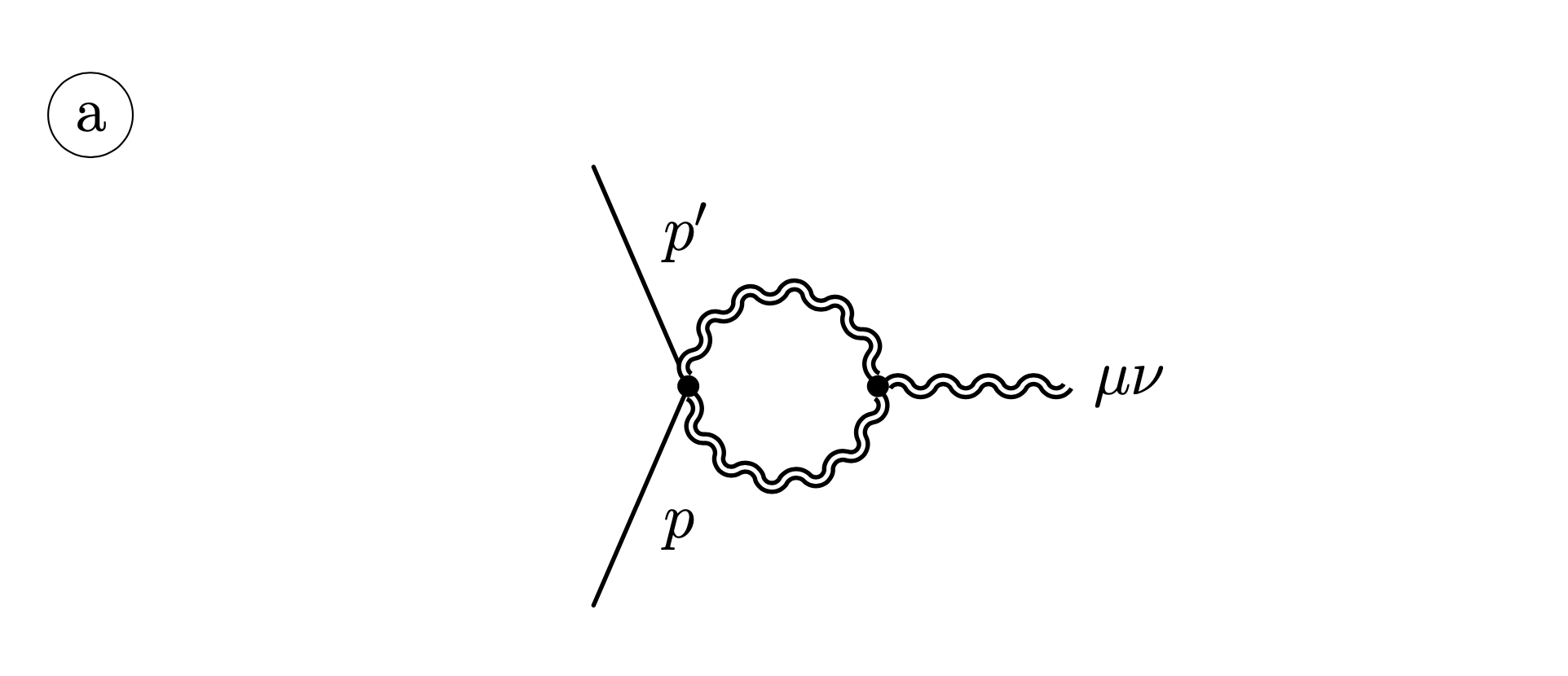} 
\includegraphics[width=0.48\textwidth, valign=c]{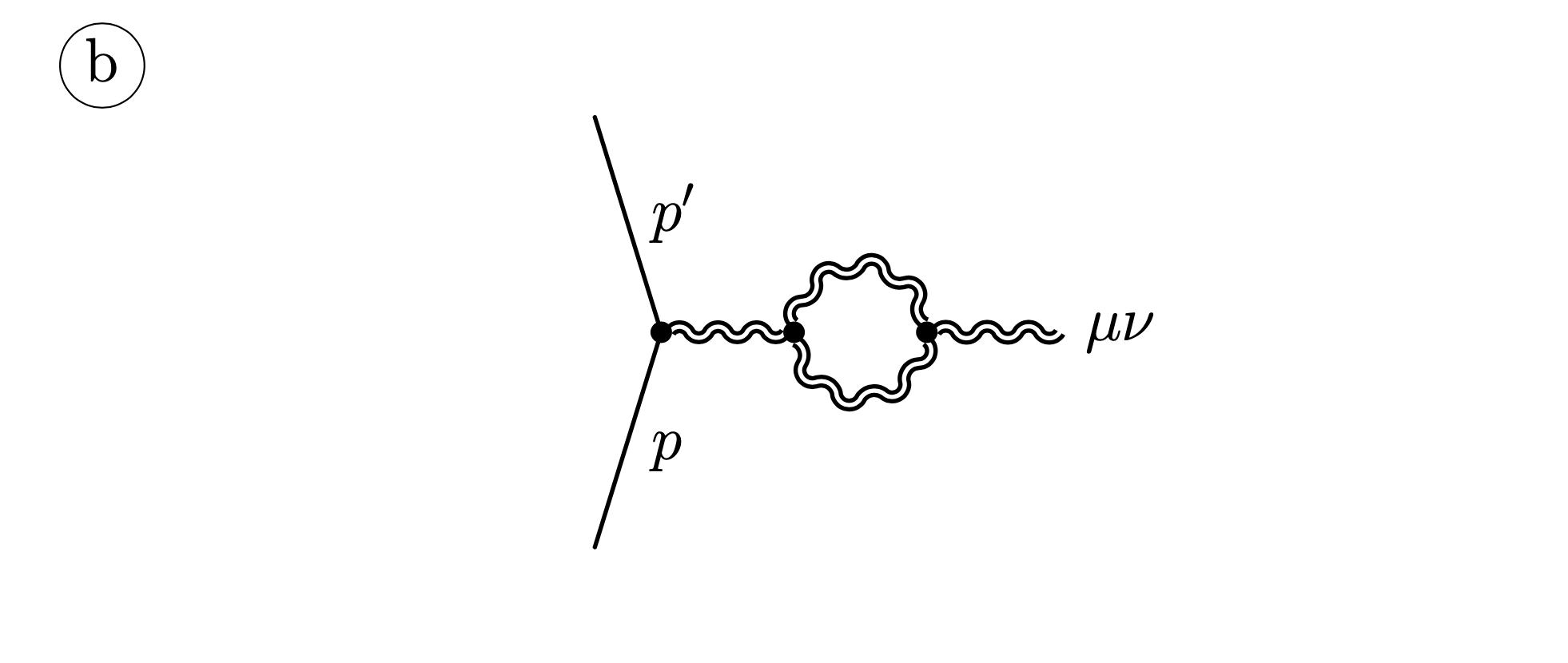} 
\includegraphics[width=0.48\textwidth, valign=c]{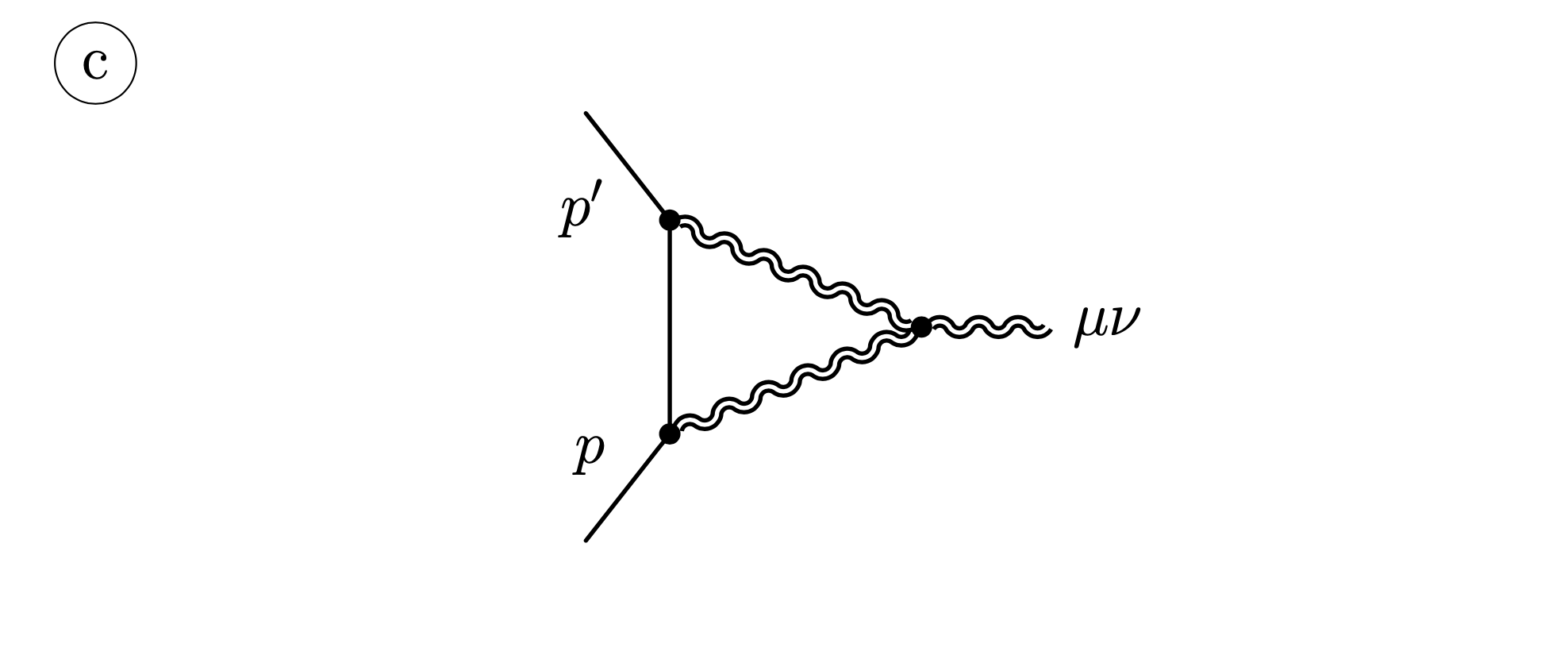} 
\includegraphics[width=0.48\textwidth, valign=c]{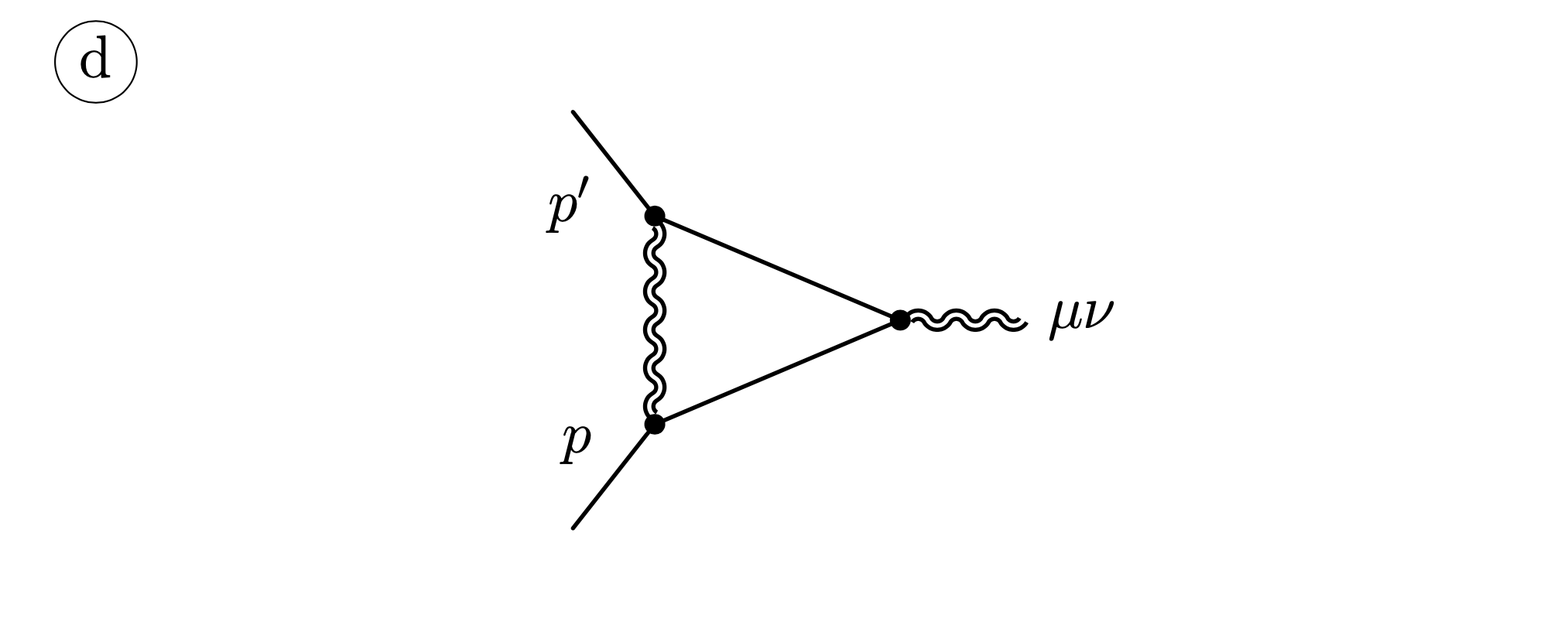} 
\caption{Some of the processes that contribute to the 1-loop correction of the 2 scalars - 1 graviton vertex.}
\label{1Loop_Example}
\end{figure}

For the 1-loop case the situation is more tricky. Since there exist many 1-loop diagrams it is not known a priori which process has a classical leading behavior. Some of the diagrams that contribute to the process are reported in figure \ref{1Loop_Example}.
Now it is not necessary to compute every single process and see at the end whether or not they are classical, in fact exploiting the knowledge of how vertices and propagators depend explicitly by $\hbar$, it can be understood immediately which diagram has a purely quantum leading behavior. For what concerns diagram (a), the behavior of the stress-energy tensor is
\begin{equation}
\frac{1}{\hbar^3}T_{\mu\nu}^{(\text{a}) 1-loop}\propto \frac{\hbar \hbar}{\hbar^3 \hbar} \Rightarrow T_{\mu\nu}^{(\text{a}) 1-loop} \propto \hbar\ ,
\end{equation}
which shows that such contribution will vanish in the classical limit, and therefore it is not needed if we are interested only in classical phenomenology. For diagram (b) the same procedure reads 
\begin{equation}
\frac{1}{\hbar^3}T_{\mu\nu}^{(\text{b}) 1-loop}\propto \frac{\hbar \hbar \hbar}{\hbar^3 \hbar \hbar} \Rightarrow T_{\mu\nu}^{(\text{b}) 1-loop} \propto \hbar\ ,
\end{equation}
which lead to the same conclusions. The process (c) instead behaves like 
\begin{equation}
\frac{1}{\hbar^3}T_{\mu\nu}^{(\text{c}) 1-loop}\propto \frac{\hbar^2 \hbar \hbar}{\hbar^3 \hbar^3 \hbar} \Rightarrow T_{\mu\nu}^{(\text{c}) 1-loop} \propto \hbar^0\ ,
\end{equation}
which means that the leading contribution is entirely classical and it is needed for reconstructing classical phenomenology. 
Different analysis for the diagram (d), which since turns out to be independent from the transferred momentum $\vec{\tilde q}$, gives rise to a local contribution \cite{Donoghue:1994dn, Bjerrum-Bohr:2002fji}. So then this argument shows that among the processes in figure \ref{1Loop_Example}, only the type (c) contributes to the classical limit of the observable, and this result can be generalized to conclude that every other diagram gives either an $\hbar$ correction or a local one, including diagrams that involve ghosts. 

\medskip

This explicit example outlines the idea behind the formulation of selection rules of Feynman diagrams, which only pick the contributions needed for classical phenomenology. So then, we can state that:
\begin{itemize}
\item The low-energy limit imposes that the only processes that contribute to the classical behavior are the ones in which internal massive lines do not mediate the interaction, meaning that such intermediate states cannot propagate the external momentum to the final states. This idea is explained in details in chapter \ref{sec:Classical_Limit_Grav_Emission_Processes}.
\item The classical limit imposes that a classical gravitational observable is recovered only from those diagrams which have an internal tree structure.
\end{itemize}
Therefore the selection rules lead to the conclusion that every classical contribution to the metric is made of $l$-loop amplitudes whose structure is composed by a single massive line which emits $l+1$ gravitons that interact in a tree-level structure ending up with a graviton emission. 
One of the most important outcomes of this analysis is that while classical electromagnetism is recovered only by tree-level processes, general relativity is reconstructed with multi-loop diagrams, from which we interpret the non-linearity of gravity as self-interaction of the gravitons even at the classical level. In fact we proved that the naive statement that says that a loop expansion is synonymous of an $\hbar$ expansion is false, and that gravity is classically reconstructed by quantum processes like loop amplitudes, which is something not deeply understood yet. 

\medskip

The actual calculation for the 1-loop contribution to the metric then follows the same steps as before. Considering then the low-energy limit of the amplitude (c), an explicit calculation leads to the form factors \cite{Donoghue:1994dn}
\begin{equation}
\begin{aligned}
F_1^{1-loop}(\vec{\tilde q}^2)=\frac{1}{16}\frac{G_N\pi m\, \tilde{q}^2}{\sqrt{-\tilde{q}^2}}+O(\hbar)=-\frac{1}{16}G_N\pi \, m\, |\vec{\tilde q}|+O(\hbar) \\ 
F_2^{1-loop}(\vec{\tilde{q}}^2)=\frac{7}{8}\frac{G_N \pi m^3}{\hbar^2\sqrt{-\tilde{q}^2}}+O(1/\hbar)=\frac{7}{8}\frac{G_N\pi m^3}{\hbar^2|\vec{\tilde q}|}+O(1/\hbar)\ ,
\end{aligned}
\end{equation}
where we used $\tilde{q}^2=-\vec{\tilde q}^2$.
Replacing the above coefficients inside eq. \eqref{Metric_with_FormFactors_Example} and using the relations \eqref{D4_Identity2} and \eqref{D4_Identity3}, it is possible to extract the classical contributions of the metric perturbation as
\begin{equation}\label{Example_SchwarzschildPerturbation_deDonder_1loop}
\begin{aligned}
\kappa \, h_{00}^{1-loop} &=2\frac{G_N^2 m^2}{r^2}+O(\hbar) \\
\kappa \, h_{i0}^{1-loop} &= O(\hbar)\\
\kappa \, h_{ij}^{1-loop} &= -\frac{G_N^2m^2}{r^2}\left(5\delta_{ij}-7\frac{x_i x_j}{r^2}\right)+O(\hbar)\ .
\end{aligned}
\end{equation}

\medskip

So at the very end, considering that the metric is reconstructed by 
\begin{equation}
g_{\mu\nu}=\eta_{\mu\nu}+\kappa\, h_{\mu\nu}=\eta_{\mu\nu}+\kappa\Bigl(h_{\mu\nu}^{tree}+h_{\mu\nu}^{1-loop}+\cdots)\ ,
\end{equation}
picking only the classical contributions one gets the Schwarzschild metric in the de Donder gauge as
\begin{equation}\label{Example_SchwarzschildMetric_deDonder_from_Amplitudes}
\begin{aligned}
g_{00} &=1-\frac{2G_N\, m}{r}+\frac{2G_N^2\, m^2}{r^2}+O(1/r^3) \\
g_{i0} &= 0\\
g_{ij} &= -\delta_{ij}-\delta_{ij}\frac{2G_N\, m}{r}-\frac{G_N^2m^2}{r^2}\left(5\delta_{ij}-7\frac{x_i x_j}{r^2}\right)+O(1/r^3)\ .
\end{aligned}
\end{equation}
Eq. \eqref{Example_SchwarzschildMetric_deDonder_from_Amplitudes} shows a very important result: the classical observable is reconstructed in terms of a post-Minkowskian expansion, which is related to the loop perturbative series of the quantum computation. In fact we can infer that an $l$-loop amplitude contributes to the $l+1$-post-Minkowskian order of the observable, which goes like $1/r^{l+1}$. In this expansion the first order is recovered from the tree-level stress-energy tensor that corresponds to what we called $T^{\mu\nu}_{\phi}$, which is zeroth-order in the post-Minkowskian series, while the multi-loop diagrams reconstruct the self-interaction stress-energy tensor $t^{\mu\nu}$.
It is important to notice that in this case the powers of the radial coordinate and the gravitational coupling constant are the same, but this is just an artifact of the fact that our theory has one coupling. 

\section{Schwarzschild metric in harmonic gauge}
\label{sec:Schwarzschild_Harmonic}

In the previous section we recovered the second post-Minkowskian order of the Schwarzschild metric in the de Donder gauge performing the classical limit of quantum gravity scattering amplitudes. Now we want to compare the solution in eq. \eqref{Example_SchwarzschildMetric_deDonder_from_Amplitudes} with the classical derivation of it and prove the consistency of the strategy outlined before. The crucial step is to take into account the gauge in which the metric is computed, in fact while in the quantum context the de Donder gauge is very useful, the classical space-time solutions in literature are often given in spherical coordinates or harmonic ones. It turns out that the de Donder gauge and the harmonic one are very similar, and it is possible to link them and convert eq. \eqref{Example_SchwarzschildMetric_deDonder_from_Amplitudes} in harmonic coordinates, and finally compare the derivation from scattering amplitudes with the completely classical one. 

\medskip

The Schwarzschild metric in spherical coordinates $(t, r, \theta, \varphi)$ has the form 
\begin{equation}
ds^2= \left(1-\frac{2G_N m}{r} \right)dt^2-\left(1-\frac{2G_N m}{r} \right)^{-1}dr^2-r^2d\Omega_{2}\ ,
\end{equation}
where $d\Omega_{2}=d\theta^2+\sin(\theta)^2d\varphi^2$ is the 2-sphere metric and $m$ is the mass of the black hole. It is possible to construct another set of coordinates $\left(t, \vec{x}\right)$, called harmonic, in which following \cite{Weinberg:1972kfs} the metric reads 
\begin{equation}\label{Example_Harmonic_Schwarzschild_notCompact}
ds^2=\left(\frac{r-G_Nm}{r+G_Nm}\right)dt^2-\left(1+\frac{G_Nm}{r}\right)^2d\vec{x}^2-\left(\frac{r+G_Nm}{r-G_Nm}\right)\frac{G_N^2m^2}{r^4}\left(\vec{x}\cdot d\vec{x}\right)\ ,
\end{equation}
where $r^2=\vec{x}\cdot\vec{x}$ and where the metric satisfies the relation 
\begin{equation}\label{Harmonic_gauge_Condition}
g^{\mu\nu}\Gamma_{\mu\nu}^{\alpha}=0\ ,
\end{equation}
which imposes the harmonic gauge. Finally, the metric can be expanded in a post-Minkowskian series component by component, which reads
\begin{equation}\label{Example_Harmonic_Schwarzschild_Expanded}
\begin{aligned}
g_{00}=&\frac{\displaystyle 1-\frac{G_Nm}{r}}{\displaystyle 1+\frac{G_Nm}{r}}=1-2\frac{G_Nm}{r}+2\frac{G_N^2m^2}{r^2}+O(1/r^3)\\
g_{i0}=&0\\
g_{ij}=&-\left(1+\frac{G_Nm}{r}\right)^2\delta_{ij}-\left(\frac{\displaystyle 1+\frac{G_Nm}{r}}{\displaystyle 1-\frac{G_Nm}{r}}\right)\frac{G_N^2m^2}{r^4}x_i x_j\\
&=-\delta_{ij}\left(1+2\frac{G_Nm}{r}+\frac{G_N^2m^2}{r^2}\right)-\frac{x_i x_j}{r^2}\frac{G_N^2m^2}{r^2}+O(1/r^3)\ .
\end{aligned}
\end{equation}

\medskip

Now in order to do the comparison we are looking for, we need to rewrite eq. \eqref{Example_SchwarzschildMetric_deDonder_from_Amplitudes} in the harmonic gauge, and compare the result with the expression \eqref{Example_Harmonic_Schwarzschild_Expanded}. In the last section we saw that a perturbative computation of the metric in terms of scattering amplitudes organizes itself in a post-Minkowskian series. Therefore it is useful to decouple the Einstein equations from the beginning as
\begin{equation}\label{Einstein_eq_NoGauge_PMexpansion_tree}
G_{\mu\nu}^{tree}=\frac{\kappa^2}{4}T_{\mu\nu}^{tree}
\end{equation}
\begin{equation}\label{Einstein_eq_NoGauge_PMexpansion_1loop}
G_{\mu\nu}^{1-loop}=\frac{\kappa^2}{4}T_{\mu\nu}^{1-loop}\ .
\end{equation}
Then imposing the metric decomposition in terms of a post-Minkowskian expansion as well
\begin{equation}
g_{\mu\nu}=\eta_{\mu\nu}+\kappa h_{\mu\nu}^{tree}+\kappa h_{\mu\nu}^{1-loop}+\cdots\ ,
\end{equation}
the harmonic gauge condition, which is non-linear in the graviton field, reads
\begin{equation}
\begin{aligned}
0=g^{\mu\nu}\Gamma_{\mu\nu}^\alpha=&\kappa \Bigl(\de_\beta h^{tree\, \alpha \beta}-\frac{1}{2}\de^\alpha h^{tree}\Bigl)+\kappa\Bigl(\de_\beta h^{1-loop\, \alpha \beta}-\frac{1}{2}\de^\alpha h^{1-loop}\Bigl)\\
&-\kappa^2h^{tree\, \alpha\beta}\Bigl(\de^\mu h^{tree}_{\beta\mu}-\frac{1}{2}\de_\beta h^{tree}\Bigl) + \cdots\ .
\end{aligned}
\end{equation}
The above equation has to be satisfied order by order, and then for the first two post-Minkowskian contributions it is verified
\begin{equation}
\begin{gathered}
\kappa \Bigl(\de_\beta h^{tree\, \alpha \beta}-\frac{1}{2}\de^\alpha h^{tree}\Bigl)=0\\
\kappa\Bigl(\de_\beta h^{1-loop\, \alpha \beta}-\frac{1}{2}\de^\alpha h^{1-loop}\Bigl)-\kappa^2h^{tree\, \alpha\beta}\Bigl(\de^\mu h^{tree}_{\beta\mu}-\frac{1}{2}\de_\beta h^{tree}\Bigl)=0\ .
\end{gathered}
\end{equation}
Now we have to replace these conditions in both \eqref{Einstein_eq_NoGauge_PMexpansion_tree} and \eqref{Einstein_eq_NoGauge_PMexpansion_1loop} and obtain the analogue of eq. \eqref{Box_Graviton_Example} in the harmonic case. The tree-level harmonic gauge condition corresponds exactly to the de Donder gauge\footnote{The fact that the leading order of the harmonic gauge corresponds to the de Donder gauge, means that the graviton propagator remains unchanged since the correction quadratic in the first derivative of the field is the same.}, therefore the field equations for the graviton field at first post-Minkowskian order coincide with eq. \eqref{Box_Graviton_Example} and the metric perturbation remains unchanged in this gauge
\begin{equation}
\begin{aligned}
\kappa \, h_{00}^{tree} &= -2\frac{G_N\, m}{r}\\
\kappa \, h_{i0}^{tree} &= 0\\
\kappa \, h_{ij}^{tree} &= -2\frac{G_N\, m}{r}\, \delta_{ij}\ ,
\end{aligned}
\end{equation}
in fact we can see that the expressions above perfectly coincide with the metric in \eqref{Example_Harmonic_Schwarzschild_Expanded} at first post-Minkowskian order. 

\medskip

At the second post-Minkowskian order the two gauges differ, and we expect the field equations to be modified, in fact they read
\begin{equation}\label{Example_1loop_BoxEquation_Expanded}
\begin{aligned}
\Box\left(h_{\mu \nu}^{1-loop}-\frac{1}{2} \eta_{\mu \nu} h^{1-loop}\right)=&\frac{\kappa}{4} T_{\mu \nu}^{1-loop} +\partial_{\mu}\left(h^{tree\, \lambda \sigma}\left(\partial_{\lambda} h_{\sigma \nu}^{tree}-\frac{1}{2} \partial_{\nu} h_{\lambda \sigma}^{tree}\right)\right)\\
&+\partial_{\nu}\left(h^{tree\, \lambda \sigma}\left(\partial_{\lambda} h_{\sigma \mu}^{tree}-\frac{1}{2} \partial_{\mu} h_{\lambda \sigma}^{tree}\right)\right) \\
&-\eta_{\mu \nu} \partial^{\alpha}\left(h^{tree\, \lambda \sigma}\left(\partial_{\lambda} h_{\alpha \sigma}^{tree}-\frac{1}{2} \partial_{\alpha} h_{\lambda \sigma}^{tree}\right)\right)\ .
\end{aligned}
\end{equation}
Now we know the explicit solution of $h^{tree}_{\mu\nu}$ that it is needed to be replaced inside the above equation, which can be compactly rewritten as
\begin{equation}
h_{\mu\nu}^{tree}=f(r)\delta_{\mu\nu}\ ,
\end{equation}
where 
\begin{equation}
f(r)=-2\frac{G_N m}{r}\ ,
\end{equation}
and finally eq. \eqref{Example_1loop_BoxEquation_Expanded} can be traced and expressed as
\begin{equation}\label{Example_1loop_BoxEquation_Traced}
\Box h_{\mu\nu}^{1-loop}=-\frac{\kappa}{2}\left(T_{\mu\nu}^{1-loop}-\frac{1}{2}\eta_{\mu\nu}T^{1-loop}\right)-\de_\mu\Bigl(f(r)\de_\nu f(r)\Bigl)-\de_\nu\Bigl(f(r)\de_\mu f(r)\Bigl)\ .
\end{equation} 
Since it holds the static limit, which implies that the temporal derivative is vanishing both on the metric and on the function $f(r)$, using the relation 
\begin{multline}
-\de_\mu\Bigl(f(r)\de_\nu f(r)\Bigl)-\de_\nu\Bigl(f(r)\de_\mu f(r)\Bigl)\\
=\de_i\Bigl(f(r)\de_j f(r)\Bigl)+\de_j\Bigl(f(r)\de_i f(r)\Bigl)=4G_N^2m^2\nabla^2\left(\frac{\delta_{ij}}{r^2}-2\frac{x_ix_j}{r^4}\right)\ ,
\end{multline}
where $\Box=\de_\mu\de_\nu\eta^{\mu\nu}=\de_t^2-\nabla^2$, the field equations in \eqref{Example_1loop_BoxEquation_Traced} becomes
\begin{equation}
\begin{aligned}
\nabla^2 h_{00}^{1-loop}=&\frac{\kappa}{2}\left(T_{00}^{1-loop}-\frac{1}{2}T^{1-loop}\right)\\
\nabla^2 h_{i0}^{1-loop}=&0\\
\nabla^2 h_{ij}^{1-loop}=&\frac{\kappa}{2}\left(T_{ij}^{1-loop}+\frac{1}{2}\delta_{\mu\nu}T^{1-loop}\right)+\nabla^2 \left(4\frac{G_N^2m^2}{r^2}\left(\delta_{ij}-2\frac{x_ix_j}{r^2}\right)\right)\ .
\end{aligned}
\end{equation}
Comparing the above equations with the analogous expression in the de Donder gauge in eq. \eqref{Graviton_Example}, one concludes that in order to move to the harmonic gauge, it is sufficient to add the extra term to the de Donder gauge expression in eq. \eqref{Example_SchwarzschildPerturbation_deDonder_1loop}, obtaining 
\begin{equation}
\begin{aligned}
\kappa \, h_{00}^{1-loop} =&2\frac{G_N^2 m^2}{r^2}\\
\kappa \, h_{i0}^{1-loop} =& 0\\
\kappa \, h_{ij}^{1-loop} =& -\frac{G_N^2m^2}{r^2}\left(5\delta_{ij}-7\frac{x_i x_j}{r^2}\right)+4\frac{G_N^2m^2}{r^2}\left(\delta_{ij}-2\frac{x_ix_j}{r^2}\right)\\
&=-\frac{G_N^2m^2}{r^2}\left(\delta_{ij}+\frac{x_i x_j}{r^2}\right)\ ,
\end{aligned}
\end{equation}
which at this order matches exactly the Schwarzschild metric in harmonic gauge in eq. \eqref{Example_Harmonic_Schwarzschild_Expanded}. 

\medskip

In the end we proved that the approach of section \ref{sec:Schwarzschild_Metric_from_Graviton_Emission}, with the right gauge choice, is perfectly coherent and leads to the expected classical metric. This analysis also underlines how the de Donder gauge is much more suitable for this approach, since the procedure outlined to convert the de Donder gauge into the harmonic one can be very tedious at higher orders. To avoid this problem, further in this thesis the harmonic gauge will be no longer used, and the quantum computation will be compared directly with the observables in the de Donder gauge, whose procedure to impose it is shown in chapter \ref{chap:RNT_Classical_Solution}. 

\section{Reissner-Nordstr\"om metric from graviton emission}

In the previous two sections we recovered the Schwarzschild metric from graviton emission amplitudes in which the gravitational field interacts exclusively with discharged scalars. However there exist classical cases in which gravity is coupled to the electromagnetic field, and for a static and spherically symmetrically mass distribution, such interaction gives rise to the Reissner-Nordstr\"om metric \cite{Reissner, Nordstrom}, which in spherical coordinates reads
\begin{equation}
ds^2=\left(1-\frac{2G_N m}{r}+\frac{Q^2 G_N}{4 \pi \, r^2}\right)dt^2-\left(1-\frac{2G_N m}{r}+\frac{Q^2 G_N}{4 \pi \, r^2}\right)^{-1}dr^2-r^2d\Omega_2\ ,
\end{equation}
where $Q$ is the electric charge of the black hole. As we did for the Schwarzschild case, it is possible to express the above metric in harmonic gauge, and the post-Minkowskian expansion, truncated at the first order in $G_N$, leads to 
\begin{equation}\label{Classical_RN_Metric}
\begin{aligned}
g_{00}&=1-\frac{2G_N m}{r}+\frac{Q^2G_N}{4\pi\, r^2}+O(G_N^2)\\
g_{i0}&=0\\
g_{ij}&=-\delta_{ij}-\delta_{ij}\frac{2G_N m}{r}+\frac{Q^2G_N}{4\pi \, r^2}\frac{x_ix_j}{r^2}+O(G_N^2)\ .
\end{aligned}
\end{equation}
As we anticipated earlier, in this thesis we define the post-Minkowskian expansion as an expansion in $O(1/r)$,  which due to the presence of the charge does not coincide with an expansion in $G_N$, conversely to what happened in the Schwarzschild case. This means that the expansion in \eqref{Classical_RN_Metric} is missing some pieces in order to be second order in the post-Minkowskian expansion, and those terms are exactly the ones already encountered in the Schwarzschild case and computed in the harmonic gauge.

\medskip

\begin{figure}[h]
\centering
\includegraphics[width=0.35\textwidth, valign=c]{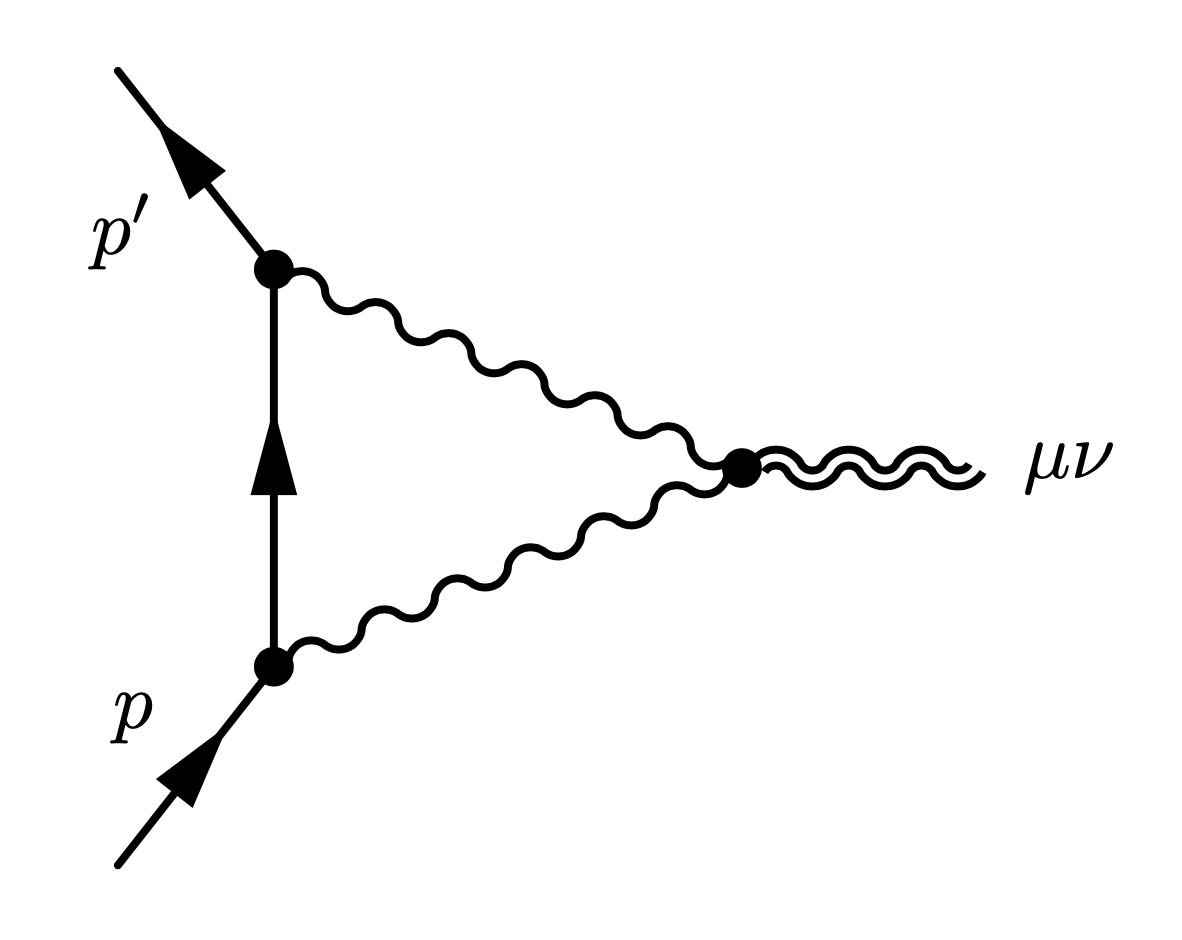} 
\caption{1-loop correction of the 2 scalars - 1 graviton vertex with photons circulating in the loop.}
\label{img:RN_Example}
\end{figure}

Now following the same steps outlined in section \ref{sec:Schwarzschild_Metric_from_Graviton_Emission}, it is possible to recover such metric from scattering amplitudes in which both photons and gravitons circulate in the loops \cite{Donoghue:1994dn}. In fact the theory associated to the present system is exactly the one studied in chapter \ref{QG_as_EFT}, and since the long-range limit imposes that the interaction is mediated by massless particles, one has to take into account all the massless particles present in the theory. This means that the Reissner-Nordstr\"om metric is recovered from eq. \eqref{Metric_with_FormFactors_Example} as well, where to the 1-loop correction of the vertex in  eq. \eqref{Generic_Example_Process_chap3} contributes also the amplitude in figure \ref{img:RN_Example}, in which two photons are emitted from the massive line and end up interacting in a 2 photons - 1 graviton vertex. Notice that here and in the following the single wiggly lines stand for photons. 
From the selection rules previously introduced, it is possible to show that up to this loop order, this process is the only one that contributes to reconstruct the term of the metric proportional to $Q^2$. From an explicit computation it is possible to extract the classical contributions of the form factors associated to such process \cite{Donoghue:2001qc}, which explicitly read
\begin{equation}\label{Example_RN_FormFactors}
\begin{aligned}
F_1^{RN}(\vec{\tilde q}^2)&=\frac{3}{64}\frac{Q^2\, \tilde{q}^2}{m\sqrt{-\tilde{q}^2}}+O(\hbar)=-\frac{3}{64}\frac{Q^2\, |\vec{\tilde q}|}{m}+O(\hbar)\\
F_2^{RN}(\vec{\tilde q}^2)&=\frac{1}{32}\frac{Q^2 m}{\hbar^2\sqrt{-\tilde {q}^2}}+O(1/\hbar)=\frac{1}{32}\frac{Q^2 m}{\hbar^2|\vec{	\tilde q}|}+O(1/\hbar)\ .
\end{aligned}
\end{equation}

\medskip

Replacing eq. \eqref{Example_RN_FormFactors} inside \eqref{Metric_with_FormFactors_Example} and using the usual Fourier transforms in appendix \ref{App:masterintegral}, the metric perturbation reads 
\begin{equation}\label{RN_Metric_Chap3}
\begin{aligned}
\kappa \, h_{00}^{RN}&=\frac{Q^2G_N}{4\pi r^2}+O(\hbar)\\
\kappa \, h_{i0}^{RN}&=0\\
\kappa \, h_{ij}^{RN}&=\frac{Q^2G_N}{4\pi r^2}\frac{x_ix_j}{r^2}+O(\hbar)\ ,
\end{aligned}
\end{equation}
which exactly corresponds to the classical metric in eq. \eqref{Classical_RN_Metric}. Interestingly in fact, in this case even if the classical calculation is done in the harmonic gauge and the quantum one in de Donder gauge, the two expressions coincide. This is traceable to the fact that the difference between the two gauges enters only in graviton self-interaction vertices, and since in the diagram in figure \ref{img:RN_Example} they are not present, the calculation is independent by this choice.

\medskip

In the end we can comment that the same analysis can be performed in order to obtain non-static space-time solutions, such as the Kerr or the Kerr-Newman ones. Considering in fact fermionic matter instead of scalar fields, it is possible to take into account the spinning dynamics of metric and describe rotating black holes. The explicit calculation in $D=4$ is performed in \cite{Bjerrum-Bohr:2002fji} where the Kerr metric is obtained up to 1-loop order from spinor-graviton interactions, while in \cite{Donoghue:2001qc} it is computed the lowest order in the electromagnetic coupling constant of the Kerr-Newman metric.

\chapter{Reissner-Nordstr\"om-Tangherlini solution in de Donder gauge}\label{chap:RNT_Classical_Solution}

The picture that has emerged so far, is that it is possible to compute the post-Minkowskian expansion of gravitational observables by exploiting the classical limit of scattering amplitudes in quantum gravity as an effective field theory. The rest of the thesis will aim to recover the Reissner–Nordstr\"om-Tangherlini solution from scattering amplitudes in which charged scalars, photons and gravitons interact with each other. 
In this chapter we will firstly review such solution, originally obtained by Tangherlini \cite{Tangherlini:1963bw} from the generalization to arbitrary dimension of the Reissner–Nordstr\"om metric, and then, following the procedure discussed in \cite{Mougiakakos:2020laz} applied to an electromagnetic coupled theory in \cite{Fragomeno, DOnofrio:2021tap}, a systematic procedure will be outlined in order to impose the de Donder gauge and express the solution in a post-Minkowskian expansion. During the derivation it is noticed how the expansion is affected by divergences in four and five space-time dimensions, and a procedure to eliminate them and obtain finite observables is carried out. 

\section{The coordinate change function}

The Reissner–Nordstr\"om-Tangherlini solution describes the metric and the electromagnetic potential of a static, spherically symmetric and electrically charged mass distribution in generic space-time dimension. 
The resulting metric in $D=d+1$ dimensions, obtained by solving the Einstein equations in spherical coordinates, has the expression 
\begin{equation}\label{Classical_Metric_FF}
ds^2 = \bigg( 1-\frac{\mu_m}{r^{d-2}}+\frac{\mu_Q}{r^{2(d-2)}} \bigg)dt^2 - \frac{dr^2}{1-\frac{\mu_m}{r^{d-2}}+\frac{\mu_Q}{r^{2(d-2)}}} - r^2 d\Omega_{d-1} \ ,
\end{equation}
where $\mu_m$ and $\mu_Q$ are respectively related to the mass and the electric charge of the black hole, by the relations 
\begin{equation} \label{mus}
    \mu_m = \frac{16 \pi G_N m }{(d-1)\Omega_{d-1}} \quad \text{and} \quad \mu_Q=\frac{8 \pi G_N Q^2}{(d-2)(d-1)\Omega_{d-1}^2} \ ,
\end{equation}
with $d\Omega_{d-1}$ the metric of the $(d-1)$-sphere, whose surface is
\begin{equation}
\Omega_{d-1}=\frac{2 \pi^{d/2}}{\Gamma(d/2)}\ ,
\end{equation}
while the electromagnetic potential of the mass distribution, obtained by solving the Maxwell equations in Lorentz gauge, has the form
\begin{equation}\label{Classical_EM_Potential_FF}
    A_\mu(r)=\delta_\mu^0 \frac{1}{(d-2)\Omega_{d-1}}\frac{Q}{r^{d-2}}\ .
\end{equation}

\medskip

Following the analysis outlined in \cite{Mougiakakos:2020laz}, we want to express such solution in the de Donder gauge, imposed by the condition 
\begin{equation}\label{deDonder_Gauge}
\eta^{\mu\nu}\Gamma^\alpha_{\mu\nu}=0\ .
\end{equation}
Given a generic metric of the form 
\begin{equation}
    ds^2 = C(r)dt^2 - \frac{dr^2}{C(r)} - r^2 d\Omega_{d-1} \ ,
\end{equation}
moving to cartesian coordinates $(t, \vec{x})$, in which $r^2=\vec{x}\cdot\vec{x}$ and considering the relations
\begin{equation}
dr=\frac{\vec{x} \cdot d\vec{x}}{r} \quad \text{and} \quad d\vec{x}^2=dr^2+r^2d\Omega_{d-1}\ ,
\end{equation}
its expression becomes
\begin{equation}\label{Generic_Metric_Cr}
     ds^2 = C(r)dt^2-d\vec{x}^2- \frac{1-C(r)}{C(r)} \frac{(\vec{x}\cdot d\vec{x})^2}{r^2} \ .
\end{equation} 
Inspired by the fact that the metric in \eqref{Generic_Metric_Cr} depends only on $r$, in order to find a reference frame that satisfies the condition \eqref{deDonder_Gauge}, we look for a transformation that involves only the radial coordinate, such as
\begin{equation}\label{Coordinate_Transformation_deDonder}
(t, \vec{x}) \rightarrow (t,f(r)\vec{x}) \quad \text{with} \quad r\rightarrow f(r)r \ ,
\end{equation}
where $f(r)$ is the coordinate change function that completely determines the transformation.
Then keeping trace of
\begin{equation}
d\vec{x}\rightarrow d\Big(f(r)\vec{x}\Big)=f(r)d\vec{x}+f'(r)\frac{\vec{x}\cdot d\vec{x}}{r}\, \vec{x}\ , 
\end{equation}
the metric in the new coordinates becomes
\begin{equation}\label{Metric_Cr_deDonder}
    ds^2 = h_0(r)dt^2-h_1(r)d\vec{x}^2- h_2(r) \frac{(\vec{x}\cdot d\vec{x})^2}{r^2} \ ,
\end{equation}
where $h_0 (r)$, $h_1 (r)$ and $h_2 (r)$ are determined in terms of $C(r)$ and $f(r)$ as 
\begin{equation}\label{Form_Metric_Cr_deDonder}
\begin{aligned}
   h_0(r) &= C(f(r)r) \\
 h_1(r) &= f(r)^2\\
 h_2(r) &= -f(r)^2+\frac{(f (r)+rf'(r))^2}{C(f(r)r)} \ . 
\end{aligned}
\end{equation}
Expanding the condition \eqref{deDonder_Gauge} in terms of the metric, it reads
\begin{equation}
-\de_jg_{00}-2\de_i g_{ij}+\de_jg_{ii}=0\ ,
\end{equation}
where we used the fact that the metric is time-independent. Imposing the above condition for 
\begin{equation}
\begin{aligned}
g_{00}&=h_0(r)\\
g_{i0}&=0\\
g_{ij}&=-h_1(r)\delta_{ij}-h_2(r)\frac{x_ix_j}{r^2}\ ,
\end{aligned}
\end{equation}
then the de Donder gauge condition leads to the equation 
\begin{equation}\label{Generic_eq_fr}
    r \frac{d}{dr} \Big( h_0(r) + (d-2) h_1(r) - h_2(r) \Big) = 2(d-1) h_2(r) \ ,
\end{equation}
by which replacing in it eq. \eqref{Form_Metric_Cr_deDonder}, it becomes a differential equation for $f(r)$.

\medskip

Since we are interested in the metric in \eqref{Classical_Metric_FF}, we want to solve eq. \eqref{Generic_eq_fr} for
\begin{equation}\label{Definition_Cr}
C(r) =   1-\frac{\mu_m}{r^{d-2}}+\frac{\mu_Q}{r^{2(d-2)}}  \ . 
\end{equation}
Following \cite{Mougiakakos:2020laz}, we define the parameter
\begin{equation}\label{Definition_rho}
    {\rho}(r) = \frac{\Gamma(\frac{d}{2}-1) \pi^{1-d/2}}{r^{d-2}} \ , 
\end{equation}
from which substituting eq. \eqref{Definition_Cr} in \eqref{Form_Metric_Cr_deDonder} one gets
\begin{equation}\label{hiRN}
\begin{aligned}
h_0(r) &=  1 - 4m G_N \frac{d-2}{d-1}\frac{\rho(r)}{f(r)^{d-2}} \bigg( 1-\frac{\alpha}{2m} \frac{\rho (r)}{f(r)^{d-2}}\bigg) \\
 h_1(r) &= f(r)^2\\
 h_2(r) &=-f(r)^2+\frac{\left(f(r)+(2-d)\rho (r) \frac{df
(r)}{d\rho}\right)^2}{ 1 - 4m G_N \frac{d-2}{d-1}\frac{\rho(r)}{f(r)^{d-2}} \left( 1-\frac{\alpha}{2m} \frac{\rho (r)}{f(r)^{d-2}}\right)} \ ,
\end{aligned}
\end{equation}
where we have introduced for convenience the fine structure constant $\alpha= \frac{Q^2}{4\pi}$ in natural units. Plugging this last expression into eq. \eqref{Generic_eq_fr} and rewriting it as an equation for the parameter $\rho$, using the relation 
\begin{equation}
r\frac{d}{dr}=(2-d)\rho(r) \frac{d}{d\rho(r)}
\end{equation}
we obtain 
\begin{equation}\label{Eq_for_fr}
(2-d) \rho(r)\frac{d}{d\rho(r)} \Big( h_0(r) + (d-2) h_1(r) - h_2(r) \Big) = 2(d-1) h_2(r) \ .
\end{equation}

\medskip

Now the idea is to solve this equation perturbatively for $f(r)$ by expanding it as a power series in $\rho$ 
\begin{equation}\label{fr_Expansion_withbees}
    f(r) = 1 + \sum_{n=1}^{\infty} \sum_{\substack{j=0\\j \ \text{even}}}^n b_{n, j}(d) \, (mG_N)^{n-j} (\alpha G_N)^{\frac{j}{2}}\,  \rho(r)^n \ , 
\end{equation}
which corresponds to its post-Minkowskian expansion\footnote{In this context we define the post-Minkowskian expansion as an expansion in $\rho (r)$.}. 
The structure of the above expression follows from dimensional reasons, in fact in natural units the dimension of the objects involved is
\begin{equation}
[\rho]=L^{2-d} \quad [G_N]=L^{d-1} \quad [\alpha]=L^{d-3} \quad [m]=L^{-1}\ ,
\end{equation}
which imply 
\begin{equation}\label{DImensional_Reasons}
[m G_N]^{n-j}[\alpha G_N]^{\frac{j}{2}}[\rho]^n=1\ ,
\end{equation}
and therefore this structure makes the coefficients $b_{n, j}(d)$ dimensionless and only dependent on the space-time dimension. Moreover for the sake of simplicity it is useful to compact eq. \eqref{fr_Expansion_withbees} in 
\begin{equation}\label{fr_Expansion}
    f(r) = 1 + \sum_{n=1}^{\infty} a_n(d) \,  \rho(r)^n \ , 
\end{equation}
in which we defined implicitly the coefficients
\begin{equation}
a_n(d)=\sum_{\substack{j=0\\j \ \text{even}}}^n b_{n, j}(d) \, (mG_N)^{n-j} (\alpha G_N)^{\frac{j}{2}}\ ,
\end{equation}
that are still functions of the space-time dimension, but clearly not dimensionless anymore. 
Finally replacing \eqref{fr_Expansion} inside eq. \eqref{Eq_for_fr} and solving the equation, up to third order in the post-Minkowskian expansion, the coordinate change function reads
\begin{equation} \label{fr_Solved}
    \begin{aligned}
    f(r) &=1+\frac{2 m G_N }{d-1}\rho-\frac{  \alpha G_N \left(d^2-4 d+3\right)+4 \left(d^2-4 d+5\right) m^2 G_N^2}{(d-4) (d-1)^2}\rho^2\\&+\frac{2 m G_N}{3 (d-1)^3 (d-4)(d-3)} \Big(\alpha  G_N\left(6 d^4-53 d^3+167 d^2-223 d+103\right) \\ &+2 \left(7 d^4-57 d^3+172 d^2-232 d+122\right) m^2 G_N^2\Big)\rho^3 +O\left(\rho^4\right) \ ,
    \end{aligned}
\end{equation}
where the explicit value of the coefficients can be easily read and the $r$-dependence of $\rho$ left understood from here on.

\medskip

Once the function $f(r)$ is known, it enables through the coordinate transformation \eqref{Coordinate_Transformation_deDonder} to determine the expression that any gravitational observable takes in the de Donder gauge. However in four and five space-time dimensions, that is when $d=3$ and $d=4$, eq. \eqref{fr_Solved} is affected by divergences. This leads to the conclusion that in these particular cases the ansatz in eq. \eqref{fr_Expansion} is no longer true and has to be changed, creating the need to properly define some adding terms inside $f(r)$ in order to cancel the divergences whenever they appear. This can be done according to the following strategy: every time there is a divergent coefficient we have to add at that order all the terms with a power of logarithm permitted by the order of the polynomial, considering a new coefficient in the expansion for each newly added term. In the end every coefficient will be fixed except for the ones associated to the divergent terms, which will remain free parameters \cite{Mougiakakos:2020laz}. 

\subsubsection*{$\bm{f(r)}$ in $\bm{d=3}$}

In the case of $d=3$, looking at eq. \eqref{fr_Solved} the only divergence that appears is at third post-Minkowksian order, which corresponds to $\rho^3$ order. Following the statement that we just gave, in the new ansatz the coordinate change function becomes 
\begin{equation}
f(r)\bigl|_{d=3} = 1+ a_1(3) \rho+ a_2(3) \rho^2 + (a_{3,2}(3) \log(\rho)^2 + a_{3,1}(3)\log(\rho) + a_3(3)) \rho^3 + O(\rho^4) \ .
\end{equation} 
Replacing it inside the differential equation \eqref{Eq_for_fr}, each coefficient of the expansion reads
\begin{equation}
\begin{gathered}
a_1(3)=G_Nm \qquad a_2(3) = 2G_N^2m^2\\
a_{3, 2}(3)=0 \qquad a_{3, 1}(3)=\frac{2}{3}\Big(2G_N^2 m \alpha-G_N^3m^3\Big) \qquad a_3(3)=\text{not fixed}\ .
\end{gathered}
\end{equation}
The fact that $a_3(3)$ is not fixed means that with this new ansatz the de Donder gauge condition is satisfied for any $a_3(3)$, introducing a redundant degree of freedom in the system. Physically this means that by changing the value of $a_3(3)$ we move to a different reference frame still preserving the de Donder condition. So then redefining the arbitrary constant in a convenient form, such that the argument of the logarithms become dimensionless, the function $f(r)\bigl|_{d=3}$ takes the expression 
\begin{equation}\label{fr_d3}
\begin{aligned}
 f(r)\bigl|_{d=3} & =1+m G_N  \rho+2  m^2G_N^2 \rho^2+\frac{2}{3} mG_N  \left(2 \alpha G_N \log \left(\frac{2 mG_N  \rho}{c}\right)\right.\\ & \left. - m^2G_N^2 \log \left(\frac{2 mG_N  \rho}{c}\right)\right) \rho^3+O\left(\rho^4\right) \ ,
 \end{aligned}
\end{equation}
where the arbitrary constant is redefined in terms of $c$.

\subsubsection*{$\bm{f(r)}$ in $\bm{d=4}$}

The same arguments hold for the $d=4$ case, in which looking at eq. \eqref{fr_Solved} we have divergences both at second and third post-Minkowskian order. Then in this case the function $f(r)$ takes the form 
\begin{equation}
\begin{aligned}
f(r)\bigl|_{d=4} =& 1+ a_1(4) \rho+ (a_{2,1}(4) \log(\rho) + a_2(4)) \rho^2 \\
&+ (a_{3,2}(4) \log(\rho)^2 + a_{3,1}(4)\log(\rho) + a_3(4)) \rho^3 + O(\rho^4) \ .
\end{aligned}
\end{equation}
Replacing it inside \eqref{Eq_for_fr} and solving the differential equation, the coefficients read
\begin{equation}
\begin{gathered}
a_1(4)=\frac{2}{3}G_Nm \qquad a_{2, 1}(4)=-\frac{1}{18}\bigl(20G_N^2m^2+3G_N\alpha\bigl) \qquad a_{2}(4)=\text{not fixed} \\
a_{3, 1}(4)=\frac{1}{9} G_N^2 m \left(3 \alpha +20 G_N m^2\right) \qquad a_{3, 2}(4)=0 \\ a_3(4)=\frac{2}{81} \left(-81 \, a_2(4) G_N m+16 G_N^3 m^3+33 \alpha  G_N^2 m\right)\ .
\end{gathered}
\end{equation}
Interestingly even if in this case the unfixed coefficients are two, namely $a_2(4)$ and $a_3(4)$, they are related to each other, which means that the free degree of freedom in the transformation is still one. Finally, conveniently manipulating the free constant similarly as before, in this case the coordinate change function reads 
\begin{equation} \label{frd=4}
\begin{aligned}
    f &(r)\bigl|_{d=4} = 1+\frac{2}{3} m G_N \rho+\frac{1}{18}  G_N \left(-20 \log \left(\frac{8 m G_N \rho}{3 c}\right) G_N m^2-3 \alpha  \log \left(\frac{8 m G_N \rho}{3 c}\right)\right) \rho^2\\&+\frac{1}{81} m G_N^2 \left(180 \log \left(\frac{8 m G_N \rho}{3 c}\right) G_N m^2+32 G_N m^2+66 \alpha +27 \alpha  \log \left(\frac{8 m G_N \rho}{3 c}\right)\right) \rho^3\\&+O\left(\rho^4\right) \ ,
\end{aligned}
\end{equation}
where again the arbitrary constant becomes $c$.

\medskip

The fact that in $d=3$ and $d=4$ the coordinate change function has logarithmic terms and it is defined up to an arbitrary constant is not a coincidence, and it is traceable to the properties of the gauge transformations. Consider an infinitesimal coordinate transformation $x^\alpha\rightarrow x^\alpha + \epsilon^\alpha(x)$, performed while the de Donder gauge is satisfied. If the condition 
\begin{equation}\label{Harmonic_Function}
\Box \epsilon^\alpha(x)=0
\end{equation}
is verified up to orders $O(\epsilon^2)$, then even the new reference frame satisfies the de Donder gauge \cite{GR_Gualtieri}. This means that the function $\epsilon^\alpha(x)$ parametrizes a redundancy inside the gauge. Explicitly, an harmonic function that satisfies eq. \eqref{Harmonic_Function} and does not involve time coordinates is 
\begin{equation}
\epsilon^i(\vec{x})=\beta(d)\frac{x^i}{r^d} \qquad  \epsilon^0(\vec{x})=0\ ,
\end{equation}
where $\beta(d)$ is a dimensional constant inserted in order to respect $[\epsilon^i(\vec{x})]=L$. In $d=3$ this last condition imposes that $[\beta(3)]=L^3$, whose using the coupling we have a disposal can only be satisfied if 
\begin{equation}
\beta(3)\thicksim (G_Nm)^3, \alpha G_N^3m\ .
\end{equation}
This means that in $d=3$ it is possible to define a redundant gauge transformation like 
\begin{equation}
\epsilon^i(\vec{x})\big|_{d=3}=\beta(3)\frac{x^i}{r^3}
\end{equation} 
inside the de Donder gauge, and this redundancy, that appear exactly at third post-Minkowskian order, is in fact reconstructed by the parameter $a_3(3)$ that we already discussed. Similarly, in $d=4$ in order to define such coordinate transformation it must be true that $[\beta(4)]=L^4$, which corresponds to 
\begin{equation}
\beta(4)\thicksim (G_Nm)^2, \alpha G_N\ .
\end{equation}
Again since the redundant transformation is possible, we expect a free parameter in the coordinate change function already at second post-Minkowskian order, which is exactly $a_2(4)$. 
Moreover this analysis shows that the fact that the free degree of freedom is just one, and the two unfixed coefficients are related, is not a coincidence but it is because the whole redundancy is controlled just by one parameter, namely $\beta(4)$. 

\medskip

Finally in $d\geq 5$ there is no combination of the couplings with an integer power that satisfy $[\beta(d)]=L^d$, therefore does not exist any harmonic function like the one in \eqref{Harmonic_Function}. This means that for $d\geq 5$ no redundant gauge transformation is possible, meaning that no free parameters can be present in the coordinate change function. This argument not only shows that we expected the divergences of $f(r)$ exactly where they are, but states also that besides $d=3$ and $d=4$ no divergences can occur \cite{Jakobsen:2020ksu}. 

\section{Metric in de Donder gauge}

Now that the coordinate transformation is completely worked out, to get a specific observable in the de Donder gauge it is sufficient to consider eq. \eqref{Coordinate_Transformation_deDonder} associated with the explicit form of $f(r)$, which we computed up to third post-Minkowskian order. In the metric case part of the work has already been done, and it is sufficient to plug eq. \eqref{fr_Solved} inside \eqref{hiRN}, which leads to
\begin{equation}\label{hid}
\begin{aligned}
h_0& (r)  =  1-\frac{4 (d-2) }{d-1}m G_N {\rho } + \left(\frac{2 (d-2)}{d-1} \alpha G_N + \frac{8 (d-2)^2  }{(d-1)^2} m^2 G_N^2\right) {{\rho }^2}
\\ &+\left(-\frac{4 (d-2)^2 (3 d-11)   }{(d-4) (d-1)^2} m \alpha G_N^2
+\frac{8 (7-3 d) (d-2)^3 }{(d-4) (d-1)^3} m^3  G_N^3\right){\rho }^3 +O( {\rho }^4)\\
h_1& (r) = 1+ \frac{4  }{d-1}m G_N {\rho }+\left(-\frac{2 (d-3) }{(d-4) (d-1)} \alpha G_N +
\frac{4 \left(-2 d^2+9 d-14\right) } {(d-4) (d-1)^2}m^2 G_N^2 \right) {\rho }^2
\\&+\left(\frac{8 (3 d^3-25 d^2+69 d-65)}{3 (d-4) (d-3) (d-1)^2} m\alpha G_N^2+\frac{8 \left(7 d^4-63 d^3+214 d^2-334 d+212\right)}{3 (d-4) (d-3) (d-1)^3}m^3 G_N^3 \right) \rho^3\\
& +O(\rho^4)\\
 h_2&(r) = \left( \frac{2 (d-2)^2 }{(d-4) (d-1)}\alpha G_N+\frac{4 (d-2)^2 (3 d-2)} {(d-4) (d-1)^2}m^2 G_N^2 \right)\rho^2\\& +\left(\frac{4 (d-2)^2  \left(-3 d^3+19   d^2-33  d+17 \right)}{(d-4) (d-3) (d-1)^3}m\alpha G_N^2  \right. \\& \left.+\frac{8 (d-2)^2 \left(-2 d^3+13 d^2-25 d+10\right) }{(d-4) (d-3) (d-1)^3}m^3 G_N^3 \right)\rho^3   +O(\rho^4) \ .
\end{aligned}
\end{equation}
The above expression has a structure that directly follows from eq. \eqref{fr_Expansion_withbees}, in which the coupling constants combination together with the associated power of the $\rho$ are dimensionless. 
As expected eq. \eqref{hid} is not well-defined in $d=3$ and $d=4$, whose cases need the proper form of the coordinate change function. In the $d=3$ case in fact, one need to consider the explicit form of $f(r)\big|_{d=3}$ given in eq. \eqref{fr_d3}, from which the metric reads
\begin{equation}\label{hi3}
\begin{aligned}
    h_0(r)\bigl|_{d=3} &= 1-\frac{2 m G_N}{r} +\left({\alpha G_N }+2 m^2 G_N^2 \right) \frac{1}{r^2}+\left(-2 m \alpha G_N^2 + 2 m^3 G_N^3 \right) \frac{1}{r^3} +O\left(\frac{1}{r^4}\right) \\
    h_1(r)\bigl|_{d=3} &=1+\frac{2 m G_N}{r}+\frac{5 m^2G_N^2 }{r^2}\\ & + \left(\frac{8}{3 } m \alpha G_N^2  \log \left(\frac{2 m G_N }{c r}\right) +\frac{4}{3} m^3 G_N^3 \left(3-\log \left(\frac{2 m G_N }{c r}\right)\right) \right)\frac{1}{r^3}+O\left(\frac{1}{r^4}\right) \\
    h_2(r)\bigl|_{d=3} &= \left(  -\alpha  G_N -7 m^2 G_N^2\right)\frac{1}{r^2} +\left(\frac{2}{3} m\alpha G_N^2 \left(-12 \log \left(\frac{2 m G_N }{c r}\right)  -7  \right)\right.\\
    & \left.+\frac{2 }{3 } m^3 G_N^3\left(6 \log \left(\frac{2 m G_N }{c r}\right)-19\right)\right) \frac{1}{r^3} + O\left(\frac{1}{r^4}\right)  \ .
\end{aligned}
\end{equation}
As we can see from the above metric, in $d=3$ logarithmic terms arise and the free constant $c$ introduces a redundant gauge freedom in the system. 
Similarly in the $d=4$ case, taking into account \eqref{frd=4}, the outcome is 
\begin{equation}\label{hi4}
\begin{aligned}
   h_0& (r)\bigl|_{d=4} = 1-\frac{8 m G_N}{3 \pi  r^2}
   +\left(\frac{4 \alpha G_N}{3 \pi ^2 } +\frac{32 m^2 G_N^2}{9 \pi ^2 }\right)\frac{1}{r^4}+\left(-\frac{8 m \alpha G_N^2}{9 \pi ^3 } \left(\log \left(\frac{8 m G_N}{3 c \pi  r^2}\right)  +4 \right)\right. \\& \left. +\frac{32 m^3 G_N^3}{27 \pi ^3 } \left(-5 \log \left(\frac{8 m G_N}{3 c \pi  r^2}\right)-3\right)\right) \frac{1}{r^6}+O\left(\frac{1}{r^8}\right) \\
   h_1 & (r)\bigl|_{d=4} = 1+\frac{4 m G_N}{3 \pi  r^2}+ 
   \left(-\frac{\alpha G_N}{3 \pi ^2 } \log \left(\frac{8  m G_N}{3 c \pi  r^2}\right)+\frac{4 m^2 G_N^2}{9 \pi ^2} \left(1-5 \log \left(\frac{8 m G_N}{3 c \pi  r^2}\right)\right) \right) \frac{1}{r^4}\\&+ \left(\frac{4 m \alpha G_N^2 }{27 \pi ^3 }  \left(3 \log \left(\frac{8 m G_N}{3 c \pi  r^2}\right)  +11 \right)+\frac{16 m^3 G_N^3 }{81 \pi ^3}\left(15 \log \left(\frac{8 m G_N}{3 c \pi  r^2}\right)+4\right)\right)\frac{1}{r^6}+O\left(\frac{1}{r^8}\right) \\
   h_2& (r)\bigl|_{d=4} =\left( \frac{2 \alpha G_N}{3 \pi ^2 } \left(2   \log \left(\frac{8 m G_N}{3 c \pi  r^2}\right)-1 \right) +\frac{40 m^2 G_N^2}{9 \pi ^2 } \left(2 \log \left(\frac{8 m G_N}{3 c \pi  r^2}\right)+1\right)\right) \frac{1}{r^4}\\ & +\left(-\frac{8 m \alpha G_N^2}{9 \pi ^3 } \left(\log \left(\frac{8 m G_N}{3 c \pi  r^2}\right)  +13  \right)+\frac{32 m^3 G_N^3 }{27 \pi ^3 } \left(-5 \log \left(\frac{8 m G_N }{3 c \pi  r^2}\right)-4\right)\right)\frac{1}{r^6} \\
   & +O\left( \frac{1}{r^8}\right) \ ,
\end{aligned}
\end{equation}
where again logarithmic terms are present as well as the free parameter $c$.

\section{Electromagnetic potential in de Donder gauge}\label{sec:EM_in_ChapRNT}

Having determined the post-Minkowskian expansion of the metric in the de Donder gauge, we now proceed to compute the same expansion for the electromagnetic potential, reported in eq. \eqref{Classical_EM_Potential_FF} in cartesian coordinates. Considering the coordinate transformation \eqref{Coordinate_Transformation_deDonder}, the potential transforms according to 
\begin{equation}
A_\mu(r)\rightarrow A_\mu^{\text{dD}}(r)=\frac{\de x^\nu}{\de x'^\mu}A_\nu(f(r)r)\ , 
\end{equation}
where $x'^\mu$ and $x^\nu$ are respectively the new and the old coordinates. Since in the cartesian reference frame only the temporal component of the electromagnetic potential is non-vanishing, then 
\begin{equation}
A_\mu^{\text{dD}}(r)=\frac{\de x^0}{\de x'^\mu}A_0(f(r)r)\ , 
\end{equation}
but since all the $\vec{x}'=f(r)\vec{x}$ are time-independent, even for the transformed potential the temporal component is the only non-vanishing one, and the transformation simply reads 
\begin{equation}
A_\mu^{\text{dD}}(r)=\delta_{\mu}^0A_0(f(r)r)=\delta_\mu^0 \frac{1}{(d-2)\Omega_{d-1}}\frac{Q}{f(r)^{d-2}r^{d-2}}\ .
\end{equation}
Finally, analogously to the metric case we can express it in terms of the parameter $\rho$ as
\begin{equation}\label{EM_dD_rho}
A^{\text{dD}}_\mu(r)=\delta_\mu^0\frac{Q\, \rho}{4 \pi }\frac{1}{f(r)^{d-2}}\ .
\end{equation}

\medskip

Now in order to obtain an explicit expression it is sufficient to replace \eqref{fr_Solved} inside the above equation, which leads to 
\begin{equation}\label{EM_Potential_deDonder}
\begin{aligned}
    A^{\rm dD}_0(r)& =\frac{Q}{4\pi} \Biggl( \rho-\frac{2(d-2) }{(d-1)}m G_N \rho ^2 \\
     & +\left(\frac{2(d-2)^2(3d-7)}{(d-4)(d-1)^2} m^2 G_N^2  + \frac{(d-3) (d-2)}{(d-4)(d-1)}\alpha G_N\right)\rho^3\Biggl)+O(\rho^4)\ .
\end{aligned}
\end{equation}
In this case the third post-Minkowskian order of the potential is only divergent in $d=4$. This is just an artifact of the expression \eqref{EM_dD_rho}, in which it is clear that for the computation of the $O(\rho^3)$ terms of the potential we only need the second post-Minkowskian order of $f(r)$, which is well-defined in $d=3$. So then in five space-time dimensions the electromagnetic field reads
\begin{equation}\label{EM_Potential_deDonderd=4}
\begin{aligned}
    A^{\rm dD}_0 & (r)\bigl|_{d=4}=\frac{Q}{4\pi}\Biggl( \frac{1}{\pi}\frac{1}{r^2}-\frac{4}{3\pi^2}G_N m\frac{1}{r^4}\\
    & + \left(\frac{4G_N^2 m^2}{9\pi^3}\left(5\log\left(\frac{8G_Nm}{3\pi  c r^2}\right)+3\right) +\frac{\alpha G_N}{3 \pi ^3 }\log \left(\frac{8 G_N m}{3 \pi  c r^2}\right) \right)\frac{1}{r^6}\Biggl)+O\left(\frac{1}{r^8}\right)\ ,
\end{aligned}
\end{equation}
in which it is coherently considered the same redefinition of the unfixed constant of the metric case.  

\chapter{Graviton emission processes}
\label{chap:Metric_from_Amplitudes}

The original work of \cite{Donoghue:1994dn, Donoghue:2001qc, Bjerrum-Bohr:2002fji}, which gave birth to the idea that it is possible to recover general relativity from the classical limit of quantum gravity as an effective field theory, inspired a series of papers whose objective was to formalize and make systematical the procedure for extracting classical physics from quantum scattering amplitudes. 

\medskip

In this chapter we will firstly review the work presented in \cite{Mougiakakos:2020laz, Bjerrum-Bohr:2018xdl}, in which the strategy to extract the classical limit from graviton emission processes is improved, and then applied to the computation of the Schwarzschild-Tangherlini metric in a completely systematical way. Then following \cite{Fragomeno, DOnofrio:2021tap} the procedure is generalized to an electromagnetically coupled theory in order to obtain the Reissner-Nordstr\"om-Tangherlini metric from scattering amplitudes with photons and gravitons circulating in the loops. After the strategy review, explicit calculations of graviton emission processes are carried out, and the perfect agreement with the classical derivation of chapter \ref{chap:RNT_Classical_Solution} is shown. Then it is noticed how in the metric divergences arise in four and five space-time dimensions, and a procedure to renormalize them and obtain finite results has to be carried out. According to \cite{Mougiakakos:2020laz}, in section \ref{sec:Renorm_Metric} in fact the non-minimal action compatible with the principles of the effective field theories is defined with the intention to use the new interaction vertices as counter-terms\footnote{In the context of the world-line formalism, these higher-derivative couplings had already been introduced in \cite{Goldberger:2004jt}, and shown not to contribute to any physical observable. The equivalence between the scattering amplitude and world-line approaches was shown in \cite{Mogull:2020sak}.}, in order to renormalize the observable \cite{Fragomeno, DOnofrio:2021tap}. Finally it is shown how such procedure gives results in complete agreement with the classical derivation in which logarithmic terms appear, and it is proved that the counter-terms needed to renormalize the electromagnetic sector of the metric are exactly the ones already found in \cite{Mougiakakos:2020laz}.
Notice that from here on it is set again $\hbar=1$.

\section{Classical limit of graviton emission processes}
\label{sec:Classical_Limit_Grav_Emission_Processes}

Inspired by chapter \ref{Classical_Chapter} we now want to recover the Reissner-Nordstr\"om-Tangherlini metric from the classical limit of graviton emission scattering amplitudes. Considering the quantum effective action defined in section \ref{sec:Quantization_of_Gravity} in space-time dimension $D=d+1$, we look for a solution of the gravitational field equation expanded in a post-Minkowskian series. Referring to the the notation of chapter \ref{chap:RNT_Classical_Solution}, in which to the $n$-th post-Minkowskian order are associated powers of $\alpha$ up to the integer part of $j/2$, the expansion of the gravitational source takes the form
\begin{equation}
T_{\mu\nu}=\sum_{n=0}^{+\infty}\sum_{\substack{j=0 \\ j \ \text{even}}}^{n}T_{\mu\nu}^{(n, j)}\ .
\end{equation}
In the same way we can expand from the beginning the graviton field in a post-Minkowskian series as
\begin{equation}
     h_{\mu\nu} =  \sum_{n=1}^{+\infty}\sum_{\substack{j=0 \\ j \ \text{even}}}^{n}h_{\mu\nu}^{(n, j)}\ ,
\end{equation}
in which the de Donder gauge condition \eqref{deDonder_Gauge} now holds order by order as
\begin{equation}
    \partial_\lambda h^{\lambda}{}_{\nu}{}^{(n, j)}-\frac{1}{2}\partial_\nu h^{(n, j)}=0\ .
\end{equation}
Then it is possible to write down the graviton field equations in the de Donder gauge, and for each order they read
\begin{equation}\label{Box_MetricPerturbation}
    \Box h_{\mu\nu}^{(n, j)}(x) = -\frac{\kappa}{2} \left(T_{\mu\nu}^{(n-1, j)}(x) - \frac{1}{d-1}\eta_{\mu\nu}T^{(n-1, j)}(x)\right)\ ,
\end{equation}  
which say that the $n-1$-th post-Minkowskian term of the stress-energy tensor is the source of the $n$-th post-Minkowskian term of the metric. As we already commented, this is due to the fact that the zeroth order of the stress-energy tensor is reconstructed by the matter sector of the theory, which gives rise to the first post-Minkowskian order of the metric, while higher orders correspond to graviton self-interaction terms \cite{Bjerrum-Bohr:2002fji}. 

\medskip

We are interested in static solutions of eq. \eqref{Box_MetricPerturbation}, which means impose $\de_t h_{\mu\nu}(x)=0$. From this condition we can explicit the graviton field as
\begin{equation}\label{h_EMT}
    h_{\mu\nu}^{(n,j)}(\vec{x}) = -\frac{\kappa}{2}\int \frac{d^d\vec{q}}{(2\pi)^d} \frac{e^{i\vec{q}\cdot \vec{x}}}{\vec{q}^2}\left(T^{(n-1,j)}_{\mu\nu}(\vec{q}^2) - \frac{1}{d-1}\eta_{\mu\nu}T^{(n-1,j)}(\vec{q}^2) \right) \ ,
\end{equation}
where $T_{\mu\nu}(\vec{q}^2)$ is the Fourier transform of the stress-energy tensor defined in eq. \eqref{StressEnergyTensor_as_MatrixElement_Example} and in which $q^0=0$ is the usual momentum-space version of the static limit. 
While the graviton field explicitated in the above equation is a quantum object, we are interested only in its classical contributions that survive at $\hbar\rightarrow 0$, such pieces are reconstructed by picking the terms of the perturbative expansion of $T_{\mu\nu}(\vec{q}^2)$ which satisfy the selection rules stated in section \ref{sec:Schwarzschild_Metric_from_Graviton_Emission}. Reproducing the results of \cite{Mougiakakos:2020laz}, the classical contributions of the observable are reconstructed from $l$-loop amplitudes in which $n-j$ gravitons and $j$ photons are emitted by a single massive line and whose internal tree structure ends up with a graviton emission as
\begin{equation}\label{Generic_Example_Process}
\includegraphics[width=0.4\textwidth, valign=c]{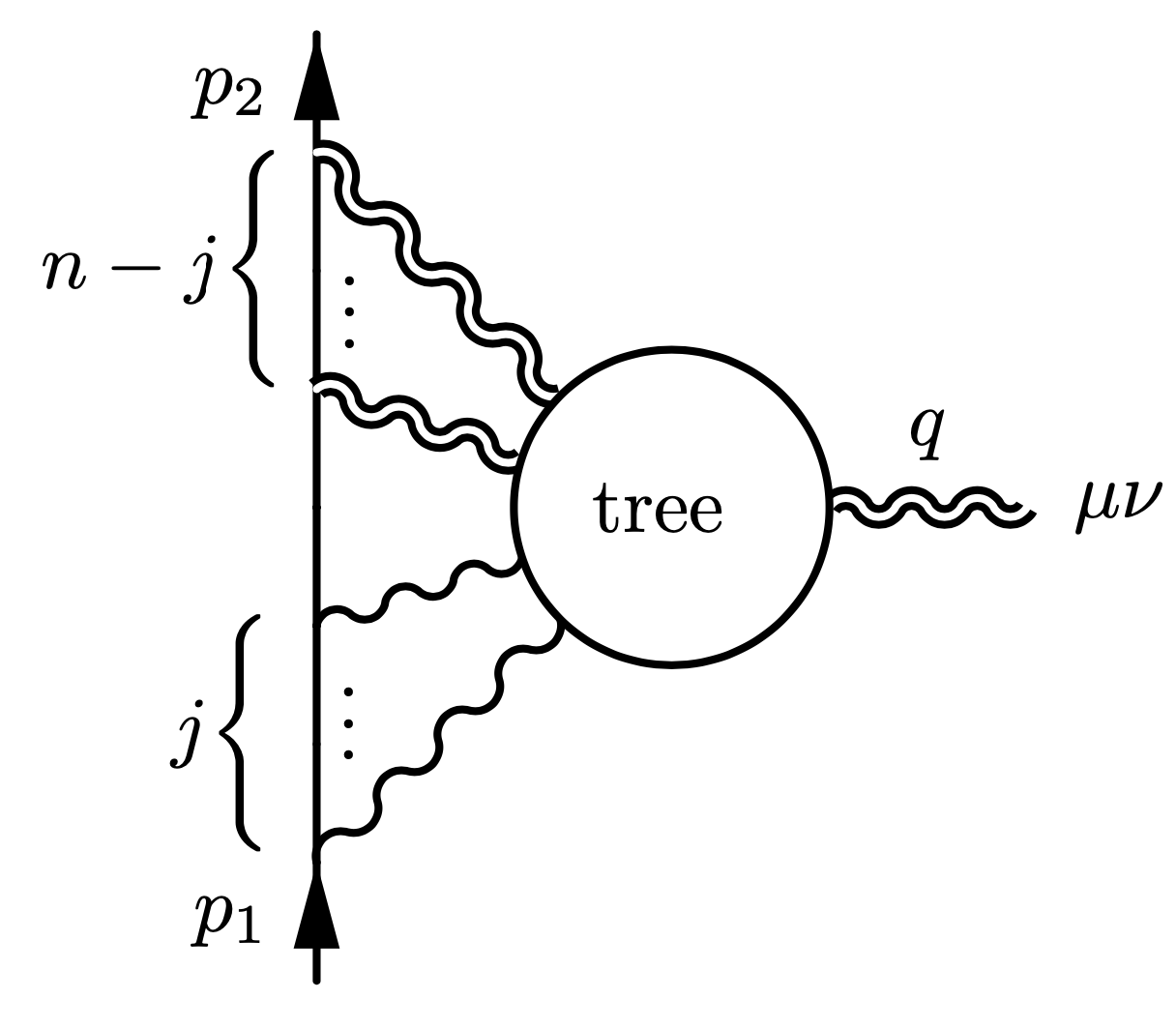}=-\frac{i\, \kappa}{2} \, \sqrt{4E_1 E_2} \, T_{\mu\nu}^{(l,j)}(\vec{q}^2)\ ,
\end{equation}
where, as already noticed in prior chapters, the post-Minkowskian expansion is related to the perturbative loop series by the relation $n=l+1$. Now since eq. \eqref{h_EMT} must reconstruct the post-Minkowskian expansion of the classical metric perturbation, it has to be true that 
\begin{equation}\label{Imposition_on_EMT}
\int \frac{d^d\vec{q}}{(2\pi)^d} \frac{e^{i\vec{q}\cdot \vec{x}}}{\vec{q}^2}\, T^{(l,j)}_{\mu\nu}(\vec{q}^2) \propto \rho^{l+1}\ ,
\end{equation}
where $\rho$ is defined in eq. \eqref{Definition_rho}. The function that satisfy the relation \eqref{Imposition_on_EMT} is the so called $l$-loop massless sunset integral (or briefly master integral), defined by 
\begin{equation}
J_{(l)}(\vec{q}^2) = \int \prod_{i=1}^{l}\frac{d^d\vec{\ell_i}}{(2\pi)^d}\frac{\vec{q}^2}{\left(\prod_{i=1}^{l}\vec{\ell_i}^2\right) \left(\vec{q}-\vec{\ell_1}-...-\vec{\ell_l}\right)^2}\ , 
\end{equation}
whose mathematical properties are listed in appendix
\ref{App:masterintegral}. Out of this analysis turns out that\footnote{Since we are interested in recovering classical physics, eq. \eqref{Tmunu_propto_J} must be understood in the sense that the contributions of the stress-energy tensor which reconstruct the classical metric must be proportional to the master integral. In principle there would be also contributions not proportional to $J_{(l)}(\vec{q}^2)$, which however we neglect since do not belong to classical phenomenology.}
\begin{equation}\label{Tmunu_propto_J}
T_{\mu\nu}^{(l,j)}(\vec{q}^2)\propto J_{(l)}(\vec{q}^2)\ ,
\end{equation}
from which exploiting the fact that the stress-energy tensor is conserved and Lorentz covariant, from dimensional reasons already discussed in eq. \eqref{DImensional_Reasons}, in terms of dimensionless form factors $c_1^{(l,j)}(d)$ and $c_2^{(l,j)} (d)$, the gravitational source in momentum space reads
\begin{equation}\label{EMT_FormFactors}
\begin{aligned}
    & T_{\mu\nu}^{(l)}(\vec{q}^2) = \sum_{\substack{j=0 \\ j \ \text{even}}}^{l+1} T_{\mu\nu}^{(l,j)} (\vec{q}^2) \\
    &\simeq    \sum_{\substack{j=0 \\ j \ \text{even}}}^{l+1}m \left(c_1^{(l, j)}(d)\delta_\mu^0\delta_\nu^0+c_2^{(l, j)}(d)\left(-\frac{q_\mu q_\nu}{\vec{q}^2}-\eta_{\mu\nu}\right)\right)(G_N m)^{l-j}(\alpha G_N)^\frac{j}{2}\, \pi^{l}\, J_{(l)}(\Vec{q}^2)\ ,
    \end{aligned}
\end{equation}
in which from here on the symbol $"\simeq"$ means that we consider only the leading-order in energy and the factor $\pi^l$ is inserted in order to make the form factors independent by it. The reason why we consider only the leading-order momentum expansion, is because since we are interested only in classical physics, non-leading-order pieces lead to quantum corrections even considering the amplitude in eq. \eqref{Generic_Example_Process}, exactly as it was in section \ref{sec:Schwarzschild_Metric_from_Graviton_Emission}. 

\medskip

Now it is possible to replace eq. \eqref{EMT_FormFactors} inside \eqref{h_EMT} obtaining the classical metric perturbation order by order in $l$ and $j$ as
\begin{equation}\label{MetricPerturb_loop_Expansion}
    \begin{aligned}
    \kappa\, & h_{\mu\nu}^{(l+1, j)}(\vec{x}) \\ &
    =-16\, \pi^{l+1} \int \frac{d^d\vec{q}}{(2\pi)^d} \frac{e^{i\vec{q}\cdot \vec{x}}}{\vec{q}^2}\left( c_1^{(l, j)}(d)\left(\delta_\mu^0\delta_\nu^0-\frac{\eta_{\mu\nu}}{d-1}\right) + c_2^{(l, j)}(d)\left(-\frac{q_\mu q_\nu}{\vec{q}^2}+\frac{\eta_{\mu\nu}}{d-1}\right)\right) \\
    & \times (G_N m)^{l+1-j}(\alpha G_N)^\frac{j}{2}J_{(l)}(\Vec{q}^2)\ .
\end{aligned}
\end{equation}
Then employing the Fourier transform in eqs. \eqref{app_J_Identity_1}
 and  \eqref{app_J_Identity_2}, one can derive the explicit post-Minkowskian expansion of the metric perturbation according to the definition \eqref{Metric_Cr_deDonder} as
\begin{equation} \label{metric_pert}
    \begin{aligned}
h_{0}^{(l+1,j)}(r) =&-\frac{16}{d-1}\left((d-2) c_{1}^{(l,j)}(d)+c_{2}^{(l,j)}(d)\right)\left(\frac{\rho}{4}\right)^{l+1}(G_N m)^{l+1-j}(\alpha G_N)^\frac{j}{2} \\
h_{1}^{(l+1,j)}(r) =&\frac{16}{d-1}\left(c_{1}^{(l,j)}(d)-\left(1+\frac{d-1}{2-l(d-2)}\right) c_{2}^{(l,j)}(d)\right)\\&\times\left(\frac{\rho}{4}\right)^{l+1}(G_N m)^{l+1-j}(\alpha G_N)^\frac{j}{2} \\
h_{2}^{(l+1,j)}(r) =&16 \frac{(d-2)(l+1)}{2-l(d-2)} c_{2}^{(l,j)}(d)\left(\frac{\rho}{4}\right)^{l+1}(G_N m)^{l+1-j}(\alpha G_N)^\frac{j}{2} \ .
\end{aligned}
\end{equation}
These expressions impose an explicit relation between the metric and the quantum amplitude calculation, in which the only object needed in order to completely recover the observable are the form factors, which can be extracted directly from amplitude computations.  

\medskip

Following \cite{Mougiakakos:2020laz}, in order to work out a systematic procedure to extract the form factors from amplitudes like \eqref{Generic_Example_Process}, we decompose such processes in 
\begin{equation}\label{Massive_Line}
\begin{aligned}
&\mathcal{N}_{\mu_1\nu_1, ..., \mu_{l-j+1}\nu_{l-j+1}, \beta_1, ..., \beta_j}(p_1, p_2, \ell_1, ..., \ell_l)=\includegraphics[width=0.4\textwidth, valign=c]{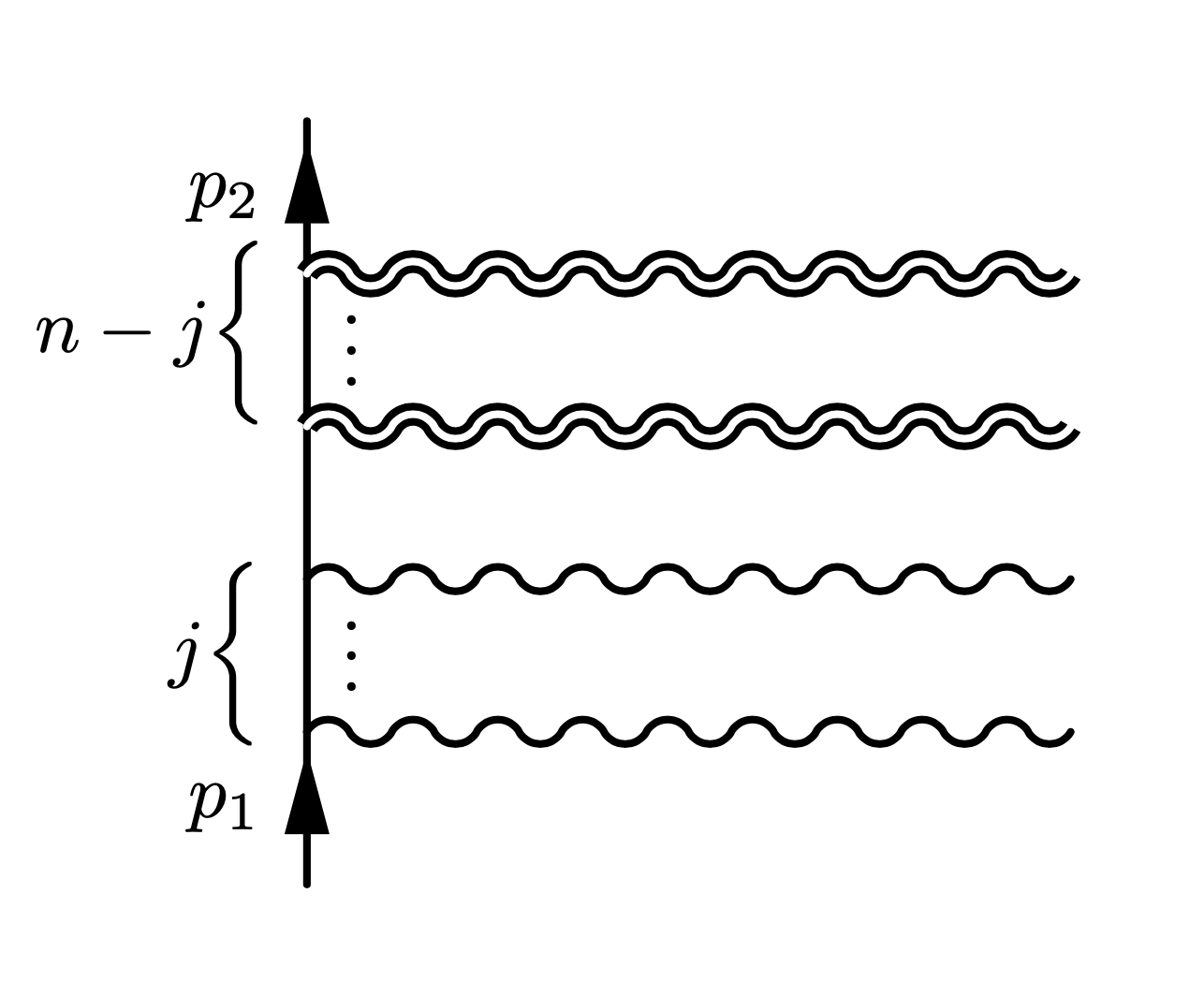}\\
&=\frac{(i)^l\left(\prod_{i=1}^{l+1-j}\left( \tau_{\phi^2 h}\right)_{\mu_i\nu_i}\right)\left(\prod_{i=1}^{j}\left( \tau_{\phi^2 A}\right)_{\beta_i}\right)}{\prod_{i=1}^l\left( \left(p_1-\sum_{k=1}^i\ell_k\right)^2-m^2+i\epsilon\right)}\ ,
\end{aligned}
\end{equation}
where we take into account all the vertices attached to the massive line, whose explicit expressions are reported in appendix \ref{Appendix_FeynRules}, and in 
\begin{equation}
\mathcal{M}_{\lambda_1\sigma_1, ..., \lambda_{l-j+1}\sigma_{l-j+1}, \alpha_1, ..., \alpha_j, \mu\nu}(p_1, p_2, \ell_1, ..., \ell_l)=\includegraphics[width=0.4\textwidth, valign=c]{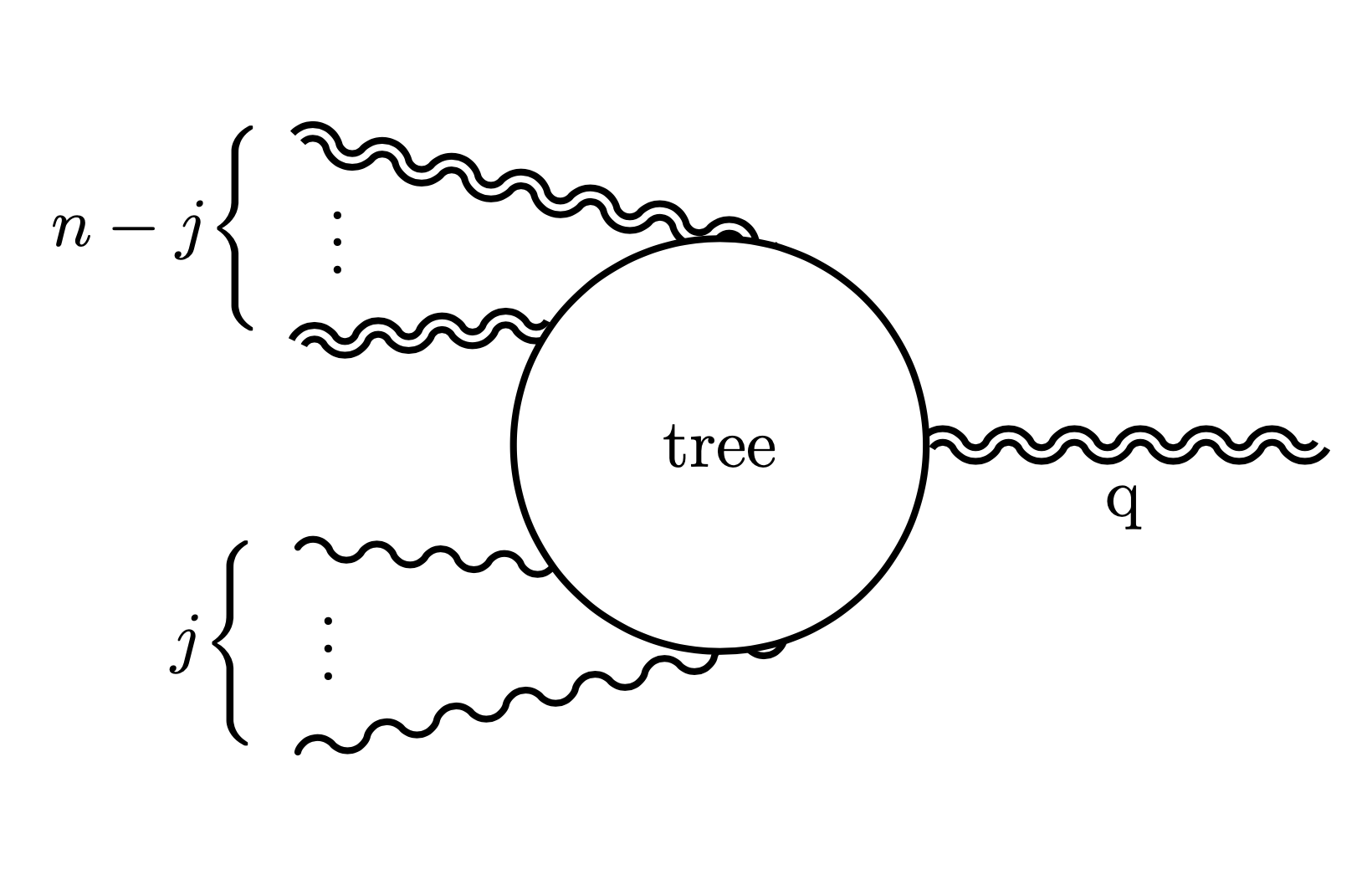}\ , 
\end{equation}
where the explicit form of the tree internal amplitude depends on the specif process. Considering the momentum conservation $\ell_1+...+\ell_{l+1}=p_1-p_2=q$, putting all together it is possible to compactly write down the amplitude in eq. \eqref{Generic_Example_Process} as
\begin{equation}
\begin{aligned}\label{EMT_Amplitude_Complete}
   & -\frac{i\, \kappa}{2}\, \sqrt{4E_1 E_2}\, T_{\mu\nu}^{(l,j)}(q^2)\\ 
   &= \int \prod_{i=1}^l\frac{d^{d+1}\ell_i}{(2\pi)^{d+1}} \mathcal{N}_{\mu_1\nu_1, ..., \beta_j} \mathcal{M}_{\lambda_1\sigma_1, ...,\alpha_j, \mu\nu}\frac{\Big(\prod_{i=1}^{l+1-j} i\, P^{\lambda_i\sigma_i, \mu_i\nu_i}\Big)\Big(\prod_{i=1}^{j} -i\, \eta^{\alpha_i, \beta_i\nu_i}\Big)}{\prod_{i=1}^{l+1} \Big(\ell_i^2+i\epsilon\Big)}\ .
\end{aligned}
\end{equation}

\medskip

Up to now the expressions of the above amplitudes are completely generic, but since we are interested only in those contributions that reconstruct the classical phenomenology, we can impose the classical limit\footnote{From hereafter with $"\text{classical limit}"$ we mean the whole set of considerations needed to extract the $\hbar$-independent long-range contributions from a loop amplitude, including the energy expansion of a certain diagram.} in order to simplify the integral. As usual this means expand the amplitude in an energy expansion and pick only the leading-order contribution considering the limit $|\vec{p_i}|\ll m$ and neglecting higher orders in the momentum whose lead to quantum corrections. Then the vertices attached to the massive line take the form
 \begin{equation} \label{static_vertex}
    \left(\tau_{\phi^2 h}\right)_{\mu\nu}\simeq -i\, \kappa \, m^2\delta_\mu^0\delta_\nu^0\quad \text{and} \quad  \left(\tau_{\phi^2 A}\right)_{\beta}\simeq -2\, i\, Q\, m\, \delta_\beta^0\ .
\end{equation}
Moreover it is possible to simplify also the expression of the massive propagator contained in eq. \ref{Massive_Line}. In fact for a generic internal massive line it is verified that
\begin{equation}
\frac{i}{\left(p_1-\sum_{k=1}^i\ell_k\right)^2-m^2+i\epsilon}\simeq \frac{i}{-2m\left(\sum_{k=1}^i\ell_k^0\right)+i\epsilon}\ ,
\end{equation}
from which it is possible to write
\begin{equation}
\begin{aligned}
\mathcal{N}_{\mu_1\nu_1, ..., \beta_j}&= (-i)^{l+1}\kappa^{l+1-j}m^{2+2l-j}2^jQ^j\left(\prod_{i=1}^{l+1-j}\delta_{\mu_i}^0\delta_{\nu_i}^0\right)\left(\prod_{i=1}^{j}\delta_{\beta_i}^0\right)\\
&\times\left(\frac{i}{-2m}\right)^{l}\prod_{i=1}^l\frac{1}{\left(\sum_{k=1}^i\ell_k^0\right)-i\epsilon}\ .
\end{aligned}
\end{equation}
Now the above expression is invariant under the exchange of any internal massless line, which means that it is possible to sum over all the permutations of the internal momenta up to weight the expression with the right combinatorial factor. In particular identifying with $\mathfrak{S}_{l+1}$ the group of $l+1$ permutations, using the relation \cite{Mougiakakos:2020laz} 
\begin{equation}
\sum_{\sigma \in \mathfrak{S}_{l+1}} \prod_{i=1}^{l+1} \frac{1}{\sum_{k=1}^{i} x_{\sigma(k)}}=\prod_{i=1}^{l+1} \frac{1}{x_{i}}\ ,
\end{equation}
exploiting the momentum conservation\footnote{It is verified that $\ell_1^0+...+\ell_{l+1}^0=q^0=0$.} in order to write 
\begin{equation}
\prod_{i=1}^{l} \frac{1}{\left(\sum_{k=1}^{i} \ell^0_{k}\right)-i\epsilon} = -i\epsilon\prod_{i=1}^{l+1} \frac{1}{\left(\sum_{k=1}^{i} \ell^0_{k}\right)-i\epsilon}\ ,
\end{equation}
we can finally get
\begin{equation}
\sum_{\sigma \in \mathfrak{S}_{l+1}} \prod_{i=1}^{l+1} \frac{1}{\left(\sum_{k=1}^{i} \ell^0_{\sigma(k)}\right)-i\epsilon}=\prod_{i=1}^{l+1} \frac{1}{\ell^0_{i}-i\epsilon}\ ,
\end{equation}
from which one obtains
\begin{equation}\label{N_Last_Expr}
\begin{aligned}
&\mathcal{N}_{\mu_1\nu_1, ..., \beta_j} = \frac{1}{(l+1)!}\sum_{\sigma \in \mathfrak{S}_{l+1}}\mathcal{N}_{\mu_1\nu_1, ..., \beta_j} \\ 
&=\frac{1}{(l+1)!}(-i)^{l+1}\kappa^{l+1-j}m^{2+2l-j}2^jQ^j\left(\prod_{i=1}^{l+1-j}\delta_{\mu_i}^0\delta_{\nu_i}^0\right)\left(\prod_{i=1}^{j}\delta_{\beta_i}^0\right)\left(\frac{i}{-2m}\right)^{l}(-i\epsilon)\prod_{i=1}^{l+1} \frac{1}{\ell^0_{i}-i\epsilon}\ .
\end{aligned}
\end{equation}

\medskip

Now inserting back this expression in eq. \eqref{EMT_Amplitude_Complete}, following \cite{Bjerrum-Bohr:2018xdl}, we can integrate out the temporal component of the internal momenta from the loop integral. In fact in the limit $\epsilon\rightarrow 0$, eq. \eqref{N_Last_Expr} vanishes unless some of the $\ell_i^0$ vanish at the same time, and since eq. \eqref{EMT_Amplitude_Complete} must be non-vanishing, one needs to integrate over $\ell_i^0$ in the loop integral picking the residue $\ell_i^0=i\epsilon$. This leads to the cancellation of the extra $-i\epsilon$ since now $\ell^0_{l+1}=0$, and simplifies the amplitude calculation as
\begin{equation}
\begin{aligned}
   & -\frac{i\, \kappa}{2}\,2m\, T_{\mu\nu}^{(l,j)}(q^2)\\ 
   &= \int \prod_{i=1}^l\frac{d^{d}\ell_i}{(2\pi)^{d}}  \mathcal{M}_{\lambda_1\sigma_1, ...,\lambda_{l+1-j}\sigma_{l+1-j}, 0, ..., 0,\mu\nu}\Big|_{\ell_i^0=0}\frac{\Big(\prod_{i=1}^{l+1-j} i\, P^{\lambda_i\sigma_i, 00}\Big) (-i)^j}{\prod_{i=1}^{l+1} \Big(-\vec{\ell_i}^2\Big)}\\
   &\times \frac{1}{(l+1)!}(-i)^{l+1}\kappa^{l+1-j}m^{2+2l-j}2^jQ^j\left(\frac{i}{-2m}\right)^{l}\ ,
\end{aligned}
\end{equation}
from which one finally obtains
\begin{equation}\label{Tmunufromamplitude}
\begin{aligned} 
    T_{\mu\nu}^{(l,j)}(\vec{q}^2)&=\frac{i}{(l+1)!}\int\prod_{i=1}^l\frac{d^d\Vec{\ell_i}}{(2\pi)^d} (-1)^{j+l+1}2^{j-l}m^{l+1-j}\kappa^{l-j}Q^j\frac{\prod_{i=1}^{l-j+1}P^{00, \lambda_i\sigma_i}}{\prod_{i=1}^{l+1}\Vec{\ell_i}^2}\\
    &\times \mathcal{M}_{\lambda_1\sigma_1, ..., \lambda_{l-j+1}\sigma_{l-j+1}, 0, ..., 0, \mu\nu}(p_1, p_2, \ell_1, ..., \ell_l)\Big|_{\ell_i^0=0}\ ,
\end{aligned}
\end{equation}
where
\begin{equation}
    P_{00, \mu\nu}=\delta_\mu^0\delta_\nu^0-\frac{\eta_{\mu\nu}}{d-1}\ .
\end{equation}

\medskip

\begin{figure}[h]
\centering
\includegraphics[width=0.3\textwidth, valign=c]{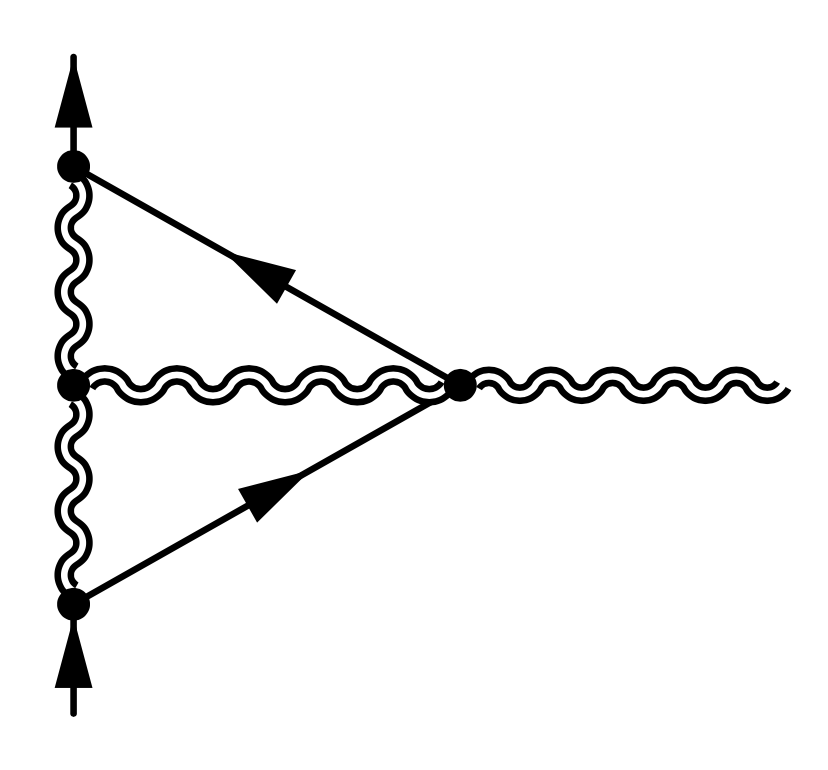}
\caption{2-loop amplitude that contributes to a quantum correction of the third post-Minkowskian order of the metric.}
\label{Example_Duff_NoLead}
\end{figure}

Then the strategy to extract the form factors needed for the computation of eq. \eqref{metric_pert} out of the above expression, will be to choose a particular process which fixes the expression of $\mathcal{M}$, considering the classical limit of such amplitude and performing the replacement $\ell_i^0=0$ once the right factors due to the residue integration are added. Then out of the whole expression in \eqref{Tmunufromamplitude} we will select only those contributions proportional to the master integral, extracting in this way the explicit form of $c_1^{(l,j)}(d)$ and $c_2^{(l,j)} (d)$ by comparison with eq. \eqref{EMT_FormFactors}. Moreover such analysis shows how eq. \eqref{Tmunufromamplitude} is equivalent to a quantum tree-graph in which the internal massive lines are cut and replaced by external sources
\begin{equation}\label{Duff_Amp}
-i\, \kappa \, m\, T_{\mu\nu}^{(l,j)}(\vec{q}^2)\Big|_{\text{classical}}=\includegraphics[width=0.45\textwidth, valign=c]{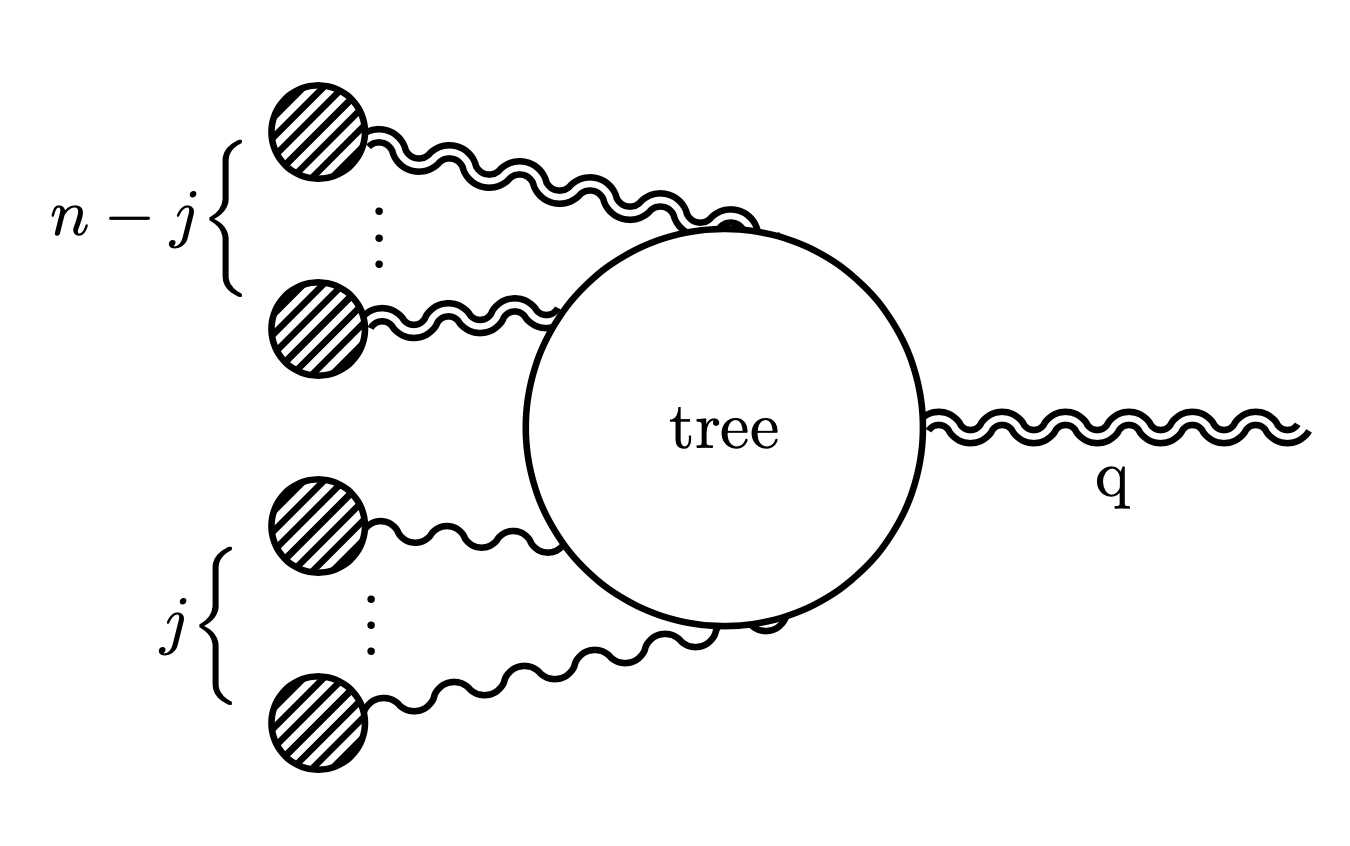}\Bigg|_{\text{leading-order}}
\ .
\end{equation}
This was first noticed by Duff in \cite{Duff:1973zz}, in which the Schwarzschild metric was recovered up to $O(G_N^3)$ by the explicit calculation of amplitudes like \eqref{Duff_Amp}. This argument motivates the selection rule given in section \ref{sec:Schwarzschild_Metric_from_Graviton_Emission}, which states that the low-energy limit imposes that massive particles cannot mediate the interaction, and in massless emission processes this means that massive states enter only as a single line that connects the two initial particles. The reason why this is true, is that since the classical limit cuts internal massive lines, the amplitude is reorganized in terms of the interaction of $l+1$ external sources like in eq. \eqref{Duff_Amp}, whose in \cite{Duff:1973zz} are proven to fully reconstruct classical phenomenology. For example the amplitude in figure \ref{Example_Duff_NoLead}, even having in principle a purely classical leading order behavior obtained from the replacement rules of section \ref{Sec_Classical_Limit}, an explicit calculation will show that such contribution does not exist and the leading-order is actually $O(\vec{q}^4)$, which correspond to quantum corrections. In the modern language we can say that the leading contribution of the diagram is not proportional to the master integral and therefore it is not classical.  

\section{Metric from amplitude calculations}
\label{sec:AmpCalc_Metric}

Now we are ready to employ the techniques elaborated in the previous section in order to explicitly compute the metric out of amplitude calculations. As anticipated earlier the strategy will be to consider processes like \eqref{Generic_Example_Process}, and extract out of them the classical part following eq. \eqref{Tmunufromamplitude}, picking up only those contributions proportional to the master integral. The calculation will be performed up to fourth post-Minkowskian order, which means up to 3 loops. Since for each loop order there exist different contributions of the stress-energy tensor associated with different values of $j$, for each pair of $l$ and $j$ will be evaluated the corresponding form factors and the associated metric contribution. 

\subsection*{Tree-level}

\begin{figure}[h]
\centering
\includegraphics[width=0.25\textwidth, valign=c]{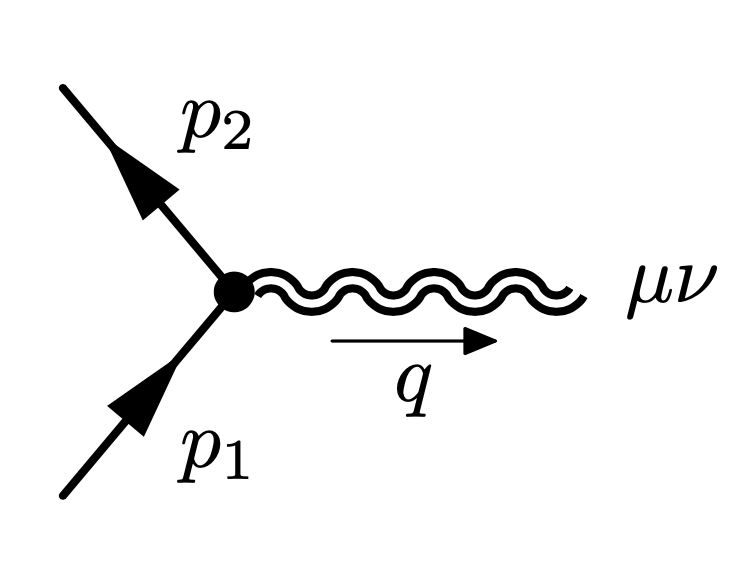}
\caption{Tree-level diagram for graviton emission.}
\label{Tree_level_GravitonEmission}
\end{figure}

We first consider the tree-level amplitude in figure \ref{Tree_level_GravitonEmission}, corresponding to eq. \eqref{Generic_Example_Process}  for $l=0$, $j=0$.
Considering on-shell particles which satisfy $p_1^2=p_2^2=m^2$, with transferred momentum $q=p_1-p_2$, the stress-energy tensor arising at tree-level is
\begin{equation}
\frac{-i\kappa}{2}\sqrt{4E_1 E_2}T_{\mu\nu}^{(0,0)}(q^2) = (\tau_{\phi^2h})_{\mu\nu}\ .
\end{equation} 
Using the relations \eqref{static_vertex} one gets
\begin{equation}
    T_{\mu\nu}^{(0,0)}(\vec{q}^2) = m \, \delta_\mu^0\delta_\nu^0\ ,
\end{equation} 
that compared with eq. \eqref{EMT_FormFactors} lead to 
\begin{equation}
\begin{aligned}
    c_1^{(0,0)} (d) &= 1 \\
    c_2^{(0,0)} (d) &= 0\ ,
\end{aligned}
\end{equation}
and from \eqref{metric_pert}  one gets the metric components 
\begin{equation}
    \begin{aligned}
        h_0^{(1,0)}(r) &= -4\frac{d-2}{d-1}G_N m\,\rho\\
        h_1^{(1,0)}(r) &= \frac{4}{d-1}G_N m\,\rho\\
        h_2^{(1,0)}(r) &= 0 \ .
    \end{aligned}
    \label{h_i^1}
\end{equation}
This result is in perfect agreement with the classical post-Minkowskian metric derived in eq. \eqref{hid} at first order. 

\subsection*{1-loop}

\begin{figure}[h]
\centering
\includegraphics[width=0.4\textwidth, valign=c]{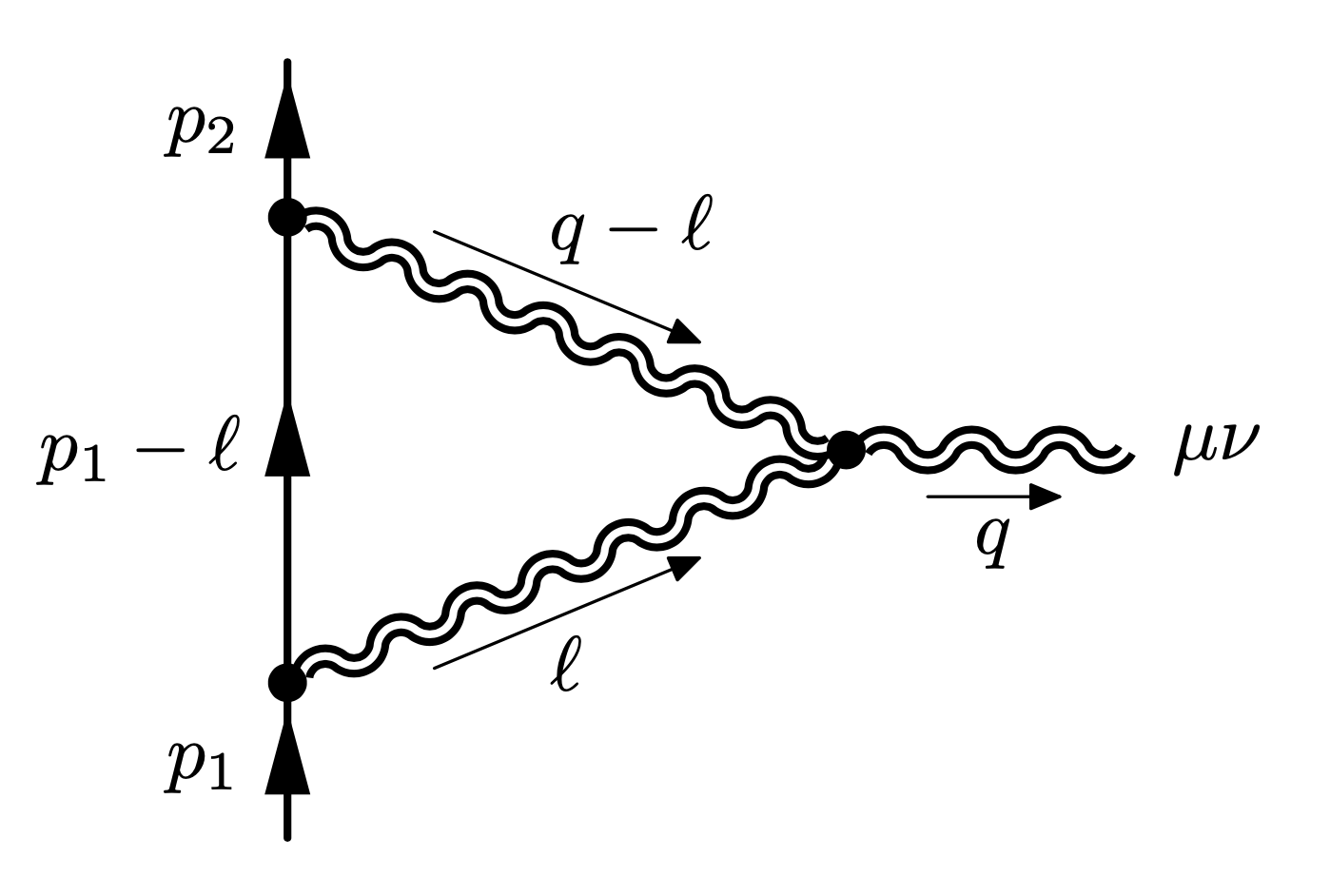}
\caption{1-loop diagram for graviton emission with 2 internal gravitons.}
\label{1Loop_GravitonEmission_2InternalGrav}
\end{figure}

We now  compute the contribution to the stress-energy tensor that arises at 1-loop order. There are two diagrams that contribute, \textit{i.e.} $l=1, \ j=0$ and $l=1, \ j=2$. The first one is obtained by evaluating the amplitude that involves a 3-gravitons vertex, as in figure \ref{1Loop_GravitonEmission_2InternalGrav}.
The amplitude associated to this diagram is 
\begin{equation}
\begin{aligned}
     \frac{-i\kappa}{2}\sqrt{4E_1 E_2}&T^{(1,0)}_{\mu\nu}(q^2)   \\
   &= \int\frac{d^{d+1}\ell}{(2\pi)^{d+1}} \frac{-i P^{\alpha\beta,\lambda\kappa}P^{\gamma\delta,\rho\sigma}  \left(\tau_{\phi^2h}\right)_{\,\alpha\beta}\,\left(\tau_{\phi^2h}\right)_{\,\gamma\delta}\,\left(\tau_{h^2h}\right)_{\mu\nu,\rho\sigma,\lambda\kappa}(\ell,q)}{(\ell^2+i\epsilon)((\ell-q)^2+i\epsilon)((\ell-p_1)^2-m^2+i\epsilon)}\ ,
    \end{aligned}
\end{equation}
where the 3 gravitons vertex $(\tau_{h^2h})$ is defined in eq. \eqref{3Graviton_Vertex_Donoghue}. In particular here we are allowed to consider the vertex with an external graviton which simplifies the calculation, although the vertex $(\tau_{h^3})$ in eq. \eqref{3Graviton_Vertex_deWitt_AppEq} would give the same result after more tedious calculations. The stress-energy tensor form factors associated to this diagram are computed in \cite{Mougiakakos:2020laz}, and they read
\begin{equation}
    \begin{aligned}
       &  c_1^{(1,0)} (d) =  -2\frac{4d^2-15d+10}{(d-1)^2}\\
       &  c_2^{(1,0)} (d) = -\frac{2(d-2)(3d-2)}{(d-1)^2} \ ,
    \end{aligned}
\end{equation}
from which by eq. \eqref{metric_pert} we get
\begin{equation}
    \begin{aligned}
        h_0^{(2,0)} (r) &= \frac{8(d-2)^2}{(d-1)^2}(G_N m)^2\rho^2 \\
        h_1^{(2,0)} (r) &= -\frac{4(2d^2 - 9d + 14)}{(d-4)(d-1)^2}(G_N m)^2\rho^2 \\
        h_2^{(2,0)} (r) &= \frac{4(d-2)^2(3d-2)}{(d-4)(d-1)^2}(G_N m)^2\rho^2 \ .
    \end{aligned}
    \label{h_ST^2}
\end{equation}
Again the above metric is in perfect agreement with eq. \ref{hid}  for the terms proportional to $ m^2G_N^2\rho^2$. In particular we notice some divergences that arise in $d=4$ (five space-time dimensions), that have to be renormalized in order to obtain finite results. 

\medskip

\begin{figure}[h]
\centering
\includegraphics[width=0.4\textwidth, valign=c]{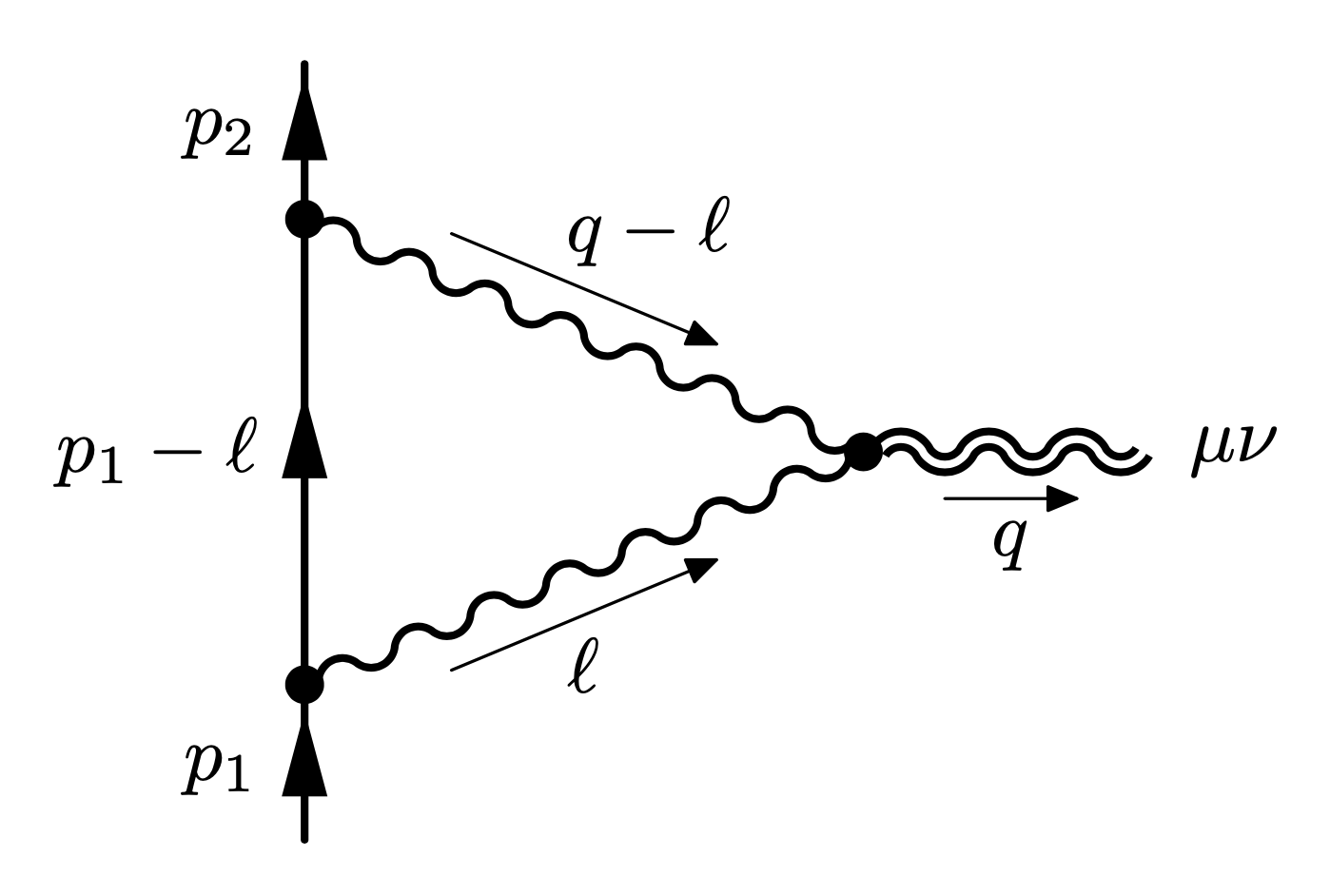}
\caption{1-loop diagram for graviton emission with 2 internal photons.}
\label{1Loop_GravitonEmission_2InternalPhot}
\end{figure}

The other contribution at 1-loop order is given by the diagram that contains two photons drawn in figure \ref{1Loop_GravitonEmission_2InternalPhot}.
The resulting contribution to the stress-energy tensor is
\begin{equation}
\begin{aligned}
    \frac{-i\kappa}{2}\sqrt{4E_1 E_2}&T^{(1,2)}_{\mu\nu}(q^2) \\ & = \int\frac{d^{d+1}\ell}{(2\pi)^{d+1}} \frac{-i\left(\tau_{\phi^2A}\right)_{\alpha}\left(\tau_{\phi^2A}\right)_{\beta}\left(\tau_{A^2h}\right){}_{\mu\nu}{}^{\alpha,\beta}(\ell,\ell-q)}{(\ell+i\epsilon)^2((\ell-q)^2+i\epsilon)((p_1-\ell)^2-m^2+i\epsilon)} \ , 
  \end{aligned}
\end{equation}
where $(\tau_{\phi^2A})$ is the 2 scalars - 1 photon vertex in \eqref{2scalars_1photon} and $(\tau_{A^2h})$ is the 2 photons - 1 graviton vertex in \eqref{2photon_1graviton}. Following the procedure of the previous section one gets
\begin{equation}
   T^{(1,2)}_{\mu\nu}(\Vec{q}^2) = i\frac{Q^2}{\kappa} \int \frac{d^{d}\ell}{(2\pi)^{d}} \frac{\left(\tau_{A^2h}\right)_{\mu\nu}{}^{0, 0}(\ell,\ell-q)\Big|_{\ell^0=0}}{\vec{\ell}^2 (\Vec{q}-\Vec{\ell})^2} \ , 
\end{equation}
where the numerator is explicitly
\begin{equation}
   \left(\tau_{A^2h}\right)_{\mu\nu}{}^{0, 0}(\ell,\ell-q)\Big|_{\ell^0=0} =i\kappa\left( \left(\delta_\mu^0\delta_\nu^0-\frac{\eta_{\mu\nu}}{2}\right) \ell\cdot (\ell-q)+\ell_\mu \ell_\nu -\frac{1}{2}(q_\mu \ell_\nu +q_\nu \ell_\mu) \right)\ .
\end{equation}

\medskip

Now that the complete loop integral reads
\begin{equation}\label{Example_Master_Extraction}
   T^{(1,2)}_{\mu\nu}(\Vec{q}^2) = -Q^2 \int \frac{d^{d}\ell}{(2\pi)^{d}} \frac{\left(\delta_\mu^0\delta_\nu^0-\frac{\eta_{\mu\nu}}{2}\right) \ell\cdot (\ell-q)+\ell_\mu \ell_\nu -\frac{1}{2}(q_\mu \ell_\nu +q_\nu \ell_\mu)}{\vec{\ell}^2 (\Vec{q}-\Vec{\ell})^2} \ ,
\end{equation}
we have to extract out of it the contributions proportional to the master integral. In order to do that one has to reshape the expression in eq. \eqref{Example_Master_Extraction} using symmetry and integral properties to obtain the integral definition of $J_{(l)}(\vec{q}^2)$. The needed loop integral reductions are listed in appendix \ref{App:LoopRed}, whose lead to 
\begin{equation}\label{Tmunu_1Loop_Explicit_For_Calculation}
\begin{gathered}
T^{(1,2)}_{00}(\vec{q}^2)= -\frac{Q^2}{4}J_{(1)}(\vec{q}^2) \\
\delta^{ij}T^{(1,2)}_{ij}(\Vec{q}^2)  =  \frac{Q^2}{4}(2-d)J_{(1)}(\vec{q}^2) \ . 
\end{gathered}
\end{equation}
The reason why we express the stress-energy tensor in the above form is because it allows us to use scalar loop integral reduction formulas like the one in appendix \ref{App:LoopRed}, instead of more complicated relations that involve free indices. Moreover the expression \eqref{Tmunu_1Loop_Explicit_For_Calculation} it is sufficient to extract the form factors. In fact expressing eq. \eqref{EMT_FormFactors} in an equivalent form as
\begin{equation}\label{Tmunu_Separated}
\begin{gathered}
T^{(l, j)}_{00}(\vec{q}^2)= m \left(c_1^{(l, j)}(d)-c_2^{(l, j)}(d)\right)(G_N m)^{l-j}(\alpha G_N)^\frac{j}{2}\, \pi^l\, J_{(l)}(\Vec{q}^2) \\
\delta^{ij}T^{(l, j)}_{ij}(\Vec{q}^2)  =  m \, c_2^{(l, j)}(d)\left(d-1\right)(G_N m)^{l-j}(\alpha G_N)^\frac{j}{2}\, \pi^l\, J_{(l)}(\Vec{q}^2) \ ,
\end{gathered}
\end{equation}
then in the present case we get
\begin{equation}
\begin{gathered}
T^{(1, 2)}_{00}(\vec{q}^2)= \left(c_1^{(1, 2)}(d)-c_2^{(1, 2)}(d)\right)\frac{Q^2}{4} J_{(1)}(\Vec{q}^2) \\
\delta^{ij}T^{(1, 2)}_{ij}(\Vec{q}^2)  =  c_2^{(1, 2)}(d)\left(d-1\right)\frac{Q^2}{4} J_{(1)}(\Vec{q}^2) \ ,
\end{gathered}
\end{equation}
from which a direct comparison with eq. \eqref{Tmunu_1Loop_Explicit_For_Calculation} leads to 
\begin{equation}
    \begin{aligned}
    & c_1^{(1,2)}(d)  = \frac{-2d+3}{d-1} \\
    & c_2^{(1,2)} ( d) = \frac{2-d}{d-1}\ .
    \end{aligned}
\end{equation}
Finally replacing the above expression inside eq. \eqref{metric_pert} one obtains
\begin{equation}
    \begin{aligned}
        h_0^{(2,2)} (r)&= -\frac{(-2d+4)}{d-1}G_N\alpha\rho^2 \\
        h_1^{(2,2)} (r)&= \frac{-2d + 6}{(d-1)(d-4)}G_N\alpha\rho^2 \\
        h_2^{(2,2)} (r)&= 2\frac{(d-2)^2}{(d-1)(d-4)} G_N\alpha\rho^2 \ .
    \end{aligned}
\end{equation}
The obtained metric is in perfect agreement with eq. \eqref{hid} for the terms proportional to $\alpha G_N\rho^2$. Moreover as expected we encounter divergences in $d=4$ that have to be renormalized.

\subsection*{2-loop}

\begin{figure}[h]
\centering
\includegraphics[width=0.3\textwidth, valign=c]{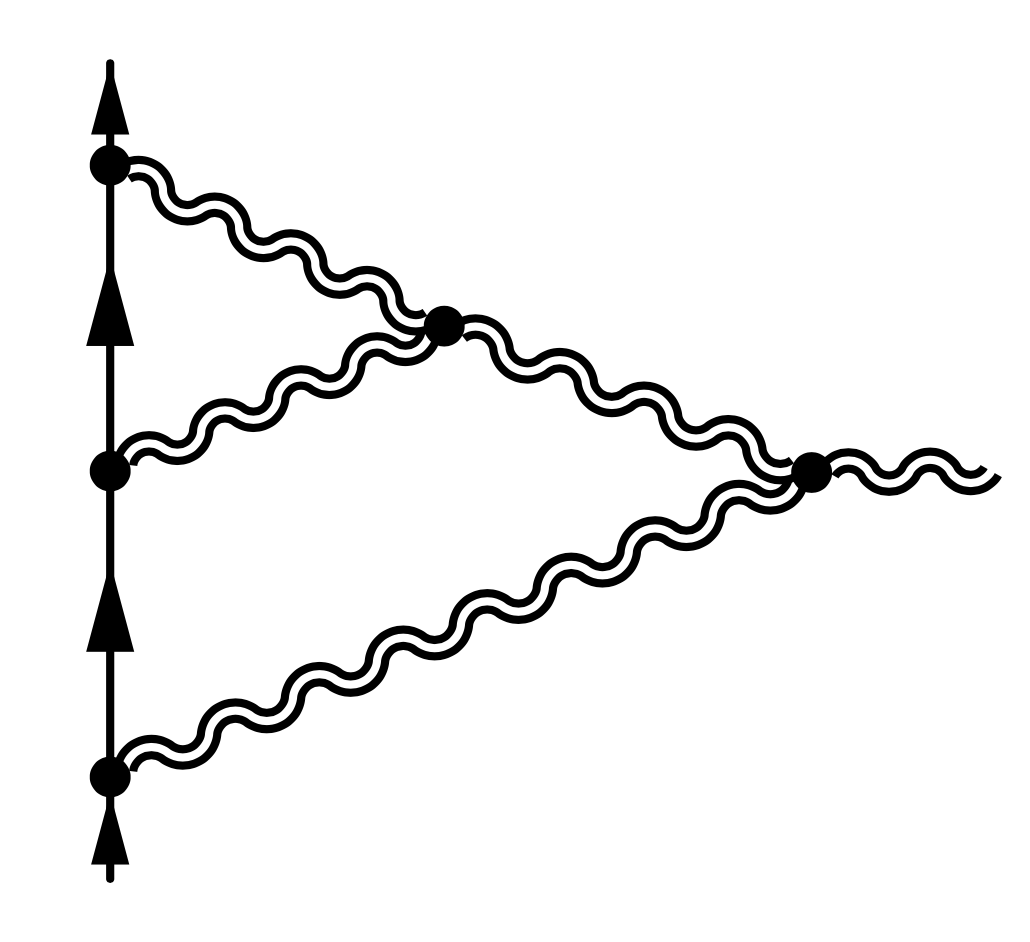}
\hspace{2cm}
\includegraphics[width=0.3\textwidth, valign=c]{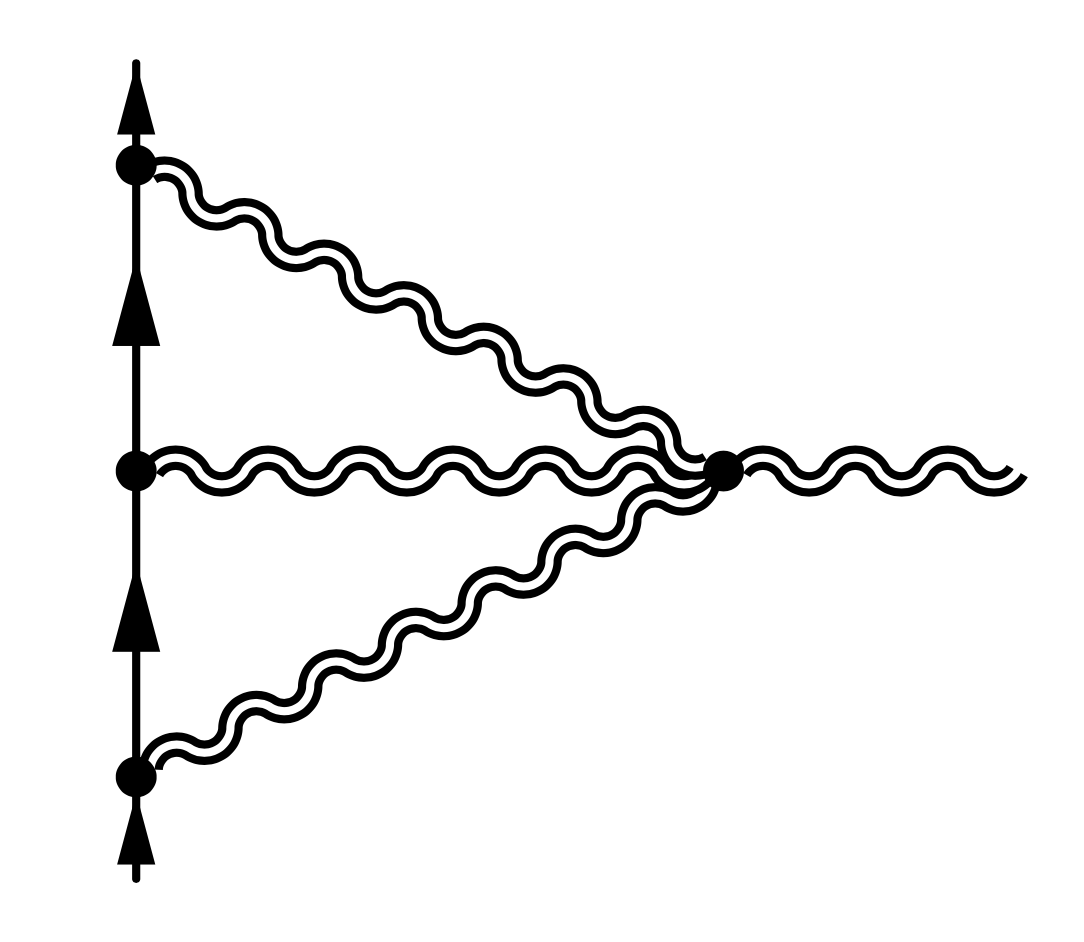}
\caption{2-loop diagrams for graviton emission with only graviton internal lines.}
\label{2Loop_GravitonEmission_OnlyGrav}
\end{figure}

The third post-Minkowskian order of the metric is given by the 2-loop diagrams of the type in \eqref{Generic_Example_Process}, which correspond to $l=2$ and are divided in $j=0$ and $j=2$.
The pure graviton case ($j=0$) corresponds to the diagrams in which there are only graviton internal lines, which are given in figure \ref{2Loop_GravitonEmission_OnlyGrav}.
The computation of these diagrams was carried out in \cite{Mougiakakos:2020laz} and we report here the corresponding form factors
\begin{equation}
\begin{aligned}
c_{1}^{(2,0)}(d)=&\frac{32}{3(d-4)(d-1)^{3}}\left(9 d^{4}-70 d^{3}+203 d^{2}-254 d+104\right)\\
c_{2}^{(2,0)}(d)=&\frac{64(d-2)}{3(d-4)(d-1)^{3}}\left(2 d^{3}-13 d^{2}+25 d-10\right) \ .
\end{aligned}
\end{equation}
From these, one computes the metric components
\begin{equation}
\begin{aligned}
h_0^{(3,0)}(r) =&\frac{8 (7-3 d) (d-2)^3 }{(d-4) (d-1)^3}m^3G_N^3 \rho^3\\
h_1^{(3,0)}(r) =& \frac{8 \left(7 d^4-63 d^3+214 d^2-334 d+212\right) }{3 (d-4) (d-3) (d-1)^3}m^3G_N^3 \rho^3\\
 h_2^{(3,0)}(r) =& \frac{8 (d-2)^2 \left(-2 d^3+13 d^2-25 d+10\right) }{(d-4) (d-3) (d-1)^3}m^3G_N^3 \rho^3 \ .
\end{aligned}
\end{equation}
This result is in agreement with the post-Minkowskian expansion of the metric \eqref{hid} for the terms proportional to $ m^3G_N^3\rho^3$. In particular, they are all divergent in five dimensions, while $h_1^{(3,0)}$ and $h_2^{(3,0)}$ are also divergent in four dimensions.

\medskip

In the case  $j=2$ we have to consider diagrams with two photons and one graviton emitted from the scalar line. 
At this loop order it is important to consider the multiplicity factor of each diagram, in fact since the classical limit of the amplitudes is independent of the exchange between photon and graviton lines emitted by the massive scalar, we have to consider all the permutations between them, and weight the amplitudes with the right factor. Since the vertices defined in appendix \ref{Appendix_FeynRules} are all written with the right multiplicity factor inside, this is the only combinatorial observation we have to do.  
Finally there are in total three diagrams that we must take into account, and we will compute the stress-energy tensor of each diagram and then sum all of them.

\medskip

\begin{figure}[h]
\centering
\includegraphics[width=0.9\textwidth, valign=c]{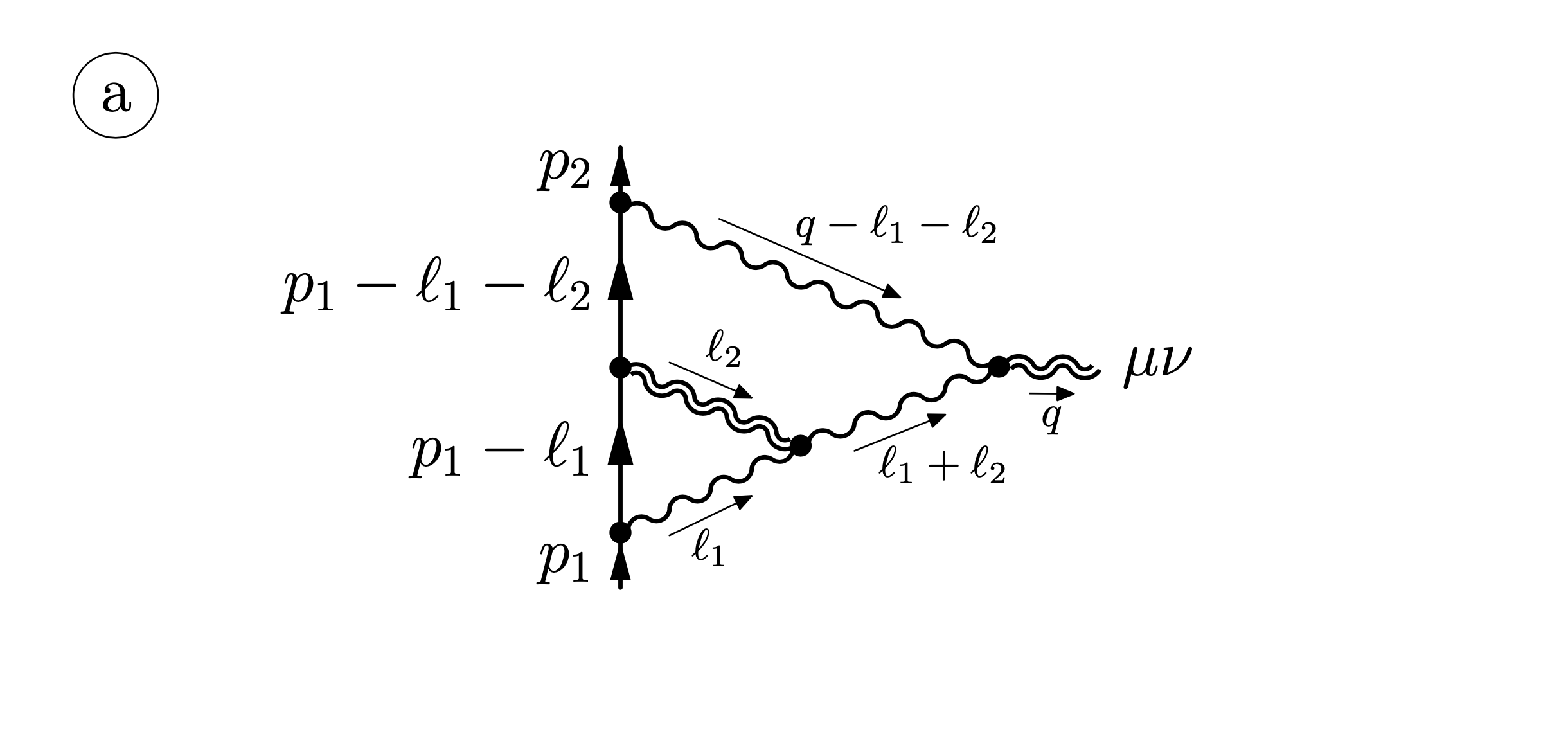}
\caption{2-loop diagram for graviton emission with 1 internal graviton and 3 internal
photons.}
\label{2Loop_GravitonEmission_a}
\end{figure}
The first diagram is given in figure \ref{2Loop_GravitonEmission_a}, and its  contribution to the stress-energy tensor is
\begin{equation}
    \begin{aligned}
    &-\frac{i\kappa}{2} \sqrt{4E_1 E_2} T^{(2,2)(\text{a})}_{\mu\nu}(q^2) \\
    & =\,6  \times\int \frac{d^{d+1}\ell_1}{(2\pi)^{d+1}}\frac{d^{d+1}\ell_2}{(2\pi)^{d+1}}\frac{\left(\tau_{\phi^2 A}\right)^\alpha
    \left(\tau_{\phi^2 A}\right)^\beta \left(\tau_{\phi^2 h}\right)^{\rho\sigma}}{\left((p_1-\ell_1)^2-m^2+i\epsilon\right)\left((p_1-\ell_1-\ell_2)^2-m^2+i\epsilon\right)} \\
    &\times\frac{P_{\rho\sigma,\eta\chi}
    \left(\tau_{A^2 h}\right)_{\mu\nu,\gamma,\alpha} (k,k')
    \left(\tau_{A^2 h}\right)^{\eta\chi}{}_{\beta}{}^\gamma (p,p')}{\left(\ell_1^2+i\epsilon\right)\left(\ell_2^2+i\epsilon\right)\left((\ell_1+\ell_2)^2+i\epsilon\right)\left((\ell_1+\ell_2-q)^2+i\epsilon\right)} 
    \ ,
    \end{aligned}
\end{equation}
where we introduced the new momenta 
\begin{equation}\label{TwoLoop_newmomenta}
\begin{aligned} 
    p&=\ell_1 &k&=\ell_1+\ell_2 \\
    p'&=\ell_1+\ell_2 &k'&=\ell_1+\ell_2-q \ ,
\end{aligned}
\end{equation}
and where the factor multiplying the integral is the multiplicity of the diagram. Then from the analysis of section \ref{sec:Classical_Limit_Grav_Emission_Processes} one obtains
\begin{equation}
\begin{aligned}\label{T22aafterl0integral}
      T^{(2,2)(\text{a})}_{\mu\nu}(\vec{q}^2) =& mQ^2\int \frac{d^d\vec{\ell}_1}{(2\pi)^d}\frac{d^d\vec{\ell}_2}{(2\pi)^d} \frac{\left(\tau_{A^2h}\right)_{\mu\nu,\gamma,0}(k,k')P_{00,\eta\chi}\left(\tau_{A^2h}\right)^{\eta\chi}{}_{0}{}^\gamma(p,p')\bigl|_{\ell_i^0=0}}{\vec{\ell}_1^2\vec{\ell}_2^2(\vec{\ell}_1+\vec{\ell}_2-\vec{q})^2(\vec{\ell}_1+\vec{\ell}_2)^2} \ ,
\end{aligned}
\end{equation}
whose numerator, using the explicit expressions of the vertices, becomes
\begin{equation}
\begin{aligned}
   \left(\tau_{A^2h}\right)_{\mu\nu,\gamma,0}(k,k')&P_{00,\eta\chi}\left(\tau_{A^2h}\right)^{\eta\chi}{}_{0}{}^\gamma(p,p')\bigl|_{\ell_i^0=0}\\&= -\kappa^2 \frac{d-2}{d-1}p\cdot p' \Biggl(k\cdot k' \left(\delta^0_\mu \delta^0_\nu - \frac{1}{2}\eta_{\mu\nu}\right) + \frac{1}{2}(k_\mu k'_\nu + k_\nu k'_\mu)\Biggl) \ .
\end{aligned}
\end{equation}
Substituting this in eq. \eqref{T22aafterl0integral} and using the identities in appendix \ref{App:LoopRed}, rewriting the stress-energy tensor according to eq. \eqref{Tmunu_Separated}, one gets
\begin{equation}
\begin{gathered}
T^{(2,2)(\text{a})}_{00}(\vec{q}^2)=\frac{1}{12}\frac{d-2}{d-1}mQ^2\kappa^2 J_{(2)}(\vec{q}^2) \\
    \delta^{ij}T^{(2,2)(\text{a})}_{ij}(\vec{q}^2) = \frac{1}{12}\frac{(d-2)^2}{d-1}mQ^2\kappa^2  J_{(2)}(\vec{q}^2) \ ,
\end{gathered}
\end{equation}
which we will add to the other 2-loop contributions in order to obtain the final form factors to be replaced inside \eqref{metric_pert}.

\medskip

\begin{figure}[h]
\centering
\includegraphics[width=0.9\textwidth, valign=c]{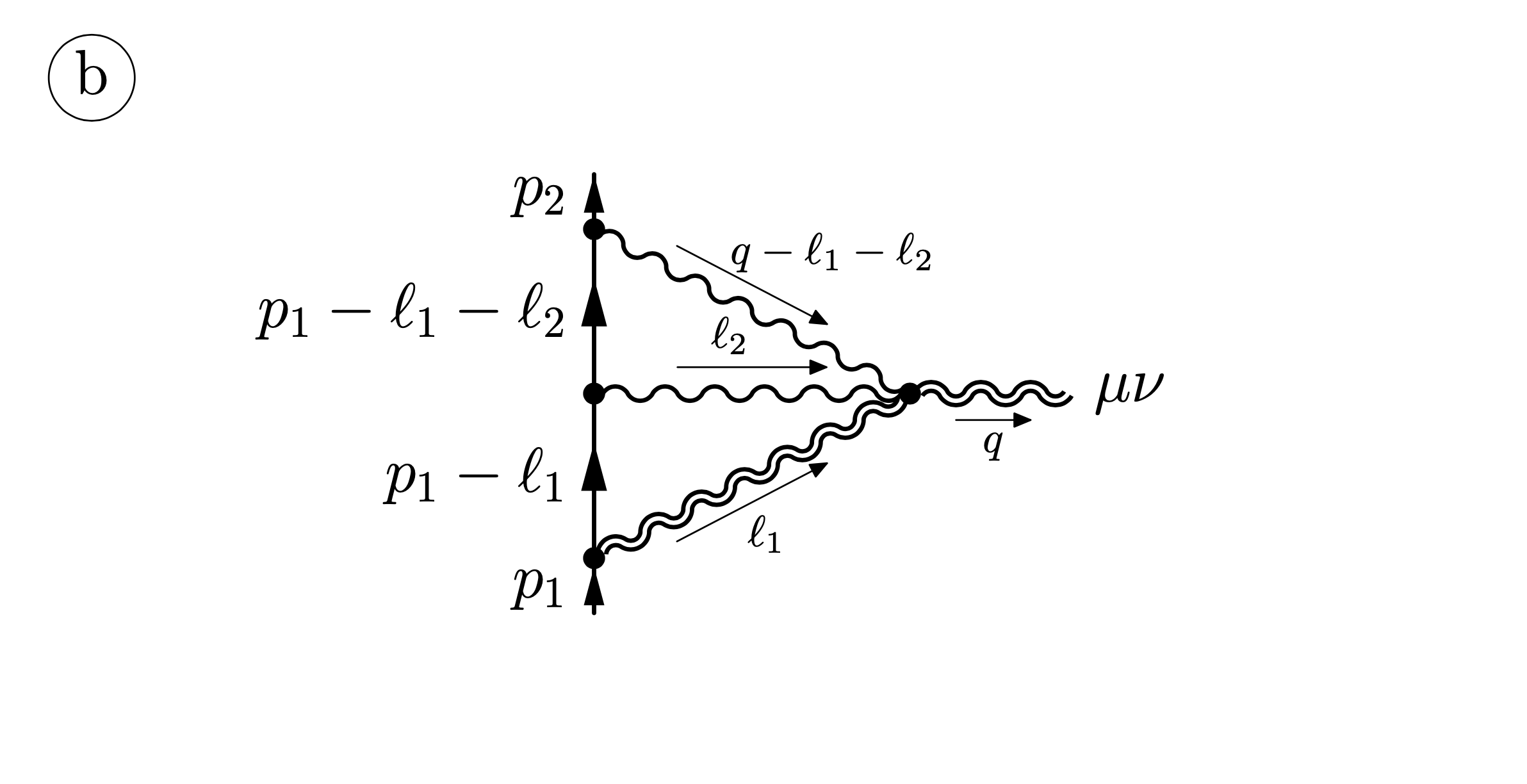}
\caption{2-loop diagram for graviton emission with 1 internal graviton and 2 internal
photons.}
\label{2Loop_GravitonEmission_b}
\end{figure}
The second diagram that we consider is the one in figure \ref{2Loop_GravitonEmission_b}, which gives the contribution to the stress-energy tensor
\begin{equation}
    \begin{aligned}
    & -\frac{i\kappa}{2}\sqrt{4E_1 E_2} T^{(2,2)(\text{b})}_{\mu\nu}(q^2) \\ &=3\times  \int \frac{d^{d+1}\ell_1}{(2\pi)^{d+1}}\frac{d^{d+1}\ell_2}{(2\pi)^{d+1}}\frac{i\left(\tau_{\phi^2 A}\right)^\alpha
    \left(\tau_{\phi^2 A}\right)^\beta \left(\tau_{\phi^2 h}\right)^{\rho\sigma}}{\left((p_1-\ell_1)^2-m^2+i\epsilon\right)\left((p_1-\ell_1-\ell_2)^2-m^2+i\epsilon\right)} \\
    &\times\frac{P_{\rho\sigma,\gamma\delta}
    \left(\tau_{A^2 h^2}\right){}_{\mu\nu}{}^{\gamma\delta}{}_{\alpha,\beta} (p,p')}{\left(\ell_1^2+i\epsilon\right)\left(\ell_2^2+i\epsilon\right)\left((\ell_1+\ell_2-q)^2+i\epsilon\right)}\ ,
\end{aligned}\end{equation}
where we defined the momenta $p=\ell_2$ and $p' = \ell_1+\ell_2-q$ and where $(\tau_{A^2h^2})$ is the 2 photons - 2 gravitons vertex in \eqref{VertA2h2}.
Imposing the classical limit to the above equation one obtains
\begin{equation}
\begin{aligned} \label{T22bafterintegration}
     T^{(2,2)(\text{b})}_{\mu\nu}(\vec{q}^2) =&\,\frac{i}{2}\,m\,Q^2\int \frac{d^d\vec{\ell}_1}{(2\pi)^d}\frac{d^d\vec{\ell}_2}{(2\pi)^d} \frac{P_{00,\gamma\delta} \left(\tau_{A^2 h^2}\right){}_{\mu\nu}{}^{\gamma\delta}{}_{0,0} (p,p')\bigl|_{\ell_i^0=0}}{\vec{\ell}_1^2\vec{\ell}_2^2(\vec{\ell}_1+\vec{\ell}_2-\vec{q})^2} \ ,
\end{aligned}
\end{equation}
from which the numerator explicitly gets the expression
\begin{equation}
    \begin{aligned}
  &P_{00,\gamma\delta}\left(\tau_{A^2 h^2}\right){}_{\mu\nu}{}^{\gamma\delta}{}_{0,0} (p,p')\bigl|_{\ell_i^0=0} \\&=-\frac{i\kappa^2}{4}\left(\frac{2(d-3)}{d-1}\left(p_\mu p'_\nu + p_\nu p'_\mu-p\cdot p' \eta_{\mu\nu}\right) + \frac{2(3d-7)}{d-1}p\cdot p' \delta^0_\mu\delta^0_\nu\right) \ .
\end{aligned}
\end{equation}
Finally, substituting the above equation in eq. \eqref{T22bafterintegration}, the resulting components of the stress-energy tensor are 
\begin{equation}
\begin{gathered}
     T^{(2,2)(\text{b})}_{00}(\vec{q}^2)  =-\frac{1}{12}\frac{d-2}{d-1} Q^2m\kappa^2J_{(2)}(\vec{q}^2) \\
         \delta^{ij}T^{(2,2)(\text{b})}_{ij}(\vec{q}^2) = -\frac{1}{24}\frac{(d-3)(d-2)}{d-1} mQ^2\kappa^2 J_{(2)}(\vec{q}^2) \ .
\end{gathered}
\end{equation}

\medskip

\begin{figure}[h]
\centering
\includegraphics[width=0.9\textwidth, valign=c]{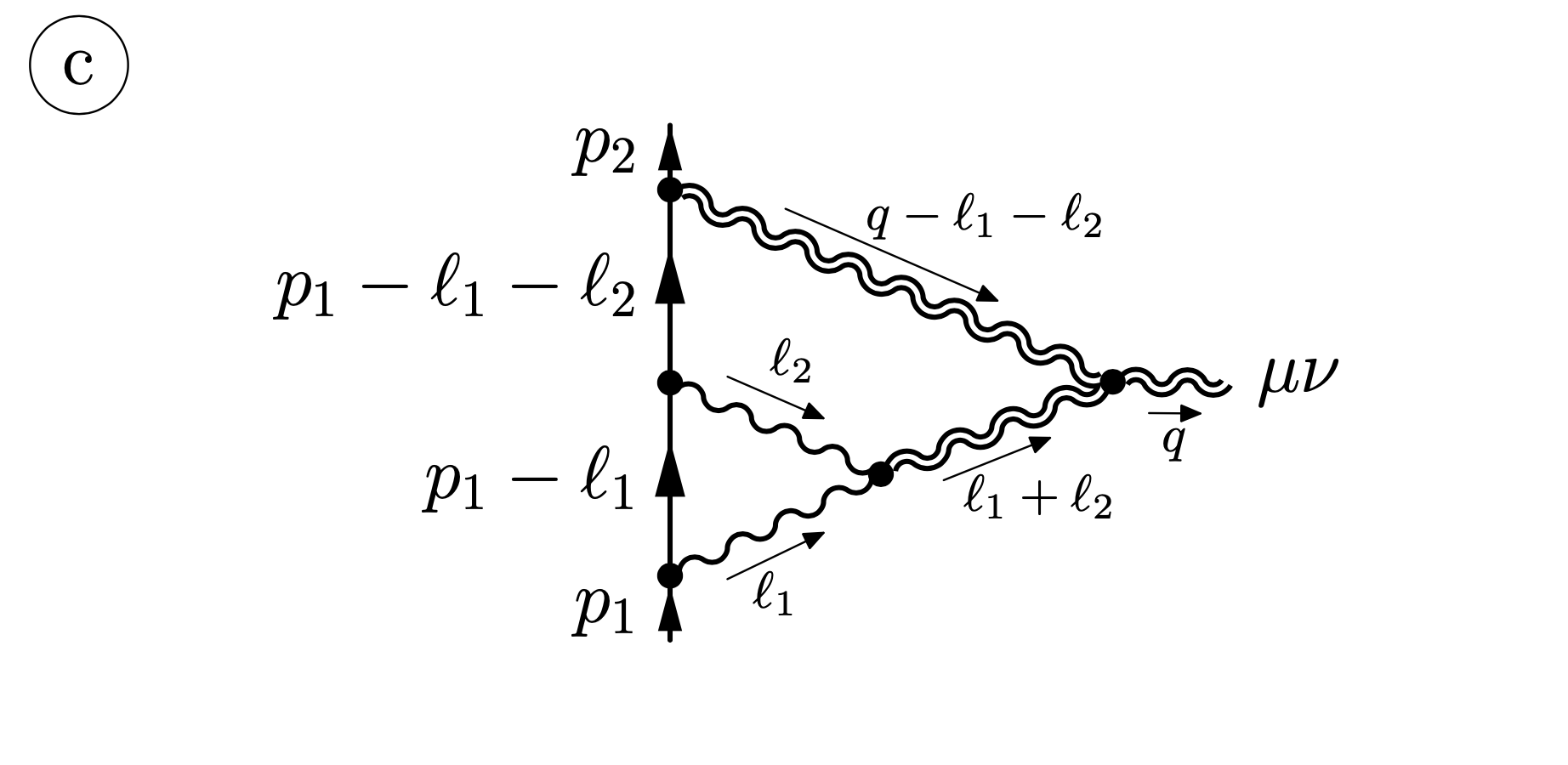}
\caption{2-loop diagram for graviton emission with 2 internal gravitons and 2 internal
photons.}
\label{2Loop_GravitonEmission_c}
\end{figure}
The last diagram for the emission of a graviton at 2-loop order is the one in figure \ref{2Loop_GravitonEmission_c}. The corresponding stress-energy tensor is 
\begin{equation}
    \begin{aligned}
    &-\frac{i\kappa}{2}\sqrt{4E_1 E_2} T^{(2,2)(\text{c})}_{\mu\nu}(q^2) \\ 
    &=3 \times  \int \frac{d^{d+1}\ell_1}{(2\pi)^{d+1}}\frac{d^{d+1}\ell_2}{(2\pi)^{d+1}}\frac{-\left(\tau_{\phi^2 A}\right)^\alpha
    \left(\tau_{\phi^2 A}\right)^\beta \left(\tau_{\phi^2 h}\right)^{\rho\sigma}}{\left((p_1-\ell_1)^2-m^2+i\epsilon\right)\left((p_1-\ell_1-\ell_2)^2-m^2+i\epsilon\right)} \\&\times
    \frac{P_{\rho\sigma}{}^{\eta\chi}P_{\theta\xi}{}^{\gamma\delta}
    \left(\tau_{A^2 h}\right)^{\theta\xi}{}_{\alpha,\beta} (p,p') \left(\tau_{h^2h}\right)_{\mu\nu,\gamma\delta,\eta\chi} (k,k')}{\left(\ell_1^2+i\epsilon\right)\left(\ell_2^2+i\epsilon\right)\left((\ell_1+\ell_2)^2+i\epsilon\right)\left((\ell_1+\ell_2-q)^2+i\epsilon\right)} 
    \ ,
    \end{aligned}
\end{equation}
where we have introduced the new momenta 
\begin{equation}\label{TwoLoop_newmomentaC}
\begin{aligned} 
    p&=\ell_1 &k&=\ell_1+\ell_2 \\
    p'&=-\ell_2 &k'&=q \ .
\end{aligned}
\end{equation}
Employing the considerations of section \ref{sec:Classical_Limit_Grav_Emission_Processes} then one obtains 
\begin{equation}\label{Twoloop_grav_EMTnumB}
\begin{aligned}
     T^{(2,2)(\text{c})}_{\mu\nu}(\vec{q}^2) =&-\frac{1}{2}mQ^2\int \frac{d^d\vec{\ell}_1}{(2\pi)^d}\frac{d^d\vec{\ell}_2}{(2\pi)^d} \frac{\left(\tau_{A^2h}\right)^{\alpha\beta}{}_{0,0}(p,p')P_{\alpha\beta}{}^{\gamma\delta}P_{00}{}^{\rho\sigma}\left(\tau_{h^2h}\right)_{\mu\nu,\gamma\delta,\rho\sigma} (k,k')\bigl|_{\ell_i^0=0}}{\vec{\ell}_1^2\vec{\ell}_2^2(\vec{\ell}_1+\vec{\ell}_2-\vec{q})^2(\vec{\ell}_1+\vec{\ell}_2)^2} \ .
\end{aligned}
\end{equation}
Introducing the vertex $\left(\tilde{\tau}_{h^2h}\right)$ defined as
\begin{equation}\label{VertTildeDef}
    \left(\tilde{\tau}_{h^2h}\right){}^{\mu\nu}{}_{\alpha\beta,\gamma\delta} = {{P}_{\alpha\beta}}^{\rho\sigma}{{P}_{\gamma\delta}}^{\eta\xi}\left(\tau_{h^2h}\right){}^{\mu\nu}{}_{\rho\sigma,\eta\xi} \ ,
\end{equation}
where detailed properties are given in \cite{Mougiakakos:2020laz, DOnofrio:2021tap}, the numerator of the integrand can be written as
\begin{equation}
    \begin{aligned}
    \left(\tau_{A^2h}\right)^{\alpha\beta}{}_{0,0}(p,p')P_{\alpha\beta}{}^{\gamma\delta}P_{00}{}^{\rho\sigma}\left(\tau_{h^2h}\right)_{\mu\nu,\gamma\delta,\rho\sigma} (k,k') = \left(\tau_{A^2h}\right)^{\alpha\beta}{}_{0,0}(p,p')\left(\tilde{\tau}_{h^2h}\right)_{\mu\nu,\alpha\beta,00} (k,k') \ ,
    \end{aligned}
\end{equation}
from which an explicit calculation leads to 
\begin{equation}
    \begin{aligned}
     &\left(\tau_{A^2h}\right)^{\alpha\beta}{}_{0,0}(p,p')\left(\tilde{\tau}_{h^2h}\right)_{00,\alpha\beta,00} (k,k')\bigl|_{\ell_i^0=0}\\ &=\frac{\kappa^2}{4}\frac{1}{d-1}\left(p\cdot p' \left(-dk'^2+ \frac{(3d-3)(d-2)}{d-1}k^2+(d-2)(k-k')^2\right)+ (4d-8) (k'\cdot p) (k'\cdot p')\right)
    \end{aligned}
\end{equation}
for the $00$-component, and to
\begin{equation}
    \begin{aligned}
     \delta^{ij}&\left(\tau_{A^2h}\right)^{\alpha\beta}{}_{0,0}(p,p')\left(\tilde{\tau}_{h^2h}\right)_{ij,\alpha\beta,00} (k,k')\bigl|_{\ell_i^0=0}\\ &=-\frac{\kappa^2}{4} \frac{d-2}{d-1} \left(p\cdot p'\left((d-2) k^2 +(d-4) (k-k')^2 + 3d k'^2\right) + 4 ( k'\cdot p) (k'\cdot p')\right)
    \end{aligned}
\end{equation}
for the trace of the spatial components.
Substituting the above expressions in \eqref{Twoloop_grav_EMTnumB}, inserting the momenta \eqref{TwoLoop_newmomentaC} and using the loop reduction identities in appendix \ref{App:LoopRed} (in particular \eqref{LR_Loop2_l1q_l2q}), we have 
\begin{equation}
\begin{gathered}
     T^{(2,2)(\text{c})}_{00}(\vec{q}^2) =\frac{(3d^2-19d+32)}{24(d-4)(d-1)} mQ^2\kappa^2J_{(2)}(\vec{q}^2) \\
          \delta^{ij}T^{(2,2)(\text{c})}_{ij}(\vec{q}^2) =\frac{(d-2)(5d^2-27d+30)}{24(d-4)(d-1)}mQ^2\kappa^2J_{(2)}(\vec{q}^2) \ .
\end{gathered}
\end{equation}

\medskip

Finally, summing all the contributions, we obtain the total stress-energy tensor 
\begin{equation}
    T^{(2,2)}_{\mu\nu}(\vec{q}^2)=\sum_{i=\text{a}, \text{b}, \text{c}}T^{(2,2)(i)}_{\mu\nu}(\vec{q}^2)\ ,
\end{equation}
that explicitly it reads
\begin{equation}\label{Twoloop_grav_EMTcomponents}
        \begin{gathered}
   T_{00}^{(2,2)}(\vec{q}^2) = \frac{(3d^2-19d+32)}{24(d-4)(d-1)}mQ^2\kappa^2J_{(2)}(\vec{q}^2)\\
   \delta^{ij}T^{(2,2)}_{ij} (\vec{q}^2)= \frac{ (d-2) (3 d^2-16 d+17)}{ 12(d-4) (d-1)}{mQ^2\kappa^2}J_{(2)}(\vec{q}^2) \ .
    \end{gathered}
\end{equation}
Then from a direct comparison with eq. \eqref{Tmunu_Separated} we can easily read the coefficients $c^{(l,j)}_i(d)$, obtaining
\begin{equation}
        \begin{aligned}
   c_1^{(2,2)}(d) &= \frac{16 \left(9 d^3-66 d^2+149 d-100\right)}{3(d-4) (d-1)^2} \\
        c_2^{(2,2)}(d) &=  \frac{32 (d-2) \left(3 d^2-16 d+17\right)}{3 (d-4) (d-1)^2} \ .
    \end{aligned}
\end{equation}
From these coefficients, using eq. \eqref{metric_pert} we finally obtain the metric components as
\begin{equation}
    \begin{aligned}
    h_0^{(3,2)} (r) =& -\frac{4(d-2)^2(3d-11)}{(d-4)(d-1)^2}m\alpha G_N^2{\rho}^3\\
    h_1^{(3,2)}(r)=&\frac{8(3d^3-25d^2+69d-65)}{3(d-4)(d-3)(d-1)^2} m\alpha G_N^2{\rho}^3\\
    h_2^{(3,2)}(r) = & -\frac{4(d-2)^2(3d^3-19d^2+33d-17)}{(d-4)(d-3)(d-1)^3}m\alpha G_N^2 {\rho}^3 \ .
    \end{aligned}
\end{equation}
We can compare this result with the part proportional to $ m\alpha G_N^2$ of \eqref{hid} and observe that they match perfectly. In particular, all these terms diverge in $d=4$, and $ h_1^{(3,2)}$ and $  h_2^{(3,2)} $ also diverge in $d=3$.

\medskip

As a summary, we have shown that  the loop computations give exactly the expression for the metric of the Reissner-Nordstr\"om-Tangherlini solution in de Donder gauge given in eq. \eqref{hid}. 
However at third post-Minkowskian order in $d=3$ arise divergences in $h_1$ and $h_2$, while in $d=4$ all the components diverge. To obtain the values of the metrics component in these dimensions  we need to perform a renormalization procedure which will be described in  the next section.

\section{Renormalization of the metric}
\label{sec:Renorm_Metric}

In the previous section we noticed the appearance of divergences in  $d=3$ and $d=4$, and now we show how such divergences are renormalized by non-minimal couplings. After regularizing the singularities that appear in the metric explicitating the poles structure, firstly we analyze the non-minimal couplings that are needed to be added to the original theory and the consequent variation of the metric, then we use these results to renormalize the divergences.

\subsection{Regularization}

The regularization procedure consists only in isolating the poles through dimensional regularization. Since the observable is already expressed in generic space-time dimension we only need to expand the parameter $d$ around the problematic values, as is usual in the regularization process of any field theory. 

\subsection*{$\bm{d=3}$ case}

In this dimension the only divergences that appear are at 2-loop order.
From the explicit expression of the 2-loop contribution of the metric in $d=3$, one can notice that $h_0^{(d=3)}\bigl|_{2-\text{loop}}$ is non divergent and can be already compared with the \eqref{hi3} with a perfect match. After the regularization process, the other two components turn out to be equal up to a multiplicative constant, in particular we have $h_2^{(d=3)}\bigl|_{2-\text{loop}}=-3h_1^{(d=3)}\bigl|_{2-\text{loop}}$, so we will renormalize only the $h_1^{(d=3)}\bigl|_{2-\text{loop}}$ component and then obtain the other one using this relation. We perform a dimensional regularization using $d=3+\epsilon$, where $\epsilon$ is a small parameter. Neglecting terms that vanish when $\epsilon \rightarrow 0$, the result is
\begin{equation}
\begin{aligned}
    h_1^{(d=3)}\bigl|_{2-\text{loop}}& =  \frac{1}{\epsilon} \left(\frac{4}{3}\frac{m\alpha G_N^2}{r^3} -\frac{2}{3}\frac{m^3G_N^3}{r^3}\right) +\frac{m^3G_N^3}{r^3}\left(2\log \left(2C_Er\right)-\frac{4}{3}\right)  \\ & -4\frac{m\alpha G_N^2}{r^3}\log \left(2C_Er\right)  \ ,
\end{aligned}
\end{equation}
where following \cite{Mougiakakos:2020laz} we defined the constant $C_E^2= \pi e^{\gamma_{EM}}$, with $\gamma_{EM}$ the Euler-Mascheroni constant. 

\subsection*{$\bm{d=4}$ case}
In $d=4$ the metric is divergent at both 1-loop and 2-loop order.
Again, from the explicit form of the metric, we can see that at 1-loop order the $h_0^{(d=4)}\bigl|_{1-\text{loop}}$ component is non-divergent and no regularization procedure has to be carried on. The $h_1^{(d=4)}\bigl|_{1-\text{loop}}$ and $h_2^{(d=4)}\bigl|_{1-\text{loop}}$ components are related by the identity $h_2^{(d=4)}\bigl|_{1-\text{loop}}=-4h_1^{(d=4)}\bigl|_{1-\text{loop}}$ and so we can focus just on one component. Regularizing in $d=4+\epsilon$, and neglecting again terms that vanish when $\epsilon \rightarrow 0$, the result at 1-loop is
\begin{equation}
\begin{aligned}
     h_1^{(d=4)}\bigl|_{1-\text{loop}} & =\frac{1}{\epsilon}\left(-\frac{2}{3}\frac{\alpha G_N}{\pi^2r^4}-\frac{40}{9}\frac{G_N^2m^2}{\pi^2r^4}\right)+\frac{m^2 G_N^2}{\pi^2r^4}\left(-\frac{4}{27}+\frac{40}{9}\log \left(C_E^2 r^2 \right)\right)  \\& +\frac{\alpha G_N}{\pi^2r^4} \left(-\frac{4}{9}+\frac{2}{3}\log \left(C_E^2 r^2 \right)\right)    \ .
\end{aligned}
\end{equation}
For what concerns the 2-loop order, we can focus just on the $h_1^{(d=4)}\bigl|_{2-\text{loop}}$ component since the relation $h_0^{(d=4)}\bigl|_{2-\text{loop}}=h_2^{(d=4)}\bigl|_{2-\text{loop}}=-2h_1^{(d=4)}\bigl|_{2-\text{loop}}$ holds, and after the regularization we obtain
\begin{equation}
\begin{aligned}
    h_1^{(d=4)}\bigl|_{2-\text{loop}}&=\frac{1}{\epsilon} \left(\frac{8}{9}\frac{m\alpha G_N^2}{\pi^3r^6} +\frac{160}{27}\frac{m^3G_N^3}{\pi^3r^6}\right) + \frac{m ^3G_N^3 }{\pi^3 r^6} \left(\frac{208}{81}-\frac{80}{9}\log \left(C_E^2 r^2 \right)  \right) \\&+ \frac{\alpha m G_N^2}{\pi^3 r^6}\left(\frac{64}{27}-\frac{4}{3}\log \left(C_E^2 r^2\right)  \right) \ .
\end{aligned}
\end{equation}

\subsection{Non-minimal couplings}

Now that the poles are isolated we need to cancel the singularities. As in any other quantum field theory this is done by the so called renormalization procedure, in which exploiting the fact that the couplings of the theory that appear in the original action are not the physical ones,  it is possible to define some counter-terms that introduce new Feynman rules by which one can eliminate the divergences. However, since we are treating an effective field theory which is non-renormalizable, we expect that it is not possible to perform such procedure only considering the original Lagrangian, and in fact do not exist such counter-terms that eliminate the poles. Nevertheless, according to the principles of effective field theories, it is possible to add some non-minimal couplings that are higher-order in energy with the purpose of canceling such divergences.
As shown in \cite{Mougiakakos:2020laz}, such non-minimal coupling terms are linear in the Riemann tensor and quadratic in the first derivative of the scalar field, whose explicit form is
\begin{equation}\label{NonMinimal_Action}
    \begin{split}
        \delta^{(\mathfrak{n})}S^{ct} =& \sum_{k=0}^{+\infty} (G_Nm)^{\frac{2(\mathfrak{n}-k)}{d-2}}(\alpha G_N)^{\frac{k}{d-2}} \int d^{d+1}x \sqrt{-g}\Bigl(\alpha^{(\mathfrak{n},k)}(d)(D^2)^{\mathfrak{n}-1}RD_{\mu}\phi D^{\mu}\phi \\
        &+\left(\beta_0^{(\mathfrak{n},k)}(d)D_{\mu}D_{\nu}(D^2)^{\mathfrak{n}-2}R + \beta_1^{(\mathfrak{n},k)}(d)(D^2)^{\mathfrak{n}-1}R_{\mu\nu}\right)D^{\mu}\phi D^{\nu}\phi\Bigr)\ ,
    \end{split}
\end{equation}
where $D_{\mu}$ is the covariant derivative, and where we consider only positive integer powers of the gravitational and electromagnetic coupling. It is important to notice that the above action is not the most generic one. In fact, here and in the following, we will only consider the addition of the non-minimal terms useful to the renormalization of the theory, since every other operator belongs to high-energy physics.

\medskip

Due to the properties of the Fourier transforms, terms with $\mathfrak{n}\geq 2$ and terms proportional to $\beta_0^{(\mathfrak{n},k)}$ and $\beta_1^{(\mathfrak{n},k)}$ do not contribute to the classical limit of the metric \cite{Mougiakakos:2020laz}, and this statement holds even for the present electromagnetically coupled theory. Defining  
\begin{equation}\label{alpha}
    \alpha^{(\mathfrak{n})}(d) = \sum_{k=0}^{+\infty} (G_Nm)^{\frac{-2k}{d-2}}(\alpha G_N)^{\frac{k}{d-2}}\alpha^{(\mathfrak{n},k)}(d)\ 
\end{equation}
for the sake of simplicity, and $\beta_0^{(\mathfrak{n})}$ and $\beta_1^{(\mathfrak{n})}$ in the same way, thus the only contribution to the non-minimal coupling action that will renormalize the metric is
\begin{equation}\label{theonlynonminimalcoupling}
\delta^{(1)}S^{ct} = (G_N m)^{\frac{2}{d-2}}\alpha^{(1)}(d)\int d^{d+1}x  \sqrt{-g}\, R\,  D_{\mu}\phi D^{\mu}\phi \ . 
\end{equation}

\medskip

In the following, we compute the Feynman rules associated to the counter-terms only in the case $\mathfrak{n}=1$, but for completeness we give the complete relations considering also the term proportional to $\beta_1^{(1)}$. Considering the interaction between two scalars and one graviton that arise from the action in eq. \eqref{NonMinimal_Action}, for $\mathfrak{n}=1$ one gets
\begin{equation}\label{Massive_CT_1}
\begin{aligned}
&\includegraphics[width=0.25\textwidth, valign=c]{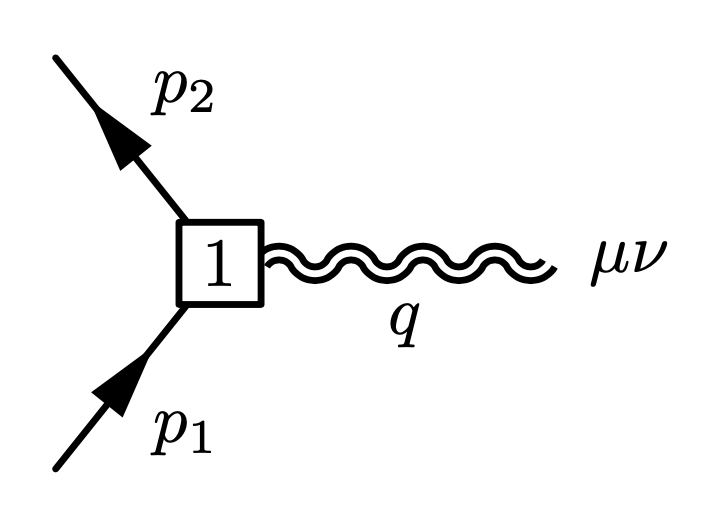}= \left(\tau_{\phi^2 h}^{ct}\right)_{\mu\nu}(q)  \\
&=(G_N m)^{\frac{2}{d-2}} i\kappa \left(\alpha^{(1)}(d)\left(-q_{\mu}q_{\nu} + \eta_{\mu\nu}q^2\right) p_1\cdot p_2 + \beta^{(1)}_1(d) \frac{q^2}{2}  p_{1\,\mu}p_{2 \, \nu} \right)\ ,
\end{aligned}
\end{equation}
whose classical limit is
\begin{equation}\label{MassiveCT}
    \left(\tau_{\phi^2 h}^{ct}\right)_{\mu\nu}(q) \simeq -(G_N m)^{\frac{2}{d-2}} i\kappa m^2 \left(\alpha^{(1)}(d)\left(q_{\mu}q_{\nu} + \eta_{\mu\nu}\vec{q}^2\right) + \beta^{(1)}_1(d) \frac{\vec{q}^2}{2}  \delta_{\mu}^{0}\delta_{\nu}^0 \right) \ .
\end{equation}
Another possible vertex is constituted by two photons and one graviton. It arises since we must take into account the electromagnetic gauge symmetry into the covariant derivative. Its contribution is then
\begin{equation}\label{Massless_CT_2}
\begin{aligned}
\includegraphics[width=0.25\textwidth, valign=c]{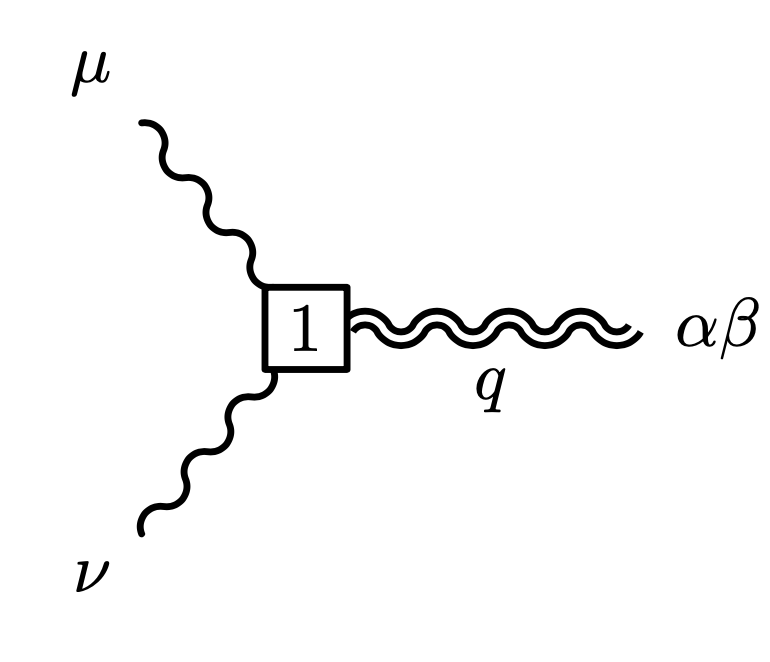}= \left(\tau_{A^2 h}^{ct}\right)_{\alpha\beta,\mu,\nu}(q) = &-(G_N m)^{\frac{2}{d-2}} Q^2  i\kappa \Bigl(\alpha^{(1)}(d)(-q_{\alpha}q_{\beta} + \eta_{\alpha\beta} q^2) 2\eta_{\mu\nu} \\
&+ \beta^{(1)}(d) \frac{q^2}{2}(\eta_{\mu\alpha} \eta_{\nu\beta} + \eta_{\nu\alpha} \eta_{\mu\beta})  \Bigr)\ .
\end{aligned}
\end{equation}
Now, we can use these counter-terms to compute the diagrams that will be necessary for the  renormalization process.

\medskip

\begin{figure}[h]
\centering
\includegraphics[width=0.25\textwidth, valign=c]{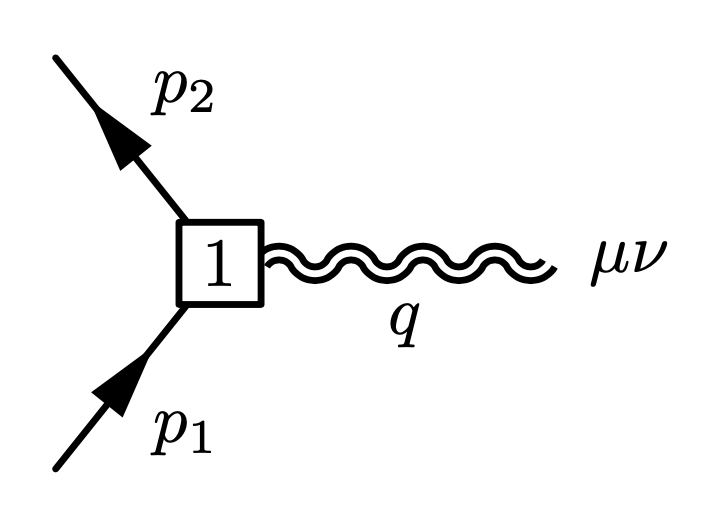}
\caption{Insertion of non-minimal coupling at tree level.}
\label{CT1}
\end{figure}

Following \cite{Mougiakakos:2020laz}, the insertion of the non-minimal coupling at tree-level gives the process in figure \ref{CT1}, and the associated contribution to the stress-energy tensor is given by
\begin{equation}
    \frac{-i\kappa}{2}\, \sqrt{4E_1E_2}\, \delta^{(1)}T_{\mu\nu}^{(0)}(q^2) = \left(\tau_{\phi^2 h}^{ct}\right)_{\mu\nu}(q) \ .
\end{equation}
Considering the relation between the stress-energy tensor and the metric perturbation, we obtain the variations of the metric components
\begin{equation}\label{CT1_Metric}
\begin{aligned}
    \delta^{(1)}h_0^{(1)}(r) &= 0 \\
    \delta^{(1)}h_1^{(1)}(r) &= \frac{16\alpha^{(1)}(d)\Gamma\left(\frac{d}{2}\right)}{\pi^{\frac{d-2}{2}}} \frac{(G_Nm)^{\frac{d}{d-2}}}{r^d} \\
    \delta^{(1)}h_2^{(1)}(r) &= \frac{-32\alpha^{(1)}(d)\Gamma\left(\frac{d+2}{2}\right)}{\pi^{\frac{d-2}{2}}} \frac{(G_Nm)^{\frac{d}{d-2}}}{r^d}\ ,
\end{aligned}
\end{equation}
where $\alpha^{(1)}(d)$ is defined in (\ref{alpha}). 
\begin{figure}[h]
\centering
\includegraphics[width=0.35\textwidth, valign=c]{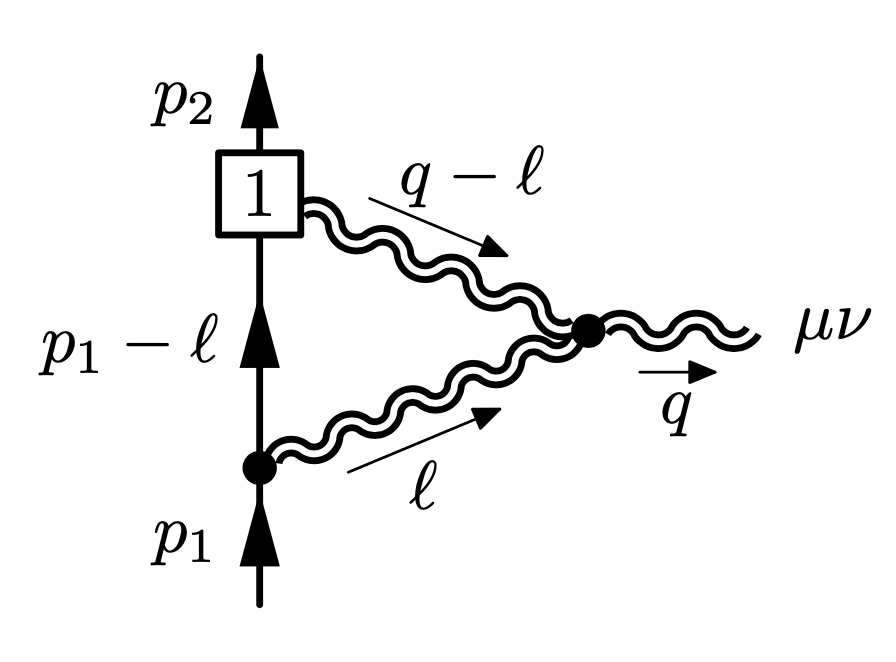}
\caption{Insertion of non-minimal coupling in the 1-loop diagram for graviton emission with 2 internal gravitons.}
\label{CT2}
\end{figure}
Then we can consider the insertion in the 1-loop diagram as in figure  \ref{CT2}, whose contribution to the stress-energy tensor is 
\begin{equation}
\begin{aligned}
    & -\frac{i\, \kappa}{2}\, \sqrt{4E_1E_2}\, \delta^{(1)}T_{\mu\nu}^{(1)}(q^2)=\\& =2\times  \int\frac{d^D\ell}{(2\pi)^D}\frac{-i\, P_{\rho\sigma,\alpha\beta}P_{\eta\xi,\gamma\delta}\left(\tau_{\phi^2h}\right)_{\rho\sigma} \left(\tau_{\phi^2 h}^{ct}\right)_{\eta\xi}(q-\ell)\left(\tau_{h^2h}\right){}^{\mu\nu}{}_{\alpha\beta,\gamma\delta}(l,q) }{((p_1-\ell)-m^2 + i\epsilon)(\ell^2+i\epsilon)((q-\ell)^2+i\epsilon)}
\end{aligned}
\end{equation}
from which using the property in \eqref{Master_CT}, we determine the variation of the metric as
\begin{align}
    \delta^{(1)}h_0^{(2)}(r) &= 64 \alpha^{(1)}(d) \frac{(d-2) \Gamma\left(\frac{d}{2}\right)^{2}}{(d-1) \pi^{d-2}}\left(\frac{\left(G_{N} m\right)^{\frac{1}{d-2}}}{r}\right)^{2(d-1)} \\
    \delta^{(1)}h_1^{(2)}(r) &= -64 \alpha^{(1)}(d) \frac{(d-2) \Gamma\left(\frac{d}{2}\right)^{2}}{(d-1) \pi^{d-2}}\left(\frac{\left(G_{N} m\right)^{\frac{1}{d-2}}}{r}\right)^{2(d-1)} \\
    \delta^{(1)}h_2^{(2)} (r)&= 128 \alpha^{(1)}(d) \frac{(d-2) \Gamma\left(\frac{d}{2}\right)^{2}}{(d-1) \pi^{d-2}}\left(\frac{\left(G_{N} m\right)^{\frac{1}{d-2}}}{r}\right)^{2(d-1)} \ .
\end{align}

\medskip

\begin{figure}[h]
\centering
\includegraphics[width=0.35\textwidth, valign=c]{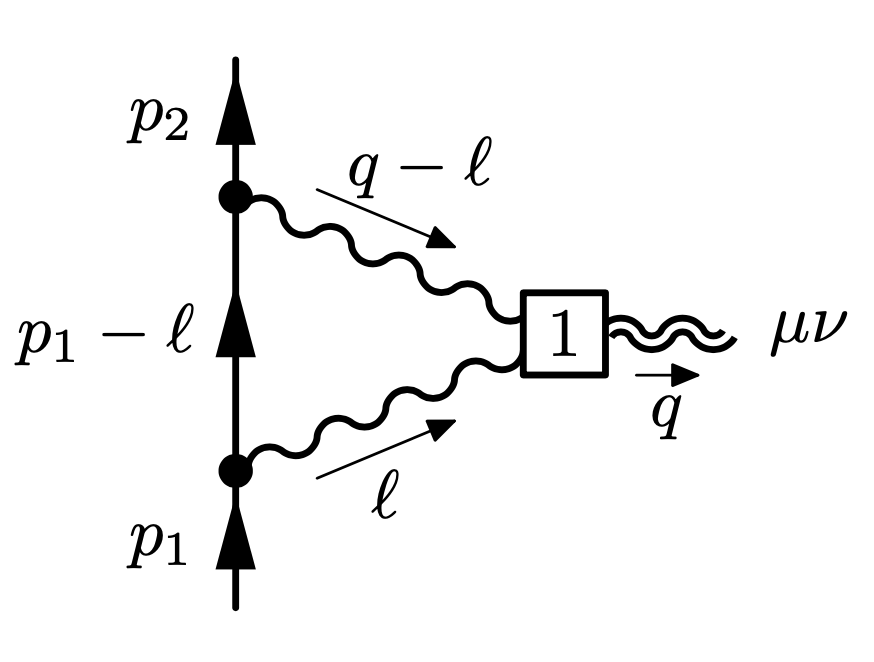}
\caption{Insertion of non-minimal coupling in the 1-loop diagram for graviton emission with 2 internal photons.}
\label{CT2_Bis}
\end{figure}

In principle there exists another 1-loop insertion given by the amplitude in figure \ref{CT2_Bis}.
However it turns out that the classical limit of the metric variation associated to such process is vanishing, and then does not contribute to the renormalization process of the classical observable. 

\subsection{Complete metric expression}

Now that we have the variations of the metric associated to the insertion of counter-terms in the amplitudes, we can use them to compute the complete metric expression in which the divergences are canceled. In order to do that we have to sum the right metric variation contribution to the regularized metric by checking the power of the coupling constants. Finally we can fix the parameter $\alpha^{(1)}(d)$, in which referred to eq. \eqref{alpha} do contribute only specific allowed values of $k$, imposing that the complete summed metric expression becomes finite. 

\subsection*{Renormalization in $\bm{d=3}$}

In four dimensions, the only values of $k$ that respect the constraints on the power of the coupling constant are $k=0,1,2$. However, the case $k=2$ does not have a match in the post-Minkowskian expansion and will not be considered.
As a consequence, eq. (\ref{alpha}) becomes
\begin{equation}\label{AlphaCTd3}
    \alpha^{(1)}(3) = \alpha^{(1,0)}(3) + \frac{\alpha G_N}{(G_Nm)^2}\alpha^{(1,1)}(3) \,.
\end{equation}
In order to cure the divergences, $\alpha^{(1)}(3)$ must have the form
\begin{equation}\label{alpha3}
    \alpha^{(1)}(3)=\frac{\omega(3)}{d-3}+\Omega(3)\ ,
\end{equation}
where $\omega(3)$ is the coefficient that we need to fix to renormalize the metric, while $\Omega(3)$ is a finite term. In $d=3$ the 2-loop corrections to the metric are given by the diagram in figure \ref{CT1} and, expanding the metric contributions (\ref{CT1_Metric}) in $d=3+\epsilon$, we find the renormalized metric component \cite{Mougiakakos:2020laz, DOnofrio:2021tap}
\begin{equation}
\begin{aligned}
    h_{1}^{(d=3)}& \bigl|^{renorm}_{2-loop}= h_{1}^{(d=3)}\bigl|_{2-loop} + \delta^{(1)}h_1^{(1)} \\& =\frac{8 G_N^3 m^3}{\epsilon r^3}\left(\omega(3) -\frac{1}{12}+\frac{\alpha}{6G_N m^2}\right)-\frac{4 G_N^3 m^3}{3 r^3}+\frac{8 G_N^3 m^3  }{r^3}\omega(3)+\frac{8 G_N^3 m^3  }{r^3}\Omega(3)\\&+\frac{2 G_N^3 m^3}{r^3} \log \left(2 C_E r\right)-\frac{4 \alpha  G_N^2 m }{r^3}\log \left(2 C_E r\right)-\frac{8 G_N^3 m^3 }{r^3}\omega(3)  \log \left(2 C_E r G_N^2 m^2 \right) \ .
\end{aligned}
\end{equation}
Imposing the above expression to be finite, we can fix $\omega(3)$ in order to cancel the divergent term in the metric component, obtaining for the divergent contribution
\begin{equation}
    \omega(3)=\frac{1}{12}-\frac{\alpha G_N}{6(G_N m)^2}\ .
\end{equation}
In the above expression we recognize the term independent of the electric charge to be equivalent to the one already found in \cite{Mougiakakos:2020laz}, while the other is the new piece due to electromagnetic interaction, which leads to
\begin{equation}
    \begin{aligned}
     h_{1}^{(d=3)}\bigl|^{renorm}_{2-loop}=&\frac{2 G_N^3 m^3}{3 r^3} \left(2 \log \left(\frac{2 C_E r}{G_N m}\right)+ 12\, \Omega(3) -1\right)\\ &-\frac{4 G_N^2 m \alpha}{3 r^3}\left(1  +2   \log \left(\frac{2 C_E r}{G_N m}\right)\right) \ .
    \end{aligned}
\end{equation}
Now it is straightforward to see that this expression matches the metric in (\ref{hi4}) up to finite terms. In fact, although we cannot fix the finite part $\Omega(3)$, since it belongs to the high-energy regime, we can keep trace of it, and imposing a strict agreement between the renormalized metric and the one computed classically, the condition
\begin{equation}
    \Omega(3)= 2 \, \omega(3)\log \left(\frac{c}{4 C_E}\right)- \omega(3)+\frac{2}{3} \ 
\end{equation}
must hold, which relates the freedom in the quantum computation with the gauge freedom in the classical one.

\subsection*{Renormalization in $\bm{d=4}$}

Referring to (\ref{NonMinimal_Action}), in five dimensions the only allowed values of $k$ are $k=0, 2$. Then the expression in (\ref{alpha}) reads
\begin{equation}
    \alpha^{(1)}(4) = \alpha^{(1,0)}(4)+\frac{\alpha G_N}{(G_Nm)^{2}}\alpha^{(1,2)}(4)\ .
\end{equation}
As we already did in the previous case, in order to cancel the divergences we write
\begin{equation}
    \alpha^{(1)}(d)=\frac{\omega(4)}{d-4}+\Omega(4)\ ,
\end{equation}
where $\omega(4)$ is the coefficient we need to fix to renormalize the divergences in $d=4$ and $\Omega(4)$ is a finite contribution. In $d=4+\epsilon$ the metric is renormalized at 1 loop adding the contribution of the diagram in figure \ref{CT1}. Then we can set the constant $\omega(4)$ in order to cancel the divergence of the metric, whose value becomes
\begin{equation}\label{omega4}
    \omega(4)=\frac{5}{18 \pi}+\frac{G_N \alpha}{(G_N m)^2}\frac{1}{24 \pi}\ ,
\end{equation}
and again we can match the purely gravitational result with \cite{Mougiakakos:2020laz}.
As done for $d=3$ we can obtain a perfect match with the classical expression \ref{hi4} by fixing the finite contribution $\Omega(4)$, finding 
\begin{equation}\label{CapOmega4}
    \Omega(4)=\frac{\omega(4)}{2}\log\left(\frac{3 \pi c}{8C_E^2}\right)+\frac{\omega(4)}{6}-\frac{4}{27\pi} \ .
\end{equation}
Finally the 1-loop renormalized metric component with this choice of coupling constants is \cite{Mougiakakos:2020laz, Fragomeno}
\begin{equation}
\begin{aligned}
    h_{1}^{(d=4)}\bigl|^{renorm}_{1-loop}= h_{1}^{(d=4)}\bigl|_{1-loop} &+ \delta^{(1)}h_1^{(1)} =\frac{56 G_N^2 m^2}{27 \pi ^2 r^4}-\frac{\alpha  G_N}{9 \pi ^2 r^4} +\frac{16 G_N^2 m^2 }{\pi  r^4} \Omega(4)
   \\&+ \frac{20 G_N^2 m^2}{9 \pi ^2 r^4} \log \left(\frac{C_E^2 r^2}{G_N m}\right)+\frac{\alpha  G_N}{3 \pi ^2 r^4}  \log \left(\frac{C_E^2 r^2}{G_N m}\right)\ ,
\end{aligned}
\end{equation}
which perfectly match the corresponding order of eq. \eqref{hi4}.
For the 2-loop computation the steps are the same as the above case, but using the 1-loop diagram in figure \ref{CT2}. The computations are independent from the previous case and lead to the same constants $\omega(4)$ and $\Omega(4)$ defined above.
At this order the renormalized metric component is found to be \cite{Mougiakakos:2020laz, DOnofrio:2021tap}
\begin{equation}
\begin{aligned}
    h_{1}^{(d=4)}\bigl|^{renorm}_{2-loop}= h_{1}^{(d=4)}\bigl|_{2-loop} &+ \delta^{(1)}h_1^{(2)} =-\frac{112 G_N^3 m^3}{81 \pi ^3 r^6}+\frac{16 \alpha  G_N^2 m}{9 \pi ^3 r^6} -\frac{64 G_N^3 m^3 \Omega(4) }{3 \pi ^2 r^6}
    \\&-\frac{80 G_N^3 m^3}{27 \pi ^3 r^6} \log \left(\frac{C_E^2 r^2}{G_N m}\right)-\frac{4 \alpha  G_N^2 m}{9 \pi ^3 r^6} \log \left(\frac{C_E^2 r^2}{G_N m}\right) \ ,
\end{aligned}
\end{equation}
which again shows a perfect agreement with the classical expression in eq. \eqref{hi4}.

\medskip

Summarizing, we obtain complete renormalized expressions for the metric that are in complete agreement with the classical calculation of chapter \ref{chap:RNT_Classical_Solution}. We showed that while the logarithmic terms coincide naturally, in order to get a strict correspondence between the quantum and the classical derivation, which involves also finite parts, one must impose a relation that links the freedom in the quantum computation, parametrized by $\Omega(d)$, with the classical gauge freedom contained in the parameter $c$, and this unique definition is coherent with every independent object of the theory. Moreover it is important to observe that in the quantum context it is possible to prove that the divergences that we renormalized are the only ones that can appear, reproducing the classical result of chapter \ref{chap:RNT_Classical_Solution}. In fact for $\mathfrak{n}=1$, $d=3$ and $d=4$ are the only dimensions in which the coupling constant structures in eq. \eqref{NonMinimal_Action} have non-negative integer powers. This means that in other dimensions it would not exist any counter-term that renormalizes the observable, therefore cannot exist any divergence in those cases. 

\chapter{Photon emission processes}
\label{chap:Photon_Emission_Processes}

In the previous chapters we reviewed the literature on the state-of-the-art techniques for extracting classical physics from quantum gravity loop amplitudes, which culminate with the reconstruction of the Reissner-Nordstr\"om-Tangherlini metric from graviton emission processes. Since in chapter \ref{chap:RNT_Classical_Solution} it is showed how the complete black hole solution requires also the knowledge of the electromagnetic potential, we can ask ourselves if this can be recovered from scattering amplitudes too. 

\medskip

In this chapter, where all the results have to be considered original, we will first show how along the same picture of section \ref{sec:Classical_Limit_Grav_Emission_Processes}, it is possible to formalize a systematic procedure in order to obtain the electromagnetic potential from photon emission scattering amplitudes. As in the metric case, an explicit calculation of the potential shows the appearance of divergences in five dimensions, which as in the metric case have to be renormalized. In the end it is shown how the procedure to cancel the singularities leads to the very same counter-terms of chapter \ref{chap:Metric_from_Amplitudes}, which underlines the extraordinary coherence of the theory. 

\section{Classical limit of photon emission processes}
\label{sec:Classical_Limit_Photon_Emission}

Following the procedure outlined in section \ref{sec:Classical_Limit_Grav_Emission_Processes}, considering the usual quantum action defined in section \ref{sec:Quantization_of_Gravity}, we want to solve the field equations
\begin{equation}
\Box A_\mu(x)=j_\mu(x)
\end{equation}
of the electromagnetic potential in the Feynman gauge and space-time dimensions $D=d+1$, perturbatively in the post-Minkowskian expansion, where $j_\mu(x)$ is the electromagnetic source current. The idea then, is that the photon field that appears in the above equation, is the quantum counterpart of the classical observable we want to recover up to consider the classical limit in which $\hbar\rightarrow 0$. Considering the post-Minkowskian expansion of both the potential and the source, with the usual notation which employs the indices $n$ and $j$, we can write\begin{equation}
j_{\mu}=\sum_{n=0}^{+\infty}\sum_{\substack{j=0 \\ j \ \text{even}}}^{n}j_{\mu}^{(n, j)} \quad \text{and} \quad A_{\mu}=\sum_{n=1}^{+\infty}\sum_{\substack{j=0 \\ j \ \text{even}}}^{n}A_{\mu}^{(n, j)}\ ,
\end{equation}
and the field equations at each order become
\begin{equation}\label{Eq_Motion_A}
    \Box A_\mu^{(n, j)}(x) = j_\mu^{(n-1, j)} (x) \ ,
\end{equation}
where again the $n$-th post-Minkowskian order of the potential is generated from the $n-1$-th order of the source. 

\medskip

As usual we are interested in static solutions of eq. \eqref{Eq_Motion_A}, from which imposing such condition we obtain  
\begin{equation}\label{ClassicalEM_potential_withCurrent}
    A_\mu^{(n,j)}(\vec{x}) = \int \frac{d^dq}{(2\pi)^d}\frac{e^{i\vec{q}\cdot \vec{x}}}{\vec{q}^2}j_\mu^{(n-1,j)}(\vec{q}^2) \ ,
\end{equation}
where $j_\mu^{(n-1,j)}(\vec{q}^2)$ is the Fourier transform of the electromagnetic current that corresponds to the matrix element
\begin{equation}
    j_{\mu}(q^2)=\bra{p_2}j_{\mu}(x=0)\ket{p_1}\ , 
\end{equation}
in which $p_1$ and $p_2$ are the momenta of two external on-shell massive particles and where $q=p_1-p_2$ is the momentum transferred.
So then, the problem of calculating the electromagnetic potential is moved to the problem of compute the matrix element of the source, and since we are interested in classical phenomenology, we pick only those contributions we know will not vanish in the classical limit, which are given exactly by the selection rules of section \ref{sec:Schwarzschild_Metric_from_Graviton_Emission}. The outcome then, is that the classical contributions of $j_\mu(\vec{q}^2)$ are reconstructed by $l$-loop photon emission amplitudes, in which a single massive line emits $n-j-1$ gravitons and $j+1$ photons, ending up in a tree internal interaction structure as 
\begin{equation}\label{SpecificLoop_PhotonEmission_Process}
\includegraphics[width=0.4\textwidth, valign=c]{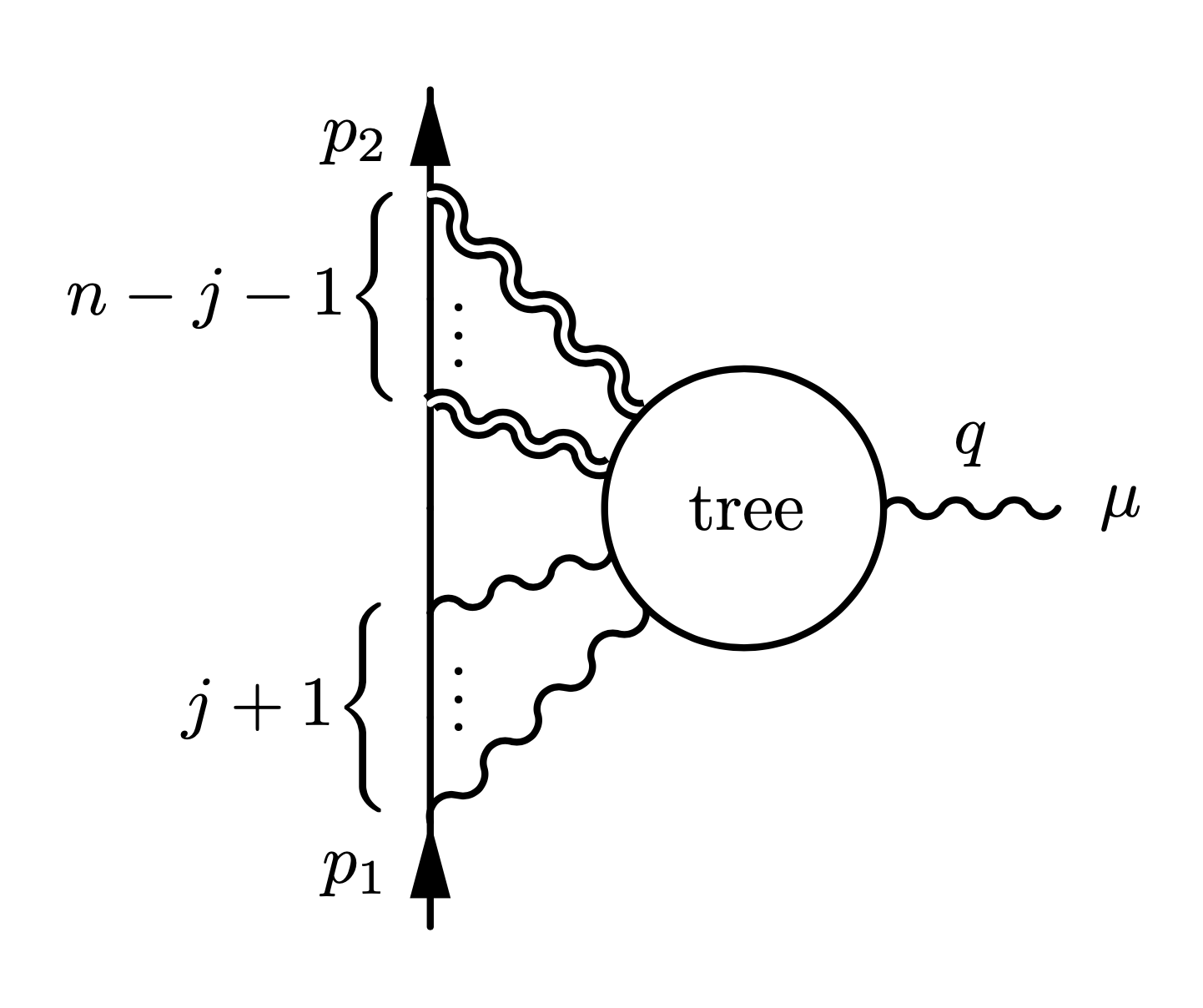} = -i \, \sqrt{4E_1 E_2}\, j_\mu{}^{(l,j)}(q^2)\ ,
\end{equation}
where the number of loops and the post-Minkowskian order are related through the usual relation $n=l+1$. Now analogously to the metric case, since eq. \eqref{ClassicalEM_potential_withCurrent} is meant to reconstruct the classical electromagnetic potential, we are interested only in those contributions of the electromagnetic source whose
\begin{equation}
j_\mu(\vec{q}^2)^{(l, j)}\propto J_{(l)}(\vec{q}^2)\ ,
\end{equation}
from which using the Lorentz covariance and the fact that the current is conserved, an expansion in terms of dimensionless form factors $c^{(l, j)}(d)$ leads to 
\begin{equation}\label{Electro_Current_FormFactor}
    j_\mu^{(l)}(\vec{q}^2)=\sum_{\substack{j=0 \\ j \ \text{even}}}^{l+1} j_\mu^{(l, j)}(\vec{q}^2) \simeq\sum_{\substack{j=0 \\ j \ \text{even}}}^{l+1} Q \,  \delta_\mu^0 \, c^{(l, j)}(d)  (G_N m)^{l-j}(\alpha G_N)^\frac{j}{2} J_{(l)}(\Vec{q}^2)\ .
\end{equation}
Then using the Fourier transform of the master integral in eq. \eqref{app_J_Identity_1}, one finally  obtains 
\begin{equation}\label{EM_Pot_FormFact_NotUseful}
    A_\mu^{(l+1, j)}(r) = Q\, \delta_\mu^0c^{(l, j)}(d)  (G_N m)^{l-j}(\alpha G_N)^\frac{j}{2}\left(\frac{\rho}{4 \pi}\right)^{l+1} \ ,
\end{equation}
where we observe that only the temporal component of the potential is non-vanishing. 
This analysis outlines the explicit relation between the classical electromagnetic potential and the amplitude calculations, by which the form factors\footnote{Notice that in the expansion in eq. \eqref{Electro_Current_FormFactor} we miss a $\pi^l$ factor which makes the form factors dependent by it.} can be directly extracted. However, with respect to the metric case, the above expressions are much simpler to manage, therefore it is not needed to extract the form factors and $j_\mu^{(l, j)}(\vec{q}^2)$ can be directly computed. This means that the form factor expansion in eq. \eqref{EM_Pot_FormFact_NotUseful} will be not used in actual computations, nevertheless shows that the temporal component is the only non-vanishing one, reproducing the classical result of chapter \ref{chap:RNT_Classical_Solution}. 

\medskip

At this point we have to follow section \ref{sec:Classical_Limit_Grav_Emission_Processes} in order to extract the contributions proportional to the master integral out of amplitudes like \eqref{SpecificLoop_PhotonEmission_Process}. The procedure is exactly the same and it is not needed to repeat it. Decouple the amplitude in a massive line emitting massless particles, parametrized by $\mathcal{N}$, and in an internal tree structure that depends on the specific process, parametrized by $\mathcal{M}$, then the classical limit of the electromagnetic current reads 
\begin{equation}\label{Electro_Current_Amplitudes}
\begin{aligned}
    j_{\mu}^{(l,j)}(\vec{q}^2)=\frac{i}{(l+1)!}\int\prod_{i=1}^l\frac{d^d\Vec{\ell_i}}{(2\pi)^d} (-1)^{l+j}2^{j-l}m^{l-j}\kappa^{l-j}Q^{j+1}\frac{\prod_{i=1}^{l-j}P^{00, \lambda_i\sigma_i}}{\prod_{i=1}^{l+1}\Vec{\ell_i}^2}\\
    \times \mathcal{M}_{\lambda_1\sigma_1, ..., \lambda_{l-j+1}\sigma_{l-j+1}, 0, ..., 0, \mu}(p_1, p_2, \ell_1, ..., \ell_l)\Bigl|_{\ell_i^0=0}\ .
\end{aligned}
\end{equation}
It is worth stressing again the fact that classical physics is recovered from extracting out of the above expression only the contributions whose leading-order in energy is proportional to the master integral, while all the other pieces give rise to quantum corrections. This statement can be summarized by rewriting eq. \eqref{Electro_Current_Amplitudes} in terms of quantum tree-graphs, adapting the formalism developed by Duff in \cite{Duff:1973zz} to photon emission processes as
\begin{equation}\label{Duff_Amp_Phot}
-i\, 2m\, j_{\mu}^{(l,j)}(\vec{q}^2)\Big|_{\text{classical}}=\includegraphics[width=0.45\textwidth, valign=c]{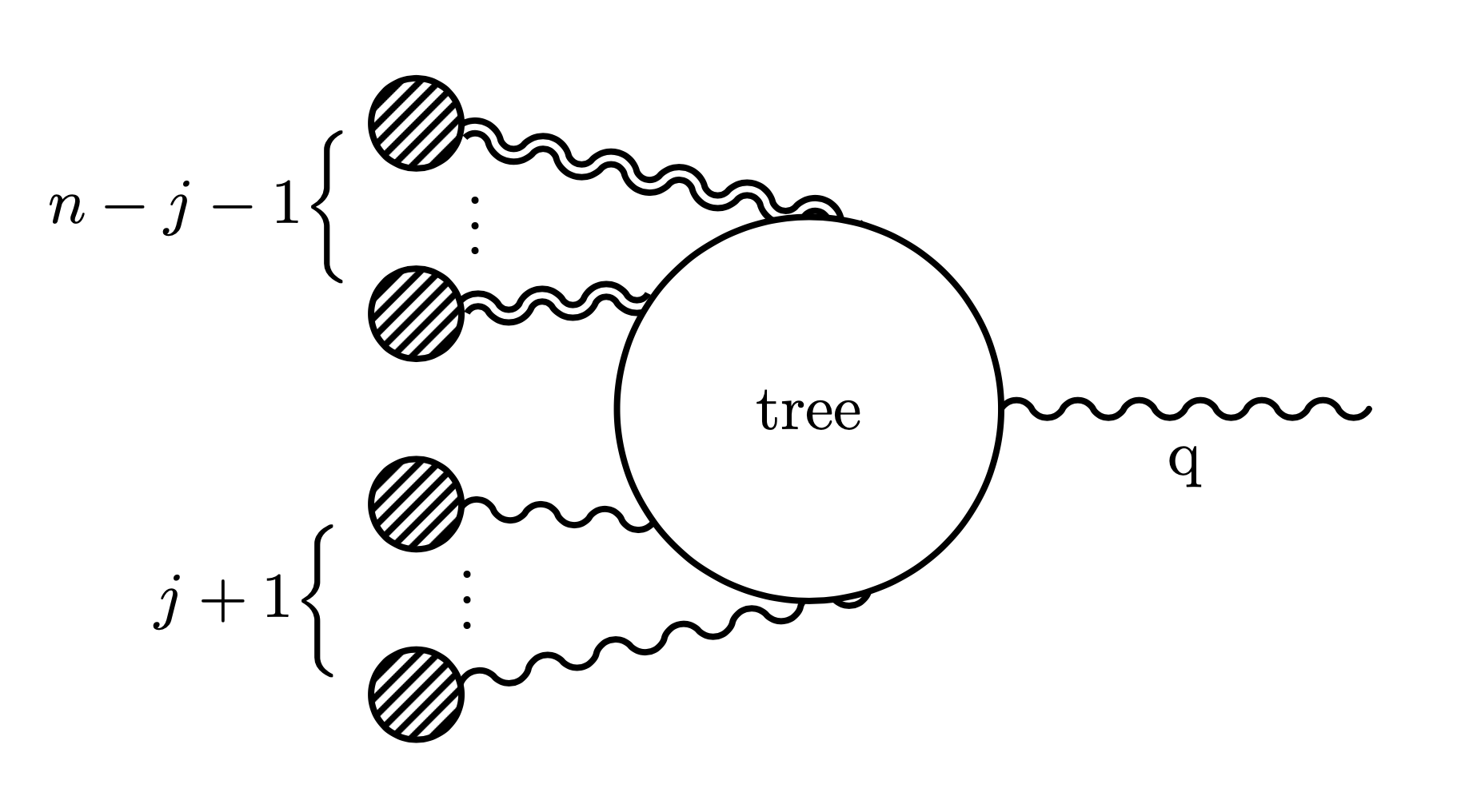}\Bigg|_{\text{leading-order}}
\ .
\end{equation}

\section{Potential from amplitude calculations}
\label{sec:AmpCalc_Pot}

Now we can perform the actual computation of the photon emission process in \eqref{SpecificLoop_PhotonEmission_Process} up to 2-loop order, and explicitly recover the expression of the electromagnetic potential by directly evaluating eq. \eqref{ClassicalEM_potential_withCurrent}. Then
we will show the appearance of divergences at 2 loops in $d=4$, which will be treated in detail in the next section. As already noticed previously, only the time component of the gauge potential is non-vanishing, so in the following it will be the only component taken into account. 

\subsection*{Tree level}

As for the metric case, we start the analysis at tree level for completeness. In the notation of section \ref{sec:Classical_Limit_Photon_Emission}, the tree-level diagram corresponds to the term $l=0,\ j=0$ in the expansion of eq. \eqref{SpecificLoop_PhotonEmission_Process}.
\begin{figure}[h]
\centering
\includegraphics[width=0.25\textwidth, valign=c]{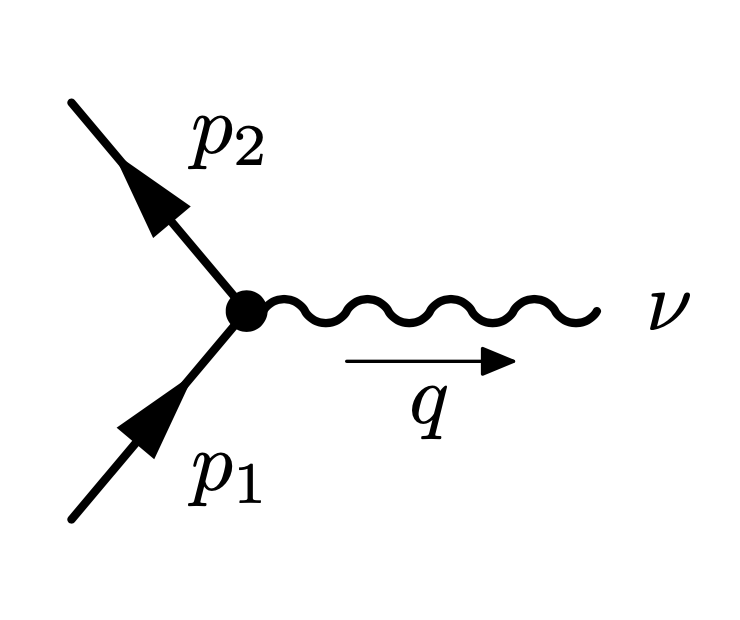}
\caption{Tree-level diagram for photon emission.}
\label{figure:TreeLevel_PhotonEmission}
\end{figure}
Evaluating the matrix element of the current due to the diagram in figure \ref{figure:TreeLevel_PhotonEmission}, and using the Feynman rules in section \ref{Appendix_FeynRules}, one obtains 
\begin{equation}
    -i \, \sqrt{4E_1 E_2}\, j_\nu^{(0, 0)}(\vec{q}^2) = \left(\tau_{\phi^2 A}\right)_\nu\ .
\end{equation}
Considering the classical limit, the tree-level electromagnetic current is simply the charge,
\begin{equation}
    j_0^{(0, 0)}(\vec{q}^2) = Q\ ,
\end{equation}
and replacing it inside eq. \eqref{ClassicalEM_potential_withCurrent} and using the Fourier transform identities of the master integral in appendix \ref{App:masterintegral}, we get
\begin{equation}
    A_0^{(1, 0)}(r) =\frac{Q}{4\pi}\rho\ ,
\end{equation}
which is exactly the first order of the expression in \eqref{EM_Potential_deDonder}.

\subsection*{1-loop}

\begin{figure}[h]
\centering
\includegraphics[width=0.4\textwidth, valign=c]{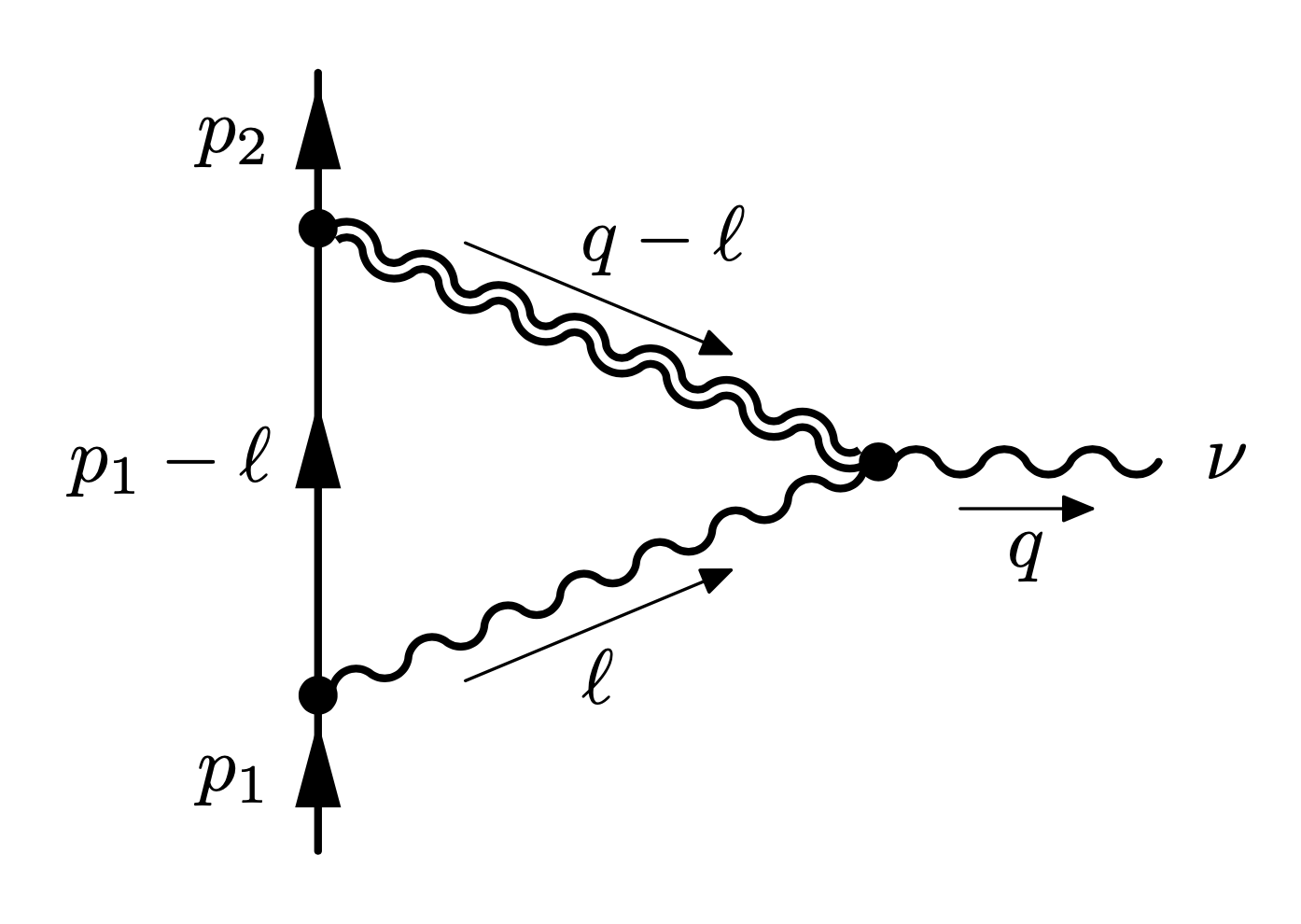}
\caption{1-loop diagram for photon emission.}
\label{figure:1Loop_PhotonEmission}
\end{figure}

At this loop order we can only have $j=0$. So the 1-loop process is given by the amplitude in figure \ref{figure:1Loop_PhotonEmission}.
The electromagnetic current associated to this process reads
\begin{equation}
-i \, \sqrt{4E_1E_2}\, j_{\nu}^{(1, 0)}(q^2) = 2\times \int \frac{d^{d+1}\ell}{(2\pi)^{d+1}} \frac{i\, P_{\alpha\beta, \sigma\rho}\left(\tau_{\phi^{2} A}\right)^{\mu}\left(\tau_{\phi^{2} h}\right)^{\alpha\beta}\left(\tau_{A^{2} h}\right)^{\sigma\rho}{}_{\mu,\nu}\left(\ell, q\right)}{\left((p_1-\ell)^2-m^2+i\epsilon\right)\left(\ell^2+i\epsilon\right)\left((q-\ell)^2+i\epsilon\right)}\ ,
\end{equation}
where a factor $"2"$ is implied in order to consider the multiplicity of the diagram. Performing the classical limit to obtain an expression like eq. \eqref{Electro_Current_Amplitudes}, one gets
\begin{equation}
j_{0}^{(1, 0)}(\vec{q}^2) = -\frac{i}{2} \kappa Q m \int \frac{d^d\ell}{(2\pi)^d} \frac{P_{00, \alpha \beta}\left(\tau_{A^{2} h}\right)^{\alpha \beta}{}_{0,0}\left(\ell, q\right)\big|_{\ell^0=0}}{\vec{\ell}^2(\vec{q}-\vec{\ell})^2}\ ,
\end{equation}
from which considering the tensorial contraction 
\begin{equation}
    P_{00, \alpha \beta}\left(\tau_{A^{2} h}\right)^{\alpha \beta}{}_{0,0}\left(\ell, q\right)\big|_{\ell^0=0}=-i\kappa\frac{d-2}{d-1}\Vec{\ell}\cdot\Vec{q}\ ,
\end{equation}
we can express the current in terms of the master integral through the identities in appendix \ref{App:LoopRed}. In particular using (\ref{LR_1Loop_1}) one end up with the compact expression
\begin{equation}
    j_0^{(1, 0)}(\vec{q}^2)= -\frac{1}{4}m Q\kappa^2\frac{d-2}{d-1}J_{(1)}(\Vec{q}^2)\ ,
\end{equation}
recovering the structure of eq. \eqref{Electro_Current_FormFactor}.
Replacing the above expression inside the (\ref{ClassicalEM_potential_withCurrent}), exploiting the property \eqref{app_J_Identity_1}, one obtains
\begin{equation}\label{1-Loop_EM_POT}
A_0^{(2, 0)}(r) = - \frac{Q}{4\pi} \frac{2(d-2)}{d-1}m G_N \rho^2\ ,
\end{equation}
which shows, at this order, a perfect agreement with the expression \eqref{EM_Potential_deDonder}.

\subsection*{2-loop}

\begin{figure}[h]
\centering
\includegraphics[width=0.9\textwidth, valign=c]{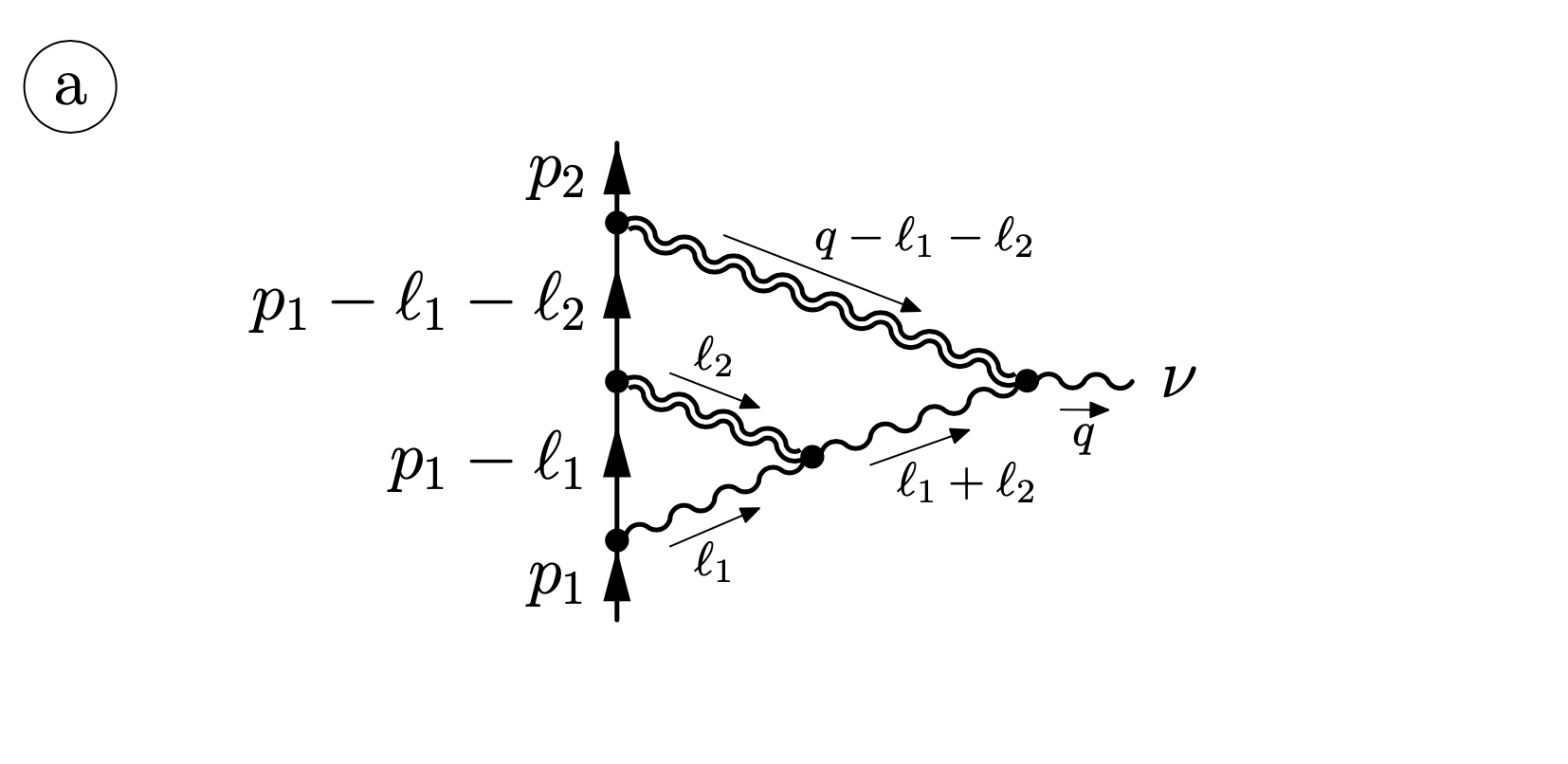}
\caption{2-loop diagram for photon emission with 2 internal photons and 2 internal gravitons.}
\label{fig:PhotonEMission_2Loop_2Photon2Gravitons}
\end{figure}

At 2 loops we can have either $j=0$ or $j=2$. We first discuss the case $j=0$, in which contribute three different diagrams that differ by the internal tree structure. As we already did in the metric case, after computing the current for each process we sum all the contributions in order to get the complete electromagnetic potential.
The first amplitude we consider is the one in figure \ref{fig:PhotonEMission_2Loop_2Photon2Gravitons}, which has 2 internal photons and 2 internal gravitons. 
The associated electromagnetic current to this process is then
\begin{multline}
    -i \, \sqrt{4E_1E_2}\, j_\nu^{(2, 0)(\text{a})}(q^2)= 6\times \int \frac{d^{d+1} \ell_1}{(2\pi)^{d+1}}\frac{d^{d+1} \ell_2}{(2\pi)^{d+1}} \frac{-P_{\sigma\rho, \alpha\beta}P_{\gamma\delta, \eta\chi}\left(\tau_{\phi^2A}\right)_{\lambda}\left(\tau_{\phi^2h} \right)^{\eta\chi}\left(\tau_{\phi^2h}\right)^{\sigma\rho}}{\left(\ell_1^2+i\epsilon\right)\left(\ell_2^2+i\epsilon\right)\left((q-\ell_1-\ell_2)^2+i\epsilon\right)} \\
    \times\frac{\left(\tau_{A^2h}\right)^{\gamma\delta,\mu,\lambda}(\ell_1, \ell_1+\ell_2)\left(\tau_{A^2h} \right)^{\alpha\beta}{}_{\mu,\nu}(\ell_1+\ell_2, q)}{\left((\ell_1+\ell_2)^2+i\epsilon\right)((p_1-\ell_1)^2-m^2+i\epsilon)((p_1-\ell_1-\ell_2)^2-m^2+i\epsilon)}\ .
\end{multline}
Exploiting the procedure outlined in section \ref{sec:Classical_Limit_Photon_Emission}, one gets
\begin{equation}
\begin{aligned}
    j_0^{(2, 0)(\text{a})}(\vec{q}^2)&=-\frac{\kappa^2 Q m^2}{4}\int \frac{d^d \ell_1}{(2\pi)^d}\frac{d^d \ell_2}{(2\pi)^d}\frac{P^{\alpha \beta, 00}P^{\sigma \rho, 00}}{\vec{\ell_1}^2\vec{\ell_2}^2(\vec{q}-\vec{\ell_1}-\vec{\ell_2})^2(\vec{\ell_1}+\vec{\ell_2})^2}\\
    &\times(\tau_{A^2h})_{0\mu, \alpha \beta}(\ell_1, \ell_1+\ell_2)(\tau_{A^2h})^{0\mu}{}_{\sigma \rho}(\ell_1+\ell_2, q)\bigl|_{\ell_i^0=0}\ ,
\end{aligned}
\end{equation}
where the tensor contraction gives
\begin{equation}
\begin{aligned}
   P^{\alpha \beta, 00}&P^{\sigma \rho, 00}(\tau_{A^2h})_{0,\mu, \alpha \beta}(\ell_1, \ell_1+\ell_2)(\tau_{A^2h})^{0,\mu}{}_{\sigma \rho}(\ell_1+\ell_2q)\bigl|_{\ell_i^0=0}\\&=-\kappa^2\frac{(d-2)^2}{(d-1)^2}(\Vec{\ell_1}+\Vec{\ell_2})\cdot\Vec{q}\,(\Vec{\ell_1}+\Vec{\ell_2})\cdot \Vec{\ell_1}\ .
   \end{aligned}
\end{equation}
Then using the reduction identities in appendix \ref{App:LoopRed}, one gets the current in terms of the master integral as
\begin{equation}
    j_0^{(2, 0)(\text{a})}(\vec{q}^2)= \frac{\kappa^4Qm^2}{12}\frac{(d-2)^2}{(d-1)^2}J_{(2)}(\Vec{q}^2)\ , 
\end{equation}
from which, due to the usual relations of the Fourier transform of the master integral, one recovers 
\begin{equation}
    A_0^{(3, 0)(\text{a})}(r)=\frac{Q}{4\pi}m^2G_N^2\frac{16}{3}\frac{(d-2)^2}{(d-1)^2}\rho^3\ .
\end{equation}

\medskip

\begin{figure}[h]
\centering
\includegraphics[width=0.9\textwidth, valign=c]{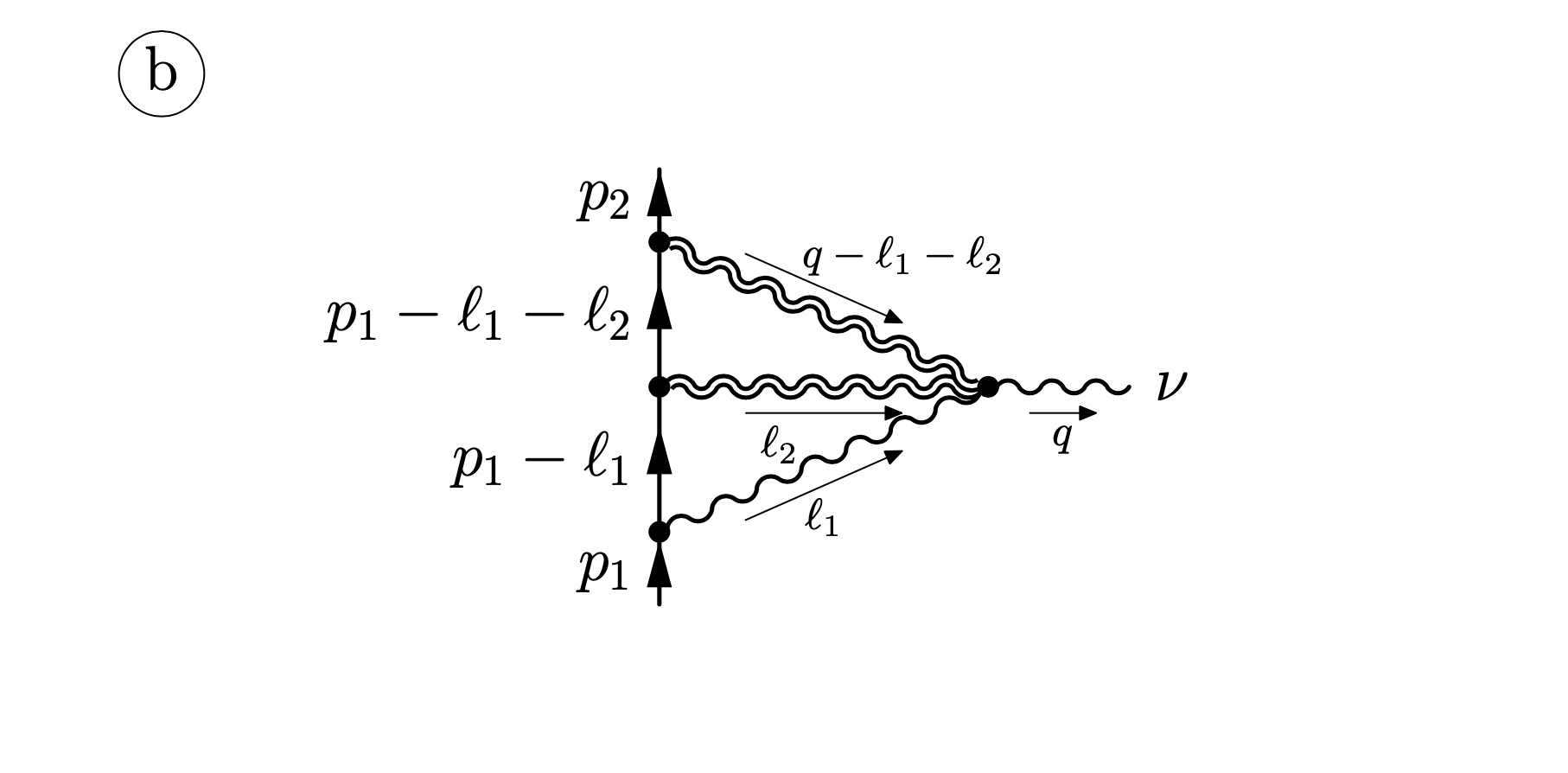}
\caption{2-loop diagram for photon emission with 1 internal photon and 2 internal gravitons.}
\label{fig:PhotonEMission_2Loop_1Photon2Gravitons}
\end{figure}

The second amplitude we consider is the one in figure \ref{fig:PhotonEMission_2Loop_1Photon2Gravitons}, which has 1 internal photon and 2 internal gravitons.
The electromagnetic current associated to the diagram is 
\begin{equation}
\begin{aligned}
    -i \, \sqrt{4E_1E_2}\, &j_\nu^{(2, 0)(\text{b})}(q^2)= 3\times \int \frac{d^{d+1} \ell_1}{(2\pi)^{d+1}}\frac{d^{d+1} \ell_2}{(2\pi)^{d+1}} \frac{-iP_{\sigma \rho, \alpha\beta}P_{\gamma\delta, \eta\chi}\left(\tau_{\phi^2A}\right)^{\mu}\left(\tau_{\phi^2h} \right)^{\eta\chi}\left(\tau_{\phi^2h}\right)^{\sigma \rho}}{\left(\ell_1^2+i\epsilon\right)\left(\ell_2^2+i\epsilon\right)\left((q-\ell_1-\ell_2)^2+i\epsilon\right)}\\
    &\times\frac{\left(\tau_{A^2h^2}\right)^{\alpha\beta, \gamma\delta}{}_{\mu,\nu}(\ell_1, q)}{((p_1-\ell_1)^2-m^2+i\epsilon)((p_1-\ell_1-\ell_2)^2-m^2+i\epsilon)}\ .
    \end{aligned}
\end{equation}
Performing the classical limit, one gets
\begin{equation}
    j_0^{(2, 0)(\text{b})}(\vec{q}^2)= \frac{i}{8}\kappa^2m^2Q \int \frac{d^d\ell_1}{(2\pi)^d} \frac{d^d\ell_2}{(2\pi)^d} \frac{P_{00, \mu\nu}P_{00, \alpha\beta}(\tau_{A^2h^2})^{0,0, \mu\nu, \alpha\beta}(\ell_1, q)\bigl|_{\ell_i^0=0}}{\vec{\ell_1}^2 \vec{\ell_2}^2\left(\vec{q}-\vec{\ell_1}-\vec{\ell_2}\right)^2}\ ,
\end{equation}
where the tensorial contraction of the numerator leads to 
\begin{equation}
    P_{00, \mu\nu}P_{00, \alpha\beta}(\tau_{A^2h^2})^{0,0, \mu\nu, \alpha\beta}(\ell_1, q)\bigl|_{\ell_i^0=0}=i\frac{\kappa^2}{2}\frac{(d-2)(3d-7)}{(d-1)^2}\Vec{\ell_1}\cdot\Vec{q} \ .
\end{equation}
Then using the relation (\ref{LiteRed_2Loop_2}), the current expressed in terms of the master integral reads
\begin{equation}
    j_0^{(2, 0)(\text{b})}(\vec{q}^2)= -\frac{m^2Q\kappa^4}{48}\frac{(d-2)(3d-7)}{(d-1)^2} J_{(2)}(\Vec{q}^2)\ ,
\end{equation}
from which considering the usual Fourier transform one gets
\begin{equation}
    A_0^{(3, 0)(\text{b})}(r)=-\frac{Q}{4\pi}m^2G_N^2\frac{4}{3}\frac{(d-2)(3d-7)}{(d-1)^2}\rho^3\ .
\end{equation}

\medskip

\begin{figure}[h]
\centering
\includegraphics[width=0.9\textwidth, valign=c]{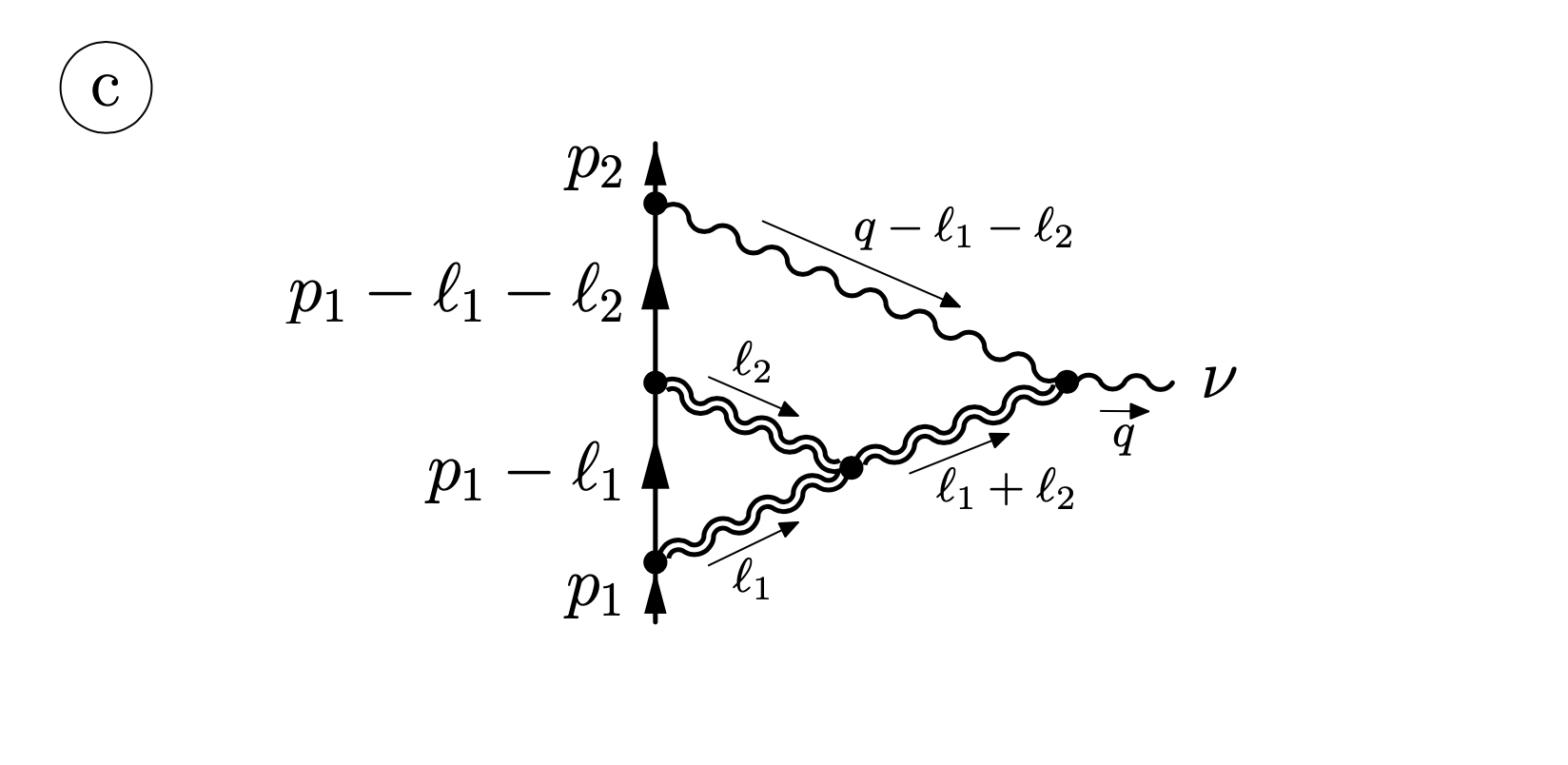}
\caption{2-loop diagram for photon emission with 1 internal photon and 3 internal gravitons.}
\label{fig:PhotonEMission_2Loop_1Photon3Gravitons}
\end{figure}

The last diagram that contributes to the process with $l=2, \ j=0$ is the one with an internal 3 gravitons vertex, as shown in figure \ref{fig:PhotonEMission_2Loop_1Photon3Gravitons}. 
Since the 3 gravitons vertex is internal, we must use the expression given in  \eqref{3Graviton_Vertex_deWitt_AppEq}. Then the current associated to this diagram is 
\begin{equation}
\begin{aligned}
    & -i \, \sqrt{4E_1E_2}\, j_\nu^{(2, 0)(\text{c})}(q^2)= 3\times \int \frac{d^{d+1} \ell_1}{(2\pi)^{d+1}}\frac{d^{d+1} \ell_2}{(2\pi)^{d+1}} \frac{P_{\alpha_1\beta_1, \gamma \delta}P_{\alpha_2\beta_2, \eta\chi}P_{\alpha_3\beta_3, \sigma\rho}\left(\tau_{\phi^2A}\right)^{\mu}}{\left(\ell_1^2+i\epsilon\right)\left(\ell_2^2+i\epsilon\right)\left((q-\ell_1-\ell_2)^2+i\epsilon\right)} \\
    & \times\frac{\left(\tau_{\phi^2h} \right)^{\gamma\delta}\left(\tau_{\phi^2h}\right)^{\eta\chi}\left(\tau_{h^3}\right)^{\alpha_1\beta_1,\alpha_2\beta_2, \alpha_3\beta_3}(\ell_1, \ell_2, -\ell_1-\ell_2)\left(\tau_{A^2h} \right)^{\sigma\rho}{}_{\mu, \nu}(q-\ell_1-\ell_2, q)}{\left((\ell_1+\ell_2)^2+i\epsilon\right)((p_1-\ell_1)^2-m^2+i\epsilon)((p_1-\ell_1-\ell_2)^2-m^2+i\epsilon)}\ .
    \end{aligned}
\end{equation}
Referring to eq. \eqref{Electro_Current_Amplitudes}, one gets 
\begin{equation}
\begin{aligned}
    j_0^{(2, 0)(\text{c})}(\vec{q}^2)& = \frac{1}{8}\kappa^2Qm^2\int \frac{d^d \ell_1}{(2\pi)^d}\frac{d^d \ell_2}{(2\pi)^d}\frac{P_{00, \mu\nu}P_{00, \alpha\beta}P_{\gamma \delta, \sigma \rho}}{\vec{\ell_1}^2\vec{\ell_2}^2(\vec{q}-\vec{\ell_1}-\vec{\ell_2})^2(\vec{\ell_1}+\vec{\ell_2})^2}\\
    &\times \left(\tau_{h^3}\right)^{\mu\nu, \alpha\beta, \gamma\delta}(\ell_1, \ell_2, -\ell_1-\ell_2)\left(\tau_{A^2h} \right)^{\sigma \rho}{}_{0,0}(q-\ell_1-\ell_2, q)\bigl|_{\ell_i^0=0}\ ,
    \end{aligned}
\end{equation}
where considering from the beginning the fact that all the expressions that contain internal momenta are symmetric under the exchange of $\ell_1\leftrightarrow \ell_2$, the tensor contraction at the numerator reads
\begin{equation}
    \begin{aligned}
    &P_{\gamma\delta, \sigma\rho}P_{\mu\nu, 00}P_{\alpha\beta, 00}\left(\tau_{h^3}\right)^{\mu\nu, \alpha\beta, \gamma\delta}(\ell_1, \ell_2, -\ell_1-\ell_2)\left(\tau_{A^2h} \right)^{\sigma \rho}{}_{0,0}(q-\ell_1-\ell_2, q)\bigl|_{\ell_i^0=0}\\
    &=-\frac{\kappa^2}{4}\frac{d-2}{(d-1)^2}\Biggl(4(d-1)\Vec{\ell_1}\cdot (\Vec{q}-\vec{\ell_1}-\vec{\ell_2}) \ \vec{\ell_1}\cdot \vec{q}+2(d-1)\Vec{\ell_1}\cdot (\Vec{q}-\vec{\ell_1}-\vec{\ell_2}) \ \vec{\ell_2}\cdot \vec{q} \\
    &+(d-5)\Vec{q}\cdot (\Vec{q}-\vec{\ell_1}-\vec{\ell_2}) \ \vec{\ell_1}\cdot \vec{\ell_2}\Biggl)\ .
    \end{aligned}
\end{equation}
Exploiting the expressions in appendix \ref{App:LoopRed}, the current  finally  reads
\begin{equation}
    j_0^{(2, 0)(\text{c})}(\vec{q}^2) = \frac{\kappa^4m^2Q}{96}\frac{(d-2)(34-29d+7d^2)}{(d-4)(d-1)^2}J_{(2)}(\Vec{q}^2)\ ,
\end{equation}
from which we obtain
\begin{equation}
    A_0^{(3, 0)(\text{c})}(r) = \frac{Q}{4\pi} m^2G_N^2 \frac{2}{3}\frac{(d-2)(34-29d+7d^2)}{(d-4)(d-1)^2}\rho^3\ .
\end{equation}
At the end we can sum up all the $j=0$  pieces to get 
\begin{equation}
    A_0^{(3, 0)}(r)=\sum_{i=\text{a}, \text{b}, \text{c}}A_0^{(3, 0)(i)}(r)=\frac{Q}{4\pi} m^2G_N^2 \frac{2(d-2)^2(3d-7)}{(d-1)^2(d-4)}\rho^3\ ,
\end{equation}
which is exactly the contribution that appears in (\ref{EM_Potential_deDonder}). In this last expression a divergence in $d=4$ arises, and as we did in the metric case it has to be renormalized. 

\medskip

\begin{figure}[h]
\centering
\includegraphics[width=0.6\textwidth, valign=c]{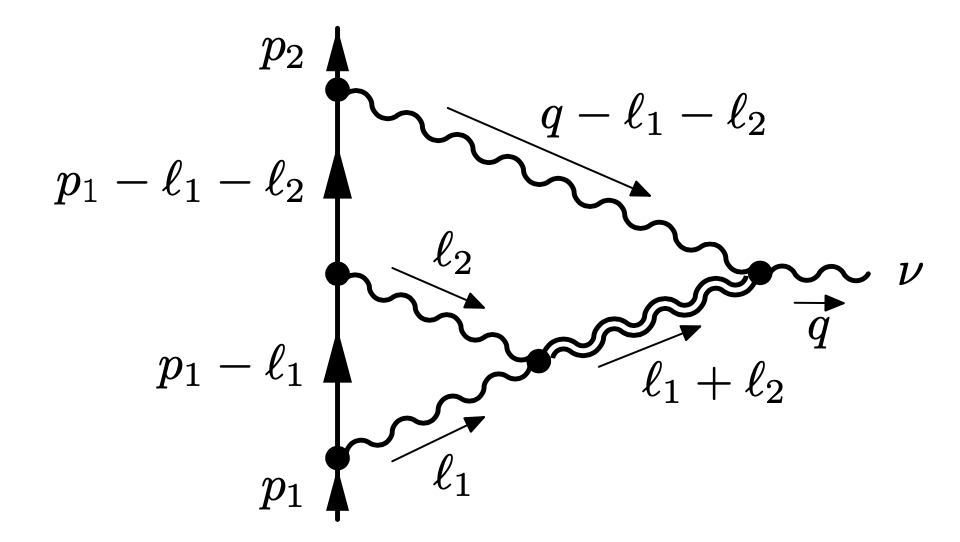}
\caption{2-loop diagram for photon emission with 3 internal photons and 1 internal graviton.}
\label{fig:PhotonEMission_2Loop_3Photon1Gravitons}
\end{figure}

Finally, the only process that contributes at $l=2, \ j=2$ is the one which has 3 internal photons and 1 internal graviton, shown in figure \ref{fig:PhotonEMission_2Loop_3Photon1Gravitons}.
The electromagnetic current associated to this process is then
\begin{equation}
    \begin{aligned}
    -i \, \sqrt{4E_1E_2}\, &j_\nu^{(2, 2)}(q^2)= 3\times \int \frac{d^{d+1} \ell_1}{(2\pi)^{d+1}}\frac{d^{d+1} \ell_2}{(2\pi)^{d+1}} \frac{P_{\gamma\delta, \sigma\alpha}\left(\tau_{\phi^2A}\right)^{\beta}\left(\tau_{\phi^2A}\right)^{\sigma}\left(\tau_{\phi^2A}\right)^{\rho}}{\left(\ell_1^2+i\epsilon\right)\left(\ell_2^2+i\epsilon\right)\left((q-\ell_1-\ell_2)^2+i\epsilon\right)} \\
    &\times\frac{\left(\tau_{A^2h}\right)^{\sigma\alpha}{}_{\beta,\sigma}(\ell_1,-\ell_2)\left(\tau_{A^2h} \right)^{\gamma\delta}{}_{\rho,\nu}(q-\ell_1-\ell_2, q)}{\left((\ell_1+\ell_2)^2+i\epsilon\right)((p_1-\ell_1)^2-m^2+i\epsilon)((p_1-\ell_1-\ell_2)^2-m^2+i\epsilon)}\ ,
    \end{aligned}
\end{equation}
which in the classical regime reads
\begin{equation}
    j_0^{(2, 2)}(\vec{q}^2)=\frac{1}{2}Q^3\int\frac{d^d \ell_1}{(2\pi)^d}\frac{d^d \ell_2}{(2\pi)^d}\frac{P_{\mu\nu, \alpha\beta}\left(\tau_{A^2h} \right)^{\mu\nu}{}_{00}(\ell_1, -\ell_2)\left(\tau_{A^2h} \right)^{\alpha\beta}{}_{00}(q-\ell_1-\ell_2, q)\bigl|_{\ell_i^0=0}}{\vec{\ell_1}^2\vec{\ell_2}^2(\vec{q}-\vec{\ell_1}-\vec{\ell_2})^2(\vec{\ell_1}+\vec{\ell_2})^2}\ .
\end{equation}
Computing the tensor contraction in the numerator as
\begin{equation}
    \begin{aligned}
    P_{\mu\nu, \alpha\beta}&\left(\tau_{A^2h} \right)^{\mu\nu}{}_{0,0}(\ell_1, -\ell_2)\left(\tau_{A^2h} \right)^{\alpha\beta}{}_{0,0}(q-\ell_1-\ell_2, q)\bigl|_{\ell_i^0=0}\\
    &=\frac{\kappa^2}{2}\left(\frac{d-3}{d-1}\Vec{\ell_1}\cdot\Vec{\ell_2}\ \Vec{q}\cdot(\Vec{q}-\Vec{\ell_1}-\Vec{\ell_2})+2\ \Vec{q}\cdot\Vec{\ell_2}\ \Vec{\ell_1}\cdot(\Vec{q}-\Vec{\ell_1}-\Vec{\ell_2})\right)\ , 
    \end{aligned}
\end{equation}
in which again we exploited the symmetry $\ell_1\leftrightarrow \ell_2$, and using the expressions in appendix \ref{App:LoopRed}, we  obtain
\begin{equation}
    j_0^{(2, 2)}(\vec{q}^2)=\frac{Q^3 \kappa^2}{8}\frac{(d-3)(d-2)}{(d-4)(d-1)}J_{(2)}(\Vec{q}^2)\ .
\end{equation}
From this we  get the contribution to the electromagnetic potential
\begin{equation}
    A_0^{(3, 2)}(r)=\frac{Q}{4\pi}\alpha G_N\frac{(d-3)(d-2)}{(d-4)(d-1)}\rho^3\ .
\end{equation}
This last expression matches perfectly the corresponding term of eq. (\ref{EM_Potential_deDonder}), and as before we notice a divergence for $d=4$ which must be renormalized. 

\medskip

To summarize, we have shown that from the loop computations one recovers exactly the expression for the electromagnetic potential of the Reissner-Nordstr\"om-Tangherlini solution in de Donder gauge given in eq. \eqref{EM_Potential_deDonder}. However a renormalization procedure has to be discussed since at third post-Minkowskian order appear a divergence in $d=4$, as expected from the analysis in chapter \ref{chap:RNT_Classical_Solution}.

\section{Renormalization of the potential}
\label{sec:Renormalization_Potential}

In this section we renormalize the electromagnetic potential following the strategy already outlined in section \ref{sec:Renorm_Metric}. First we regularize the potential by isolating the pole, and then, considering the non-minimal action defined in eq. \eqref{NonMinimal_Action}, we compute the potential variation that arises by the insertion of counter-terms in photon emission amplitudes. In the end we sum the regularized potential with its variation in order to obtain the complete renormalized observable, which will be compared with the one in eq. \eqref{EM_Potential_deDonderd=4}.

\medskip

In the present case the only divergence that appears is for $d=4$ at 2-loop order. Then we just need to regularize the expression given by 
\begin{equation}
A_0\big|_{\text{2-loop}}=A_0^{(3, 0)}+A_0^{(3, 2)}\ ,
\end{equation}
from which expanding it around $d=4+\epsilon$, neglecting terms that vanish for $\epsilon\rightarrow 0$, we obtain 
\begin{equation}\label{Reg_A}
\begin{aligned}
    A_0^{(d=4)}\bigl|_{2-\text{loop}}&= \frac{1}{\epsilon}\left(\frac{10}{9}\frac{Q(G_Nm)^2}{\pi^4r^6}+\frac{1}{6}\frac{QG_N \alpha}{\pi^4 r^6}\right)
    -\frac{1}{27}\frac{Q(G_Nm)^2}{\pi^4r^6}\left(-28+45\log(C_E^2r^2)\right)\\ & -\frac{1}{36}\frac{QG_N\alpha}{\pi^4r^6}\left(-7+9\log(C_E^2r^2)\right) \ .
    \end{aligned}
\end{equation} 

\medskip

\begin{figure}[h]
\centering
\includegraphics[width=0.35\textwidth, valign=c]{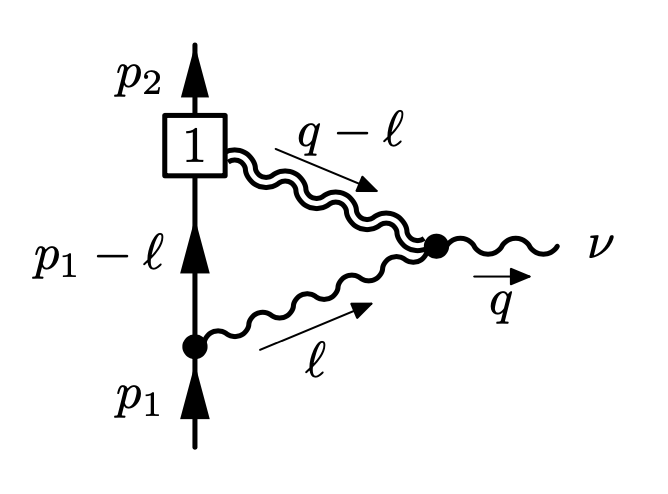}
\caption{Insertion of an internal counter-term vertex at 1-loop with an external photon.}
\label{fig:Ext_Photon_CT_1}
\end{figure}

Now we have to consider the potential variation that arises from the insertion of the counter-terms vertices reported in eqs. \eqref{Massive_CT_1} and \eqref{Massless_CT_2} into photon emission amplitudes. At tree-level no insertion can be performed since does not exists a 2 scalars - 1 photon counter-term vertex. This means that the first amplitude in which we can insert the counter-term is at 1-loop order, leading to the fact that no divergences can arise at that order. So then, the fact that eq. \eqref{1-Loop_EM_POT} is perfectly finite it is not a coincidence, but derives from the consideration by which there would be no counter-term to cancel the singularity, and this protects the 1-loop contribution of the potential to be well-defined. Moreover from the same argument can be derived that we expect divergences in $d=3$ only starting from the 3-loop order.
For what concerns the insertion in the 1-loop amplitude, the first process to take into account is the one in figure \ref{fig:Ext_Photon_CT_1}, whose contribution to the electromagnetic current is
\begin{equation}
\begin{aligned}
&-i\, \sqrt{4E_1E_2}\, \delta^{(1)}j_{\nu}^{(1)}(q^2)\\ &=2 \times \int \frac{d^{d+1} \ell}{(2 \pi)^{d+1}} \frac{iP_{\alpha \beta \gamma \delta} \ \left(\tau_{\phi^{2} A}\right)^{\mu}\left(\tau_{\phi^{2} h}^{ct}\right)^{\gamma \delta}\left(q-\ell\right)\left(\tau_{A^{2} h}\right)^{\alpha \beta}{}_{\mu \nu}(\ell, q)}{(\left(p_{1}-\ell\right)^{2}-m^{2}+i \epsilon)(\ell^{2}+i\epsilon)(\left(q-\ell\right)^{2}+i\epsilon)} \ .
\end{aligned}
\end{equation}
Considering the classical limit we get
\begin{equation}
    -i\, 2m\, \delta^{(1)}j_{0}^{(1)}(\vec{q}^2)=-iQ \int\frac{d^d\ell}{(2\pi)^d}\frac{P_{\mu\nu, \alpha\beta}\left(\tau_{\phi^{2} h}^{ct}\right)^{\mu\nu}\left(q-\ell\right)\left(\tau_{A^{2} h}\right)^{\alpha \beta}{}_{00}(\ell, q)\bigl|_{\ell^0=0}}{\Vec{\ell}^2(\Vec{q}-\Vec{\ell})^2}\ ,
\end{equation}
from which computing
\begin{equation}
\begin{aligned}
    P_{\mu\nu, \alpha\beta}&\left(\tau_{\phi^{2} h}^{ct}\right)^{\mu\nu}\left(q-\ell\right)\left(\tau_{A^{2} h}\right)^{\alpha \beta}{}_{00}(\ell, q)\bigl|_{\ell^0=0}\\
    &=-\frac{\kappa^2}{2(d-1)}(G_Nm)^{\frac{2}{d-2}}\Bigl(\beta^{(1)}(d)(d-2)(\Vec{\ell}^2\, \Vec{\ell}\cdot\Vec{q}-2\Vec{\ell}\cdot\Vec{q}\, \Vec{\ell}\cdot\Vec{q}+\Vec{\ell}\cdot\Vec{q}\, \Vec{q}^2)\\
    &-2\alpha^{(1)}(d)m^2\Vec{\ell}\cdot\vec{q}(\Vec{q}^2+\Vec{\ell}^2)-2\alpha^{(1)}(d)(d-3)m^2\Vec{\ell}\cdot\Vec{q}\, \Vec{\ell}\cdot\vec{q}\Bigl)\ ,
    \end{aligned}
\end{equation}
and exploiting eqs. \eqref{LR_1Loop_2} and \eqref{LR_1Loop_3}, one finally obtains the contribution to the electromagnetic current
\begin{equation}
    \delta^{(1)}j_{0}^{(1)}(\vec{q}^2)=\frac{1}{4}Q\kappa^2m\alpha^{(1)}(d)(G_Nm)^{\frac{2}{d-2}}\vec{q}^2J_{(1)}(\Vec{q}^2)\ .
\end{equation}
Then using eq. (\ref{Master_CT}), we obtain the variation to the electromagnetic potential as 
\begin{equation}\label{deltaA}
    \delta^{(1)}A_0^{(2)}(r) =  -4 G_N m (G_Nm)^{\frac{2}{d-2}} Q\alpha^{(1)}(d)\frac{\Gamma\left(\frac{d}{2}\right)^{2}}{\pi^{d-1} r^{2(d-1)}} \ . 
\end{equation}

\medskip

\begin{figure}[h]
\centering
\includegraphics[width=0.35\textwidth, valign=c]{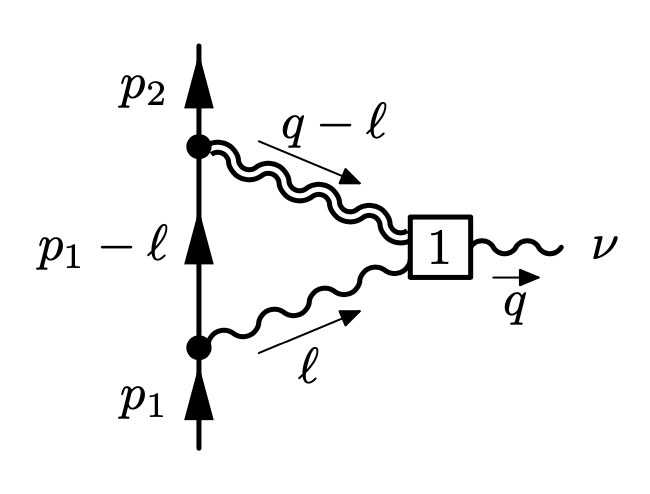}
\caption{Insertion of an external counter-term vertex at 1-loop with an external photon.}
\label{fig:Ext_Photon_CT_2}
\end{figure}

Furthermore, likely to the metric case, there should be another counter-term insertion given by the vertex in eq. \eqref{Massless_CT_2}, drawn in figure \ref{fig:Ext_Photon_CT_2}.
However also in this case an explicit calculation shows that the classical limit of the electromagnetic potential associated to such process is vanishing. We conclude then, that up to 2-loop order, the pure massless counter-term \eqref{Massless_CT_2} never contributes to the renormalization of the theory. It is interesting to observe that there is no known reason by which this should happen, so then cannot be excluded the possibility for it to be necessary for higher loop-order calculations. 

\medskip

In the end it is possible to obtain the complete renormalized potential by considering the addition of eq. \eqref{deltaA}, which must be expanded around $d=4+\epsilon$, to the regularized expression in \eqref{Reg_A}. One of the most important outcome is that the counter-term that is needed to be fixed in order to obtain finite results, is the same as the one already encountered in section \ref{sec:Renorm_Metric}. In fact, reporting here for clarity 
\begin{equation}
    \alpha^{(1)}(d)=\frac{\omega(4)}{d-4}+\Omega(4)\ ,
\end{equation}
where
\begin{equation}
    \omega(4)=\frac{5}{18 \pi}+\frac{G_N \alpha}{(G_N m)^2}\frac{1}{24 \pi}\ ,
\end{equation}
the 2-loop renormalized potential reads 
\begin{multline}\label{EM_Renorm}
    A_0^{(d=4)}\bigl|^{renorm}_{2-loop}=A_0^{(d=4)}\bigl|_{2-loop}+\delta^{(1)}A_0^{(2)}=\frac{Q}{4\pi}\Biggl(-\frac{8}{27}\frac{(G_Nm)^2}{\pi^3r^6}+\frac{1}{9}\frac{G_N\alpha}{\pi^3r^6}\\
    -\frac{16(G_Nm)^2}{\pi^2 r^6}\Omega(4)+\frac{20}{9}\frac{(G_Nm)^2}{\pi^3r^6}\log\left(\frac{G_Nm}{C_E^2r^2}\right)+\frac{1}{3}\frac{G_N\alpha}{\pi^3r^6}\log\left(\frac{G_Nm}{C_E^2r^2}\right)\Biggl)\ .
\end{multline}
Finally using the relation (\ref{CapOmega4}) the renormalized electromagnetic potential perfectly matches the classical calculation in eq. \eqref{EM_Potential_deDonderd=4}.

\medskip

In conclusion, we proved that from photon emission scattering amplitudes, in which both photons and gravitons circulate in the loops, it is possible to recover the electromagnetic potential associated to the Reissner-Nordstr\"om-Tangherlini solution. We showed the divergences that arise in $d=4$ are renormalized by the same counter-terms already  employed in the metric case, and that such procedure gives rise to logarithmic terms that perfectly reproduce the classical solution.

\chapter{Conclusions}
\label{chap:Conclusions}

In this thesis we have shown that the Reissner-Nordstr\"om-Tangherlini solution, describing the metric and the potential generated by a static and spherically symmetric object of charge $Q$ and mass $m$ in any dimension, can be derived from scattering amplitudes describing the emission of either a graviton or a photon from a scalar field with the same charge and mass. This generalizes the work of \cite{Mougiakakos:2020laz} in which the chargeless case was considered up to 3-loop order. The present work extends the original papers \cite{Donoghue:2001qc, Bjerrum-Bohr:2002fji} where amplitude computations were applied to derive the metric of all black hole solutions in four dimensions up to second post-Minkowskian order. 

\medskip

In chapter \ref{QG_as_EFT} we reconstructed the scalar electrodynamics action by gauging the global $\text{U}(1)$ symmetry, and we discussed the Faddeev-Popov procedure to quantize it. Then we applied the same considerations in order to get general relativity by gauging the global translational symmetry, recovering the Einstein-Hilbert action as the kinetic term of such theory. Moreover we concluded that since the general coordinate transformations are intrinsically non-abelian, general relativity can be considered a non-abelian gauge theory. Following the Faddeev-Popov procedure an attempt to quantize general relativity has been made, but since the canonical mass dimension of the gravitational coupling constant is negative for all the dimensions greater than two, we recover the well-known result by which such theory is non-renormalizable. Then, we applied the tools of effective field theories in order to rewrite such theory by means of an infinite series of all the possible energy ordered operators compatible with the symmetry of the original action. Finally in section \ref{Sec_Classical_Limit} we showed how systematically restore the $\hbar$'s in the theory with the idea to make the classical limit in which $\hbar\rightarrow 0$ more simple. An important outcomes is that the effective theory allows us to separate the effects of massive and massless particles, which contribute to the classical limit in different ways.

\medskip

In chapter \ref{Classical_Chapter} we made an explicit example of a classical gravitational observable recovered from the classical limit of quantum computations. Reviewing the original work of \cite{Donoghue:1994dn, Donoghue:2001qc, Bjerrum-Bohr:2002fji}, we showed how to compute static metrics in four space-time dimensions from graviton emission scattering amplitudes. Starting with the Schwarzschild metric, considering the tree-level graviton emission from a scalar particle, in eq. \eqref{Tree_Level_Chap3_Calc} it is recovered the first post-Minkowskian order of the metric in the de Donder gauge. In fact, since in section \ref{sec:Quantization_of_Gravity} we chose to employ the de Donder gauge for the gravitational sector (since the graviton field equations acquire a simple expression), then all the quantum computations we have made in this thesis are expressed in this gauge. For the 1-loop case the number of processes to take into account proliferate, and it is necessary to state some selection rules that pick only those contributions which are non-vanishing in the classical limit. These rules then lead to eq. \eqref{Example_SchwarzschildPerturbation_deDonder_1loop}, and the comparison with the classical Schwarzschild metric is done in section \ref{sec:Schwarzschild_Harmonic}, in which a procedure to convert the de Donder gauge in the harmonic one, which is more used in literature, has been provided. Finally a perfect match between the two computations is found. In the end we consider another static metric, namely the Reissner-Nordstr\"om one, which describes a static electrically charged black hole solution. In order to recover such metric, it is shown how the addition of photons in the graviton emission loop amplitude is needed, and from the 1-loop computation it is recovered the term proportional to the electric charge of the metric in the harmonic gauge in eq. \eqref{RN_Metric_Chap3}.

\medskip

Then the attention is focused to recover the Reissner-Nordstr\"om-Tangherlini solution from scattering amplitudes, and as a preliminary step, since we are interested in compare classical and quantum derivations, in chapter \ref{chap:RNT_Classical_Solution} we reviewed the classical solution provided in \cite{Tangherlini:1963bw}. In particular, following the strategy outlined in \cite{Mougiakakos:2020laz}, we recovered the classical solution in the de Donder gauge in terms of a post-Minkowskian expansion of both the metric and the electromagnetic potential associated to the black hole. The core of the procedure was find the coordinate change function in eq. \eqref{fr_Solved}, that completely determines the reference frame transformation from spherical coordinates to de Donder coordinates. Substituting this expression inside the metric and the potential, in eqs. \eqref{hid} and \eqref{EM_Potential_deDonder} we reach our claim. However the procedure gives rise to divergences in four and five space-time dimensions, that we classically proved to be traceable to a redundant gauge freedom in those dimensions. Once modified the ansatz of the coordinate change function, we found its well-defined form for $d=3$ and $d=4$, which consequently lead to the expressions of the metric and the potential in those dimensions. In the end we observed how in such expressions logarithmic terms arise in the post-Minkowksian expansion, as well as free constants, which parametrize the redundant gauge freedom.

\medskip

In chapter \ref{chap:Metric_from_Amplitudes} actual computations of graviton emission processes are carried out with the aim to recover the Reissner-Nordstr\"om-Tangherlini metric. First we outlined the general procedure to extract from the quantum amplitudes only the terms which will reconstruct the classical phenomenology. Following \cite{Mougiakakos:2020laz}, this is accomplished by considering only the amplitudes in which a single massive line emits $n-j$ gravitons and $j$ photons, whose internal-tree structure ends up with a graviton emission, and extracting out from their energy expansion the term proportional to the master integral $J_{(l)}(\vec{q}^2)$. In order to do that the strategy followed is the one presented in \cite{Bjerrum-Bohr:2018xdl}, in which exploiting the low-energy and static limit it is possible to integrate out the temporal component of the internal momenta, and reduce the amplitudes in an interaction of external sources reproducing the original work of \cite{Duff:1973zz}. In fact this result can be read backward, and state that the only quantum processes that contribute to reconstruct classical physics are the ones in which performed the classical limit, that corresponds to cut the massive lines, reproduce the quantum tree-graphs of \cite{Duff:1973zz}. Then in section \ref{sec:AmpCalc_Metric} calculations are carried out, and we showed, as expected, the appearance of divergences in four and five dimensions. These divergences
are renormalized by the insertion of the non-minimal action in eq. \eqref{NonMinimal_Action}, in which are taken into account both gravitational and electromagnetic coupling constants. Then we commented how from Fourier transform properties, it is possible to show that in the renormalization process of the metric only eq. \eqref{theonlynonminimalcoupling} contributes, and decoupling the counter-term \eqref{alpha} in a finite and a divergent part, following \cite{Mougiakakos:2020laz, Fragomeno, DOnofrio:2021tap}, a renormalized expression of the metric has obtained, showing that the same logarithmic terms of the classical solution arise. We commented that this procedure only needs to fix the divergent part of the counter-term, since the finite one belongs to high-energy physics. Moreover, even if the logarithmic terms match naturally, it is possible to impose a strict relation between the finite parts, which leads to a relation between the constant that parametrize the redundant classical gauge freedom and the unfixed high-energy parameter. We finally showed that the exact same relation is coherent for both the first and second post-Minkowskian order expansion of the metric.

\medskip

Finally the original calculations in which we proved that the electromagnetic potential associated to the Reissner-Nordstr\"om-Tangherlini solution is recovered from photon emission scattering amplitudes are presented. In particular we use the exact same scheme we employed in the metric case, in which the classical contributions of the electromagnetic current in momentum space are given by the terms of the amplitudes proportional to the master integral, in which a single massive line emits $n-j-1$ gravitons and $j+1$ photons, ending up in a photon emission through an internal tree-structure interaction. Then we proved that such structure leads to the conclusion that the classical limit is again reconstructed by quantum-tree graphs, which using the formalism of Duff, are expressed in eq. \eqref{Duff_Amp_Phot}. Exploiting again the program outlined in \cite{Bjerrum-Bohr:2018xdl}, integrating out the temporal component of the internal momenta, explicit calculations are provided and we showed a perfect agreement with the classical potential up to the problematic case in which a divergence appears, which in particular happens at third post-Minkowksian order for $d=4$. One of the most important results then, is that such divergence is canceled by the very same count-terms that we already defined in the metric case, leading to eq. \eqref{EM_Renorm}, from which using the relation \eqref{CapOmega4} a perfect match with the classical solution can be observed, also in the finite pieces. The aforementioned observations gave rise to the publication of a paper in which all the results obtained in this thesis are presented \cite{DOnofrio:2022cvn}.

\medskip

\begin{figure}[h]
\centering
\includegraphics[width=0.4\textwidth, valign=c]{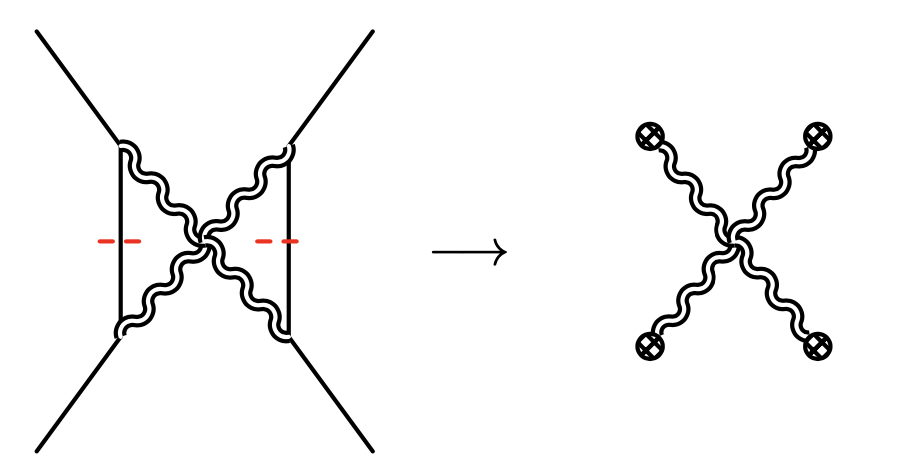}
\hspace{1cm}
\includegraphics[width=0.4\textwidth, valign=c]{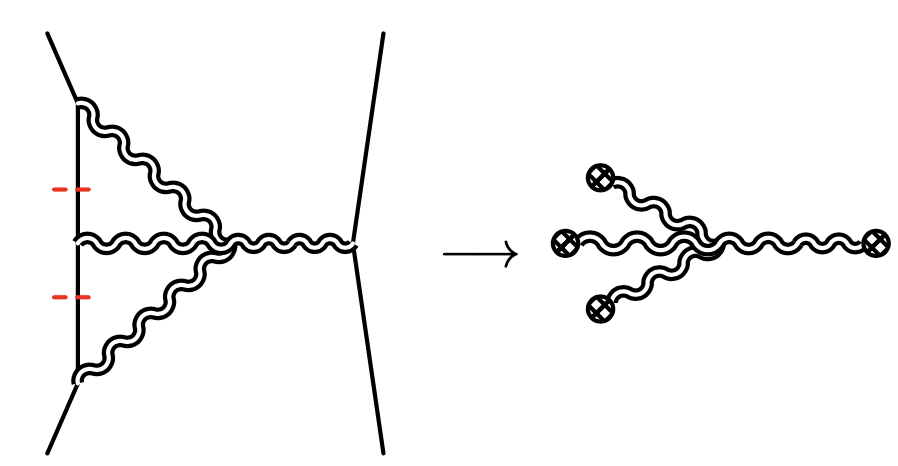}
\caption{Example of 2-loop diagrams that contributes to the classical limit of the 2-to-2 scattering amplitudes of scalar massive particles taken from \cite{Bjerrum-Bohr:2022blt}. In the picture it is shown how in the classical limit, that corresponds to cut the internal massive lines, the diagrams correspond to external sources (represented by the blobs) that interact through a quantum-tree graph.}
\label{fig:Conclusion}
\end{figure}

In the end, we conclude that the present strategy for extracting classical gravitational physics from quantum scattering amplitudes is proved to be perfectly coherent and self-consistent. In particular all the general results that can be recovered classically have a quantum counter-part, which makes this calculation method completely equivalent to a classical derivation. For instance we showed that classically the reason why we expect divergences in the de Donder gauge is because in $d=3$ and $d=4$ there exists a redundant gauge transformation. The quantum approach to the problem shows that it is possible to arrive at the same conclusion just by looking at the non-minimal action that renormalizes the theory, and since it would not be possible to define eq. \eqref{theonlynonminimalcoupling} in those dimensions, there would be no divergences except the ones we encounter. Similarly in section \ref{sec:EM_in_ChapRNT} we showed that only the temporal component of the classical potential is non-vanishing, and in eq. \eqref{EM_Pot_FormFact_NotUseful} we arrive to the same conclusion through a completely quantum approach. 
This extraordinary correspondence is not present only for these observables, but in every application of this approach. In fact many considerations we did, were not specific for the metric or the electromagnetic potential, but hold for every gravitational observable computed with the present program. For example if we focus on the derivation of the Newton potential between two massive objects, even non static ones, then it is possible to reconstruct it in a post-Minkowskian expansion by means of 2-to-2 scattering amplitudes of massive particles mediated by gravitons. In this case, even if the procedure to extract the classical contributions from the amplitudes is more involved than the one we discussed in this thesis, the general conclusions regarding the phenomenology of the classical limit hold. In fact, performing the classical limit of such amplitudes, it leads to cut the internal massive lines and consider the interaction of external source like in figure \ref{fig:Conclusion}. 
Such quantum tree-graphs are identical to the ones that arise from the world-line formalism \cite{Jakobsen:2022fcj, Kalin:2020fhe, Mogull:2020sak, Jakobsen:2021smu, Jakobsen:2021zvh, Dlapa:2021npj}, and show how many results obtained in this thesis are universal, since we got the same kind of diagrams for the graviton and photon emission amplitudes. 

\medskip

There are many ways in which this work can be extended. 
As mentioned in chapters \ref{chap:Metric_from_Amplitudes} and \ref{chap:Photon_Emission_Processes}, the counter-term vertex in eq.  \eqref{Massless_CT_2} does not contribute to the renormalization of the theory up to 2-loop calculations. However, it is interesting to observe that there are no known reasons why this should happen, and therefore an higher order calculation could show whether this vertex contributes or does not. If the vertex will contribute at higher-order, then what happens in this work is proved to be a coincidence. Otherwise if it will not contribute even beyond 2 loops, this would seem an insight of a general behavior.
It would be also very interesting to perform this higher loop computation in a generalized hybrid gauge as in \cite{Jakobsen:2020ksu}. In fact, since in the reference it is shown that up to 1-loop order there exists a particular gauge choice in which the graviton emission amplitudes do not diverge, and consequently the  Schwarzschild-Tangherlini metric is completely finite, it could be nice to see if this special gauge makes the amplitudes divergenceless also in the electromagnetically coupled case, and if it is still true at higher orders. Moreover, since the non-minimal couplings are gauge invariant, they still contribute, and then we expect that in this gauge the variations of the observables that come from the counter-terms disappear in the classical limit.
Finally one of the most exciting extension is the possibility to generalize the present work to graviton and photon emission processes from fermionic matter,  extending the four-dimensional 1-loop analysis of \cite{Donoghue:2001qc, Bjerrum-Bohr:2002fji}, finding the post-Minkowskian expansion of a rotating black hole metric in arbitrary dimension. The reason why this is important is because the literature on spinning black holes in arbitrary dimensions is much less abundant than the one that treat static space-time solutions, and this program could give many insights in order to explore the richer structure of black hole solutions in higher dimensions \cite{Emparan:2008eg}.

\medskip

The techniques to extract classical contributions from quantum processes discussed in thesis then, are not only interesting on their own, but have many applications. In fact they have been widely applied in the last few years to determine the dynamics of binary systems from two-body scattering amplitudes \cite{Cheung:2018wkq,Bern:2019crd,Bern:2019nnu,Bern:2021yeh, Chung:2019yfs,Cheung:2020gyp,Bjerrum-Bohr:2019kec,Mogull:2020sak,Guevara:2018wpp,Guevara:2019fsj,Guevara:2020xjx,Bautista:2021wfy,Moynihan:2019bor,Emond:2020lwi}. As we did our analysis in generic space-time dimensions, also these computations have been extended to higher-dimensions \cite{Cristofoli:2020uzm,KoemansCollado:2018hss} (for a review see   \cite{Bjerrum-Bohr:2022blt}), for two reasons mainly. First, it is known that gravity has a much richer phenomenology in higher dimensions, which means that hidden theoretical structures could emerge in $D\neq 4$, which would give clues for other problems. Second, as it is clear from this thesis work, even if we calculate classical physics out of quantum computations, the amplitudes are affected by divergences that have to be renormalized. However, in every renormalization process one has to impose a regularization procedure, which usually corresponds to the dimensional one. So then, following this approach, it is worth to impose a generic space-time dimension from the beginning. 

\medskip

The reason why the scientific community is so interested in the dynamics of massive objects though, and in particular in bound binary systems, is because it can be linked to the gravitational-wave emission of a coalescence process. In fact, since this research program can be applied to make systematic computations for the post-Minkowskian dynamics of binary bound systems, it could lead to high-precision predictions in the context of gravitational-wave emission \cite{Buonanno:2022pgc}. 
Moreover one of the most relevant feature of the proposed techniques, is that they allow the computations of gravitational observables in a complete analytic form, which means that such program, not only has the power to significantly improve the state-of-the-art precision of the predictions of gravitational-wave emission templates, but it is candidate to be an analytical alternative to numerical methods. In fact since up to now the only way to theoretically model the merger phase of a binary coalescence is by employing numerical relativity, it seems to be possible to extend these techniques on a curved space-time in order to obtain analytic predictions of such process \cite{Buonanno:2022pgc}.
This last research line, and all the other ones, are far from being completely understood or worked out, and this is one of the reasons why it is worth to continue to explore the methods to extract classical gravitational observables from scattering amplitudes.

\appendix

\chapter{Feynman rules}\label{Appendix_FeynRules}

In this appendix are listed all the Feynman rules used in the thesis. In the first part are given the propagators in the interested gauges and in generic space-time dimension $D=d+1$, while in the second part are reported the interaction vertices between photons, gravitons and massive scalars. Notice that the single wiggly lines stand for photons, the double wiggly lines stand for gravitons and the solid lines stand for charged scalars. 

\begin{itemize}

\item Scalar propagator of mass $m$:
\begin{equation}
\includegraphics[width=0.15\textwidth, valign=c]{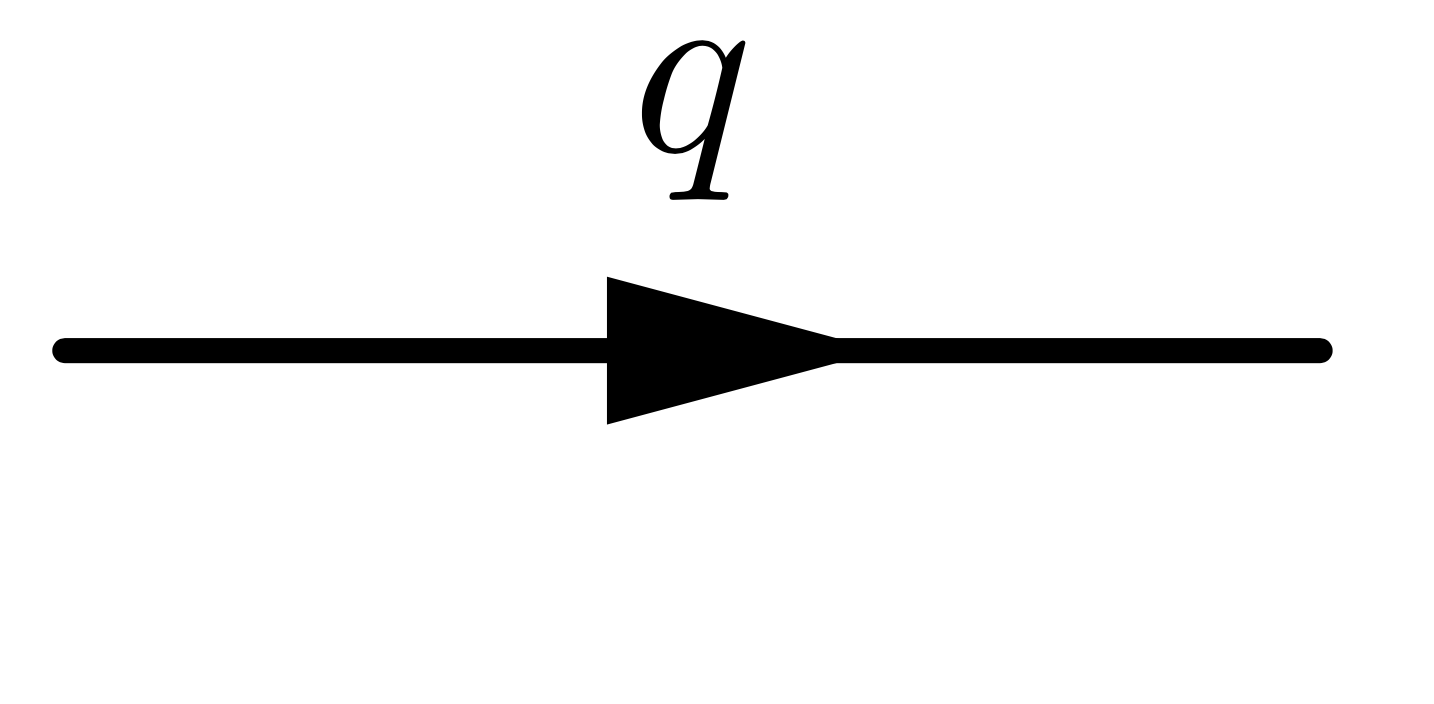}=\frac{i}{q^2-m^2+i\epsilon}\ .
\end{equation}

\item Photon propagator in Feynman gauge: 
\begin{equation}
\includegraphics[width=0.18\textwidth, valign=c]{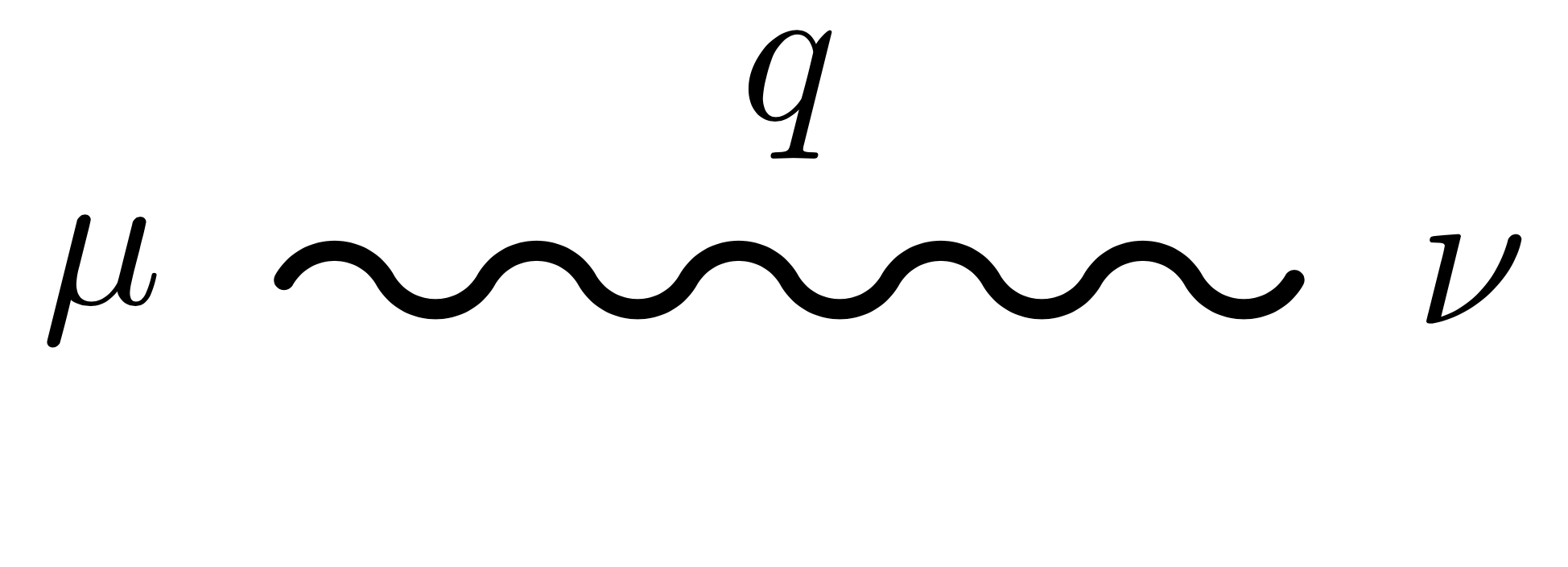}=-\frac{i\eta^{\mu\nu}}{q^2 +i\epsilon}\ .
\end{equation}

\item Graviton propagator in de Donder gauge: 
\begin{equation}
\includegraphics[width=0.20\textwidth, valign=c]{Diagrams/Graviton_Propagator.png}=i\frac{P_{\alpha\beta,\mu\nu}}{q^2 +i\epsilon} \ ,
\end{equation}
with $P_{\alpha\beta,\mu\nu}$ defined by
\begin{equation}\label{PropGraviton}
    P_{\alpha\beta,\mu\nu} =  \frac{1}{2} \Big( \eta_{\mu\alpha}\eta_{\nu\beta}+\eta_{\mu\beta}\eta_{\nu\alpha}-\frac{2}{d-1}\eta_{\mu\nu}\eta_{\alpha\beta} \Big) \ .
\end{equation}

\item 2 scalars - 1 photon vertex: 
\begin{equation}\label{2scalars_1photon}
\includegraphics[width=0.25\textwidth, valign=c]{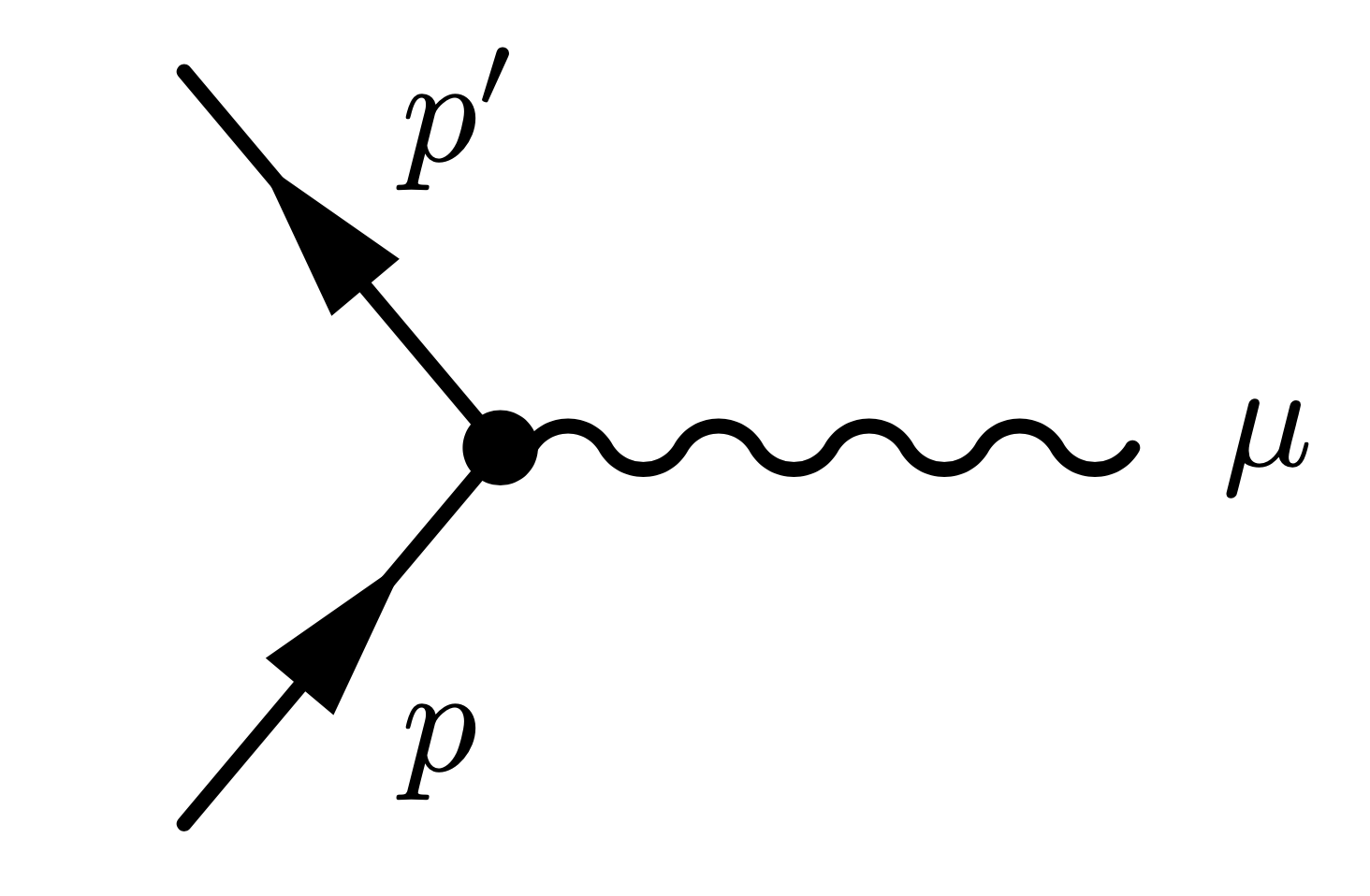}=\left(\tau_{\phi^2A}\right)^\mu(p,p')=-iQ\left(p+p'\right)^\mu \ . 
\end{equation}

\item 2 scalars - 1 graviton vertex \cite{Bjerrum-Bohr:2014lea}: 
\begin{equation}\label{2scalars_1graviton}
\includegraphics[width=0.25\textwidth, valign=c]{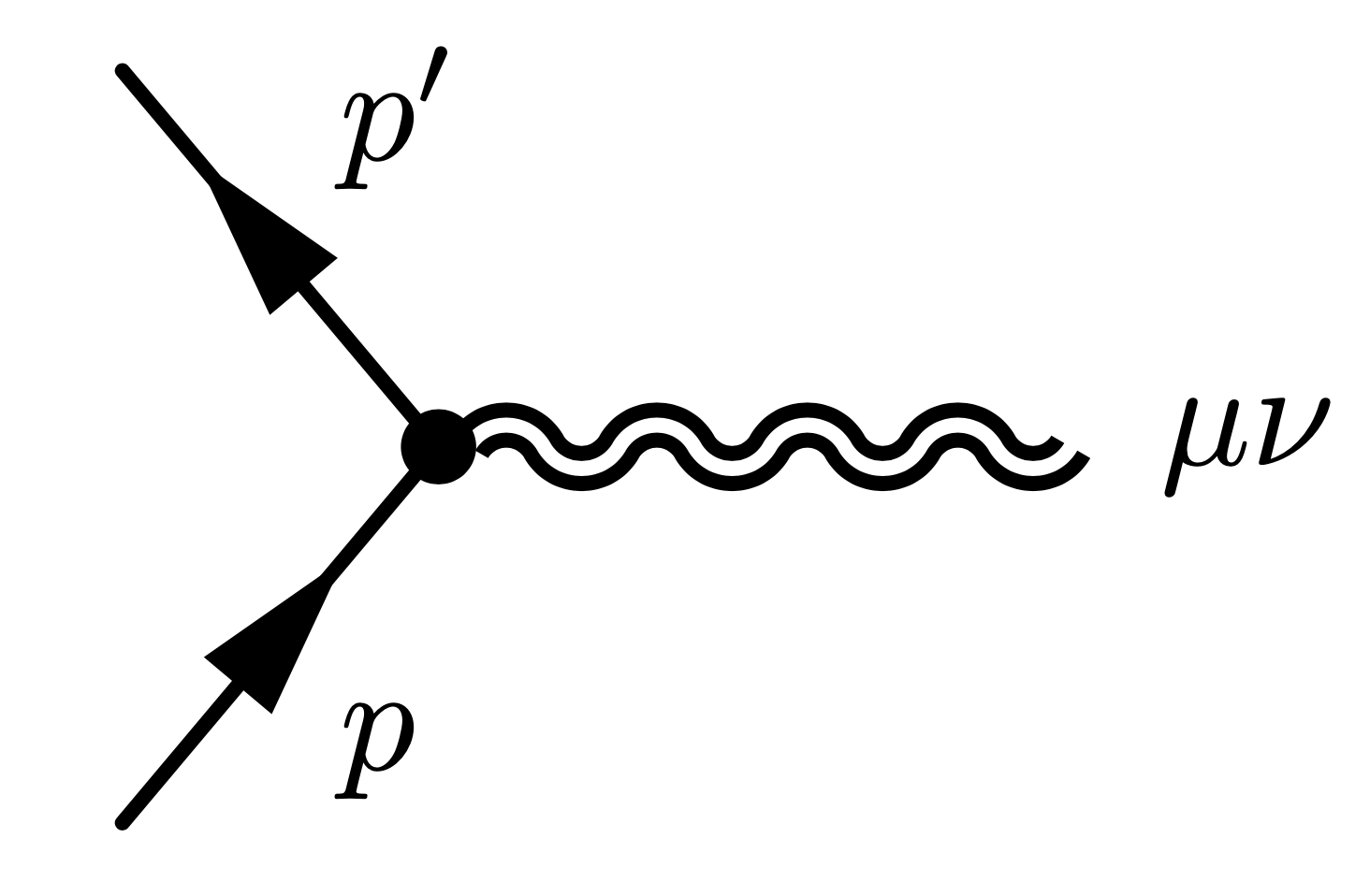}=\left(\tau_{\phi^2 h}\right) ^{\mu\nu}(p,p')=-\frac{i\kappa}{2}\Big(\big( p ^\mu p'{}^\nu + p ^\nu p'{}^\mu\big) - \eta ^{\mu\nu}\big(p\cdot p' - m^2 \big)\Big) \ . 
\end{equation}

\item 2 photons - 1 graviton vertex \cite{Bjerrum-Bohr:2014lea}: 
\begin{equation*}
\includegraphics[width=0.25\textwidth, valign=c]{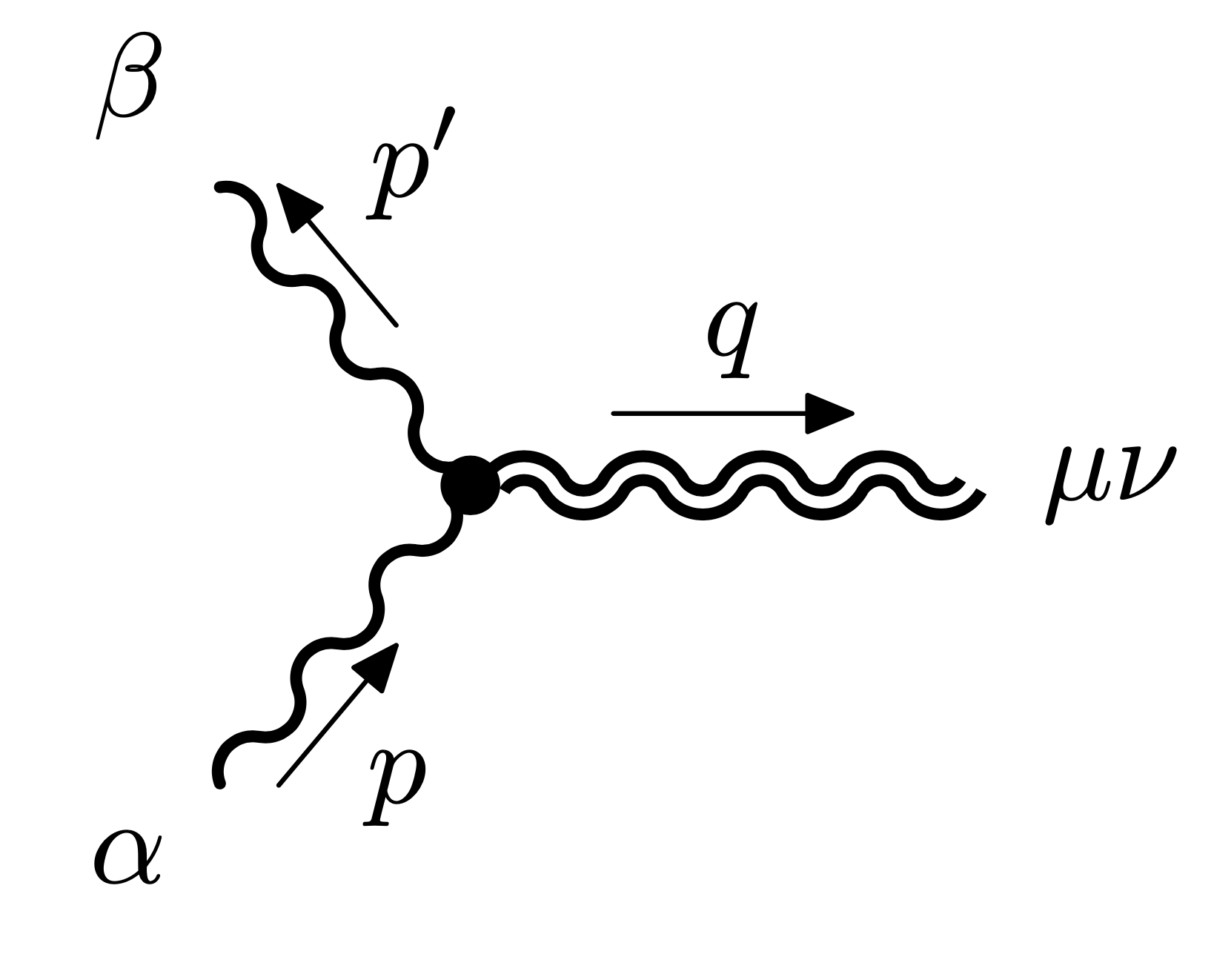}=
\left(\tau_{A^2h}\right) ^{\mu\nu,\alpha,\beta}(p,p')
\end{equation*}
\begin{equation}\label{2photon_1graviton}
\begin{split}
=i\kappa&\left( P_{(4)}  ^{\mu\nu,\alpha\beta} (p\cdot p')+ \frac{1}{2} \left( \frac{1}{2} \eta ^{\mu\nu} (p ^\alpha p'{}^\beta + p ^\beta p' {}^\alpha) + \eta ^{\alpha\beta} (p^ \mu p'{}^ \nu + p^ \nu p'{}^ \mu) \right.\right.\\ &\left.\left. -\frac{1}{2}\left(
    \eta^ {\beta\nu} (p^ \mu p'{}^ \alpha + p^ \alpha p'{}^ \mu) + \eta^ {\alpha\nu} (p^ \mu p'{}^ \beta + p^ \beta p'{}^ \mu)+ \eta^ {\alpha\mu} (p^ \nu p'{}^\beta + p^\beta p'{}^\nu) \right.\right.\right.\\&\left.\left.\left.+ \eta^{\beta\mu} (p^\nu p'{}^\alpha + p^\alpha p'{}^\nu)
    \right)\right)\right) \  ,
\end{split}
\end{equation}
where $P_{(4)}^{\mu\nu, \alpha\beta}$ is \eqref{PropGraviton} for $d=3$.

\item 2 photons - 2 gravitons vertex \cite{Bjerrum-Bohr:2014lea}: 
\begin{equation*}
\includegraphics[width=0.25\textwidth, valign=c]{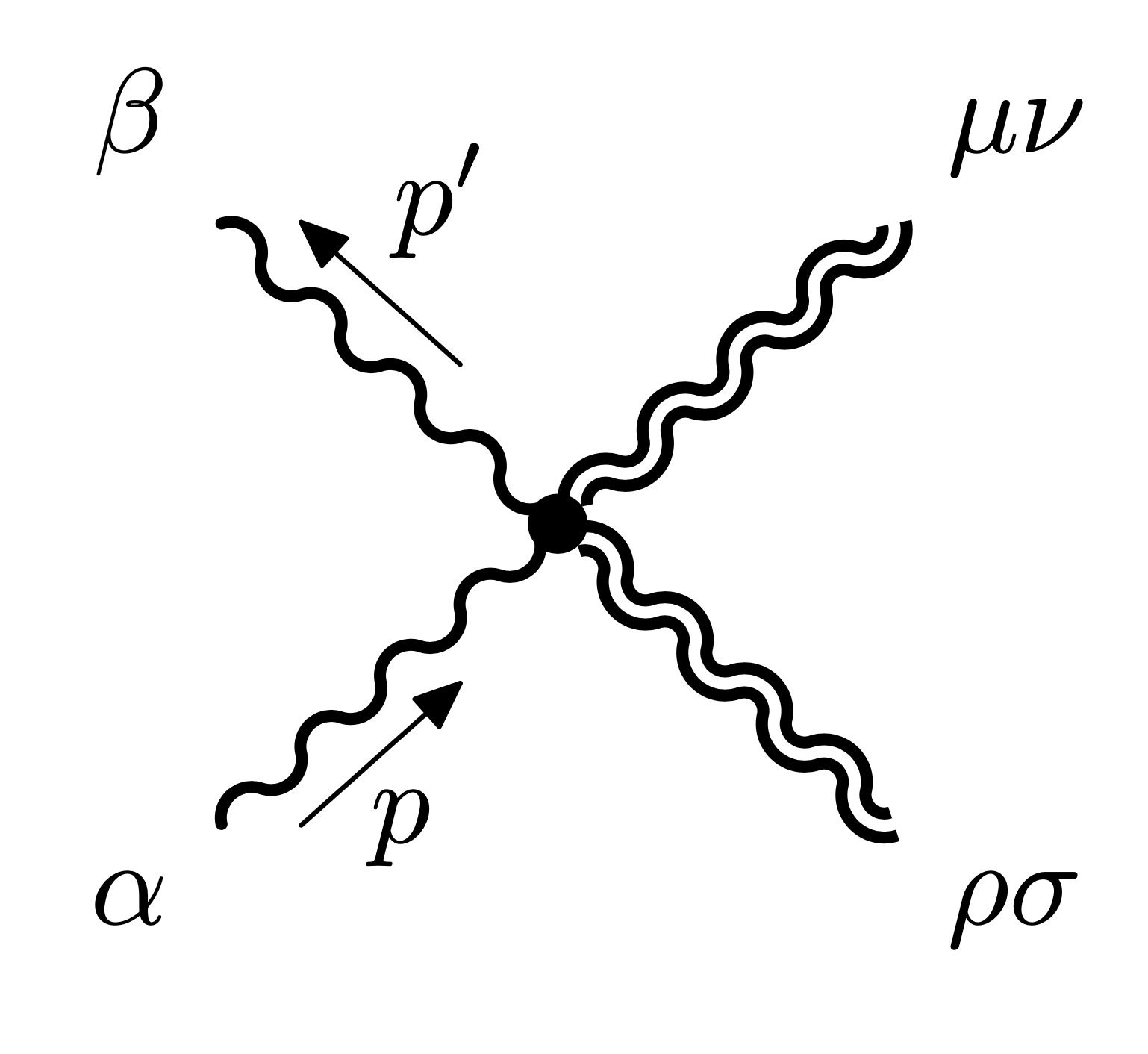}=\left({\tau}_{A^2h^2}\right){}_{\mu\nu,}{}_{\rho\sigma,\alpha,\beta}(p,p')
\end{equation*}
\begin{equation}\label{VertA2h2}
\begin{aligned} 
\quad=&- \frac{i\kappa^{2}}{4}\left( \left( p_{ \beta} p'_{ \alpha}-\eta_{\alpha \beta}p\cdot p'\right)\left(\eta_{\mu \rho} \eta_{\nu \sigma}+\eta_{\mu \sigma} \eta_{\nu \rho}-\eta_{\mu \nu} \eta_{\rho \sigma}\right)+\eta_{\mu \rho}\left(\eta_{\alpha \beta}\left(p_{ \nu} p'_{ \sigma}+p_{ \sigma} p'_{ \nu}\right) \right. \right. \\
&\left.-\eta_{\alpha \nu} p_{ \beta} p'_{ \sigma}-\eta_{\beta \nu} p_{ \sigma} p'_{ \alpha}
-\eta_{\beta \sigma} p_{ \nu} p'_{ \alpha}-\eta_{\alpha \sigma} p_{ \beta} p'_{ \nu}+p\cdot p'\left(\eta_{\alpha \nu} \eta_{\beta \sigma}+\eta_{\alpha \sigma} \eta_{\beta \nu}\right)\right) \\
&+\eta_{\mu \sigma}\left(\eta_{\alpha \beta}\left(p_{ \nu} p'_{ \rho}+p_{ \rho} p'_{ \nu}\right)-\eta_{\alpha \nu} p_{ \beta} p'_{ \rho}-\eta_{\beta \nu} p_{ \rho} p'_{ \alpha} -\eta_{\beta \rho} p_{ \nu} p'_{ \alpha}-\eta_{\alpha \rho} p_{ \beta} p'_{ \nu}\right.\\
&\left.+p\cdot p' \left(\eta_{\alpha \nu} \eta_{\beta \rho}+\eta_{\alpha \rho} \eta_{\beta \nu}\right)\right) +\eta_{\nu \rho}\left(\eta_{\alpha \beta}\left(p_{ \mu} p'_{ \sigma}+p_{ \sigma} p'_{ \mu}\right)-\eta_{\alpha \mu} p_{ \beta} p'_{ \sigma}-\eta_{\beta \mu} p_{ \sigma} p'_{ \alpha}\right.\\
&\left.-\eta_{\beta \sigma} p_{ \mu} p'_{ \alpha}-\eta_{\alpha \sigma} p_{ \beta} p'_{ \mu}+p\cdot p'\left(\eta_{\alpha \mu} \eta_{\beta \sigma}+\eta_{\alpha \sigma} \eta_{\beta \mu}\right)\right) +\eta_{\nu \sigma}\left(\eta_{\alpha \beta}\left(p_{ \mu} p'_{ \rho}+p_{ \rho} p'_{ \mu}\right)\right.\\
&\left. -\eta_{\alpha \mu} p_{ \beta} p'_{ \rho}-\eta_{\beta \mu} p_{ \rho} p'_{ \alpha}\right.\left.-\eta_{\beta \rho} p_{ \mu} p'_{ \alpha}-\eta_{\alpha \rho} p_{ \beta} p'_{ \mu}+p\cdot p'\left(\eta_{\alpha \mu} \eta_{\beta \rho}+\eta_{\alpha \rho} \eta_{\beta \mu}\right)\right)\\
& -\eta_{\mu \nu}\left(\eta_{\alpha \beta}\left( p_{ \rho} p'_{ \sigma}+p_{ \sigma} p'_{ \rho} \right)-\eta_{\alpha \rho} p_{ \beta} p'_{ \sigma}-\eta_{\beta \rho} p_{ \sigma} p'_{ \alpha}-\eta_{\beta \sigma} p_{ \rho} p'_{ \alpha}-\eta_{\alpha \sigma} p_{ \beta} p'_{ \rho}\right.\\
&\left.+p\cdot p'\left(\eta_{\alpha \rho} \eta_{\beta \sigma}+\eta_{\beta \rho} \eta_{\alpha \sigma}\right)\right) -\eta_{\rho \sigma}\left(\eta_{\alpha \beta}\left(p_{ \mu} p'_{ \nu}+p_{ \nu} p'_{ \mu}\right)-\eta_{\alpha \mu} p_{ \beta} p'_{ \nu}-\eta_{\beta \mu} p_{ \nu} p'_{ \alpha}\right.\\
&\left.-\eta_{\beta \nu} p_{ \mu} p'_{ \alpha}-\eta_{\alpha \nu} p_{ \beta} p'_{ \mu}+p\cdot p'\left(\eta_{\alpha \mu} \eta_{\beta \nu}+\eta_{\beta \mu} \eta_{\alpha \nu}\right)\right) \\
&+\left(\eta_{\alpha \rho} p_{ \mu}-\eta_{\alpha \mu} p_{ \rho}\right)\left(\eta_{\beta \sigma} p'_{ \nu}-\eta_{\beta \nu}p'_{\sigma}\right)
+\left(\eta_{\alpha \sigma} p_{ \nu}-\eta_{\alpha \nu} p_{ \sigma}\right)\left(\eta_{\beta \rho} p'_{ \mu}-\eta_{\beta \mu}p'_{\rho}\right)
\\
&+\left.\left(\eta_{\alpha \sigma} p_{ \mu}-\eta_{\alpha \mu} p_{ \sigma}\right)\left(\eta_{\beta \rho} p'_{ \nu}-\eta_{\beta \nu}p'_{\sigma}\right)
+\left(\eta_{\alpha \rho} p_{ \nu}-\eta_{\alpha \nu} p_{ \rho}\right)\left(\eta_{\beta \sigma} p'_{ \mu}-\eta_{\beta \mu}p'_{\sigma}\right)
\right) \ . 
\end{aligned}
\end{equation}

\item 3 gravitons vertex \cite{DeWitt:1967uc, Sannan:1986tz}: 
\begin{equation*}
\includegraphics[width=0.3\textwidth, valign=c]{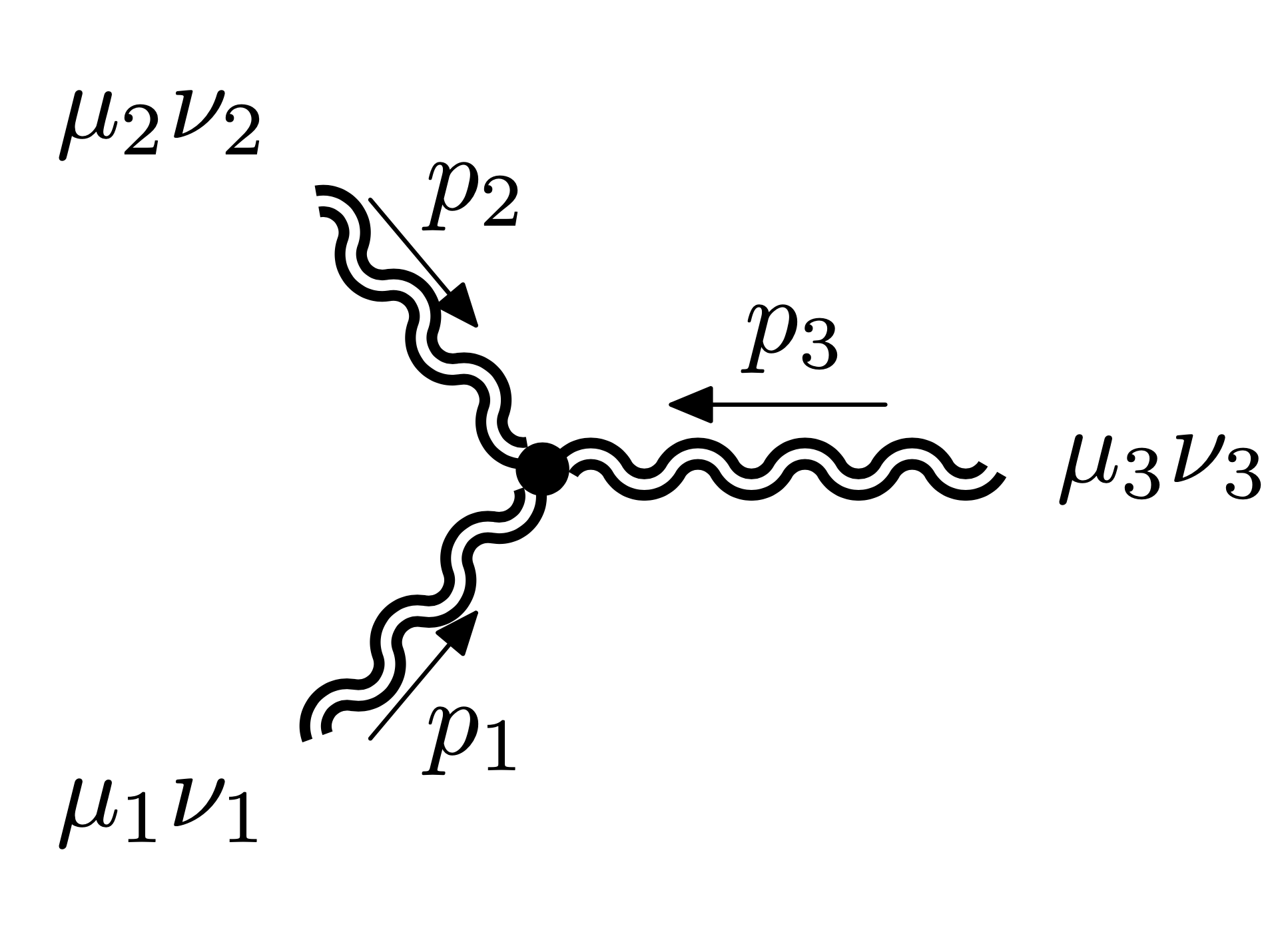}=\left(\tau_{h^3}\right)^{\mu_1\nu_1,\mu_2\nu_2,\mu_3\nu_3}(p_1,p_2,p_3)
\end{equation*}
\begin{equation}\label{3Graviton_Vertex_deWitt_AppEq}
\begin{aligned}
     = -2i\kappa\,  \text{Sym} \ \Big( &-\frac{1}{4}P_3\big(p_1 \cdot p_2 \eta^{\mu_1\nu_1}\eta^{\mu_2\nu_2}\eta^{\mu_3\nu_3} \big) -\frac{1}{4}P_6\big(p_1^{\mu_2} p_1^{\nu_2} \eta^{\mu_1\nu_1}\eta^{\mu_3\nu_3} \big) \\&+ \frac{1}{4}P_3\big(p_1 \cdot p_2 \eta^{\mu_1\mu_2}\eta^{\nu_1\nu_2}\eta^{\mu_3\nu_3} \big)
      +\frac{1}{2}P_6\big(p_1 \cdot p_2 \eta^{\mu_1\nu_1}\eta^{\mu_2\mu_3}\eta^{\nu_2\nu_3} \big) \\ &+P_3\big(p_1^{\mu_2} p_1^{\nu_3} \eta^{\mu_1\nu_1}\eta^{\nu_2\mu_3} \big)-\frac{1}{2}P_3\big(p_1^{\nu_2} p_2^{\mu_1} \eta^{\nu_1\mu_2}\eta^{\mu_3\nu_3} \big)
     \\ &+\frac{1}{2}P_3\big(p_1^{\mu_3} p_2^{\nu_3}\eta^{\mu_1\mu_2}\eta^{\nu_1\nu_2} \big) +\frac{1}{2}P_6\big(p_1^{\mu_3} p_1^{\nu_3}\eta^{\mu_1\mu_2}\eta^{\nu_1\nu_2} \big) \\ &+
     P_6\big(p_1^{\mu_2} p_2^{\nu_3}\eta^{\nu_2\mu_1}\eta^{\nu_1\mu_3} \big) +P_3\big(p_1^{\mu_2} p_2^{\mu_1}\eta^{\nu_2\mu_3}\eta^{\nu_3\nu_1} \big) \\ &-P_3\big(p_1 \cdot p_2 \eta^{\nu_1\mu_2}\eta^{\nu_2\mu_3}\eta^{\nu_3\mu_1} \big) \Big) \ .
\end{aligned}
\end{equation}
Here $"\text{Sym}"$ indicates that a symmetrization has to be performed on each index pair, while $"P"$ indicates that a summation has to be performed on all the distinct permutations of the momentum-index triplets (the subscript gives the number of terms).\footnote{For example: $$\text{Sym}\big[P_3\big(p_1^{\mu_2} p_2^{\mu_1}\eta^{\nu_2\mu_3}\eta^{\nu_3\nu_1} \big)\big] = p_1^{\mu_2} p_2^{\mu_1}\eta^{\nu_2\mu_3}\eta^{\nu_3\nu_1} + p_1^{\mu_3} p_3^{\mu_1}\eta^{\nu_3\mu_2}\eta^{\nu_2\nu_1} + p_2^{\mu_3} p_3^{\mu_2}\eta^{\nu_3\mu_1}\eta^{\nu_1\nu_2} \ .$$}

\item 3 gravitons vertex with two quantum fields and an external field \cite{Donoghue:1994dn}:
\begin{equation*}
\includegraphics[width=0.25\textwidth, valign=c]{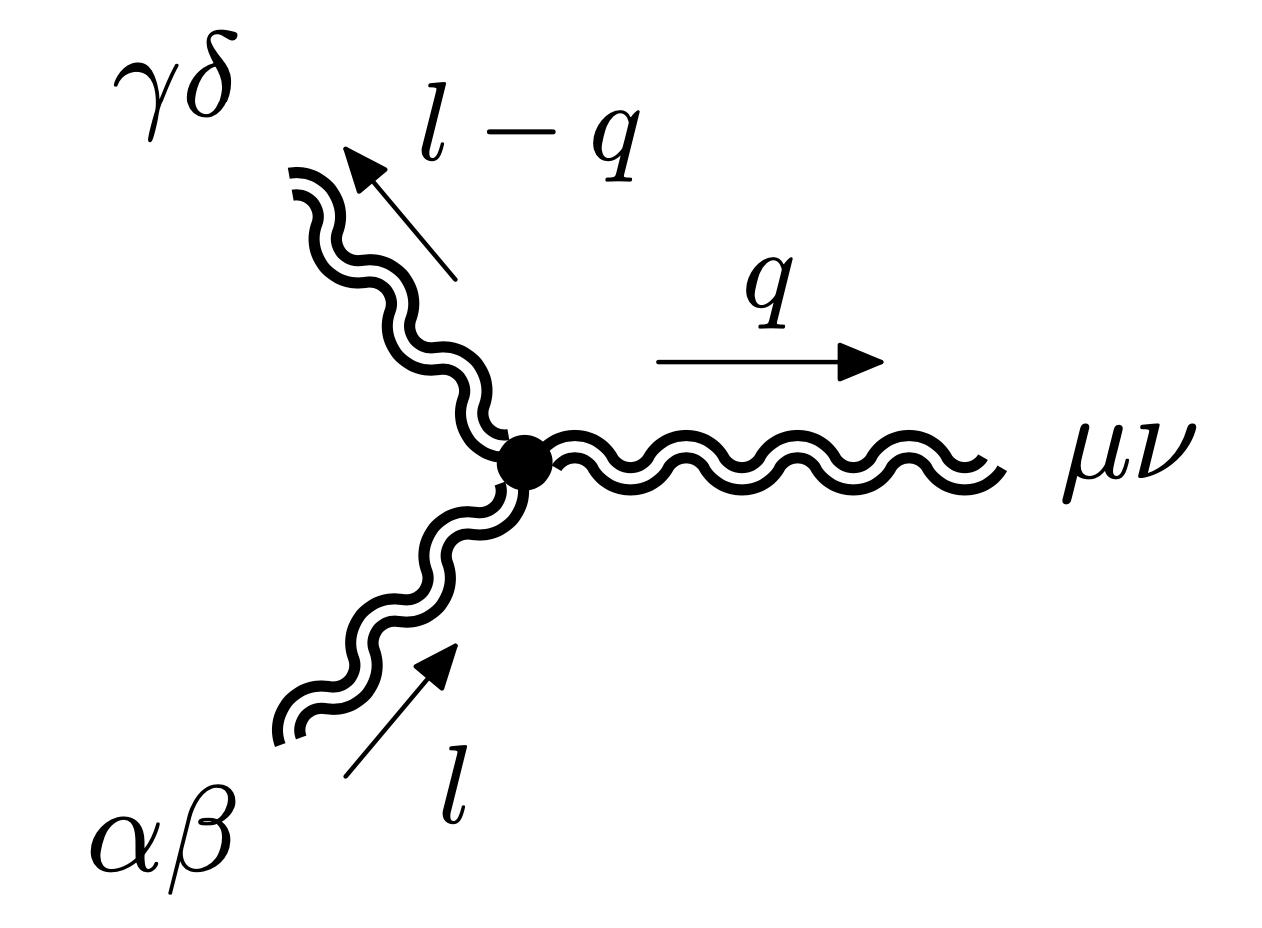}=\left(\tau_{h^2h}\right){}^{\mu\nu}{}_{\alpha\beta,\gamma\delta}(l,q) 
\end{equation*}
\begin{equation}\label{3Graviton_Vertex_Donoghue}
\begin{aligned}
=\frac{i \kappa}{2}&\bigg( P^{(4)}_{\alpha \beta,  \gamma \delta}\left(l^{\mu} l^{\nu}+(l-q)^{\mu}(l-q)^{\nu}+q^{\mu} q^{\nu}-\frac{3}{2} \eta^{\mu \nu} q^{2}\right) \\
&+2 q_{\lambda} q_{\sigma}\left({I_{\alpha \beta}}^{\lambda \sigma} {I_{\gamma \delta}}^{\mu \nu}+{I_{\gamma \delta}}^{\lambda \sigma} {I_{\alpha \beta}}^{\mu \nu}-{I_{\alpha \beta}}^{\lambda \mu }{I^{\sigma\nu}}_{\gamma \delta}-{I_{\alpha \beta}}^{\sigma \nu} {I_{\gamma \delta}}^{\lambda \nu}\right) \\
&+q_{\lambda} q^{\mu}\left(\eta_{\alpha \beta} I^{\lambda \nu}{ }_{\gamma \delta}+\eta_{\gamma \delta} {I^{\lambda \nu}}_{\alpha \beta}\right)+q_{\lambda} q^{\nu}\left(\eta_{\alpha \beta} {I_{\gamma \delta}}^{\lambda \mu}+\eta_{\gamma \delta} {I_{\alpha \beta}}^{\lambda \mu}\right)\\
&-q^{2}\left(\eta_{\alpha \beta} {I_{\gamma \delta}}^{\mu \nu}+\eta_{\gamma \delta} {I_{\alpha \beta}}^{\mu \nu}\right)-\eta^{\mu \nu} q^{\lambda} q^{\sigma}\left(\eta_{\alpha \beta} {I_{\gamma \delta \lambda \sigma}}+\eta_{\gamma \delta} I_{\alpha \beta \lambda \sigma}\right) \\
&+2 q ^ { \lambda } \left({I_{\alpha \beta}}^{\sigma \lambda} I_{\gamma \delta \sigma \nu}(l-q)^{\mu}+{I_{\alpha \beta}}^{\lambda\sigma} I_{\gamma \delta \sigma}{}^\mu(l-q)^{\nu} -{I_{\gamma \delta}}^{\lambda \sigma} I_{\alpha \beta \sigma}{}^\nu l^{\mu}-{I_{\gamma \delta}}^{\lambda \sigma} I_{\alpha \beta \sigma}{}^\mu l^{\nu}\right) \\
&+q^{2}\left({I_{\alpha \beta}}^{\sigma \mu} {I_{\gamma \delta ,\sigma}}^{\nu}+I_{\alpha \beta, \sigma}{ }^\nu {I_{\alpha \delta}}^{\sigma \mu}\right)+\eta^{\mu \nu} q^{\lambda} q_{\sigma}\left({I_{\gamma \delta}}^{\rho \sigma} I_{\alpha \beta, \lambda \rho}+{I_{\alpha \beta}}^{\rho \sigma} I_{\gamma \delta, \lambda \rho}\right) \\
&+\left(l^{2}+(l-q)^{2}\right)\left({I_{\alpha \beta}}^{\sigma \mu} {I_{\gamma \delta, \sigma}}^{\nu}+I^{\sigma \nu}{ }_{\alpha \beta} {I_{\gamma \delta, \sigma}}^{\mu}-\frac{1}{2} \eta^{\mu \nu} P^{(4)}_{\alpha \beta, \gamma \delta}\right) \\
&-l^{2} \eta_{\gamma \delta} {I_{\alpha\beta}}^{\mu \nu}-(l-q)^{2} \eta_{\alpha \beta} {I_{\gamma \delta}}^{\mu \nu}\bigg) \ ,
\end{aligned}
\end{equation}
with $I_{\mu\nu, \alpha\beta}=\frac{1}{2}(\eta_{\mu\alpha}\eta_{\nu\beta}+\eta_{\mu\beta}\eta_{\nu\alpha})$, and where the $\mu\nu$ graviton correspond to the external one, meaning that the expression is symmetric only under the exchange of the lines $\alpha\beta\leftrightarrow \gamma\delta$.
\end{itemize}

\chapter{Master integral and Fourier transforms}\label{App:masterintegral}

Here we review the definition of the master integral introduced in section \ref{sec:Classical_Limit_Grav_Emission_Processes} and we list its $d$-dimensional Fourier transform identities. In a $d$-dimensional Euclidean space in which $\vec{q}^2=q_1^2+...+q_d^2$, the $l$-loop massless sunset integral is defined by 
\begin{equation}
J_{(l)}(\vec{q}^2)=\int\prod_{i=1}^l\frac{d^d\vec{\ell}_i}{(2\pi)^d}\frac{\vec{q}^2}{\left(\prod_{i=1}^l\vec{\ell}_i^2\right)\left(\vec{q}-\sum_{i=1}^l\vec{\ell}_i\right)^2}\ .
\end{equation}
It can be explicitly computed \cite{Vanhove:2014wqa} as 
\begin{equation}\label{explicit_J_definition}
        J_{(l)}(\vec{q}^2)=\frac{\Gamma\left( l+1-\frac{ld}{2}\right)\Gamma\left(\frac{d-2}{2}\right)^{l+1}}{(4\pi)^{\frac{ld}{2}}\Gamma\left(\frac{(l+1)(d-2)}{2}\right)} (\vec{q}^2)^{\frac{l(d-2)}{2}}\ ,
\end{equation}
from which it is clear the presence of poles for $\frac{(l+1)(d-2)}{2}=0, -1, -2$. Then computing the Fourier transforms\footnote{A proof of such Fourier transforms can be found in \cite{DOnofrio:2022cvn}.} needed for the amplitude calculation of chapters \ref{chap:Metric_from_Amplitudes} and \ref{chap:Photon_Emission_Processes}, they read
\begin{equation}\label{app_J_Identity_1}
    \int \frac{d^d\vec{q}}{(2\pi)^d}  \frac{J_{(l)}(\vec{q}^2)}{\vec{q}^2}e^{i\vec{q}\cdot\vec{x}} = \left(\frac{\rho}{4\pi}\right)^{l+1} 
\end{equation}
\begin{equation}\label{app_J_Identity_2}
        \int \frac{d^d\vec{q}}{(2\pi)^d} \frac{q_iq_j}{\vec{q}^2} \frac{J_{(l)}(\vec{q}^2)}{\vec{q}^2}e^{i\vec{q}\cdot\vec{x}} =  \frac{1}{2-l(d-2)} \left(\delta_{ij} + (l+1)(2-d)\frac{x_ix_j}{r}\right) \left(\frac{\rho}{4\pi}\right)^{l+1}\ ,
\end{equation}
where the $\rho$ parameter is the one defined in \eqref{Definition_rho}. One can notice how eqs. \eqref{app_J_Identity_1} and \eqref{app_J_Identity_2} are perfectly finite, which means that the divergences in the explicit form of the master integral are canceled by the Fourier transforms. Then it is possible to compute a specific relation needed for sections \ref{sec:Renorm_Metric} and \ref{sec:Renormalization_Potential}, which reads \cite{Mougiakakos:2020laz}
\begin{equation}\label{Master_CT}
    \int \frac{d^{d} \vec{q}}{(2 \pi)^{d}}J_{(1)}(\vec{q}^{2}) e^{i \vec{q} \cdot \vec{x}} =-\frac{\Gamma\left(\frac{d}{2}\right)^{2}}{2 \pi^{d} r^{2(d-1)}} \ .
\end{equation}

\medskip

Finally from the above identities it is possible to recover the particular cases needed for the computations in chapter \ref{Classical_Chapter}, recovering the Fourier transforms listed in \cite{Donoghue:2001qc} and \cite{Bjerrum-Bohr:2002fji}. Since $J_{(0)}(\vec{q}^2)=1$, in $d=3$ and $l=0$ eq. \eqref{app_J_Identity_1} becomes
\begin{equation}\label{D4_Identity1}
\int\frac{d^3\vec{q}}{(2\pi)^3}\frac{1}{\vec{q}^2}e^{i\vec{q}\cdot\vec{x}} = \frac{1}{4 \pi \, r} \ ,
\end{equation}
while for $l=1$, the master integral explicitly reads
\begin{equation}
J_{(1)}(\vec{q}^2)\Big|_{d=3}=\frac{1}{8}|\vec{q}|\ ,
\end{equation}
whose replaced inside eqs. \eqref{app_J_Identity_1} and \eqref{app_J_Identity_2} gives
\begin{equation}\label{D4_Identity2}
\int\frac{d^3\vec{q}}{(2\pi)^3}\frac{1}{|\vec{q}|}e^{i\vec{q}\cdot\vec{x}}=\frac{1}{2 \pi^2\, r^2}
\end{equation}
\begin{equation}\label{D4_Identity3}
\int\frac{d^3\vec{q}}{(2\pi)^3}\frac{q_iq_j}{\vec{q}^3}e^{i\vec{q}\cdot\vec{x}}=\frac{1}{2 \pi^2\, r^2}\left(\delta_{ij}-2\frac{x_ix_j}{r^2}\right)\ .
\end{equation}
  
\chapter{Loop integral reduction}\label{App:LoopRed}

The classical limit of both the stress-energy tensor and the electromagnetic current, arise by extracting from the loop integrals the contributions proportional to the master integral and neglecting all the higher powers of the momentum transferred in the energy expansion, which correspond to the quantum corrections. In this appendix we list such contributions for all the loop integrals that occur in the amplitude computations in sections \ref{sec:AmpCalc_Metric} and \ref{sec:AmpCalc_Pot}. Making the notation clear, when an integral below is set to be vanish, it means that no term in the energy expansion is proportional to the master integral. Notice that for some of these relations we made use of the \texttt{LiteRed} package of Mathematica \cite{LiteRed}. 

\subsubsection*{1-loop}
\begin{equation}
    \int \frac{d^d \ell}{(2\pi)^d} \frac{\vec{\ell}^2}{\vec{\ell}^2(\vec{q}-\vec{\ell})^2} = 0\ , 
\end{equation}
\begin{equation}\label{LR_1Loop_1}
    \int \frac{d^d \ell}{(2\pi)^d} \frac{\vec{\ell}\cdot \vec{q}}{\vec{\ell}^2(\vec{q}-\vec{\ell})^2} = \frac{1}{2}J_{(1)}(\vec{q}^2)\ , 
\end{equation}
\begin{equation}\label{LR_1Loop_2}
    \int \frac{d^d \ell}{(2\pi)^d} \frac{\vec{\ell}\cdot \vec{q}\, \Vec{\ell}^2}{\vec{\ell}^2(\vec{q}-\vec{\ell})^2} = 0 \ , 
\end{equation}
\begin{equation}\label{LR_1Loop_3}
    \int \frac{d^d \ell}{(2\pi)^d} \frac{\vec{\ell}\cdot \vec{q}\, \vec{\ell}\cdot \vec{q}}{\vec{\ell}^2(\vec{q}-\vec{\ell})^2} = \frac{1}{4}\Vec{q}^2\, J_{(1)}(\vec{q}^2)\ , 
\end{equation}

\subsubsection*{2-loop}

\begin{equation}\label{LiteRed_2Loop_2}
      \int \frac{d^d\ell_1}{(2\pi)^d} \frac{d^d\ell_2}{(2\pi)^d} \frac{\vec{\ell_1}\cdot \vec{q}}{\vec{\ell_1}^2 \vec{\ell_2}^2\left(\vec{q}-\vec{\ell_1}-\vec{\ell_2}\right)^2}=\frac{1}{3}J_{(2)}\left(\vec{q}^{2}\right)
\end{equation}

\begin{equation}
      \int \frac{d^d\ell_1}{(2\pi)^d} \frac{d^d\ell_2}{(2\pi)^d} \frac{\vec{\ell_1}\cdot \vec{\ell_2}}{\vec{\ell_1}^2 \vec{\ell_2}^2\left(\vec{q}-\vec{\ell_1}-\vec{\ell_2}\right)^2}=\frac{1}{6}J_{(2)}\left(\vec{q}^{2}\right)
\end{equation}

\begin{equation}
      \int \frac{d^d\ell_1}{(2\pi)^d} \frac{d^d\ell_2}{(2\pi)^d} \frac{\vec{\ell_1}\cdot\Vec{q}\ \Vec{\ell_1}^2}{\vec{\ell_1}^2 \vec{\ell_2}^2\left(\vec{q}-\vec{\ell_1}-\vec{\ell_2}\right)^2\left(\vec{\ell_1}+\vec{\ell_2}\right)^{2}}=0
\end{equation}

\begin{equation}
      \int \frac{d^d\ell_1}{(2\pi)^d} \frac{d^d\ell_2}{(2\pi)^d} \frac{\vec{\ell_2}\cdot\Vec{q}\ \Vec{\ell_1}^2}{\vec{\ell_1}^2 \vec{\ell_2}^2\left(\vec{q}-\vec{\ell_1}-\vec{\ell_2}\right)^2\left(\vec{\ell_1}+\vec{\ell_2}\right)^{2}}=0
\end{equation}

\begin{equation}
      \int \frac{d^d\ell_1}{(2\pi)^d} \frac{d^d\ell_2}{(2\pi)^d} \frac{\vec{\ell_1}\cdot\Vec{\ell_2}\ \Vec{q}\cdot\Vec{\ell_1}}{\vec{\ell_1}^2 \vec{\ell_2}^2\left(\vec{q}-\vec{\ell_1}-\vec{\ell_2}\right)^2\left(\vec{\ell_1}+\vec{\ell_2}\right)^{2}}=\frac{1}{6}J_{(2)}\left(\vec{q}^{2}\right)
\end{equation}

\begin{equation}
      \int \frac{d^d\ell_1}{(2\pi)^d} \frac{d^d\ell_2}{(2\pi)^d} \frac{\vec{\ell_1}^2\ \Vec{q}^2}{\vec{\ell_1}^2 \vec{\ell_2}^2\left(\vec{q}-\vec{\ell_1}-\vec{\ell_2}\right)^2\left(\vec{\ell_1}+\vec{\ell_2}\right)^{2}}=0
\end{equation}

\begin{equation}
      \int \frac{d^d\ell_1}{(2\pi)^d} \frac{d^d\ell_2}{(2\pi)^d} \frac{\vec{\ell_1}\cdot\Vec{\ell_2}\ \Vec{q}^2}{\vec{\ell_1}^2 \vec{\ell_2}^2\left(\vec{q}-\vec{\ell_1}-\vec{\ell_2}\right)^2\left(\vec{\ell_1}+\vec{\ell_2}\right)^{2}}=\frac{1}{2}J_{(2)}\left(\vec{q}^{2}\right)
\end{equation}

\begin{equation}\label{LR_Loop2_l1q_l2q}
      \int \frac{d^d\ell_1}{(2\pi)^d} \frac{d^d\ell_2}{(2\pi)^d} \frac{\vec{\ell_1}\cdot\Vec{q}\ \vec{\ell_2}\cdot\Vec{q}}{\vec{\ell_1}^2 \vec{\ell_2}^2\left(\vec{q}-\vec{\ell_1}-\vec{\ell_2}\right)^2\left(\vec{\ell_1}+\vec{\ell_2}\right)^{2}}=\frac{2d-7}{6(d-4)}J_{(2)}\left(\vec{q}^{2}\right)
\end{equation}

\begin{equation}
      \int \frac{d^d\ell_1}{(2\pi)^d} \frac{d^d\ell_2}{(2\pi)^d} \frac{\vec{\ell_1}\cdot\Vec{q}\ \vec{\ell_1}\cdot\Vec{q}}{\vec{\ell_1}^2 \vec{\ell_2}^2\left(\vec{q}-\vec{\ell_1}-\vec{\ell_2}\right)^2\left(\vec{\ell_1}+\vec{\ell_2}\right)^{2}}=\frac{d-3}{3(d-4)}J_{(2)}\left(\vec{q}^{2}\right) \ .
\end{equation}

\backmatter
\cleardoublepage
\phantomsection 
\addcontentsline{toc}{chapter}{\bibname}
\bibliographystyle{JHEP} 
\bibliography{Thesis_Library} 

\providecommand{\href}[2]{#2}\begingroup\raggedright\begin{thebibliography}{10}

\bibitem{Feynman:1963ax}
R.~P. Feynman, \emph{{Quantum theory of gravitation}}, {\emph{Acta Phys.
  Polon.} {\bf 24} (1963) 697--722}.

\bibitem{DeWitt:1967yk}
B.~S. DeWitt, \emph{{Quantum Theory of Gravity. 1. The Canonical Theory}},
  \href{http://dx.doi.org/10.1103/PhysRev.160.1113}{\emph{Phys. Rev.} {\bf 160}
  (1967) 1113--1148}.

\bibitem{DeWitt:1967ub}
B.~S. DeWitt, \emph{{Quantum Theory of Gravity. 2. The Manifestly Covariant
  Theory}}, \href{http://dx.doi.org/10.1103/PhysRev.162.1195}{\emph{Phys. Rev.}
  {\bf 162} (1967) 1195--1239}.

\bibitem{DeWitt:1967uc}
B.~S. DeWitt, \emph{{Quantum Theory of Gravity. 3. Applications of the
  Covariant Theory}},
  \href{http://dx.doi.org/10.1103/PhysRev.162.1239}{\emph{Phys. Rev.} {\bf 162}
  (1967) 1239--1256}.

\bibitem{tHooft:1974toh}
G.~'t~Hooft and M.~J.~G. Veltman, \emph{{One loop divergencies in the theory of
  gravitation}}, {\emph{Ann. Inst. H. Poincare Phys. Theor. A} {\bf 20} (1974)
  69--94}.

\bibitem{Donoghue:1993eb}
J.~F. Donoghue, \emph{{Leading quantum correction to the Newtonian potential}},
  \href{http://dx.doi.org/10.1103/PhysRevLett.72.2996}{\emph{Phys. Rev. Lett.}
  {\bf 72} (1994) 2996--2999}, [\href{http://arxiv.org/abs/gr-qc/9310024}{{\tt
  gr-qc/9310024}}].

\bibitem{Donoghue:1994dn}
J.~F. Donoghue, \emph{{General relativity as an effective field theory: The
  leading quantum corrections}},
  \href{http://dx.doi.org/10.1103/PhysRevD.50.3874}{\emph{Phys. Rev. D} {\bf
  50} (1994) 3874--3888}, [\href{http://arxiv.org/abs/gr-qc/9405057}{{\tt
  gr-qc/9405057}}].

\bibitem{Donoghue:2001qc}
J.~F. Donoghue, B.~R. Holstein, B.~Garbrecht and T.~Konstandin, \emph{{Quantum
  corrections to the Reissner-Nordstr\"om and Kerr-Newman metrics}},
  \href{http://dx.doi.org/10.1016/S0370-2693(02)01246-7}{\emph{Phys. Lett. B}
  {\bf 529} (2002) 132--142}, [\href{http://arxiv.org/abs/hep-th/0112237}{{\tt
  hep-th/0112237}}].

\bibitem{Bjerrum-Bohr:2002fji}
N.~E.~J. Bjerrum-Bohr, J.~F. Donoghue and B.~R. Holstein, \emph{{Quantum
  corrections to the Schwarzschild and Kerr metrics}},
  \href{http://dx.doi.org/10.1103/PhysRevD.68.084005}{\emph{Phys. Rev. D} {\bf
  68} (2003) 084005}, [\href{http://arxiv.org/abs/hep-th/0211071}{{\tt
  hep-th/0211071}}].

\bibitem{Donoghue:1996mt}
J.~F. Donoghue and T.~Torma, \emph{{On the power counting of loop diagrams in
  general relativity}},
  \href{http://dx.doi.org/10.1103/PhysRevD.54.4963}{\emph{Phys. Rev. D} {\bf
  54} (1996) 4963--4972}, [\href{http://arxiv.org/abs/hep-th/9602121}{{\tt
  hep-th/9602121}}].

\bibitem{Iwasaki:1971vb}
Y.~Iwasaki, \emph{{Quantum theory of gravitation vs. classical theory. -
  fourth-order potential}},
  \href{http://dx.doi.org/10.1143/PTP.46.1587}{\emph{Prog. Theor. Phys.} {\bf
  46} (1971) 1587--1609}.

\bibitem{Holstein:2004dn}
B.~R. Holstein and J.~F. Donoghue, \emph{{Classical physics and quantum
  loops}}, \href{http://dx.doi.org/10.1103/PhysRevLett.93.201602}{\emph{Phys.
  Rev. Lett.} {\bf 93} (2004) 201602},
  [\href{http://arxiv.org/abs/hep-th/0405239}{{\tt hep-th/0405239}}].

\bibitem{Bjerrum-Bohr:2018xdl}
N.~E.~J. Bjerrum-Bohr, P.~H. Damgaard, G.~Festuccia, L.~Plant\'e and
  P.~Vanhove, \emph{{General Relativity from Scattering Amplitudes}},
  \href{http://dx.doi.org/10.1103/PhysRevLett.121.171601}{\emph{Phys. Rev.
  Lett.} {\bf 121} (2018) 171601}, [\href{http://arxiv.org/abs/1806.04920}{{\tt
  1806.04920}}].

\bibitem{Kosower:2018adc}
D.~A. Kosower, B.~Maybee and D.~O'Connell, \emph{{Amplitudes, Observables, and
  Classical Scattering}},
  \href{http://dx.doi.org/10.1007/JHEP02(2019)137}{\emph{JHEP} {\bf 02} (2019)
  137}, [\href{http://arxiv.org/abs/1811.10950}{{\tt 1811.10950}}].

\bibitem{Cheung:2018wkq}
C.~Cheung, I.~Z. Rothstein and M.~P. Solon, \emph{{From Scattering Amplitudes
  to Classical Potentials in the Post-Minkowskian Expansion}},
  \href{http://dx.doi.org/10.1103/PhysRevLett.121.251101}{\emph{Phys. Rev.
  Lett.} {\bf 121} (2018) 251101}, [\href{http://arxiv.org/abs/1808.02489}{{\tt
  1808.02489}}].

\bibitem{Bjerrum-Bohr:2004qcf}
N.~E. Bjerrum-Bohr, \emph{{Quantum gravity, effective fields and string
  theory}}.
\newblock PhD thesis, Bohr Inst., 2004.
\newblock \href{http://arxiv.org/abs/hep-th/0410097}{{\tt hep-th/0410097}}.

\bibitem{Duff:1973zz}
M.~J. Duff, \emph{{Quantum Tree Graphs and the Schwarzschild Solution}},
  \href{http://dx.doi.org/10.1103/PhysRevD.7.2317}{\emph{Phys. Rev. D} {\bf 7}
  (1973) 2317--2326}.

\bibitem{Mougiakakos:2020laz}
S.~Mougiakakos and P.~Vanhove, \emph{{Schwarzschild-Tangherlini metric from
  scattering amplitudes in various dimensions}},
  \href{http://dx.doi.org/10.1103/PhysRevD.103.026001}{\emph{Phys. Rev. D} {\bf
  103} (2021) 026001}, [\href{http://arxiv.org/abs/2010.08882}{{\tt
  2010.08882}}].

\bibitem{Mougiakakos:2021ngd}
S.~Mougiakakos, \emph{{Scattering Amplitudes in effective gravitational
  theories}}.
\newblock PhD thesis, U. Paris-Saclay, 2021.

\bibitem{Tangherlini:1963bw}
F.~R. Tangherlini, \emph{{Schwarzschild field in $n$ dimensions and the
  dimensionality of space problem}},
  \href{http://dx.doi.org/10.1007/BF02784569}{\emph{Nuovo Cim.} {\bf 27} (1963)
  636--651}.

\bibitem{Goldberger:2004jt}
W.~D. Goldberger and I.~Z. Rothstein, \emph{{An Effective field theory of
  gravity for extended objects}},
  \href{http://dx.doi.org/10.1103/PhysRevD.73.104029}{\emph{Phys. Rev. D} {\bf
  73} (2006) 104029}, [\href{http://arxiv.org/abs/hep-th/0409156}{{\tt
  hep-th/0409156}}].

\bibitem{Jakobsen:2020ksu}
G.~U. Jakobsen, \emph{{Schwarzschild-Tangherlini Metric from Scattering
  Amplitudes}},
  \href{http://dx.doi.org/10.1103/PhysRevD.102.104065}{\emph{Phys. Rev. D} {\bf
  102} (2020) 104065}, [\href{http://arxiv.org/abs/2006.01734}{{\tt
  2006.01734}}].

\bibitem{Jakobsen:2020diz}
G.~U. Jakobsen, \emph{{General Relativity from Quantum Field Theory}},
  Master's thesis, 2020.

\bibitem{Fragomeno}
F.~Fragomeno, \emph{The \uppercase{R}eissner-\uppercase{N}ordstr\"om metric in
  any dimensions from quantum field theory amplitudes},  Master's thesis, 2021.

\bibitem{DOnofrio:2021tap}
S.~D'Onofrio, \emph{Classical metrics from quantum amplitudes: The
  \uppercase{R}eissner-\uppercase{N}ordstr\"om-\uppercase{T}angherlini metric
  from the graviton emission of charged scalars},  Master's thesis, 2022.

\bibitem{DOnofrio:2022cvn}
S.~D'Onofrio, F.~Fragomeno, C.~Gambino and F.~Riccioni, \emph{{The
  Reissner-Nordstr\"om-Tangherlini solution from scattering amplitudes of
  charged scalars}},
  \href{http://dx.doi.org/10.1007/JHEP09(2022)013}{\emph{JHEP} {\bf 09} (2022)
  013}, [\href{http://arxiv.org/abs/2207.05841}{{\tt 2207.05841}}].

\bibitem{Buonanno:2022pgc}
A.~Buonanno, M.~Khalil, D.~O'Connell, R.~Roiban, M.~P. Solon and M.~Zeng,
  \emph{{Snowmass White Paper: Gravitational Waves and Scattering Amplitudes}},
   \href{http://arxiv.org/abs/2204.05194}{{\tt 2204.05194}}.

\bibitem{Bjerrum-Bohr:2022blt}
N.~E.~J. Bjerrum-Bohr, P.~H. Damgaard, L.~Plante and P.~Vanhove, \emph{{The
  SAGEX Review on Scattering Amplitudes, Chapter 13: Post-Minkowskian expansion
  from Scattering Amplitudes}},  \href{http://arxiv.org/abs/2203.13024}{{\tt
  2203.13024}}.

\bibitem{https://doi.org/10.48550/arxiv.1702.00786}
P.~Amaro-Seoane, H.~Audley, S.~Babak, J.~Baker, E.~Barausse, P.~Bender et~al.,
  \emph{Laser interferometer space antenna},  2017.
\newblock 10.48550/ARXIV.1702.00786.

\bibitem{Saleem:2021iwi}
M.~Saleem et~al., \emph{{The science case for LIGO-India}},
  \href{http://dx.doi.org/10.1088/1361-6382/ac3b99}{\emph{Class. Quant. Grav.}
  {\bf 39} (2022) 025004}, [\href{http://arxiv.org/abs/2105.01716}{{\tt
  2105.01716}}].

\bibitem{Reitze:2019iox}
D.~Reitze et~al., \emph{{Cosmic Explorer: The U.S. Contribution to
  Gravitational-Wave Astronomy beyond LIGO}}, {\emph{Bull. Am. Astron. Soc.}
  {\bf 51} (2019) 035}, [\href{http://arxiv.org/abs/1907.04833}{{\tt
  1907.04833}}].

\bibitem{Punturo_2010}
M.~Punturo, M.~Abernathy, F.~Acernese, B.~Allen, N.~Andersson, K.~Arun et~al.,
  \emph{The einstein telescope: a third-generation gravitational wave
  observatory},
  \href{http://dx.doi.org/10.1088/0264-9381/27/19/194002}{\emph{Classical and
  Quantum Gravity} {\bf 27} (sep, 2010) 194002}.

\bibitem{Nagar:2018zoe}
A.~Nagar et~al., \emph{{Time-domain effective-one-body gravitational waveforms
  for coalescing compact binaries with nonprecessing spins, tides and self-spin
  effects}}, \href{http://dx.doi.org/10.1103/PhysRevD.98.104052}{\emph{Phys.
  Rev. D} {\bf 98} (2018) 104052}, [\href{http://arxiv.org/abs/1806.01772}{{\tt
  1806.01772}}].

\bibitem{Varma:2018mmi}
V.~Varma, S.~E. Field, M.~A. Scheel, J.~Blackman, L.~E. Kidder and H.~P.
  Pfeiffer, \emph{{Surrogate model of hybridized numerical relativity binary
  black hole waveforms}},
  \href{http://dx.doi.org/10.1103/PhysRevD.99.064045}{\emph{Phys. Rev. D} {\bf
  99} (2019) 064045}, [\href{http://arxiv.org/abs/1812.07865}{{\tt
  1812.07865}}].

\bibitem{Varma:2019csw}
V.~Varma, S.~E. Field, M.~A. Scheel, J.~Blackman, D.~Gerosa, L.~C. Stein
  et~al., \emph{{Surrogate models for precessing binary black hole simulations
  with unequal masses}},
  \href{http://dx.doi.org/10.1103/PhysRevResearch.1.033015}{\emph{Phys. Rev.
  Research.} {\bf 1} (2019) 033015},
  [\href{http://arxiv.org/abs/1905.09300}{{\tt 1905.09300}}].

\bibitem{Gamba:2021ydi}
R.~Gamba, S.~Ak\c{c}ay, S.~Bernuzzi and J.~Williams, \emph{{Effective-one-body
  waveforms for precessing coalescing compact binaries with post-Newtonian
  twist}}, \href{http://dx.doi.org/10.1103/PhysRevD.106.024020}{\emph{Phys.
  Rev. D} {\bf 106} (2022) 024020},
  [\href{http://arxiv.org/abs/2111.03675}{{\tt 2111.03675}}].

\bibitem{LIGOScientific:2014pky}
{\scshape LIGO Scientific} collaboration, J.~Aasi et~al., \emph{{Advanced
  LIGO}}, \href{http://dx.doi.org/10.1088/0264-9381/32/7/074001}{\emph{Class.
  Quant. Grav.} {\bf 32} (2015) 074001},
  [\href{http://arxiv.org/abs/1411.4547}{{\tt 1411.4547}}].

\bibitem{VIRGO:2014yos}
{\scshape VIRGO} collaboration, F.~Acernese et~al., \emph{{Advanced Virgo: a
  second-generation interferometric gravitational wave detector}},
  \href{http://dx.doi.org/10.1088/0264-9381/32/2/024001}{\emph{Class. Quant.
  Grav.} {\bf 32} (2015) 024001}, [\href{http://arxiv.org/abs/1408.3978}{{\tt
  1408.3978}}].

\bibitem{KAGRA:2020agh}
{\scshape KAGRA} collaboration, T.~Akutsu et~al., \emph{{Overview of KAGRA:
  Calibration, detector characterization, physical environmental monitors, and
  the geophysics interferometer}},
  \href{http://dx.doi.org/10.1093/ptep/ptab018}{\emph{PTEP} {\bf 2021} (2021)
  05A102}, [\href{http://arxiv.org/abs/2009.09305}{{\tt 2009.09305}}].

\bibitem{Favata:2013rwa}
M.~Favata, \emph{{Systematic parameter errors in inspiraling neutron star
  binaries}},
  \href{http://dx.doi.org/10.1103/PhysRevLett.112.101101}{\emph{Phys. Rev.
  Lett.} {\bf 112} (2014) 101101}, [\href{http://arxiv.org/abs/1310.8288}{{\tt
  1310.8288}}].

\bibitem{Samajdar:2018dcx}
A.~Samajdar and T.~Dietrich, \emph{{Waveform systematics for binary neutron
  star gravitational wave signals: effects of the point-particle baseline and
  tidal descriptions}},
  \href{http://dx.doi.org/10.1103/PhysRevD.98.124030}{\emph{Phys. Rev. D} {\bf
  98} (2018) 124030}, [\href{http://arxiv.org/abs/1810.03936}{{\tt
  1810.03936}}].

\bibitem{Purrer:2019jcp}
M.~P\"urrer and C.-J. Haster, \emph{{Gravitational waveform accuracy
  requirements for future ground-based detectors}},
  \href{http://dx.doi.org/10.1103/PhysRevResearch.2.023151}{\emph{Phys. Rev.
  Res.} {\bf 2} (2020) 023151}, [\href{http://arxiv.org/abs/1912.10055}{{\tt
  1912.10055}}].

\bibitem{Huang:2020pba}
Y.~Huang, C.-J. Haster, S.~Vitale, V.~Varma, F.~Foucart and S.~Biscoveanu,
  \emph{{Statistical and systematic uncertainties in extracting the source
  properties of neutron star - black hole binaries with gravitational waves}},
  \href{http://dx.doi.org/10.1103/PhysRevD.103.083001}{\emph{Phys. Rev. D} {\bf
  103} (2021) 083001}, [\href{http://arxiv.org/abs/2005.11850}{{\tt
  2005.11850}}].

\bibitem{Gamba:2020wgg}
R.~Gamba, M.~Breschi, S.~Bernuzzi, M.~Agathos and A.~Nagar, \emph{{Waveform
  systematics in the gravitational-wave inference of tidal parameters and
  equation of state from binary neutron star signals}},
  \href{http://dx.doi.org/10.1103/PhysRevD.103.124015}{\emph{Phys. Rev. D} {\bf
  103} (2021) 124015}, [\href{http://arxiv.org/abs/2009.08467}{{\tt
  2009.08467}}].

\bibitem{Srednicki}
M.~Srednicki, \emph{{Quantum field theory}}.
\newblock Cambridge University Press, 1, 2007.

\bibitem{Faddeev:1967fc}
L.~D. Faddeev and V.~N. Popov, \emph{{Feynman Diagrams for the Yang-Mills
  Field}}, \href{http://dx.doi.org/10.1016/0370-2693(67)90067-6}{\emph{Phys.
  Lett. B} {\bf 25} (1967) 29--30}.

\bibitem{Donoghue:2017pgk}
J.~F. Donoghue, M.~M. Ivanov and A.~Shkerin, \emph{{EPFL Lectures on General
  Relativity as a Quantum Field Theory}},
  \href{http://arxiv.org/abs/1702.00319}{{\tt 1702.00319}}.

\bibitem{Ortin}
T.~Ortin, \emph{{Gravity and strings}}.
\newblock Cambridge Monographs on Mathematical Physics. Cambridge Univ. Press,
  3, 2004,
  \href{http://dx.doi.org/10.1017/CBO9780511616563}{10.1017/CBO9780511616563}.

\bibitem{GR_Gualtieri}
V.~Ferrari, L.~Gualtieri and P.~Pani, \emph{{General Relativity and its
  Applications}}.
\newblock CRC Press, Taylor \& Francis Group, 2020.

\bibitem{Hamber}
H.~W. Hamber, \emph{{Quantum gravitation: The Feynman path integral approach}}.
\newblock Springer, Berlin, 2009,
  \href{http://dx.doi.org/10.1007/978-3-540-85293-3}{10.1007/978-3-540-85293-3}.

\bibitem{Bjerrum-Bohr:2014lea}
N.~E.~J. Bjerrum-Bohr, B.~R. Holstein, L.~Plant\'e and P.~Vanhove,
  \emph{{Graviton-Photon Scattering}},
  \href{http://dx.doi.org/10.1103/PhysRevD.91.064008}{\emph{Phys. Rev. D} {\bf
  91} (2015) 064008}, [\href{http://arxiv.org/abs/1410.4148}{{\tt 1410.4148}}].

\bibitem{Stelle:1977ry}
K.~S. Stelle, \emph{{Classical Gravity with Higher Derivatives}},
  \href{http://dx.doi.org/10.1007/BF00760427}{\emph{Gen. Rel. Grav.} {\bf 9}
  (1978) 353--371}.

\bibitem{Goroff:1985th}
M.~H. Goroff and A.~Sagnotti, \emph{{The Ultraviolet Behavior of Einstein
  Gravity}}, \href{http://dx.doi.org/10.1016/0550-3213(86)90193-8}{\emph{Nucl.
  Phys. B} {\bf 266} (1986) 709--736}.

\bibitem{Brodsky:2010zk}
S.~J. Brodsky and P.~Hoyer, \emph{{The $\hbar$ Expansion in Quantum Field
  Theory}}, \href{http://dx.doi.org/10.1103/PhysRevD.83.045026}{\emph{Phys.
  Rev. D} {\bf 83} (2011) 045026}, [\href{http://arxiv.org/abs/1009.2313}{{\tt
  1009.2313}}].

\bibitem{Weinberg:1972kfs}
S.~Weinberg, \emph{{Gravitation and Cosmology}: {Principles and Applications of
  the General Theory of Relativity}}.
\newblock John Wiley and Sons, New York, 1972.

\bibitem{Reissner}
H.~Reissner, \emph{Über die eigengravitation des elektrischen feldes nach der
  einsteinschen theorie},
  \href{http://dx.doi.org/https://doi.org/10.1002/andp.19163550905}{\emph{Annalen
  der Physik} {\bf 355} (1916) 106--120}.

\bibitem{Nordstrom}
G.~{Nordstr{\"o}m}, \emph{{On the Energy of the Gravitation field in Einstein's
  Theory}}, {\emph{Koninklijke Nederlandse Akademie van Wetenschappen
  Proceedings Series B Physical Sciences} {\bf 20} (Jan., 1918) 1238--1245}.

\bibitem{Mogull:2020sak}
G.~Mogull, J.~Plefka and J.~Steinhoff, \emph{{Classical black hole scattering
  from a worldline quantum field theory}},
  \href{http://dx.doi.org/10.1007/JHEP02(2021)048}{\emph{JHEP} {\bf 02} (2021)
  048}, [\href{http://arxiv.org/abs/2010.02865}{{\tt 2010.02865}}].

\bibitem{Jakobsen:2022fcj}
G.~U. Jakobsen and G.~Mogull, \emph{{Conservative and Radiative Dynamics of
  Spinning Bodies at Third Post-Minkowskian Order Using Worldline Quantum Field
  Theory}}, \href{http://dx.doi.org/10.1103/PhysRevLett.128.141102}{\emph{Phys.
  Rev. Lett.} {\bf 128} (2022) 141102},
  [\href{http://arxiv.org/abs/2201.07778}{{\tt 2201.07778}}].

\bibitem{Kalin:2020fhe}
G.~K\"alin, Z.~Liu and R.~A. Porto, \emph{{Conservative Dynamics of Binary
  Systems to Third Post-Minkowskian Order from the Effective Field Theory
  Approach}},
  \href{http://dx.doi.org/10.1103/PhysRevLett.125.261103}{\emph{Phys. Rev.
  Lett.} {\bf 125} (2020) 261103}, [\href{http://arxiv.org/abs/2007.04977}{{\tt
  2007.04977}}].

\bibitem{Jakobsen:2021smu}
G.~U. Jakobsen, G.~Mogull, J.~Plefka and J.~Steinhoff, \emph{{Classical
  Gravitational Bremsstrahlung from a Worldline Quantum Field Theory}},
  \href{http://dx.doi.org/10.1103/PhysRevLett.126.201103}{\emph{Phys. Rev.
  Lett.} {\bf 126} (2021) 201103}, [\href{http://arxiv.org/abs/2101.12688}{{\tt
  2101.12688}}].

\bibitem{Jakobsen:2021zvh}
G.~U. Jakobsen, G.~Mogull, J.~Plefka and J.~Steinhoff, \emph{{SUSY in the sky
  with gravitons}},
  \href{http://dx.doi.org/10.1007/JHEP01(2022)027}{\emph{JHEP} {\bf 01} (2022)
  027}, [\href{http://arxiv.org/abs/2109.04465}{{\tt 2109.04465}}].

\bibitem{Dlapa:2021npj}
C.~Dlapa, G.~K\"alin, Z.~Liu and R.~A. Porto, \emph{{Dynamics of binary systems
  to fourth Post-Minkowskian order from the effective field theory approach}},
  \href{http://dx.doi.org/10.1016/j.physletb.2022.137203}{\emph{Phys. Lett. B}
  {\bf 831} (2022) 137203}, [\href{http://arxiv.org/abs/2106.08276}{{\tt
  2106.08276}}].

\bibitem{Emparan:2008eg}
R.~Emparan and H.~S. Reall, \emph{{Black Holes in Higher Dimensions}},
  \href{http://dx.doi.org/10.12942/lrr-2008-6}{\emph{Living Rev. Rel.} {\bf 11}
  (2008) 6}, [\href{http://arxiv.org/abs/0801.3471}{{\tt 0801.3471}}].

\bibitem{Bern:2019crd}
Z.~Bern, C.~Cheung, R.~Roiban, C.-H. Shen, M.~P. Solon and M.~Zeng,
  \emph{{Black Hole Binary Dynamics from the Double Copy and Effective
  Theory}}, \href{http://dx.doi.org/10.1007/JHEP10(2019)206}{\emph{JHEP} {\bf
  10} (2019) 206}, [\href{http://arxiv.org/abs/1908.01493}{{\tt 1908.01493}}].

\bibitem{Bern:2019nnu}
Z.~Bern, C.~Cheung, R.~Roiban, C.-H. Shen, M.~P. Solon and M.~Zeng,
  \emph{{Scattering Amplitudes and the Conservative Hamiltonian for Binary
  Systems at Third Post-Minkowskian Order}},
  \href{http://dx.doi.org/10.1103/PhysRevLett.122.201603}{\emph{Phys. Rev.
  Lett.} {\bf 122} (2019) 201603}, [\href{http://arxiv.org/abs/1901.04424}{{\tt
  1901.04424}}].

\bibitem{Bern:2021yeh}
Z.~Bern, J.~Parra-Martinez, R.~Roiban, M.~S. Ruf, C.-H. Shen, M.~P. Solon
  et~al., \emph{{Scattering Amplitudes, the Tail Effect, and Conservative
  Binary Dynamics at $O(G^4)$}},
  \href{http://dx.doi.org/10.1103/PhysRevLett.128.161103}{\emph{Phys. Rev.
  Lett.} {\bf 128} (2022) 161103}, [\href{http://arxiv.org/abs/2112.10750}{{\tt
  2112.10750}}].

\bibitem{Chung:2019yfs}
M.-Z. Chung, Y.-T. Huang and J.-W. Kim, \emph{{Kerr-Newman stress-tensor from
  minimal coupling}},
  \href{http://dx.doi.org/10.1007/JHEP12(2020)103}{\emph{JHEP} {\bf 12} (2020)
  103}, [\href{http://arxiv.org/abs/1911.12775}{{\tt 1911.12775}}].

\bibitem{Cheung:2020gyp}
C.~Cheung and M.~P. Solon, \emph{{Classical gravitational scattering at $ O
  (G^{3})$ from Feynman diagrams}},
  \href{http://dx.doi.org/10.1007/JHEP06(2020)144}{\emph{JHEP} {\bf 06} (2020)
  144}, [\href{http://arxiv.org/abs/2003.08351}{{\tt 2003.08351}}].

\bibitem{Bjerrum-Bohr:2019kec}
N.~E.~J. Bjerrum-Bohr, A.~Cristofoli and P.~H. Damgaard,
  \emph{{Post-Minkowskian Scattering Angle in Einstein Gravity}},
  \href{http://dx.doi.org/10.1007/JHEP08(2020)038}{\emph{JHEP} {\bf 08} (2020)
  038}, [\href{http://arxiv.org/abs/1910.09366}{{\tt 1910.09366}}].

\bibitem{Guevara:2018wpp}
A.~Guevara, A.~Ochirov and J.~Vines, \emph{{Scattering of Spinning Black Holes
  from Exponentiated Soft Factors}},
  \href{http://dx.doi.org/10.1007/JHEP09(2019)056}{\emph{JHEP} {\bf 09} (2019)
  056}, [\href{http://arxiv.org/abs/1812.06895}{{\tt 1812.06895}}].

\bibitem{Guevara:2019fsj}
A.~Guevara, A.~Ochirov and J.~Vines, \emph{{Black-hole scattering with general
  spin directions from minimal-coupling amplitudes}},
  \href{http://dx.doi.org/10.1103/PhysRevD.100.104024}{\emph{Phys. Rev. D} {\bf
  100} (2019) 104024}, [\href{http://arxiv.org/abs/1906.10071}{{\tt
  1906.10071}}].

\bibitem{Guevara:2020xjx}
A.~Guevara, B.~Maybee, A.~Ochirov, D.~O'connell and J.~Vines, \emph{{A
  worldsheet for Kerr}},
  \href{http://dx.doi.org/10.1007/JHEP03(2021)201}{\emph{JHEP} {\bf 03} (2021)
  201}, [\href{http://arxiv.org/abs/2012.11570}{{\tt 2012.11570}}].

\bibitem{Bautista:2021wfy}
Y.~F. Bautista, A.~Guevara, C.~Kavanagh and J.~Vines, \emph{{From Scattering in
  Black Hole Backgrounds to Higher-Spin Amplitudes: Part I}},
  \href{http://arxiv.org/abs/2107.10179}{{\tt 2107.10179}}.

\bibitem{Moynihan:2019bor}
N.~Moynihan, \emph{{Kerr-Newman from Minimal Coupling}},
  \href{http://dx.doi.org/10.1007/JHEP01(2020)014}{\emph{JHEP} {\bf 01} (2020)
  014}, [\href{http://arxiv.org/abs/1909.05217}{{\tt 1909.05217}}].

\bibitem{Emond:2020lwi}
W.~T. Emond, Y.-T. Huang, U.~Kol, N.~Moynihan and D.~O'Connell,
  \emph{{Amplitudes from Coulomb to Kerr-Taub-NUT}},
  \href{http://dx.doi.org/10.1007/JHEP05(2022)055}{\emph{JHEP} {\bf 05} (2022)
  055}, [\href{http://arxiv.org/abs/2010.07861}{{\tt 2010.07861}}].

\bibitem{Cristofoli:2020uzm}
A.~Cristofoli, P.~H. Damgaard, P.~Di~Vecchia and C.~Heissenberg,
  \emph{{Second-order Post-Minkowskian scattering in arbitrary dimensions}},
  \href{http://dx.doi.org/10.1007/JHEP07(2020)122}{\emph{JHEP} {\bf 07} (2020)
  122}, [\href{http://arxiv.org/abs/2003.10274}{{\tt 2003.10274}}].

\bibitem{KoemansCollado:2018hss}
A.~Koemans~Collado, P.~Di~Vecchia, R.~Russo and S.~Thomas, \emph{{The
  subleading eikonal in supergravity theories}},
  \href{http://dx.doi.org/10.1007/JHEP10(2018)038}{\emph{JHEP} {\bf 10} (2018)
  038}, [\href{http://arxiv.org/abs/1807.04588}{{\tt 1807.04588}}].

\bibitem{Sannan:1986tz}
S.~Sannan, \emph{{Gravity as the Limit of the Type {II} Superstring Theory}},
  \href{http://dx.doi.org/10.1103/PhysRevD.34.1749}{\emph{Phys. Rev. D} {\bf
  34} (1986) 1749}.

\bibitem{Vanhove:2014wqa}
P.~Vanhove, \emph{{The physics and the mixed Hodge structure of Feynman
  integrals}}, \href{http://dx.doi.org/10.1090/pspum/088/01455}{\emph{Proc.
  Symp. Pure Math.} {\bf 88} (2014) 161--194},
  [\href{http://arxiv.org/abs/1401.6438}{{\tt 1401.6438}}].

\bibitem{LiteRed}
R.~N. Lee, \emph{{LiteRed} 1.4: a powerful tool for reduction of multiloop
  integrals},
  \href{http://dx.doi.org/10.1088/1742-6596/523/1/012059}{\emph{Journal of
  Physics: Conference Series} {\bf 523} (jun, 2014) 012059}.

\end{thebibliography}\endgroup

\end{document}